\documentclass[a4paper,11pt,twoside]{book}
\usepackage{amsmath}
\usepackage{amssymb}
\usepackage{amsthm}
\usepackage{fancyhdr}
\usepackage{graphicx}
\usepackage{verbatim}
\usepackage{upgreek}
\usepackage{latexsym}
\usepackage{type1cm}

\makeatletter
\def\cleardoublepage{\clearpage\if@twoside \ifodd\c@page\else%
    \hbox{}%
    \thispagestyle{empty}
    \newpage%
    \if@twocolumn\hbox{}\newpage\fi\fi\fi}
\makeatother

\usepackage{sectsty}
\allsectionsfont{\usefont{OT1}{cmss}{bx}{n}\selectfont}

\usepackage[titles]{tocloft}
\renewcommand{\cftchapfont}{%
  \fontsize{12}{14}\usefont{OT1}{cmss}{bx}{n}\selectfont
}

\DeclareMathAlphabet{\mathbf}{OT1}{pnc}{b}{n}

\newcommand{\be}{\begin{equation}}
\newcommand{\ee}{\end{equation}}
\newcommand{\bea}{\begin{eqnarray}}
\newcommand{\eea}{\end{eqnarray}}
\newcommand{\bal}{\begin{align}}
\newcommand{\eal}{\end{align}}
\newcommand{\nn}{\nonumber}
\newcommand{\dd}{\mathrm{d}}
\newcommand{\ii}{\mathrm{i}}
\newcommand{\sss}{\mathrm{s}}

\newcommand{\eV}{\textrm{ eV}}
\newcommand{\GeV}{\textrm{ GeV}}
\newcommand{\TeV}{\textrm{ TeV}}
\newcommand{\cm}{\textrm{ cm}}

\DeclareMathOperator{\Det}{\textrm{Det}}
\DeclareMathOperator{\cch}{\textrm{ch}}
\DeclareMathOperator{\ccl}{\textrm{c}}
\DeclareMathOperator{\diag}{\textrm{diag}}
\DeclareMathOperator{\fslash}{\!\not\!}

\DeclareMathOperator{\ind}{\textrm{ind}}

\DeclareMathOperator{\Str}{\textrm{Str}}
\DeclareMathOperator{\tr}{\textrm{tr}}
\DeclareMathOperator{\Tr}{\textrm{Tr}}

\DeclareMathOperator{\MA}{\mathcal{A}}

\DeclareMathOperator{\MD}{\mathcal{D}}

\DeclareMathOperator{\MF}{\mathcal{F}} 
\DeclareMathOperator{\MG}{\mathcal{G}}
\DeclareMathOperator{\MH}{\mathcal{H}}

\DeclareMathOperator{\ML}{\mathcal{L}} 
\DeclareMathOperator{\MM}{\mathcal{M}}

\DeclareMathOperator{\MP}{\mathcal{P}}
\DeclareMathOperator{\MQ}{\mathcal{Q}} 
\DeclareMathOperator{\MR}{\mathcal{R}} 
\DeclareMathOperator{\MS}{\mathcal{S}}
\DeclareMathOperator{\MT}{\mathcal{T}}

\DeclareMathOperator{\MV}{\mathcal{V}}

\DeclareMathOperator{\MZ}{\mathcal{Z}}

\DeclareMathOperator{\RA}{\mathrm{A}}

\DeclareMathOperator{\RI}{\mathrm{I}}

\DeclareMathOperator{\MBFC}{\mathbf{C}}

\DeclareMathOperator{\MBFG}{\mathbf{G}}

\DeclareMathOperator{\MBFS}{\mathbf{S}} 
\DeclareMathOperator{\MBFT}{\mathbf{T}}

\DeclareMathOperator{\MBBR}{\mathbb{R}} 
\DeclareMathOperator{\MBBZ}{\mathbb{Z}} 

\begin{document}

\frontmatter
\pagestyle{empty}

\begin{titlepage}
\begin{center}

\vspace*{2cm}

{\cftchapfont{\Huge Anomaly-Free Supergravities\\in Six Dimensions}}

\vspace{5.5cm}

\textsf{\LARGE Ph.D. Thesis}

\vspace{5.5cm}


\bigskip
\textsf{\Large Spyros D. Avramis}

\bigskip
\textsf{\large National Technical University of Athens\\ School of Applied Mathematics and Natural Sciences\\ Department of Physics}

\end{center}
\end{titlepage}

\begin{titlepage}
\begin{center}

\vspace*{2cm}

{\Large Spyros D. Avramis}

\vspace{2cm}

{\LARGE{\sc Anomaly-Free Supergravities\\in Six Dimensions}}

\vspace{2cm}

Dissertation submitted to the Department of Physics of the National Technical University of Athens in partial fulfillment of the requirements for the degree of Doctor of Philosophy in Physics.

\vspace{2cm}

\begin{tabular}{ll}
Thesis Advisor: & Alex Kehagias\\
\\
Thesis Committee: & Alex Kehagias\\ 
& Elias Kiritsis\\
& George Zoupanos\\
& K. Anagnostopoulos\\
& A.B. Lahanas\\
& E. Papantonopoulos\\
& N.D. Tracas
\end{tabular}

\vspace{3cm}

Athens, February 2006
\end{center}
\end{titlepage}

\cleardoublepage

\setcounter{page}{0} 
\pagenumbering{roman}

\chapter*{Abstract\markboth{Abstract}{Abstract}}

This thesis reviews minimal $N=2$ chiral supergravities coupled to matter in six dimensions with emphasis on anomaly cancellation. In general, six-dimensional chiral supergravities suffer from gravitational, gauge and mixed anomalies which, being associated with the breakdown of local gauge symmetries, render the theories inconsistent at the quantum level. Consistency of the theory is restored if the anomalies of the theory cancel via the Green-Schwarz mechanism or generalizations thereof, in a similar manner as in the case of ten-dimensional $N=1$ supergravities. The anomaly cancellation conditions translate into a certain set of constraints for the gauge group of the theory as well as on its matter content. For the case of ungauged theories these constraints admit numerous solutions but, in the case of gauged theories, the allowed solutions are remarkably few. In this thesis, we examine these anomaly cancellation conditions in detail and we present all solutions to these conditions under certain restrictions on the allowed gauge groups and representations, imposed for practical reasons. The central result of this thesis is the existence of two more anomaly-free gauged models in addition to the only one that was known until recently. We also examine anomaly cancellation in the context of Ho\v rava-Witten--type compactifications of minimal seven-dimensional supergravity on $\mathbf{S}^1 / \mathbb{Z}_2$. Finally, we discuss some basic aspects of the 4D phenomenology of the gauged models.

\chapter*{Acknowledgements\markboth{Acknowledgements}{Acknowledgements}}

I thank my advisor Prof. Alex Kehagias for his guidance during my Ph.D. years. I would also like to thank the members of my advisory committee Profs. Elias Kiritsis and George Zoupanos and the members of my evaluation committee Profs. K. Anagnostopoulos, A.B. Lahanas, E. Papantonopoulos and N.D. Tracas. I am also grateful to all my teachers during my M.Sc. and Ph.D. time.

I thank Prof. Seif Randjbar-Daemi for sharing with me his expertise on six-dimensional supergravity, for inviting me to ICTP, and for a very interesting collaboration.

I thank the Theory Division of CERN, the High Energy, Cosmology and Astroparticle Physics Section of ICTP and the Centre de Physique Th\'eorique of the \'Ecole Polytechnique for hospitality during various stages of this work.

I thank my colleague Constantina Mattheopoulou for our various discussions, for help with bibliography and for proofreading my thesis. I also thank my colleague and collaborator Dimitrios Zoakos for useful discussions.

This thesis was financially supported from the ``Heraclitus'' program cofunded by the Greek State and the European Union.

Finally, I have to thank my parents Dimitrios and Olympia and my brother Alexandros for their constant support throughout these years.

\vskip 0.3cm

\tableofcontents

\cleardoublepage

\mainmatter

\setcounter{page}{0} 
\pagenumbering{arabic}
\pagestyle{fancy}

{\allowdisplaybreaks

\chapter{Introduction}

\section{String Theory and Supergravity}

Despite the impressive successes of the Standard Model of elementary particle physics, there still remain many fundamental problems unanswered up to date. The most important such problem is finding an appropriate quantum-mechanical formulation of gravity since, from the usual viewpoint of quantum field theory, the gravitational interaction is non-renormalizable. Another problem, of more theoretical interest, is finding a unified description of all forces in Nature. Up to now, the most successful attempts to solve these problems have been made within the framework of string theory \cite{Green:1987sp,Green:1987mn,Polchinski:1998rq,Polchinski:1998rr}.

String theory was initially developed with the hope that it would explain certain properties of the strong interactions. Namely, the soft behavior of high-energy hadronic scattering amplitudes in the Regge limit could be approximated by a sum over an infinity of resonant states whose spectrum closely resembles that of a vibrating string. This picture turned out to be problematic, mainly because string theory (i) contains a massless spin--2 state that does not appear in the hadronic spectrum and (ii) cannot properly account for the pointlike structure of hadrons discovered in deep inelastic scattering experiments, and was abandoned after the discovery of QCD. However, motivated by the existence of the massless spin--2 state, Scherk and Schwarz proposed a drastic reinterpretation of string theory as a candidate for a unified theory of all interactions, including gravity. The striking property of this interpretation was that, due to the soft behavior of string amplitudes, the ultraviolet behavior of the theory was well under control, unlike ordinary quantum gravity. Moreover, there are also massless string excitations that correspond to other field theory particles, including spin 1/2 fermions and spin-1 gauge bosons with Standard-Model--like gauge symmetries. From the above, it appeared that string theory has the potential to resolve both problems mentioned in the first paragraph, a viewpoint that is pursued up to now.

One key concept in the development of string theory is the notion of supersymmetry, a non-trivial extension of the Poincar\'e spacetime symmetry that mixes bosonic and fermionic fields. Field theories with global supersymmetry are Standard-Model extensions involving extra fields accompanying the Standard-Model ones. Field theories with local supersymmetry are furthermore required to incorporate gravity and are known as supergravity theories; despite their improved ultraviolet properties, they are still non-renormalizable. With regard to string theory, the prevailing view is that supersymmetry is actually required in order for string theories to be consistent; string theories with supersymmetry are called superstring theories.

The interactions of superstring theories in the low-energy regime where the massive modes decouple are described by effective field theories containing supergravity and, in the heterotic and Type I cases, super Yang-Mills theories. In this context, the problem of non-renormalizability of these theories is automatically resolved by the ultraviolet completion provided by string theory.

A feature of string theory is that its internal consistency requires the presence of extra spacetime dimensions in addition to the observed four; in particular, superstring theory requires 10 spacetime dimensions. In order for the theory to make sense, the extra dimensions must be unobservable for some reason. For example, the extra dimensions may form a compact manifold whose size is small enough so that it cannot be probed at present-day energies; this is called compactification and has the important feature that it gives rise to extra gauge symmetries arising from the components of the gravitational field along the extra dimensions. Compactification can also be described in the effective-field-theory viewpoint of supergravity, in which case it leads to lower-dimensional supergravity theories. One of the most known examples is the compactification of $N=1$ supergravity on Calabi-Yau 3--folds (six-dimensional K\"ahler manifolds of $\mathrm{SU}(3)$ holonomy), which leads to $N=1$ supergravities in four dimensions. More involved examples are based on compactifications of the ten-dimenwsional theories on internal product manifolds; the theories resulting from the first steps of the compactification are then described by supergravity theories in intermediate dimensions, $4<D<10$ and many of the features of the underlying theory can be studied in the framework of these lower-dimensional theories. This thesis will be concerned with a class of theories of this type, namely $N=2$ supergravities in six dimensions.

\section{Six-dimensional Supergravity}

Six-dimensional supergravity theories have been attracted a lot of interest over the years and have been studied in a variety of phenomenological contexts. Some physically interesting setups motivating the study of such theories are the following.
\begin{enumerate}
\item \emph{Self-tuning of the cosmological constant}. A major and long-standing problem in theoretical physics is the search for an explanation of the extremely small positive value of the cosmological constant that does not rely on fine tuning. In the context of theories with extra dimensions, one is thus instructed to search for classical solutions of the higher-dimensional theory that force a 4D Minkowski space upon us. Ten-dimensional superstring theories and eleven-dimensional supergravity admit solutions of this kind, the internal spaces being Calabi-Yau 3--folds and $G_2$--holonomy manifolds respectively. However, the choice of the internal manifold in these solutions is far from being unique and, moreover, there also exist other types of solutions of these theories that also contain 4D spacetime as a factor. On the other hand, there is a special class of six-dimensional theories, namely $N=2$ gauged supergravities, that admit a unique solution of the required type. These theories are severely constrained by supersymmetry, which forbids a bare 6D cosmological constant but requires a potential for the scalars of the theory. This potential is positive-definite and has a unique minimum, in which it reduces to a positive effective 6D cosmological constant. Introducing the most general ansatz for the metric and the other background fields that is compatible with a maximally symmetric 4D spacetime and a compact regular 2D internal space, it has been shown that the unique solution of the equations of motion is given by 4D Minkowski space times an internal $\MBFS^2$. Since the vanishing of the 4D cosmological constant is imposed by the equations of motion, this is referred to as self-tuning, in contrast to fine tuning. Relaxing the requirement for a regular internal space opens up other possibilities that include brane solutions with internal singularities of the conical type.

\item \emph{Flat codimension--2 branes}. Another motivation for six-dimensional theories comes from the fact that, in a bulk-brane theory with vanishing bulk cosmological constant, a codimension--2 $p$--brane is always Ricci-flat, regardless of its tension. This is most easily seen by considering the trace-reversed Einstein equations in the $(p+3)$--dimensional bulk,
\be
R_{MN} = \frac{1}{M^{p+1}} \left( T_{MN} - \frac{1}{p+1} g_{MN} T \right),
\ee
where $M$ is the $(p+3)$--dimensional Planck mass. Noting that the stress tensor for a $p$--brane is given by
\be
T_{\mu\nu} = - \rho g_{\mu\nu} ,\qquad T_{mn} = 0,
\ee
where $\mu,\nu=0,\ldots,p$ denote the directions tangent to the brane and $m,n=1,2$ denote the transverse directions, we immediately find
\be
R_{\mu\nu} = 0.
\ee
This means that, independent of any effects giving rise to a vacuum energy on the brane, the brane worldvolume is always Ricci-flat and thus the brane cosmological constant vanishes; all vacuum-energy effects are absorbed into the internal geometry. Although for a compact internal space this latter fact reintroduces fine-tuning in the guise of requiring a special value for the brane energy density, this can be remedied by considering non-compact internal spaces of finite volume, both singular and non-singular. For the physically interesting case of a 3--brane, the above scenario requires a six-dimensional bulk.

\item \emph{The hierarchy problem and sub-millimeter extra dimensions}. An attractive scenario in theories with extra dimensions is the possibility of the resolution of the hierarchy problem resulting from the discrepancy between the Planck and the TeV scale through the introduction of large extra dimensions. Considering a $(4+n)$--dimensional gravitational theory compactified down to 4 dimensions, the 4--dimensional Planck mass $M_P$ is derived from the $(4+n)$--dimensional Planck mass $M$ by
\be
M_P^2 = M^{2+n} R^n,
\ee
where $R$ is the average size of the internal dimensions. To remove the hierarchy problem, one can set the fundamental scale $M_{4+n}$ to be of the order of $1 \TeV$. Then, the fact that $M_P \sim 10^{19} \GeV$ implies that the size of the compact manifold equals $10^{15} \cm$ for $n=1$, $10^{-2} \cm$ for $n=2$, $10^{-6} \cm$ for $n=3$ and so on. We see that the six-dimensional case ($n=2$) is quite special in that it requires the internal dimensions to be in the sub-millimeter range. Moreover, the mass scale set by the internal dimensions is in this case $M_{KK} \sim 10^2 \cm^{-1} \sim 10^{-3} \eV$, which is close to the expected scale of the neutrino masses and to the mass scale set by the observed cosmological constant.
\end{enumerate}

On the more theoretical side, further motivation for studying six-dimensional supergravities is provided by (i) their relation to ten-dimensional $N=1$ supergravities, (ii) their connection, in the gravity-decoupling limit, to the much-studied $N=2$ supersymmetric gauge theories in four dimensions, (iii) the existence of self-dual strings in six dimensions and (iv) the anomaly constraints that appear in six-dimensional theories. This latter aspect will be elucidated in the following section.

\section{Anomaly Cancellation}

If any of the scenarios mentioned above is to be realized in the context of a six-dimensional supergravity theory, the theory must be consistent quantum-mechanically. Consistency of these theories at the quantum level is threatened by non-renormalizability and anomalies. Although the problem of non-renormalizability in such contexts is always assumed to be resolved by an appropriate ultraviolet completion of the theory, the problem of anomalies is an infrared problem which must be resolved within the low-energy effective theory. So, a fundamental problem is to examine the conditions for anomaly freedom and to identify the anomaly-free theories among the available possibilities.

Again, the case of six dimensions is quite special in that gravitational and mixed nonabelian anomalies are present on top of the usual gauge and mixed abelian anomalies that appear in any even number of dimensions. As a result, the requirement of anomaly cancellation leads to quite stringent constraints that single out a relatively small number of consistent models. This is in sharp contrast to the case of the Standard Model/MSSM where the anomaly cancellation conditions are weak and can be satisfied by a vast number of models. Moreover, adopting the viewpoint that the theories under consideration are effective theories arising as long-wavelength limits of fundamental theories, identifying the possible anomaly-free models can potentially enable us to infer information about the high-energy aspects of the underlying theory through low-energy considerations.

In this respect, it is instructive to recall the basic facts in the case of 10D supergravity. The string-derived chiral 10D supergravities known before the explicit calculation of higher-dimensional anomalies were Type IIB $N=2$ supergravity (realized in terms of closed strings) and $N=1$ supergravity coupled to $\textrm{SO}(N)$ Yang-Mills (realized in terms of Type I strings). At the time, it was believed that these theories were anomalous, signaling an internal inconsistency of string theory. This issue was partially settled when the calculation of 10D gravitational anomalies led to the striking result that the anomalies of the Type IIB theory completely cancel; on the other hand, $N=1$ supergravity was found to be anomalous. However, Green and Schwarz discovered that $N=1$ theory can also be made anomaly-free through a coupling of the 2--form of the supergravity multiplet to a certain 8--form constructed out of curvature invariants. The necessary and sufficient condition for anomaly cancellation was that the anomaly polynomial must factorize. This can happen only for a gauge group of dimension 496 with no sixth-order Casimirs, in which case the factorization coefficients are uniquely determined and result in a further constraint on certain group-theoretical coefficients. The obvious candidate was $\textrm{SO}(32)$ which indeed satisfied all the above requirements and the corresponding string theory was subsequently shown to also satisfy the RR tadpole cancellation condition. However, surprisingly enough, these requirements were also satisfied by the $E_8 \times E_8$ group which at that time lacked a string-theoretical interpretation, as well as by the physically uninteresting $E_8 \times \textrm{U}(1)^{248}$ and $\textrm{U}(1)^{496}$ groups; the above four groups exhaust all possibilities. The discovery of the heterotic string provided a string realization of the $E_8 \times E_8$ model which turned out to be the most phenomenologically relevant string unification model at the time. These developments made clear that anomaly cancellation not only seriously constrains the particle spectrum of a theory but can also point, from the effective-field-theory point of view, to new consistent models that may be realized through a more fundamental theory.

Anomaly cancellation via the Green-Schwarz mechanism carries over to lower-dimensional chiral theories like the minimal 6D supergravities. Here, however, things are more complicated mainly due to the existence of the massless hypermultiplets that may transform in arbitrary representations of the gauge group. A consequence of this is that the anomaly cancellation conditions are somewhat weaker than those in the 10D case. First, the condition for the cancellation of irreducible gravitational anomalies does not uniquely fix the dimension of the gauge group but, instead, it simply sets an upper bound on the number of non-singlet hypermultiplets. Second, in the case that the gauge group has fourth-order Casimirs, cancellation of the corresponding irreducible gauge anomaly leads to an equality constraint for the numbers of hypermultiplets. Finally, the factorization condition does not determine how the highest-order traces in the gauge anomaly must factorize but instead leads to two weaker constraints. The conditions mentioned above admit a large number of solutions for the gauge group and the possible hypermultiplet representations and, in fact, a complete classification is a very complicated task.

\section{Outline of the Thesis}

The objective of this thesis is the study of anomaly cancellation in minimal $N=2$ chiral supergravities in six dimensions and the search for more anomaly-free models than those already known. In the main part of the thesis, we examine the relevant anomaly cancellation conditions in detail and we conduct an exhaustive search for anomaly-free ungauged and gauged models under some restrictions on the allowed gauge groups and representations.

This thesis is structured as follows. Chapters 2, 3 and 4 contain review material, included to make the discussion as self-contained as possible. In Chapter 2, we discuss the general structure and features of supergravity theories. In Chapter 3, we do a detailed study of anomalies and their cancellation in gauge and gravitational theories. Finally, in Chapter 4 we discuss anomaly cancellation in the context of superstring theories as well as in the context of the heterotic M-theory of Ho\v rava and Witten. Chapters 5 and 6 contain the main part of this thesis and are organized as follows. In Chapter 5, we present a detailed review of anomaly cancellation in six dimensions and we present the results of a detailed search of anomaly-free models of this type. We also examine anomaly cancellation in the context of Ho\v rava-Witten--type compactifications of minimal seven-dimensional supergravity on the $\mathbf{S}^1 / \mathbb{Z}_2$ orbifold. Finally, in Chapter 6 we do a brief study of four-dimensional compactifications of the 6D supergravities under consideration. Our conventions and many useful results needed in this thesis are summarized in four appendices.

The original material contained in this thesis is presented in Chapters 5 and 6. In Sections \ref{sec-5-4} to \ref{sec-5-7}, we present the results of a search for anomaly-free $D=6$, $N=2$ supergravities initiated in \cite{Avramis:2005qt} (in collaboration with A. Kehagias and S. Randjbar-Daemi) and completed in \cite{Avramis:2005hc} (in collaboration with A. Kehagias). In Section \ref{sec-6-1}, we examine anomaly cancellation on the six-dimensional fixed points of $D=7$, $N=2$ supergravity on $\mathbf{S}^1 / \mathbb{Z}_2$, as in \cite{Avramis:2004cn} (in collaboration with A. Kehagias). Finally, Section \ref{sec-7-3} includes discussions about the 4D compactifications of the new gauged models found here, as in \cite{Avramis:2005qt}.

\chapter{Supersymmetry and Supergravity}
\label{chap-2}

\section{Supersymmetry} 
\label{sec-2-1}

Supersymmetry is an extension of the Poincar\'e spacetime symmetry of physical theories which involves fermionic generators that mix boson and fermion fields. Since its discovery, it has played a prominent role both in 4--dimensional particle physics as well as in higher-dimensional unified theories of particle physics and gravity, namely string theories and their effective supergravities.

The history of supersymmetry can be traced back to 1969, where Coleman and Mandula \cite{Coleman:1967ad} proved a no-go theorem stating that, under a set of assumptions, the spacetime and internal parts of the symmetries of the S-matrix of a field theory do not mix. The strongest assumption underlying the theorem was that the S-matrix symmetries are described by ordinary Lie algebras involving bosonic generators. During the development of string theory in the early seventies, it was recognized that the two-dimensional worldsheet theory of strings with fermionic degrees of freedom possess a supersymmetry relating bosons and fermions. Although this was not of direct relevance to the four-dimensional problem earlier posed, soon after that Haag, Lopusz\`anski and Sohnius \cite{Haag:1975qh} evaded the Coleman-Mandula result by relaxing the restriction to bosonic generators and constructed the most general form of the symmetry group. It was shown that, in such a case, there exists indeed the possibility of spacetime supersymmetry, which is an extension of the bosonic Poincar\'e symmetry of field theories involving fermionic generators that mix non-trivially with the Poincar\'e group. Subsequently, supersymmetry emerged also as a spacetime symmetry of string theory, and one that is crucial for its self-consistency. Due to its attractive properties, both in the context of particle physics and string theory, supersymmetry has evolved into a field of intense study despite the fact that it still awaits experimental verification.

In this section, we will describe some elementary aspects of supersymmetry relevant to the supergravity theories under consideration. Specifically, we will review the supersymmetry algebras in diverse dimensions and the construction of their massless representations, which are the building blocks of supergravity theories.
 
\subsection{The supersymmetry algebra} 
\label{sec-2-1-1}

To develop the basic principles of supersymmetry, our starting point is the Coleman-Mandula theorem stated earlier on. To be more specific, this theorem states that, for a four-dimensional field theory where (i) there exists a non-trivial S-matrix, (ii) the vacuum is non-degenerate, (iii) there are no massless particles and (iv) the continuous symmetries are described by Lie algebras, the symmetry group of the S-matrix is necessarily given by the \emph{direct} product of the spacetime Poincar\'e group with commutation relations
\bea
\label{e-2-1-1} 
\left[  P_a,P_b \right] &=& 0, \nn \\
\left[  P_a,M_{bc} \right] &=& \eta_{ac} P_b - \eta_{ab} P_c, \nn\\
\left[  M_{ab},M_{cd} \right] &=& \eta_{ad} M_{bc} + \eta_{bc} M_{ad} - \eta_{ac} M_{bd} - \eta_{bd} M_{ac},
\eea
with an internal symmetry group with commutation relations
\be
\label{e-2-1-2}
\left[  T_I,T_J \right] = C_{IJ}^{\phantom{IJ} K} T_K.
\ee
By ``direct'', we mean that the Poincar\'e and internal symmetries commute,
\be
\label{e-2-1-3}
\left[  P_a,T_I \right] = \left[  M_{ab},T_I \right] = 0.
\ee
The Coleman-Mandula theorem has a simple intuitive explanation. Considering a general two-body scattering process in field theory, we know that Lorentz invariance fixes the incoming and outgoing trajectories to lie all on the same plane and translation invariance fixes the energies of the outgoing particles. So, the only unknown in this process is the scattering angle. Allowing for an internal symmetry that mixes with the Poincar\'e symmetry would further constrain the kinematics of the process and would result in a quantization of the scattering angle. Analyticity of the S-matrix would then imply that the transition element is independent of the angle, i.e. that the theory is trivial. 

To evade the Coleman-Mandula theorem and the restrictions it imposes on the symmetries of a theory, one may relax the assumption that these symmetries are described by ordinary Lie algebras, allowing for the possibility of \emph{graded Lie algebras} or \emph{superalgebras} \cite{Kac:1977em,Frappat:1996pb}. This was recognized by Golf'and and Likhtman \cite{Golfand:1971iw} as early as 1971 and the most general form of such an algebra was constructed by Haag, Lopusz\`anski and Sohnius \cite{Haag:1975qh} in 1975 for the four-dimensional case. In what follows, we will sketch the construction of such a superalgebra in four dimensions.

We start by introducing a $\MBBZ_2$ grading that divides the algebra elements into two classes, even and odd. The structure of the superalgebra is such that two even elements close on an even element by commutation, one even and one odd element close on an odd element by commutation and two odd elements close on an even element by anticommutation. The Jacobi identities are modified in a fairly obvious way. The even generators are assumed to be $P_a$, $M_{ab}$ and $T_I$, while the odd generators are called supercharges and are written as $Q^{\alpha i}$ where $\alpha$ is a Lorentz index and $i=1,\ldots,N$ is an internal index labelling the different supercharges. The commutation relations (\ref{e-2-1-1})-(\ref{e-2-1-3}) of the even generators must remain intact due to the Coleman-Mandula theorem. The remaining (anti)commutation relations involving the odd generators are determined by writing down their most general form and invoking the Jacobi identities. Starting from the $[ M,Q ]$ commutator, it must have the form
\be
\label{e-2-1-4}
\left[  M_{ab},Q^{\alpha i} \right] = (m_{ab})^\alpha_{\phantom{\alpha} \beta} Q^{\beta i},
\ee
and the $MMQ$ Jacobi identity requires that the matrices $m_{ab}$ satisfy the commutation relations of the Lorentz algebra, i.e. that $Q^{\alpha i}$ form a representation of this algebra. Due to the anticommuting property of $Q^{\alpha i}$'s, this representation must be fermionic, and the simplest choice is the spinor representation; actually, it can be shown that this is the \emph{only} possible representation. So, the $Q^{\alpha i}$'s transform as a Lorentz spinor and, since we are in four dimensions, we can impose the Majorana condition to simplify things. The $[ M,Q ]$ commutator is thus
\be
\label{e-2-1-5}
\left[ M_{ab},Q^{\alpha i} \right] = \frac{1}{2} (\Gamma_{ab})^\alpha_{\phantom{\alpha} \beta} Q^{\beta i}.
\ee
Turning to the $[ P,Q ]$ commutator, its general form is
\be
\label{e-2-1-6}
\left[ P_a,Q^{\alpha i} \right] = \left[  p_1 (\Gamma_a)^\alpha_{\phantom{\alpha} \beta} + p_2 (\Gamma_5 \Gamma_a)^\alpha_{\phantom{\alpha} \beta} \right] Q^{\beta i},
\ee
and the $PPQ$ Jacobi identity, combined with the Majorana condition fixes $p_1=p_2=0$, i.e.
\be
\label{e-2-1-7}
\left[ P_a,Q^{\alpha i} \right] = 0
\ee
For the $\{ Q,Q \}$ anticommutator, we first impose for simplicity the condition that it does not involve any new bosonic generators (this excludes the possibility of \emph{central charges}). Then, the requirement for symmetry under simultaneous interchange of $(\alpha,i)$ with $(\beta,j)$, the symmetry property (\ref{e-b-2-14}) of the $C \Gamma^{a_1 \ldots a_n}$ products and the $MQQ$ Jacobi identity fix its form to 
\be
\label{e-2-1-8}
\{  Q^{\alpha i},Q^{\beta j}  \} = (\Gamma^a C^{-1})^{\alpha \beta} P_a p^{ij} + m (\Gamma^{ab} C^{-1})^{\alpha \beta} M_{ab} \delta^{ij},
\ee
where $p^{ij}$ is a symmetric matrix. By a suitable rotation of the supercharges, one may set $p^{ij} = p \delta^{ij}$ while preserving the Majorana condition. Doing so and invoking the $PQQ$ Jacobi identity fixes $m=0$, while $p$ can be fixed by a choice of scale. Setting $p=1$, we write
\be
\label{e-2-1-9}
\{  Q^{\alpha i},Q^{\beta j}  \} = (\Gamma^a C^{-1})^{\alpha \beta} P_a \delta^{ij}.
\ee
Finally, turning to the $[ T,Q ]$ commutator, one can switch from a Majorana to a Weyl basis and write 
\be
\label{e-2-1-10}
\left[  T_I,Q^{\alpha i} \right] = (t_I)^i_{\phantom{i} j} Q^{\alpha j},
\ee
where $(t_I)^i_{\phantom{i} j}$ are complex matrices. Then, the $TQ\bar{Q}$ Jacobi identity implies that these matrices are antihermitian and hence the $Q^{\alpha i}$ ($\bar{Q}^\alpha_i$) transform in the $\mathbf{N}$ ($\mathbf{\overline{N}}$) of $\mathrm{U}(N)$. This type of symmetry is called R-symmetry.

We will now discuss the generalization of the above results to other dimensions \cite{Nahm:1977tg}, where the available spinors are of different types. First of all, Eqs. (\ref{e-2-1-1}), (\ref{e-2-1-5}) and (\ref{e-2-1-7}) hold for all types of spinors and so the supersymmetry algebra contains the relations
\bea
\label{e-2-1-11}
\left[ P_a,P_b \right] &=& 0, \nn\\
\left[ P_a,M_{bc} \right] &=& \eta_{ac} P_b - \eta_{ab} P_c, \nn\\
\left[ M_{ab},M_{cd} \right] &=& \eta_{ad} M_{bc} + \eta_{bc} M_{ad} - \eta_{ac} M_{bd} - \eta_{bd} M_{ac}, \nn\\
\left[ P_a,Q^{\alpha i} \right] &=& 0, \nn\\
\left[ M_{ab},Q^{\alpha i} \right] &=& \frac{1}{2} (\Gamma_{ab})^\alpha_{\phantom{\alpha} \beta} Q^{\beta i}.
\eea
The $\{ Q,Q \}$ anticommutator and the R-symmetry group both depend on the type of available spinors for a given spacetime dimension. Using the reality properties of spinors from Table \ref{t-b-2} we see that we can choose our conventions so that the minimal spinor is Majorana or Weyl for $D=0,4 \mod 8$, (pseudo)Majorana for $D=1,3 \mod 8$, symplectic Majorana for $D=5,7 \mod 8$, Majorana-Weyl for $D=2 \mod 8$ and symplectic Majorana-Weyl for $D=6 \mod 8$. The results are as follows.
\begin{itemize}
\item $D=0,4 \mod 8$. Here, we have Majorana supercharges, satisfying
\be
\label{e-2-1-12}
\{  Q^{\alpha i},Q^{\beta j}  \} = (\Gamma^a C^{-1})^{\alpha \beta} P_a \delta^{ij} .
\ee
Alternatively, we can use a Weyl basis where the supercharges are $Q^{\alpha i\pm}$ and satisfy
\be
\label{e-2-1-13}
\{  Q^{\alpha i+},Q^{\beta j-}  \} = ( P^+ \Gamma^a C^{-1})^{\alpha \beta} P_a \delta^{ij}.
\ee
Here, the $\pm$ superscripts indicate the chirality of the supercharges and $P^\pm = \frac{1\pm\Gamma_{D+1}}{2}$.
The R-symmetry group is $\mathrm{U}(N) \cong \mathrm{SU}(N) \times \mathrm{U}(1)$ with the chiral components transforming in $\mathbf{N}_1 + \mathbf{\overline{N}}_{-1}$.
\item $D=1,3 \mod 8$. Here, we have Majorana supercharges satisfying
\be
\label{e-2-1-14}
\{  Q^{\alpha i},Q^{\beta j}  \} = (\Gamma^a C^{-1})^{\alpha \beta} P_a \delta^{ij} .
\ee
The R-symmetry group is $\mathrm{SO}(N)$ with the supercharges transforming in $\mathbf{N}$.
\item $D=5,7 \mod 8$. Here, we have symplectic Majorana supercharges, satisfying
\be
\label{e-2-1-15}
\{  Q^{\alpha i},Q^{\beta j}  \} = ( \Gamma^a C^{-1})^{\alpha \beta} P_a \Omega^{ij},.
\ee
where $\Omega^{ij}$ is the symplectic metric and where $N$ must be even. The R-symmetry group is $\mathrm{USp}(N)$ with the supercharges transforming in $\mathbf{N}$. The number $N$ must be even.
\item $D=2 \mod 8$. Here, we have Majorana-Weyl supercharges. Due to this property, the $N$ supercharges can be decomposed into $N_+$ supercharges of positive chirality and $N_-$ supercharges of negative chirality which satisfy the algebra
\be
\label{e-2-1-16}
\{  Q^{\alpha i \pm},Q^{\beta j \pm}  \} = (P^\pm \Gamma^a C^{-1})^{\alpha \beta} P_a \delta^{ij} ,
\ee
which shows that, in contrast to the $D=0,4 \mod 8$ case, supercharges of opposite chirality do not mix. The R-symmetry group is $\mathrm{SO}(N_+) \times \mathrm{SO}(N_-)$ with the supercharges transforming $(\mathbf{N_+},\mathbf{1}) + (\mathbf{1},\mathbf{N_-})$. The numbers $N_\pm$ must be even.
\item $D=6 \mod 8$. Here, we have symplectic Majorana-Weyl supercharges satisfying the algebra
\be
\label{e-2-1-17}
\{  Q^{\alpha i \pm},Q^{\beta j \pm}  \} = (P^\pm \Gamma^a C^{-1})^{\alpha \beta} P_a \Omega^{ij} ,
\ee
where again supercharges of opposite chirality do not mix. The R-symmetry group is $\mathrm{USp}(N_+) \times \mathrm{USp}(N_-)$ and the supercharges transform in $(\mathbf{N_+},\mathbf{1}) + (\mathbf{1},\mathbf{N_-})$. The numbers $N_\pm$ must be even.
\end{itemize}
For reviews on supersymmetry algebras in diverse dimensions, the reader is referred to \cite{Strathdee:1987jr,deWit:1997sz,West:1998ey,VanProeyen:1999ni}.

\subsection{Massless representations}
\label{sec-2-1-2}

Here, we discuss the construction of massless representations of supersymmetry allgebras in diverse dimensions. We start by discussing the automorphisms of supersymmetry algebras in diverse dimension. Next, we examine some general properties of the massless representations, which lead to an upper bound on the number of possible supersymmetries in a given dimension. Finally, we turn to the actual construction of such representations for the cases $D=11,10,7,6$.

\subsubsection{Automorphism groups}

The $\{ Q,Q\}$ anticommutation relation of the supersymmetry algebra possesses a set of symmetries that form the \emph{automorphism group} of the algebra. Since the existence of such a group implies that all representations of the supersymmetry algebra fall into its irreducible representations, we need to identify this automorphism group for all possible cases. 

The automorphism group naturally splits into a spacetime and an internal part. The spacetime part is the little group of the Lorentz group, i.e. the subgroup that leaves the $D$--momentum $P_a$ invariant. For the massless representations of interest, we can pass to a lightlike frame where the $D$--momentum is $P_a = (-E,E,\mathbf{0})$ and the little group is identified with the $\mathrm{SO}(D-2)$ that rotates its zero components. To find the $\mathrm{SO}(D-2)$ representations of the supercharges, we note that the latter have been defined as spinors of $\mathrm{SO}(D-1,1)$ and so we have to consider the decomposition $\mathrm{SO}(D-1,1) \to \mathrm{SO}(D-2) \times \mathrm{SO}(1,1)$. Under this decomposition, a Weyl spinor (in the even-dimensional case) and a Dirac spinor (in the odd-dimensional case) decompose as follows
\bea
\label{e-2-1-18}
D=2k+2 &:&\qquad \mathbf{2^k_\pm} \to (\mathbf{2^{k-1}_\pm})_{1/2} + (\mathbf{2^{k-1}_\mp})_{-1/2}, \nn\\ 
D=2k+3 &:&\qquad \mathbf{2^{k+1}} \to (\mathbf{2^k})_{1/2} + (\mathbf{2^k})_{-1/2}.
\eea 
where the subscripts $\pm 1/2$ correspond to the $\mathrm{SO}(1,1)$ ``helicity'' eigenvalues $\lambda_0 = \pm 1/2$. The internal part is the R-symmetry group, corresponding to ``rotations'' of the vector formed by the $N$ supercharges that preserve the supersymmetry algebra. From the previous subsection, we know that the supercharges transform in the $\mathbf{N}_1 + \mathbf{\overline{N}}_{-1}$ of $\mathrm{SU}(N) \times \mathrm{U}(1)$ for $D=0,4 \mod 8$, the $\mathbf{N}$ of $\mathrm{SO}(N)$ for $D=1,3 \mod 8$, the $\mathbf{N}$ of $\mathrm{USp}(N)$ for $D=5,7 \mod 8$, the $(\mathbf{N_+},\mathbf{1}) + (\mathbf{1},\mathbf{N_-})$ of $\mathrm{SO}(N_+) \times \mathrm{SO}(N_-)$ for $D=2 \mod 8$ and the $(\mathbf{N_+},\mathbf{1}) + (\mathbf{1},\mathbf{N_-})$ of $\mathrm{USp}(N_+) \times \mathrm{USp}(N_-)$ for $D=6 \mod 8$. 

\subsubsection{General properties of massless representations}

We will now consider the general properties of the massless representations of supersymmetry algebras. Considering, for simplicity and without loss of generality, the Majorana case, we write the $\{ Q,Q\}$ anticommutator in the form
\be
\label{e-2-1-19}
\{  Q^i, Q^{\dag j}  \} = \Gamma^a \Gamma^0 P_a \delta^{ij},
\ee
and evaluating it explicitly in our lightlike reference frame, we easily find
\be
\label{e-2-1-19a}
\{  Q^i, Q^{\dag j}  \} = E (1 - \Gamma^1 \Gamma^0) \delta^{ij} = E (1 + 2S_0) \delta^{ij}.
\ee 
Therefore, in the $s$--representation considered in Appendix C, we have
\be
\label{e-2-1-20}
\{  Q^i_\lambda, Q^{\dag j}_{\lambda'}  \} = 2E \delta_{\lambda_0 ,1/2} \delta_{\lambda'_0 ,1/2} \delta_{\boldsymbol{\lambda},\boldsymbol{\lambda'}} \delta^{ij},
\ee
where $\boldsymbol{\lambda}=(\lambda_1,\ldots,\lambda_k)$ stands for the vector of $\mathrm{SO}(D-2)$ helicities. Taking the expectation of this expression, for $\lambda_0 = \lambda'_0 = -1/2$, $\boldsymbol{\lambda} = \boldsymbol{\lambda'}$ and $i=j$, in a state $| \psi \rangle$ we get
\be
\label{e-2-1-21}
0 = \langle \psi | \{  Q^i_{-1/2,\boldsymbol{\lambda}} , Q^{\dag i}_{-1/2,\boldsymbol{\lambda}} \} | \psi \rangle = \| Q^{\dag i}_{-1/2,\boldsymbol{\lambda}}  | \psi \rangle \|^2 + \| Q^i_{-1/2,\boldsymbol{\lambda}}  | \psi \rangle \|^2.
\ee
That is, the $s_0 = -1/2$ components of the supercharges create zero-norm states and can be safely set to zero. So, in constructing the representations, we must consider only the nonzero $s_0=1/2$ components which we will collectively denote as $Q_{1/2}$. The number of real nonzero components is found using Eq. (\ref{e-b-2-27}) for the real dimension of a minimal spinor in $D$ dimensions, multiplying by the number $N$ of supersymmetries and dividing by $2$ to remove the zero components. This way, we are left with $2 \Delta$ real components where
\be
\label{e-2-1-22}
\Delta = 2^{\lfloor \frac{D}{2} \rfloor - 1 - W - M} N.
\ee
These components can be grouped into $\Delta$ pairs of lowering and raising operators $(b^I,b^{\dag I})$, where $I=1,\ldots,\Delta$. They can be normalized so as to satisfy the algebra of $\Delta$ uncoupled fermionic oscillators,
\be
\label{e-2-1-23}
\{  b^I , b^{\dag J} \} = \delta^{IJ} ,\qquad \{ b^I , b^J \} = \{ b^{\dag I} , b^{\dag J} \} = 0,
\ee
and can be chosen so that the lowering (raising) operators have weight $-\frac{1}{2}$ ($+\frac{1}{2}$) with respect to one of the $\mathrm{SO}(D-2)$ helicity operators, say $S_1$,
\be
\label{e-2-1-24}
[ S_1 , b^I ] = -\frac{1}{2} b^I ,\qquad [  S_1 , b^{\dag I} ] = \frac{1}{2} b^{\dag I}.
\ee
To obtain the representations of the algebra (\ref{e-2-1-19}), we consider a Clifford vacuum, characterized by its helicity and annihilated by all lowering operators,
\be
\label{e-2-1-25}
S_i | \boldsymbol{\lambda} \rangle = \lambda_i | \boldsymbol{\lambda} \rangle, \qquad b^I | \boldsymbol{\lambda} \rangle = 0,
\ee
and transforming in an $n$--dimensional irreducible representation of the automorphism group. Acting on this vacuum with the raising operators, we obtain a representation whose states have the form $b^{\dag I_1} \ldots b^{\dag I_\delta} | \boldsymbol{\lambda} \rangle$, where each value of $\delta$ corresponds to ${\Delta \choose \delta} n$ states of helicity $\lambda_1 + \frac{\delta}{2}$. In total, the representation space is comprised of $2^\Delta n$ states. The lowest representation is obtained from a Clifford vacuum that is a singlet of the automorphism group and contains $2^\Delta$ states forming a spinor of $\mathrm{SO}(2\Delta)$. 

From this construction, we can obtain an upper bound on the possible number of supersymmetries in any given dimension as well as an upper bound on the spacetime dimensionality for which supersymmetry can exist. The above results imply that the highest-helicity state for a given Clifford vacuum has helicity $\lambda_1 + \frac{\Delta}{2}$. However, since fields of spin higher than $2$ cannot be consistently coupled to gravity, physically relevant representations should not contain states of helicity higher than $2$ or lower than $-2$. Thus, we are led to the restriction
\be
\label{e-2-1-26}
\Delta \leqslant 8.
\ee  
Applying this bound to Eq. (\ref{e-2-1-22}), we see that it implies an upper bound on the number $N$ of supersymmetries that we may have in a given dimension. Moreover, for $D > 11$ this equation has no solution for $N$, that is, \emph{supersymmetry requires at most 11 spacetime dimensions}. 

\begin{table}[!t]
\label{t-2-1}
\begin{center} 
\begin{tabular}{|r||l|l|l|l|}
\hline
$D$  & Automorphism group & Representation of $Q_{1/2}$ & $2\Delta$ & Allowed $N$\\
\hline
$2$  & $\mathrm{SO}(N_+) \times \mathrm{SO}(N_-)$ & $(\mathbf{N_+},\mathbf{1}) + (\mathbf{1},\mathbf{N_-})$ & $N_+ + N_-$ & $1,2,\ldots,16$\\
$3$  & $\mathrm{SO}(N)$ & $\mathbf{N}$ & $N$ & $1,2,\ldots,16$ \\
$4$  & $\mathrm{SO}(2) \times \mathrm{SU}(N) \times \mathrm{U}(1)$ & $\mathbf{N}_{1/2,1} + \mathbf{\overline{N}}_{-1/2,-1}$ & $2 N$ & $1,2,\ldots,8$\\
$5$  & $\mathrm{SO}(3) \times \mathrm{USp}(N)$ & $(\mathbf{2},\mathbf{N})$ & $2 N$ & $2,4,6,8$ \\
$6$  & $\mathrm{SO}(4) \times \mathrm{USp}(N_+) \times \mathrm{USp}(N_-)$ & $(\mathbf{2_+},\mathbf{N_+},\mathbf{1}) + (\mathbf{2_-},\mathbf{1},\mathbf{N_-})$ & $2 (N_+ + N_-)$ & $2,4,6,8$\\
$7$  & $\mathrm{SO}(5) \times \mathrm{USp}(N)$ & $(\mathbf{4},\mathbf{N})$ & $4 N$ & $2,4$\\
$8$  & $\mathrm{SO}(6) \times \mathrm{SU}(N) \times \mathrm{U}(1)$ & $(\mathbf{4_+}, \mathbf{N})_1 + (\mathbf{4_-},\mathbf{\overline{N}})_{-1}$ & $8 N$ & $1,2$\\
$9$  & $\mathrm{SO}(7) \times \mathrm{SO}(N)$ & $(\mathbf{4},\mathbf{N})$ & $8 N$ & $1,2$\\
$10$ & $\mathrm{SO}(8) \times \mathrm{SO}(N_+) \times \mathrm{SO}(N_-)$ &$(\mathbf{8_+},\mathbf{N_+},\mathbf{1}) + (\mathbf{8_-},\mathbf{1},\mathbf{N_-})$ & $8 (N_+ + N_-)$ & $1,2$\\
$11$ & $\mathrm{SO}(9)$ & $\mathbf{16}$ & $16$ & $1$\\
\hline
\end{tabular}
\end{center}
\caption{Automorphism groups, representations of the nonzero supercharges, their real dimensions, and the allowed values of $N$ for supersymmetry algebras in diverse dimensions.}
\end{table}

The results of this section up to now are summarized in Table \ref{t-2-1}, which contains the various properties of the supercharges and the allowed numbers of supersymmetries in all spacetime dimensions from $D=2$ to $D=11$.

\subsubsection{Massless representations of minimal supersymmetry in $D=11,10,7,6$}

Using the above results, we can build the representations of any supersymmetry algebra in any spacetime dimension. The easiest procedure for obtaining the representations consists of the following steps: (i) consider the lowest representation which transforms as a spinor of $\mathrm{SO}(2\Delta)$, (ii) embed the automorphism group into $\mathrm{SO}(2\Delta)$ and read off the particle content of the lowest representation by considering the decomposition of this spinor representation and (iii) construct higher representations by taking the tensor product of the lowest representation with irreducible representations of the automorphism group. Since this procedure is best illustrated by considering specific examples, we will next examine the detailed construction of the representations of the minimal supersymmetry algebras in $D=11,10,7$ and $6$.

\begin{itemize}
\item $D=11$. In eleven dimensions, the unique supersymmetry algebra is the $N=1$ algebra which has $2 \Delta = 16$ real nonzero supercharges transforming as a spinor $\mathbf{16}$ of the little group $\mathrm{SO}(9)$. The fact that $\Delta=8$ implies that there exists only one representation of the supersymmetry algebra. So, acting with the $\Delta=8$ raising operators on the unique $\mathrm{SO}(9)$--singlet Clifford vacuum, we obtain a representation space of $2^8=256$ states transforming as a Dirac spinor $\mathbf{256}$ of $\mathrm{SO}(16)$.  To find the field content of this representation space, we must arrange these $\mathrm{SO}(16)$ states into irreps of the little group $\mathrm{SO}(9)$. To do so, we first decompose the $\mathrm{SO}(16)$ Dirac spinor into the two irreducible Weyl spinors $\mathbf{128_+}$ and $\mathbf{128_-}$. Next, we consider the embedding $\mathrm{SO}(16) \supset \mathrm{SO}(9)$, under which these Weyl spinors decompose as $\mathbf{128_+} \to \mathbf{44} + \mathbf{84}$ and $\mathbf{128_-} \to \mathbf{128}$. So, the representation contains the bosons $\mathbf{44}$ (graviton) and $\mathbf{84}$ ($3$-form) plus the fermion $\mathbf{128}$ (gravitino). This is the $D=11$ gravity multiplet
\be
\label{e-2-1-27}
\text{Gravity multiplet} \quad:\quad (g_{\mu\nu}, A_{\mu\nu\rho},\psi_\mu). 
\ee

\item $D=10$. In ten dimensions, the minimal supersymmetry algebras have $(N_+,N_-)=(1,0)$ and $(0,1)$. Considering for definiteness the second case, we have $2 \Delta = 8$ real nonzero supercharges transforming as a spinor $\mathbf{8_-}$ of the little group $\mathrm{SO}(8)$. Since $\Delta=4$, there are two possible representations. Acting with the 4 raising operators on a $\mathrm{SO}(8)$--singlet Clifford vacuum, we obtain $16$ states transforming as a reducible spinor $\mathbf{16}$ of $\mathrm{SO}(8)$. We note that in this case, the automorphism group coincides with the group corresponding to the Clifford algebra of the ladder operators.  To find their field content, we must arrange the reducible spinor $\mathbf{16}$ of $\mathrm{SO}(8)$ into two irreducible representations of the same group, one bosonic and one fermionic. The desired decomposition is $\mathbf{16} \to \mathbf{8_v} + \mathbf{8_+}$. So, the first representation of the supersymmetry algebra contains the boson $\mathbf{8_v}$ (gauge field) and the fermion $\mathbf{8_+}$ (positive-chirality fermion). This is the $D=10$, $N=1$ vector multiplet
\be
\label{e-2-1-28}
\text{Vector multiplet} \quad:\quad ( A_\mu, \lambda^+ ).
\ee
The second representation is obtained by tensoring the above representation with one of the 8--dimensional irreducible representations $\mathbf{8_v}$, $\mathbf{8_+}$ and $\mathbf{8_-}$ of $\mathrm{SO}(8)$. The relevant representation for $N=1$ supersymmetry turns out to be $\mathbf{8_v}$. Taking the tensor product according to
\be
\label{e-2-1-29}
\mathbf{8_v} \times (\mathbf{8_v} + \mathbf{8_+}) = \mathbf{35_v} + \mathbf{28} + \mathbf{1} + \mathbf{56_+} + \mathbf{8_-}
\ee
we obtain the bosons $\mathbf{35_v}$ (graviton), $\mathbf{28}$ ($2$-form) and $\mathbf{1}$ (dilaton) and the fermions $\mathbf{56}_+$ (positive-chirality gravitino) and $\mathbf{8_-}$ (negative-chirality fermion). We thus arrive at the $D=10$, $N=1$ gravity multiplet
\be
\label{e-2-1-30}
\text{Gravity multiplet} \quad:\quad ( g_{\mu\nu} , B_{\mu\nu} \textrm{ or } A_{\mu\nu\rho\sigma\tau\upsilon} , \phi , \psi^+_\mu , \chi^-),
\ee
where we noted that the 2--form potential can be traded for a dual 6--form potential.

\item $D=7$. In seven dimensions, the minimal supersymmetry algebra is the $N=2$ algebra, which has $2 \Delta = 8$ real nonzero supercharges transforming in the $(\mathbf{4};\mathbf{2})$ of the $\mathrm{SO}(5) \times \mathrm{USp}(2)$ automorphism group. Since $\Delta=4$, there are two possible representations. Acting with the 4 raising operators on a $\mathrm{SO}(5) \times \mathrm{USp}(2)$--singlet Clifford vacuum, we obtain $16$ states transforming as a reducible spinor $\mathbf{16}$ of $\mathrm{SO}(8)$. To find their field content, we first arrange the $\mathbf{16}$ of $\mathrm{SO}(8)$ into the irreducible representations $\mathbf{8_+}$ and $\mathbf{8_v}$. We next consider the embedding $\mathrm{SO}(8) \supset \mathrm{SO}(5) \times \mathrm{USp}(2)$, under which these representations decompose as $\mathbf{8_+} \to (\mathbf{5},\mathbf{1})+(\mathbf{1},\mathbf{3})$ and $\mathbf{8_v} \to (\mathbf{4},\mathbf{2})$. So, the first representation of the supersymmetry algebra contains the bosons $(\mathbf{5},\mathbf{1})$ (gauge field) and $(\mathbf{1},\mathbf{3})$ (real triplet of scalars) and the fermions $(\mathbf{4},\mathbf{2})$ (fermion doublet). This is the $D=7$, $N=2$ vector multiplet,
\be
\label{e-2-1-31}
\text{Vector multiplet} \quad:\quad ( A_\mu, A^A_{\phantom{A} B}, \lambda^A ), 
\ee
where $A=1,2$ is a $\mathrm{USp}(2)$ doublet index. The second representation is obtaining by taking the tensor product with $(\mathbf{5},\mathbf{1})$ according to
\be
\label{e-2-1-32}
(\mathbf{5},\mathbf{1}) \times \left( (\mathbf{5},\mathbf{1})+(\mathbf{1},\mathbf{3})+(\mathbf{4},\mathbf{2}) \right) = (\mathbf{14},\mathbf{1}) + (\mathbf{10},\mathbf{1})+(\mathbf{5},\mathbf{3})+(\mathbf{1},\mathbf{1})+(\mathbf{16},\mathbf{2})+(\mathbf{4},\mathbf{2}),
\ee
and contains the bosons $(\mathbf{14},\mathbf{1})$ (graviton), $(\mathbf{10},\mathbf{1})$ (2--form) $(\mathbf{5},\mathbf{3})$ (gauge field triplet) and $(\mathbf{1},\mathbf{1})$ (dilaton) plus the fermions $(\mathbf{16},\mathbf{2})$ (gravitino doublet) and $(\mathbf{4},\mathbf{2})$ (fermion doublet). This is the $D=7$, $N=2$ gravity multiplet
\be
\label{e-2-1-33}
\text{Gravity multiplet} \quad:\quad ( g_{\mu\nu}, B_{\mu\nu}\textrm{ or } A_{\mu\nu\rho}, A^{\phantom{\mu} A}_{\mu \phantom{A} B}, \phi, \psi^A_\mu, \chi^A ).
\ee
where we noted that the 2--form potential can be traded for its dual 3--form potential.

\item $D=6$. In six dimensions, the minimal supersymmetry algebras have $(N_+,N_-)=(2,0)$ and $(0,2)$. Considering for definiteness the second case, we have $2 \Delta = 4$ real nonzero supercharges transforming in the $(\mathbf{2_-};\mathbf{2})$ of the $\mathrm{SO}(4) \times \mathrm{USp}(2)$ automorphism group; equivalently, one may write the automorphism group as $\mathrm{SU}(2)_- \times \mathrm{SU}(2)_+ \times \mathrm{USp}(2)$ and the representation of the supercharges as $(\mathbf{1},\mathbf{2};\mathbf{2})$. Since $\Delta=2$, there are four possible representations. Acting with the $2$ raising operators on a $\mathrm{SU}(2)_- \times \mathrm{SU}(2)_+ \times \mathrm{USp}(2)$--singlet Clifford vacuum, we obtain $4$ states transforming as a reducible spinor $\mathbf{4}$ of $\mathrm{SO}(4)$.  To find their field content, we first arrange the $\mathbf{4}$ of $\mathrm{SO}(4)$ into the irreducible spinors $\mathbf{2_+}$ and $\mathbf{2_-}$. We next embed $\mathrm{SU}(2)_- \times \mathrm{USp}(2)$ in $\mathrm{SO}(4)$ in such a way that $\mathbf{2_-}$ and $\mathbf{2_+}$ correspond, respectively, to the fundamentals of $\mathrm{USp}(2)$ and $\mathrm{SU}(2)_+$. So, the first representation of the supersymmetry algebra contains the bosons $(\mathbf{1},\mathbf{1};\mathbf{2})$ (complex doublet of scalars) and the fermion $(\mathbf{1},\mathbf{2};\mathbf{1})$ (positive-chirality fermion). This is the $D=6$, $N=2$ hypermultiplet
\be
\label{e-2-1-34}
\text{Hypermultiplet} \quad:\quad ( 4 \varphi , 2 \psi^{+} ),
\ee
where each $\varphi$ is understood as a real scalar and each $\psi^{+}$ as a symplectic Majorana spinor. The remaining representations of the algebra are obtained by tensoring this representation with the three smallest irreducible representations of $\mathrm{SU}(2)_- \times \mathrm{SU}(2)_+ \times \mathrm{USp}(2)$. Taking the tensor product with $(\mathbf{2},\mathbf{1};\mathbf{1})$ according to
\be
\label{e-2-1-35}
(\mathbf{2},\mathbf{1};\mathbf{1}) \times \left( (\mathbf{1},\mathbf{1};\mathbf{2}) + (\mathbf{1},\mathbf{2};\mathbf{1}) \right) = (\mathbf{2},\mathbf{2};\mathbf{1}) + (\mathbf{2},\mathbf{1};\mathbf{2}),
\ee
we obtain the boson $(\mathbf{2},\mathbf{2};\mathbf{1})$ (gauge field) and the fermions $(\mathbf{2},\mathbf{1};\mathbf{2})$ (negative-chirality fermion doublet). This is the $D=6$, $N=2$ vector multiplet,
\be
\label{e-2-1-36}
\text{Vector multiplet} \quad:\quad ( A_\mu, \lambda^{A-} ).
\ee
Likewise, taking the tensor product with $(\mathbf{1},\mathbf{2};\mathbf{1})$ according to
\be
\label{e-2-1-37}
(\mathbf{1},\mathbf{2};\mathbf{1}) \times \left( (\mathbf{1},\mathbf{1};\mathbf{2}) + (\mathbf{1},\mathbf{2};\mathbf{1}) \right) = (\mathbf{1},\mathbf{3};\mathbf{1}) + (\mathbf{1},\mathbf{1};\mathbf{1}) + (\mathbf{1},\mathbf{2};\mathbf{2}),
\ee
we obtain the bosons $(\mathbf{1},\mathbf{3};\mathbf{1})$ (anti-self-dual 2--form) and $(\mathbf{1},\mathbf{1};\mathbf{1})$ (dilaton) and the fermions $(\mathbf{1},\mathbf{2};\mathbf{2})$ (positive-chirality fermion doublet). This is the $D=6$, $N=2$ tensor multiplet,
\be
\label{e-2-1-38}
\text{Tensor multiplet} \quad:\quad ( B^-_{\mu\nu} , \phi , \chi^{A+} ).
\ee
Finally, taking the tensor product with $(\mathbf{3},\mathbf{2};\mathbf{1})$ according to
\be
\label{e-2-1-39}
(\mathbf{3},\mathbf{2};\mathbf{1}) \times \left( (\mathbf{1},\mathbf{1};\mathbf{2}) + (\mathbf{1},\mathbf{2};\mathbf{1}) \right) = (\mathbf{3},\mathbf{3};\mathbf{1}) + (\mathbf{3},\mathbf{1};\mathbf{1}) +
(\mathbf{3},\mathbf{2};\mathbf{2}),
\ee
we obtain the bosons $(\mathbf{3},\mathbf{3};\mathbf{1})$ (graviton), $(\mathbf{3},\mathbf{1};\mathbf{1})$ (self-dual 2--form) and the fermions $(\mathbf{3},\mathbf{2};\mathbf{2})$ (negative-chirality gravitino doublet). This is the $D=6$, $N=2$ gravity multiplet,
\be
\label{e-2-1-40}
\text{Gravity multiplet} \quad:\quad ( g_{\mu\nu} , B^+_{\mu\nu} , \psi^{A-}_\mu ).
\ee
\end{itemize}

\section{Supergravity}
\label{sec-2-2}

As with many types of symmetry, supersymmetry may be treated either as a rigid symmetry, where the transformation parameters are constant, or as a local gauge symmetry, where the transformation parameters are spacetime-dependent. However, in contrast to ordinary internal symmetries, supersymmetry possesses the special property that it mixes with the spacetime Poincar\'e symmetry. So, treating supersymmetry as a local symmetry actually requires promoting the full super-Poincar\'e symmetry to a local one and thus leads to theories containing gravity. These theories are known by the name of \emph{supergravity}. The gauge field of local supersymmetry is the vector-spinor gravitino appearing in all gravity multiplets.

Supergravity theories were originally formulated with the hope that they would tame the ultraviolet divergences that plague four-dimensional quantum gravity. However, although these theories exhibit better ultraviolet behavior than ordinary quantum gravity, it turns out that they are still non-renormalizable, failing to realize this initial hope. Nevertheless, supergravity theories emerged back to life, not as fundamental theories of gravity, but rather as effective field theories describing the low-energy degrees of freedom of underlying theories, such as superstring theories and M-theory; in this viewpoint, the ultraviolet divergences are regarded as low-energy artifacts that are guaranteed to cancel in the complete theory.

During the years, there has been developed a substantial machinery for the construction of supergravity theories. Some of the popular methods are the following.
\begin{itemize}
\item \emph{Noether method}: Start from an incomplete theory (e.g. a linearized or rigidly supersymmetric theory) and appropriately modify the Lagrangian and supersymmetry variations of the fields term by term until the theory becomes locally supersymmetric.
\item \emph{Superspace methods}: Introduce fermionic spacetime coordinates in addition to the usual bosonic ones and formulate general relativity on this extended space by replacing the vielbein, the torsion and the curvature by the supervielbein, the supertorsion and the supercurvature. The only fundamental difference of this construction from ordinary general relativity is that supertorsions and supercurvatures must be subject to constraints to eliminate spurious degrees of freedom.
\item \emph{Gauging the super-Poincar\'e algebra}: Promote the super-Poincar\'e symmetry to a local symmetry by introducing appropriate gauge fields (bosonic for even generators and fermionic for odd generators) and impose suitable constraints to eliminate non-propagating degrees of freedom.
\end{itemize}
Here, we will choose the last method and illustrate it for the case of simple ($N=1$), pure (no matter multiplets) supergravity in 4 dimensions, i.e. the theory describing the minimal gravity multiplet $(g_{\mu\nu},\psi_\mu)$. The Noether method will be employed in other parts of this thesis. For more information on the construction of supergravities in diverse dimensions, the reader is referred to the classic review \cite{VanNieuwenhuizen:1981ae}, the more recent reviews \cite{Tanii:1998px,deWit:2002vz} and the collection of original papers \cite{Salam:1989fm}. 

\subsection{Gauge theory of the super-Poincar\'e group}
\label{sec-2-2-1}

Here, we will construct simple supergravity in 4 dimensions using the McDowell-Mansouri \cite{MacDowell:1977jt} formalism (see also \cite{Townsend:1977fz,vanNieuwenhuizen:2004rh}) which treats the theory as a gauge theory based on the super-Poincar\'e group. To begin, let us recall some elementary facts about gauge theories. Consider a field $\phi$ and an infinitesimal transformation under a symmetry group, 
\be
\label{e-2-2-1}
\delta_\lambda \phi = - \lambda \phi, 
\ee
where $\lambda = \lambda^I T_I$ is the gauge parameter, expressed in terms of the generators $T_I$ of the symmetry group. Gauging the symmetry group, i.e. promoting the global symmetry to a local one, is achieved by introducing a gauge field $\omega_\mu = \omega^I_\mu T_I$ with the transformation law
\be
\label{e-2-2-2}
\delta_\lambda \omega_\mu = \partial_\mu \lambda + \left[ \omega_\mu,\lambda \right]
\ee
so that the covariant derivative of $\phi$, defined as 
\be
\label{e-2-2-3}
\MD_\mu \phi = \partial_\mu \phi - \delta_{\omega_\mu} \phi,
\ee
transforms in the same way as $\phi$. The curvature $R_{\mu\nu}$ of $\omega_\mu$ is defined through
\be
\label{e-2-2-4}
\left[ \MD_\mu, \MD_\nu  \right] = - \delta_{R_{\mu\nu}},
\ee
has the explicit form
\be
\label{e-2-2-5}
R_{\mu\nu} = R_{\mu\nu}^{\phantom{{\mu\nu}} I} T_I = \partial_\mu \omega_\nu - \partial_\nu \omega_\mu + \left[ \omega_\mu , \omega_\nu \right],
\ee
and transforms covariantly so that it can be used to build invariant actions.

To construct supergravity theories, we need to apply the above principles for the super-Poincar\'e group, generated by the translation generators $P_a$, the Lorentz generators $M_{ab}$ and the supersymmetry generators $Q^\alpha$. However, to make the analogy with gauge theory more complete, we actually need to generalize the supersymmetry algebra to the super-anti-de Sitter algebra, given by
\bea
\label{e-2-2-6}
\left[ P_a,P_b \right] &=& m^2 M_{ab}, \nn\\
\left[ P_a,M_{bc} \right] &=& \eta_{ac} P_b - \eta_{ab} P_c, \nn\\
\left[ M_{ab},M_{cd} \right] &=& \eta_{ad} M_{bc} + \eta_{bc} M_{ad} - \eta_{ac} M_{bd} - \eta_{bd} M_{ac}, \nn\\
\left[ P_a,Q^\alpha \right] &=& - m ( \Gamma_a Q )^\alpha , \nn\\
\left[ M_{ab},Q^\alpha \right] &=& \frac{1}{2} ( \Gamma_{ab} Q )^\alpha , \nn\\
\{ Q^\alpha,\bar{Q}^\beta \} &=& - ( \Gamma^a )^{\alpha \beta} P_a - m (\Gamma^{ab})^{\alpha\beta} M_{ab}.
\eea
where $m$ is a free parameter. The super-Poincar\'e algebra is recovered through the In\"on\"u-Wigner contraction $\mathrm{SO}(3,2) \to \mathrm{ISO}(3,1)$, effected by the limit $m \to 0$.

We begin by considering a transformation of the form (\ref{e-2-2-1}), where the parameter $\Lambda$ is expressed in terms of the generators as
\be
\label{e-2-2-7}
\lambda = \frac{1}{2} \lambda^{ab} M_{ab} + \xi^a P_a + \bar{\epsilon}_\alpha Q^\alpha .
\ee
To promote the symmetry to a gauge symmetry, we introduce the gauge field
\be
\label{e-2-2-8}
\omega_\mu = \frac{1}{2} \omega_\mu^{\phantom{\mu} ab} M_{ab} + e^a_\mu P_a + \bar{\psi}_{\mu\alpha} Q^\alpha ,
\ee
whose individual components $\omega_\mu^{\phantom{\mu} ab}$ (spin connection), $e^a_\mu$ (vielbein) and $\psi_\mu^\alpha$ (gravitino) are associated with Lorentz transformations, translations and supersymmetry transformations respectively. Inserting (\ref{e-2-2-7}) and (\ref{e-2-2-8}) into the transformation law (\ref{e-2-2-2}) and using the algebra (\ref{e-2-2-6}), we find that these components transform according to 
\bea
\label{e-2-2-9}
\delta_\lambda e^a_\mu &=& \partial_\mu \xi^a + \omega_{\mu \phantom{a} b}^{\phantom{\mu} a} \xi^b - e^b_\mu \lambda^a_{\phantom{a} b} + \bar{\epsilon} \Gamma^a \psi_\mu ,\nn\\
\delta_\lambda \omega_\mu^{\phantom{\mu}ab} &=& \partial_\mu \lambda^{ab} + \omega_{\mu \phantom{a} c}^{\phantom{\mu} a} \lambda^{cb} + \omega_{\mu \phantom{b} c}^{\phantom{\mu} b} \lambda^{ac} - m \bar{\psi}_\mu \Gamma^{ab} \epsilon + 2 m^2 e^{[a}_\mu \xi^{b]} ,\nn\\
\delta_\lambda \psi_{\mu} &=&  \partial_\mu \epsilon + \frac{1}{4} \omega_\mu^{\phantom{\mu} ab} \Gamma_{ab} \epsilon - \frac{1}{4} \lambda^{ab} \Gamma_{ab} \psi_\mu + m \xi^a \Gamma_a \psi_\mu - m e^a_\mu \Gamma_a \epsilon.
\eea
The super-Poincar\'e covariant derivative is defined as
\be
\label{e-2-2-10}
\MD_\mu \phi = \partial_\mu \phi + \frac{1}{2} \omega_\mu^{\phantom{\mu} ab} M_{ab} \phi + e^a_\mu P_a \phi +  \bar{\psi}_\mu^\alpha Q_\alpha \phi.
\ee
The curvature of $\omega_\mu$ can be computed using either (\ref{e-2-2-4}) or (\ref{e-2-2-5}). Decomposing it as
\be
\label{e-2-2-11}
R_{\mu\nu} = \frac{1}{2} R_{\mu\nu}^{\phantom{\mu\nu} ab}(M) M_{ab} + R_{\mu\nu}^{\phantom{\mu\nu} a} (P) P_a + \bar{R}_{\mu\nu \alpha}(Q) Q^\alpha,
\ee
we find the individual components
\bea
\label{e-2-2-12}
R_{\mu\nu}^{\phantom{\mu\nu} ab} (M) &=& 2 \left( \partial_{[\mu} \omega_{\nu]}^{\phantom{\nu} ab} + \omega_{[\mu \phantom{a} c}^{\phantom{\mu} a} \omega_{\nu]}^{\phantom{\nu} cb} \right) + m \bar{\psi}_\mu \Gamma^{ab} \psi_\nu + 2 m^2 e^a_{[\mu} e^b_{\nu]}, \nn\\
R_{\mu\nu}^{\phantom{\mu\nu} a} (P) &=& 2 \left( \partial_{[\mu} e^a_{\nu]} + \omega_{[\mu \phantom{a} b}^{\phantom{[\mu} a} e^b_{\nu]} \right) + \bar{\psi}_\mu \Gamma^a \psi_\nu, \nn\\
R_{\mu\nu}^{\phantom{\mu\nu} \alpha} (Q) &=& 2 \left(  \partial_{[\mu} \psi_{\nu]}^\alpha + \frac{1}{4} \omega_{[\mu}^{\phantom{\mu} ab} ( \Gamma_{ab} \psi_{\nu]} )^\alpha \right) + m e^a_{[\mu} ( \Gamma_a \psi_{\nu]} )^\alpha.
\eea
In standard terminology, $R_{\mu\nu}^{\phantom{\mu\nu} ab} (M)$ is a true curvature while $R_{\mu\nu}^{\phantom{\mu\nu} a} (P)$ and $R_{\mu\nu \alpha} (Q)$ are torsions.

Having described the building blocks of the theory, we are now faced with the problem of constructing an invariant action. To do so, it is instructive to first consider the case of gravity and then generalize to supergravity.

\subsection{Pure gravity}
\label{sec-2-2-2}

In the case of pure gravity, the available gauge fields are $\omega_\mu^{\phantom{\mu} ab}$ and $e^a_\mu$ with curvatures
\be
\label{e-2-2-13}
R_{\mu\nu}^{\phantom{\mu\nu} ab} (M) = R_{\mu\nu}^{\phantom{\mu\nu} ab} + 2 m^2 e^a_{[\mu} e^b_{\nu]}, \qquad
R_{\mu\nu}^{\phantom{\mu\nu} a} (P) = T_{\mu\nu}^{\phantom{\mu\nu} a},
\ee
where $R_{\mu\nu}^{\phantom{\mu\nu} ab}$ and $T_{\mu\nu}^{\phantom{\mu\nu} a}$ are the usual curvature and torsion 2--forms of general relativity. To find the appropriate action for the theory, the usual gauge-theory arguments suggest that we search for the simplest possible quantity that is quadratic in the curvatures and respects the symmetries of the theory. Making the assumption that the indices of the curvatures are contracted with constant tensors, the only possible term involves the curvature $R_{\mu\nu}^{\phantom{\mu\nu} ab} (M)$ and leads to the action
\be
\label{e-2-2-14}
S_2 = - \frac{1}{64 \kappa^2 m^2} \int \dd^4 x \epsilon^{\mu\nu\rho\sigma} \epsilon_{abcd} R_{\mu\nu}^{\phantom{\mu\nu} ab} (M) R_{\rho\sigma}^{\phantom{\rho\sigma} cd} (M),
\ee
where the value of the coefficient will be justified below. Eq. (\ref{e-2-2-14}) is, of course, not the action for Einstein gravity. However, substituting the first of (\ref{e-2-2-13}) for $R_{\mu\nu}^{\phantom{\mu\nu} ab} (M)$ and expanding, we find
\be
\label{e-2-2-15}
S_2 = \frac{1}{\kappa^2} \int \dd^4 x \left( - \frac{1}{64 m^2} \epsilon^{\mu\nu\rho\sigma} \epsilon_{abcd} R_{\mu\nu}^{\phantom{\mu\nu} ab} R_{\rho\sigma}^{\phantom{\rho\sigma} cd} + \frac{1}{4} e R + \frac{3}{2} m^2 e \right).
\ee
Now, the first term in (\ref{e-2-2-15}) is a Gauss-Bonnet term that may be safely ignored. The second term is just the well-known Einstein-Hilbert action. Finally, the third term is a cosmological constant term, as expected in the gauging of the anti-de Sitter group. Applying the contraction $m \to 0$, only the second term survives and we are thus left with the Einstein-Hilbert action,
\be
\label{e-2-2-16}
S_2 = \frac{1}{4 \kappa^2} \int \dd^4 x e R.
\ee 

Although this action is the desired one, in the present formulation (\emph{first-order} formalism)  $e^a_\mu$ and $\omega_{\mu a b}$ appear as independent fields that are both to be varied in the action. In contrast, in the ordinary formulation of gravity (\emph{second-order} formalism), one has the torsion-free condition
\be
\label{e-2-2-17}
R_{\mu\nu}^{\phantom{\mu\nu} a} (P) = 0,
\ee
which fixes the form of the spin connection to
\be
\label{e-2-2-18}
\omega_{\mu ab} = \omega_{\mu ab} (e) \equiv \frac{1}{2} e^c_\mu (\Omega_{bac} + \Omega_{bca} + \Omega_{cab}) ;\qquad \Omega_{abc} = e^\mu_a e^\nu_b (\partial_\mu e_{c \nu} - \partial_\nu e_{c \mu}),
\ee
so that only $e^a_\mu$ is to be varied in the action. So, one has to show that the first-order formalism is equivalent to the second-order one, i.e. that the equations of motion in the first-order formalism are the constraint equation $\omega_{\mu a b} = \omega_{\mu a b}(e)$ and the Einstein equation. To do so, we write  the Einstein-Hilbert Lagrangian as $\ML = \frac{1}{4 \kappa^2} e R (\omega,e)$, emphasizing that $R$ is a functional of both $e^a_\mu$ and $\omega_{\mu ab}$. To obtain the equations of motion, we first use an integration by parts to write
\be
\label{e-2-2-19}
e R (\omega,e) = 2 e e^\mu_{[a} e^\nu_{b]} ( \partial_\mu \omega_\nu^{\phantom{\nu} ab}  + \omega_{\mu \phantom{a} c}^{\phantom{\mu} a} \omega_\nu^{\phantom{\nu} cb} ) = 2 \left[  - \partial_\mu ( e e^\mu_{[a} e^\nu_{b]} ) \omega_\nu^{\phantom{\nu} ab} + e e^\mu_{[a} e^\nu_{b]} \omega_{\mu \phantom{a} c}^{\phantom{\mu} a} \omega_\nu^{\phantom{\nu} cb} \right]
\ee
We see thus that the equation of motion of $\omega_{\mu ab}$ is algebraic, as expected for a non-propagating field. It is given by
\bea
\label{e-2-2-20}
\partial_\mu ( e e^\mu_{[a} e^\nu_{b]} ) - e ( e^\mu_{c} e^\nu_{[a} - e^\nu_{c} e^\mu_{[a} ) \omega_{\mu b]}^{ \phantom{\mu b]} c} = 0,
\eea
and it is a matter of simple algebra to show that its solution is indeed $\omega_{\mu a b} = \omega_{\mu a b}(e)$. To vary with respect to $e^a_\mu$, we write $e R (\omega,e) = e e^\mu_a e^\nu_b R_{\mu\nu}^{\phantom{\mu\nu} ab} (\omega)$ and so we need only vary the quantity $e e^\mu_a e^\nu_b$. Making use of the identity $\delta e = -e e^a_\mu \delta e^\mu_a$, we indeed obtain the Einstein equation
\be
\label{e-2-2-21}
R_\mu^{\phantom{\mu} a} - \frac{1}{2} e^a_\mu R = 0.
\ee 
This completes the demonstration of the equivalence of the two formalisms.

To clarify the meaning of the torsion-free condition in the present context, we consider a GCT with parameter $\xi^\mu$ and a local Poincar\'e transforation with parameter $\Lambda = - \frac{1}{2} \xi^\mu \omega_\mu^{\phantom{\mu} ab} M_{ab} + \xi^\mu e^a_\mu P_a$ acting on the vielbein. Subtracting the two relations, we find
\be
\label{e-2-2-22}
( \delta_\Lambda - \delta_{\mathrm{gct}, \xi^\mu} ) e^a_\mu = R_{\mu\nu}^{\phantom{\mu\nu} a} (P) \xi^\nu. 
\ee
Therefore, if the constraint (\ref{e-2-2-17}) holds, general coordinate invariance becomes indeed equivalent to local Poincar\'e invariance and so a gauge theory of the Poincar\'e group leads to gravity. Another consequence of this constraint is that the correct covariant derivatives to use when coupling fields to gravity are
\be
\label{e-2-2-23}
D_\mu \phi = \partial_\mu \phi + \frac{1}{2} \omega_\mu^{\phantom{\mu} ab} M_{ab} \phi,
\ee
plus its appropriate generalizations with Christoffel symbols for fields with curved-space indices. The commutator of two covariant derivatives is now given by
\be
\label{e-2-2-24}
\left[  D_\mu, D_\nu  \right] = \frac{1}{2} R_{\mu\nu}^{\phantom{\mu\nu} ab} M_{ab},
\ee
that is, it involves only the Riemann curvature.

To summarize, we have managed to construct gravity in a manner completely analogous to gauge theories. The only departures from the construction of ordinary gauge theories are that (i) the starting point must be the anti-de Sitter algebra, which is to be contracted to the Poincar\'e algebra in the end and (ii) the torsion-free constraint must be satisfied to eliminate spurious degrees of freedom.

\subsection{Simple supergravity}
\label{sec-2-2-3}

To generalize the above construction to supergravity, we start from the full expressions (\ref{e-2-2-12}), which we write as
\bea
\label{e-2-2-25}
R_{\mu\nu}^{\phantom{\mu\nu} ab} (M) &=& R_{\mu\nu}^{\phantom{\mu\nu} ab} + m \bar{\psi}_\mu \Gamma^{ab} \psi_\nu + 2 m^2 e^a_{[\mu} e^b_{\nu]}, \nn\\
R_{\mu\nu}^{\phantom{\mu\nu} a} (P) &=& T_{\mu\nu}^{\phantom{\mu\nu} a} + \bar{\psi}_\mu \Gamma^a \psi_\nu, \nn\\
R_{\mu\nu}^{\phantom{\mu\nu} \alpha} (Q) &=& 2 D_{[\mu} \psi_{\nu]}^\alpha + m e^a_{[\mu} (\Gamma_a \psi_{\nu]})^\alpha.
\eea
To find the appropriate action, we need again the simplest quantities quadratic in the curvatures and consistent with all symmetries. Neglecting supersymmetry for the moment, there is now one more such quantity, involving the curvature $R_{\mu\nu}^{\phantom{\mu\nu} \alpha} (Q)$. The complete action is given by
\be
\label{e-2-2-26}
S = - \frac{1}{64 \kappa^2 m^2} \int \dd^4 x \epsilon^{\mu\nu\rho\sigma} \left[  \epsilon_{abcd} R_{\mu\nu}^{\phantom{\mu\nu} ab} (M) R_{\rho\sigma}^{\phantom{\rho\sigma} cd} (M) + \ii \alpha \bar{R}_{\mu\nu} (Q) \Gamma_5 R_{\rho\sigma} (Q) \right],
\ee
where $\alpha$ is a coefficient to be determined later. Expanding this action according to (\ref{e-2-2-25}) and simplifying the various terms using the gamma-matrix duality relations (\ref{e-b-1-7}), we find $S = S_2 + S_{3/2}$ where $S_2$ is given in (\ref{e-2-2-14}) and 
\bea
\label{e-2-2-27}
S_{3/2} &=& \frac{1}{2 \kappa^2} \int \dd^4 x \biggl[ - \frac{\alpha}{8m} e \bar{\psi}_\mu \Gamma^{\mu\nu\rho} D_\nu \psi_\rho + \frac{m}{2} \left( 1 + \frac{\alpha}{8m} \right) e \bar{\psi}_\mu \Gamma^{\mu\nu} \psi_\nu \nn\\ && \qquad \qquad \quad + \frac{\ii}{8m} \left( 1 - \frac{\alpha}{8m} \right) \epsilon^{\mu\nu\rho\sigma} R_{\mu\nu}^{\phantom{\mu\nu} ab} \bar{\psi}_\rho \Gamma_{ab} \Gamma_5 \psi_\sigma \biggr],
\eea
We observe that the unwanted third term cancels provided that the coefficient $\alpha$ is set to $\alpha=8m$. In this case, the first and second term reduce to the action
\be
\label{e-2-2-28}
S_{3/2} = \frac{1}{2\kappa^2} \int \dd^4 x e ( - \bar{\psi}_\mu \Gamma^{\mu\nu\rho} D_\nu \psi_\rho + m \bar{\psi}_\mu \Gamma^{\mu\nu} \psi_\nu )
\ee
which contains the Rarita-Schwinger term plus a gravitino ``mass'' term, the latter being a consequence of gauging the super-anti-de Sitter group. Applying the contraction $m \to 0$, only the first term survives. So, the complete action reads
\be
\label{e-2-2-29}
S = \frac{1}{\kappa^2} \int \dd^4 x e \left( \frac{1}{4} R - \frac{1}{2} \bar{\psi}_\mu \Gamma^{\mu\nu\rho} D_\nu \psi_\rho \right),
\ee
which is the Einstein/Rarita-Schwinger action.

Again, this action may either be regarded as a first-order action $S[e,\psi,\omega]$ where $e^a_\mu$, $\psi_\mu$ and $\omega_{\mu ab}$ are all to be varied or as a second-order action $S[e,\psi]$ where $\omega_{\mu ab}$ is a function of $e^a_\mu$ and $\psi_\mu$, $\omega_{\mu ab} = \omega_{\mu ab}(e,\psi)$, and only $e^a_\mu$ and $\psi_\mu$ are to be varied. In the second-order formalism, the action (\ref{e-2-2-29}) must be supplemented with a constraint that eliminates spurious degrees of freedom and the appropriate condition turns out to be again $R_{\mu\nu}^{\phantom{\mu\nu} a} (P) = 0$, since it is the only one that may be solved algebraically. This time however, the constraint implies that we have non-trivial torsion,
\be
\label{e-2-2-30}
T_{\mu\nu}^{\phantom{\mu\nu} a} = - \bar{\psi}_\mu \Gamma^a \psi_\nu.
\ee
In the first-order formalism we have to show, as before, that this constraint follows from the equation of motion for $\omega_{\mu ab}$. To do so, we expand the second term in (\ref{e-2-2-29}) writing 
\be
\label{e-2-2-31}
S = \frac{1}{\kappa^2} \int \dd^4 x e \left( \frac{1}{4} R - \frac{1}{2} \bar{\psi}_\mu \Gamma^{\mu\nu\rho} \partial_\nu \psi_\rho - \frac{1}{8} \omega_{\nu a b} \bar{\psi}_\mu \Gamma^{\mu\nu\rho} \Gamma^{ab} \psi_\rho \right) ,
\ee
where we noted that the term involving Christoffel symbols in $D_\nu \psi_\rho$ drops out. Varying with respect to $\omega_{\mu a b}$, writing $\omega_{\mu a b}$ as the sum of $\omega_{\mu a b}(e)$ and the contorsion, 
\be
\label{e-2-2-32}
\omega_{\mu ab} = \omega_{\mu a b}(e)+\kappa_{\mu ab} ,
\ee
and recalling that $\omega_{\mu ab}(e)$ does not contribute to the variation of $e R$, we end up with the equation
\be
\label{e-2-2-33}
\kappa_{[ab]}^{\phantom{[ab]} \mu} + \kappa_{[a} e^\mu_{b]} = \frac{1}{4} \bar{\psi}_\nu \Gamma^{\nu\mu\rho} \Gamma_{ab} \psi_\rho, 
\ee
where we set $\kappa_a \equiv \kappa_{ba}^{\phantom{ba} b}$. Contracting with $e^a_\mu$ and using (\ref{e-b-1-11}) and the Majorana property, we find
\be
\label{e-2-2-34}
\kappa_a = \frac{1}{4} \bar{\psi}_\mu \Gamma^{\mu\nu b} \Gamma_{ba} \psi_\nu = - \frac{1}{2} \bar{\psi}_b \Gamma^{\mu} \psi_\mu.
\ee
Then, (\ref{e-2-2-33}) can be solved for the expression
\be
\label{e-2-2-35}
K_{b c a} \equiv \kappa_{c a b} - \kappa_{b a c},
\ee
with the result
\be
\label{e-2-2-36}
K_{b c a} = \eta_{b a} \kappa_c - \eta_{c a} \kappa_b - \frac{1}{2} \bar{\psi}_\mu \Gamma^{\mu \nu}_{\phantom{\mu \nu} a} \Gamma_{b c} \psi_\nu = \bar{\psi}_b \Gamma_a \psi_c .
\ee
Due to the antisymmetry of $\kappa_{\mu ab}$, (\ref{e-2-2-35}) can be inverted to yield the contorsion
\be
\label{e-2-2-37}
\kappa_{\mu ab} = \frac{1}{2} e^c_\mu ( K_{bac} + K_{bca} + K_{cab} ) = \frac{1}{2} ( \bar{\psi}_b \Gamma_a \psi_\mu + \bar{\psi}_b \Gamma_\mu \psi_a + \bar{\psi}_\mu \Gamma_b \psi_a ),
\ee
which results in the following expression for the spin connection
\be
\label{e-2-2-38}
\omega_{\mu ab} = \omega_{\mu ab} (e) + \frac{1}{2} ( \bar{\psi}_b \Gamma_a \psi_\mu + \bar{\psi}_b \Gamma_\mu \psi_a + \bar{\psi}_\mu \Gamma_b \psi_a ).
\ee
Using the contorsion tensor (\ref{e-2-2-37}) we obtain exactly the torsion 2--form (\ref{e-2-2-30}). Hence, we have shown once again that the first- and second-order formalisms are equivalent. We have also seen that supergravity provides an explicit realization of a gravitational theory with torsion. 

What remains is to verify that the action (\ref{e-2-2-29}) is invariant under the supersymmetry transformations in (\ref{e-2-2-9}). For the vielbein and gravitino, these transformations are given by
\be
\label{e-2-2-39}
\delta e^a_\mu = \bar{\epsilon} \Gamma^a \psi_\mu ,\qquad \delta \psi_{\mu} = D_\mu \epsilon.
\ee
However, regarding the supersymmetry variation of the spin connection, we have to choose between first- and second-order formalism. In the first case, we have to compute the variations of $e^a_\mu$ and $\psi_\mu$ and the resulting variation of $\omega_{\mu ab}(e,\psi)$ and substitute in $S[e,\psi,\omega]$, while in the second case we have to compute the variations of $e^a_\mu$ and $\psi_\mu$ and substitute in $S[e,\psi]$. Although both approaches are quite cumbersome, there exists a useful trick that combines the advantages of the two formalisms, appropriately called \emph{1.5 order} formalism. To explain it, we use the chain rule to write the variation of the second-order action in terms of the first-order action as follows
\bea 
\label{e-2-2-40}
\!\!\!\!\!\!\!\!\!\!\!\! \delta S[e,\psi] &=& \frac{\delta S[e,\psi]}{\delta e^a_\mu} \delta e^a_\mu + \frac{\delta S[e,\psi]}{\delta \psi_\mu} \delta \psi_\mu \nn\\ &=& \frac{\delta S[e,\psi,\omega]}{\delta e^a_\mu} \delta e^a_\mu + \frac{\delta S[e,\psi,\omega]}{\delta \psi_\mu} \delta \psi_\mu + \frac{\delta S[e,\psi,\omega]}{\delta \omega_{\mu ab}} \left[  \frac{\delta \omega_{\mu ab}(e,\psi)}{\delta e^c_\nu } \delta e^c_\nu + \frac{\omega_{\mu ab}(e,\psi)}{\delta \psi_\nu} \delta \psi_\nu \right].
\eea
But, at the solution $\omega_{\mu ab}(e,\psi)$, we have $\delta S[e,\psi,\omega]/\delta \omega_{\mu ab} = 0$ and we are left with
\be
\label{e-2-2-41}
\delta S[e,\psi] = \frac{\delta S[e,\psi,\omega]}{\delta e^a_\mu} \delta e^a_\mu + \frac{\delta S[e,\psi,\omega]}{\delta \psi_\mu} \delta \psi_\mu.
\ee 
Therefore, all we have to do is plug the (simple) variations of $e^a_\mu$ and $\psi_\mu$ into the (simple) first-order action and we are done. Using the 1.5 order formalism, it is quite easy to prove invariance of the action under (\ref{e-2-2-41}). We start by rewriting the action in the more convenient form 
\be
\label{e-2-2-42}
S = \frac{1}{\kappa^2} \int \dd^4 x \left( \frac{1}{4} e e^\mu_a e^\nu_b R_{\mu\nu}^{\phantom{\mu\nu} ab} + \frac{\ii}{2} \epsilon^{\mu\nu\rho\sigma} e^a_\nu \bar{\psi}_\mu \Gamma_5 \Gamma_a D_\rho \psi_\sigma \right),
\ee
where all occurences of the vielbein have been explicitly separated and where the Rarita-Schwinger term was rewritten by using gamma-matrix duality. The variation of the Einstein part is
\be
\label{e-2-2-43}
\delta S_2 = - \frac{1}{2 \kappa^2} \int \dd^4 x e \left( R_a^{\phantom{a} \mu} - \frac{1}{2} R e_a^\mu \right) \bar{\epsilon} \Gamma^a \psi_\mu .
\ee
The variation of the Rarita-Schwinger part can be written as $\delta S_{3/2} = \delta S_{3/2}^{(1)} + \delta S_{3/2}^{(2)}$ where
\be
\label{e-2-2-44}
\delta S_{3/2}^{(1)}
= \frac{1}{2 \kappa^2} \int \dd^4 x \ii \epsilon^{\mu\nu\rho\sigma} ( \bar{\psi}_\mu \Gamma_5 \Gamma_\nu D_\rho D_\sigma \epsilon - \bar{\epsilon} \Gamma_5 \Gamma_\nu D_\mu D_\rho \psi_\sigma ),
\ee
and
\be
\label{e-2-2-45}
\delta S_{3/2}^{(2)}
= \frac{1}{2 \kappa^2} \int \dd^4 x \ii \epsilon^{\mu\nu\rho\sigma} ( \bar{\psi}_\mu \Gamma_5 \Gamma_a D_\rho \psi_\sigma \bar{\epsilon} \Gamma^a \psi_\nu - \bar{\epsilon} \Gamma_5 \Gamma_a D_\rho \psi_\sigma D_\mu e^a_\nu ),
\ee
with the second terms in $\delta S_{3/2}^{(1)}$ and $\delta S_{3/2}^{(2)}$ resulting from integrating the $D_\mu \bar{\epsilon}$ term by parts. Starting from $\delta S_{3/2}^{(1)}$, we use the identity (\ref{e-b-1-7}) to write it as 
\be
\label{e-2-2-46}
\delta S_{3/2}^{(1)}
= \frac{1}{16 \kappa^2} \int \dd^4 x \ii \epsilon^{\mu\nu\rho\sigma} R_{\rho\sigma}^{\phantom{\rho\sigma} ab} \bar{\psi}_\mu \Gamma_5 \{ \Gamma_\nu , \Gamma_{ab} \} \epsilon. 
\ee
This can be further simplified using the gamma-matrix relation $\{  \Gamma_\nu , \Gamma_{ab} \} = 2 \ii \epsilon_{abcd} \Gamma_5 \Gamma^c e^d_\nu$ which follows easily from the identities of \S\ref{sec-b-1-2}. After some straightforward algebra, we find
\be
\label{e-2-2-47}
\delta S_{3/2}^{(1)}
= \frac{1}{2 \kappa^2} \int \dd^4 x \left[  e \left( R_a^{\phantom{a} \mu} - \frac{1}{2} R e_a^\mu \right) \bar{\epsilon} \Gamma^a \psi_\mu \right],
\ee
which exactly cancels $\delta S_2$. Turning to $\delta S_{3/2}^{(2)}$, we use the definition of the torsion tensor and Eq. (\ref{e-2-2-30}) to write $2 D_{[\mu} e^a_{\nu]} = T_{\mu\nu}^{\phantom {\mu\nu} a} = - \bar{\psi}_\mu \Gamma^a \psi_\nu$. This way, we find the extra two terms
\be
\label{e-2-2-48}
\delta S_{3/2}^{(2)}
= \frac{1}{2 \kappa^2} \int \dd^4 x \ii \epsilon^{\mu\nu\rho\sigma} \left( \bar{\psi}_\mu \Gamma_5 \Gamma_a D_\rho \psi_\sigma \bar{\epsilon} \Gamma^a \psi_\nu + \frac{1}{2} \bar{\epsilon} \Gamma_5 \Gamma_a D_\rho \psi_\sigma \bar{\psi}_\mu \Gamma^a \psi_\nu \right),
\ee
which cancel each other as a result of a Fierz identity. This completes the proof of supersymmetry and the construction of the theory. 

The main result is that the action
\be
\label{e-2-2-49}
S = \frac{1}{\kappa^2} \int \dd^4 x e \left[  \frac{1}{4} R (\omega(e,\psi)) - \frac{1}{2} \bar{\psi}_\mu \Gamma^{\mu\nu\rho} D_\nu (\omega(e,\psi)) \psi_\rho \right],
\ee
is invariant under the super-Poincar\'e group, including the local supersymmetry transformations 
\be
\label{e-2-2-50}
\delta e^a_\mu = \bar{\epsilon} \Gamma^a \psi_\mu ,\qquad \delta \psi_{\mu} = D_\mu (\omega(e,\psi)) \epsilon.
\ee
In Eqs. (\ref{e-2-2-49}) and (\ref{e-2-2-50}) we made explicit that the spin connection appearing in $R$ and $D_\mu$ is not the usual spin connection $\omega_{\mu ab}(e)$ of Einstein gravity, but rather the spin connection $\omega_{\mu ab}(e,\psi)$ that solves its equations of motion. If one wishes to express the action in terms of $\omega_{\mu ab}(e)$, one has to substitute the relation $\omega_{\mu ab}(e,\psi) = \omega_{\mu ab}(e)+\kappa_{\mu ab}$ in (\ref{e-2-2-49}). This results in a fairly complicated action involving extra four-fermion terms arising from the contorsion tensor.

The construction method developed in this section may be applied to more general cases, where the torsions and the supersymmetry variations are given by more complicated expressions that those encountered here. In the rest of this thesis we will state the actions of such theories neglecting four-fermion terms, which amounts to replacing $R (\omega(e,\psi))$ and $D_\mu (\omega(e,\psi))$ by $R(\omega(e))$ and $D_\mu(\omega(e))$ respectively. The only exception will be eleven-dimensional supergravity, whose action and transformation rules can be written in a compact form similar to (\ref{e-2-2-49}) and (\ref{e-2-2-50}).

\section{Scalar Coset Manifolds}
\label{sec-2-3}

A consequence of supersymmetry, in both rigid and local cases, is that certain types of scalar fields appearing in supersymmetry multiplets are required to parameterize certain coset manifolds of the form $G/H$, where the allowed forms of $G$ and $H$ depend on the particular supersymmetry algebra. Accordingly, the action describing these scalar fields has the form of a nonlinear sigma model over the coset space. Moreover, the global symmetries of the scalar manifold, a part of which corresponds to global symmetries of the supersymmetry algebra, may be promoted to local symmetries yielding gauged nonlinear sigma models. Since the structure of the scalar manifolds plays an essential role in the supergravities to be considered, we give a brief account of coset spaces and their main properties. 

\subsection{Coset spaces and their geometry}
\label{sec-2-3-1}

Given a group $G$ and a subgroup $H \subset G$, the right coset $G/H$ is defined as the set of equivalence classes of elements of $G$ under the right action of $H$, i.e. the set of points $g \in G$ modulo the identification $g \cong gh$ for all $h \in H$. To describe the symmetries of the coset space, we split the generators of $G$ into the generators $H^i$ of $H$ and the generators $K^a$ of $G/H$. Since $H$ is a subgroup, the commutator of two $H^i$'s will not involve $K^a$. By making the further assumption that that the algebra is \emph{reductive}, i.e. that $K^a$ form a representation of $H$, we can decompose the algebra of $G$ as follows
\bea
\label{e-2-3-1}
\left[  H_i , H_j  \right] &=& f_{i j}^{\phantom{i j} k} H_k, \nn\\
\left[  H_i , K_a  \right] &=& f_{i a}^{\phantom{i a} b} K_b, \nn\\
\left[  K_a , K_b  \right] &=& f_{a b}^{\phantom{a b} i} H_i + f_{a b}^{\phantom{a b} c} K_c. 
\eea 

Considering an $n$--dimensional coset manifold, we introduce coordinates $\varphi^\alpha$, $\alpha=1,\ldots,n$, where $n$ is the dimension of $G/H$, and the associated derivatives $\partial_\alpha \equiv \partial / \partial \varphi^\alpha$. From the $\varphi^\alpha$'s, we can form a $G$--valued matrix $L$, called \emph{coset representative}, and impose rigid symmetry under left multiplications with elements of $g$
\be 
\label{e-2-3-2}
L \to g^{-1} L ;\qquad g \in G,
\ee 
and local symmetry under right multiplications with elements of $H$
\be 
\label{e-2-3-3}
L \to L h (\varphi) ;\qquad h (\varphi) \in H.
\ee 
In the above construction, the choice of $L$ is not unique; one can choose many representatives for the coset space. To isolate the physical degrees of freedom, we can ``fix the gauge'' by selecting $L$ as a particular function of $\varphi$, $L=L(\varphi)$. Since the transformation (\ref{e-2-3-2}) does not preserve this gauge choice, it must be accompanied by a compensating transformation of the type (\ref{e-2-3-3}), so that the complete transformation law under $G$ reads
\be 
\label{e-2-3-4}
L(\varphi) \to L(\varphi') = g^{-1} L(\varphi) h (\varphi;g),
\ee 
where $h(\varphi;g)$ is an element of $H$ selected so that the transformation retains the functional form of $L$.

On the coset space under consideration, we can define certain geometrical quantities. From a given coset representative $L$, one can construct the $G$--valued left-invariant Maurer-Cartan form $L^{-1} \dd L$ (here $\dd=dx^\alpha \partial_\alpha$), whose elements can be expressed in terms of $H_i$ and $K_a$ as follows
\be 
\label{e-2-3-5}
L^{-1} \partial_\alpha L = \MA_\alpha + \MV_\alpha = \MA_\alpha^{\phantom{\alpha} i} H_i + \MV_\alpha^a K_a.
\ee 
From this, it is clear that $\MA_\alpha$ parameterizes the tangent-space rotations of the representative with respect to $H$ and are identified with the connection of $H$, while $V_\alpha$ defines an orthonormal frame on the coset space and is identified with the coset vielbein. This can be explicitly verified by noting that, under a local transformation (\ref{e-2-3-4}) of $L$, the Maurer-Cartan form (\ref{e-2-3-5}) transforms as
\be
\label{e-2-3-6} 
L^{-1} \partial_\alpha L \to h^{-1} ( \MA_\alpha + \partial_\alpha ) h + h^{-1} \MV_\alpha h,
\ee 
so that, under $H$, $\MA$ does indeed transform like a connection and $\MV$ like a vielbein. Now, taking the exterior derivative of the Maurer-Cartan form, we find
\be 
\label{e-2-3-7} 
\partial_{[\alpha} \left( L^{-1} \partial_{\beta]} L \right) = - \left( L^{-1} \partial_{[\alpha} L \right) \left( L^{-1} \partial_{\beta]} L \right) 
\ee 
Using (\ref{e-2-3-5}), we arrive at the relation
\be 
\label{e-2-3-8} 
\partial_\alpha \MA_\beta - \partial_\beta \MA_\alpha + \left[  \MA_\alpha , \MA_\beta  \right] + \partial_\alpha \MV_\beta - \partial_\beta \MV_\alpha + \left[  \MA_\alpha , \MV_\beta  \right] - \left[  \MA_\beta , \MV_\alpha  \right] + \left[  \MV_\alpha , \MV_\beta  \right] = 0.
\ee 
In most cases of interest, we have two simplifications. The first one is the condition 
\be 
\label{e-2-3-8b} 
f_{a b}^{\phantom{a b} c} = 0,
\ee
which defines a \emph{symmetric} coset space and implies that the commutator $[  V_\alpha , V_\beta  ]$ is an element of $H$. Then, Eq. (\ref{e-2-3-8}) can be split into the \emph{Maurer-Cartan structure equations}
\be 
\label{e-2-3-9} 
\MT_{\alpha\beta} (G/H) \equiv \partial_\alpha \MV_\beta - \partial_\beta \MV_\alpha + \left[  \MA_\alpha , \MV_\beta  \right] - \left[  \MA_\beta , \MV_\alpha  \right] = 0,
\ee 
and
\be 
\label{e-2-3-10} 
\MF_{\alpha\beta} (H) \equiv \partial_\alpha \MA_\beta - \partial_\beta \MA_\alpha + \left[  \MA_\alpha , \MA_\beta \right] = - \left[  \MV_\alpha , \MV_\beta  \right].
\ee 
The second simplification is the antisymmetry property
\be 
\label{e-2-3-11} 
f_{i a}^{\phantom{i a} b} = - f_{i b}^{\phantom{i b} a},
\ee
which can be shown to hold for any reductive space and implies that $H$ can be embedded in some $\mathrm{SO}(n-p,p)$ group. Then, using the $\mathrm{SO}(n-p,p)$--invariant tensor $\eta_{ab}$ and the vielbein, we can construct an invariant metric on $G/H$ according to
\be 
\label{e-2-3-12} 
g_{\alpha\beta} = \eta_{ab} \MV^a_\alpha \MV^b_\beta.
\ee 
Also, the connection $\MA_\alpha^{\phantom{\alpha} ab}$ resulting from the embedding of $H$ in $\mathrm{SO}(n-p,p)$ is a spin connection in the usual sense. Accordingly, the structure equations (\ref{e-2-3-9}) and (\ref{e-2-3-10}) turn into equations for the torsion of $G/H$ and the curvature of $H$ and read
\be 
\label{e-2-3-13} 
\MT_{\alpha\beta}^{\phantom{\alpha\beta} a} (G/H) \equiv \partial_\alpha \MV^a_\beta - \partial_\beta \MV^a_\alpha + \MA_{\alpha \phantom{a} b}^{\phantom{\alpha} a} \MV^b_\beta - \MA_{\beta \phantom{a} b}^{\phantom{\beta} a} \MV^b_\alpha = 0,
\ee 
and 
\be 
\label{e-2-3-14} 
\MF_{\alpha\beta}^{\phantom{\alpha\beta} ab} (H) \equiv \partial_\alpha \MA_\beta^{\phantom{\beta} ab} - \partial_\beta \MA_\alpha^{\phantom{\alpha} ab} + \MA_{\alpha \phantom{a} c}^{\phantom{\alpha} a} \MA_\beta^{\phantom{\beta} cb} - \MA_{\beta \phantom{a} c}^{\phantom{\beta} a} \MA_\alpha^{\phantom{\alpha} cb} = \MV^a_\alpha \MV^b_\beta - \MV^a_\beta \MV^b_\alpha.
\ee  

\subsection{Nonlinear sigma models and gauging}
\label{sec-2-3-2}

Based on the above preliminaries, we are in a position to describe the nonlinear sigma models that occur in supergravity theories. To this end, all we have to do is to identify the coordinates $\varphi^\alpha$ with scalar fields $\varphi^\alpha(x)$ defining a map from the spacetime manifold to the coset manifold. Accordingly, the $\varphi$--dependent transformations parameterized by $h(\varphi)$ are associated with $x$--dependent transformations parameterized by $h(x)$. Considering a coset representative $L$ we now write
\be 
\label{e-2-3-15} 
L^{-1} \partial_\mu L = \MQ_\mu + \MP_\mu. 
\ee 
where $\MQ_\mu$ and $\MP_\mu$ are called composite connection and composite vielbein respectively, and are defined as the pullbacks of $\MA_\alpha$ and $\MV_\alpha$ from the scalar to the spacetime manifold,
\be 
\label{e-2-3-16} 
\MQ_\mu = \MA_\alpha \partial_\mu \varphi^\alpha \qquad,\qquad \MP_\mu = \MV_\alpha \partial_\mu \varphi^\alpha.
\ee 
The transformation law of the Maurer-Cartan form (\ref{e-2-3-15}) is then given by
\be 
\label{e-2-3-17} 
L^{-1} \partial_\mu L \to h^{-1} ( \MQ_\mu + \partial_\mu ) h + h^{-1} \MP_\mu h.
\ee  
The Maurer-Cartan structure equations for this case can be derived as before and are given by
\be 
\label{e-2-3-18} 
\MT_{\mu\nu} (G/H) \equiv \partial_\mu \MP_\nu - \partial_\nu \MP_\mu + \left[ \MQ_\mu , \MP_\nu  \right] - \left[  \MQ_\nu , \MP_\mu  \right] = 0,
\ee 
and
\be 
\label{e-2-3-19} 
\MF_{\mu\nu} (H) \equiv \partial_\mu \MQ_\nu - \partial_\nu \MQ_\mu + \left[  \MQ_\mu , \MQ_\nu \right] = - \left[  \MP_\mu , \MP_\nu  \right].
\ee 
Now, since the coset space coordinates are now spacetime fields, one needs to write down a kinetic Lagrangian. This Lagrangian must be invariant under the usual spacetime symmetries as well as under rigid $G$ transformations and local $H$ transformations. As in gravity theories, the Lagrangian can be written in first- or second-order form. The appropriate second-order Lagrangian can be written as
\be 
\label{e-2-3-20} 
e^{-1} \mathcal{L} = - \frac{1}{2} \tr ( \MP_\mu \MP^\mu ) = - \frac{1}{2} g_{\alpha\beta}(\varphi) \partial_\mu \varphi^\alpha \partial^\mu \varphi^\beta,
\ee  
where the first form has manifest local $H$ invariance, while the second form is obtained after gauge fixing and is no longer $H$--invariant.

In the above, the group $G$ i.e. the isometry group of $G/H$ corresponds to a global symmetry of the theory. It is possible to gauge this symmetry by promoting the parameters of isometry group or a subgroup thereof to functions of the spacetime coordinates, introducing the corresponding gauge fields and coupling them in an appropriate manner to all fields that transform non-trivially under $G$. This construction is of particular importance in gauged supergravities and we illustrate it below. We start by considering an isometry transformation of the $\varphi^\alpha$, parameterized as
\be
\label{e-2-3-20a} 
\delta \varphi^\alpha = \Lambda \varphi^\alpha,
\ee
To promote this isometry to a local symmetry, we introduce dynamical gauge fields $A_\mu$ transforming as
\be
\label{e-2-3-21} 
\delta A_\mu = D_\mu \Lambda,
\ee
and we replace the ordinary derivative by the covariant derivative
\be
\label{e-2-3-22} 
\MD_\mu \varphi^\alpha = \partial_\mu \varphi^\alpha - A_\mu^I \xi^\alpha_I ,
\ee
where $g$ is the gauge coupling. Appropriately, the Maurer-Cartan form (\ref{e-2-3-15}) is replaced by its covariant version
\be
\label{e-2-3-23} 
L^{-1} \MD_\mu L = \MQ_\mu + \MP_\mu,
\ee
with
\be  
\label{e-2-3-24} 
\MQ_\mu = \MA_\alpha \MD_\mu \varphi^\alpha \qquad,\qquad \MP_\mu = \MV_\alpha \mathcal{D}_\mu \varphi^\alpha. 
\ee
The Maurer-Cartan structure equations become
\be 
\label{e-2-3-25} 
\MT_{\alpha\beta} (G/H) \equiv \partial_\mu \MP_\nu - \partial_\nu \MP_\mu + \left[ \MQ_\mu , \MP_\nu  \right] - \left[  \MQ_\nu , \MP_\mu  \right] = - \frac{1}{2} \left( L^{-1} F_{\mu\nu} L \right)_{G/H},
\ee 
and
\be 
\label{e-2-3-26} 
\MF_{\alpha\beta} (H) \equiv \partial_\mu \MQ_\beta - \partial_\nu \MQ_\mu + \left[  \MQ_\mu , \MQ_\nu \right] = - \left[  \MP_\mu , \MP_\nu  \right] - g \left( L^{-1} F_{\mu\nu} L \right)_H.
\ee 
where $F_{\mu\nu}$ is the gauge field strength. Finally, the scalar Lagrangian is
\be 
\label{e-2-3-27} 
e^{-1} \mathcal{L} = - \frac{1}{2} \tr ( \MP_\mu \MP^\mu ) = - \frac{1}{2} g_{\alpha\beta}(\varphi) \MD_\mu \varphi^\alpha \MD^\mu \varphi^\beta,
\ee   
A very detailed example of the above construction will be given in \S\ref{sec-5-1-3} for the case of the nonlinear sigma model of the hyperscalar manifold of $D=6$, $N=2$ supergravity.

\section{Supergravities in $D=11$ and in $D=10$}
\label{sec-2-4}

In the concluding section of this chapter, we will review the minimal supergravity theories in eleven and ten dimensions.

\subsection{Eleven-dimensional supergravity}
\label{sec-2-4-1}

The first case we will consider is $D=11$, which is the highest dimension in which one can construct a supergravity theory. In this dimension there exists a unique supergravity theory with $N=1$ supersymmetry, known to be related to all superstring theories and conjectured to be one of the low-energy limits of the hypothetical M-theory. 

The unique $N=1$ supersymmetry algebra in eleven dimensions admits a single representation given by the following multiplet
\be
\label{e-2-4-1}
\text{Gravity multiplet} \quad:\quad ( g_{\mu\nu}, A_{\mu\nu\rho}, \psi_\mu). 
\ee
The complete action describing the interactions of this multiplet was found by Cremmer, Julia and Scherk \cite{Cremmer:1978km} in 1977. It is given by
\bea
\label{e-2-4-2}
e^{-1} \mathcal{L} &=& \frac{1}{4} R (\omega) - \frac{1}{48} F_{\mu\nu\rho\sigma} F^{\mu\nu\rho\sigma} - \frac{1}{2} \bar{\psi}_\mu \Gamma^{\mu\nu\rho} D_\nu \left( \frac{\omega + \widehat{\omega}}{2} \right) \psi_\rho 
\nn\\ 
&&- \frac{1}{96 \sqrt{2} } \left( \bar{\psi}_\lambda \Gamma^{\lambda\mu\nu\rho\sigma\tau} \psi_\tau + 12 \bar{\psi}^\mu \Gamma^{\nu\rho} \psi^\sigma \right) \left( F_{\mu\nu\rho\sigma} + \widehat{F}_{\mu\nu\rho\sigma} \right) 
\nn\\ 
&&- \frac{1}{1728 \sqrt{2}} e^{-1} \epsilon^{\mu_1 \ldots \mu_{11}} A_{\mu_1 \mu_2 \mu_3} F_{\mu_4 \mu_5 \mu_6 \mu_7} F_{\mu_8 \mu_9 \mu_{10} \mu_{11}}	.
\eea  
and is invariant under the local supersymmetry transformations
\bea
\label{e-2-4-3}
\delta e^a_\mu &=& \bar{\epsilon} \Gamma^a \psi_\mu, 
\nn\\
\delta A_{\mu\nu\rho} &=& - \frac{1}{4 \sqrt{2}} \bar{\epsilon} \Gamma_{[\mu\nu} \psi_{\rho]},
\\
\delta \psi_\mu &=& D_\mu (\widehat{\omega}) \epsilon + \frac{1}{144 \sqrt{2}} \left( \Gamma_\mu^{\phantom{\mu} \nu\rho\sigma\tau} - 8 \delta_\mu^\nu \Gamma^{\rho\sigma\tau} \right) \widehat{F}_{\nu\rho\sigma\tau} \epsilon.\nn
\eea
Here, $\omega$ is the spin connection in the first-order formalism, which is written as
\be
\label{e-2-4-4}
\omega_{\mu a b} = \omega_{\mu a b}(e) + \frac{1}{2} \left( \bar{\psi}_\mu \Gamma_b \psi_a + \bar{\psi}_b \Gamma_a \psi_\mu + \bar{\psi}_b \Gamma_\mu \psi_a \right) - \frac{1}{4} \bar{\psi}_\nu \Gamma_{\mu a b}^{\phantom{\mu a b} \nu \rho} \psi_\rho.
\ee 
As for $\widehat{\omega}$, it is the supercovariant spin connection, given by
\be
\label{e-2-4-5}
\widehat{\omega}_{\mu a b} = \omega_{\mu a b} + \frac{1}{4} \bar{\psi}_\nu \Gamma_{\mu a b}^{\phantom{\mu a b} \nu \rho} \psi_\rho.
\ee
Also, $F_4$ and $\widehat{F}_4$ stand for the 4-form field strength and its supercovariant generalization respectively,
\be
\label{e-2-4-6}
F_4 = \dd A_3 ,\qquad\qquad \widehat{F}_{\mu\nu\rho\sigma} = F_{\mu\nu\rho\sigma} + 3 \bar{\psi}_{[\mu} \Gamma_{\nu\rho} \psi_{\sigma]}.
\ee
Also, apart from the usual continuous symmetries, this theory has a global discrete symmetry under parity/time reversal, accompanied by the sign reversal
\be
\label{e-2-4-7}
A'_{\mu\nu\rho} = - A_{\mu\nu\rho}.
\ee 
for all $\mu,\nu,\rho$ lying in the transverse directions.

The theory described above is unique in many senses. First, as already mentioned, $D=11$ is the highest dimension for supergravity and admits a unique $N=1$ multiplet. Second, it is impossible to obtain a theory described in terms of a dual 6--form potential: a duality transformation is impossible due to the presence of the Chern-Simons term and attempts to directly construct the dual theory have failed due to uncancelled supersymmetry variations \cite{Nicolai:1981kb}. Third, the theory does not admit the usual extension via the addition of a cosmological term and a gravitino mass term \cite{Bautier:1997yp}. Fourth, it does not admit any gauging \cite{Castellani:1983kd} since the only global symmetry of the theory is a discrete one. A thorough review of 11D supergravity and its solutions is given in \cite{Duff:1986hr}.

Moreover, although $D=11$ supergravity is not chiral and cannot give rise to a chiral theory upon ordinary compactification, it can yield a chiral $D=10$ theory if one defines it on the $\MBFS^1 / \MBBZ_2$ orbifold. In such a case, supersymmetry requires chiral boundary conditions for the fermions and yields chiral fermions on the orbifold fixed points. Anomaly considerations require that the theories on the fixed points be coupled to $D=10$ vector multiplets with gauge group $E_8$. This is the heterotic M-theory of Ho\v rava and Witten \cite{Horava:1996qa,Horava:1996ma}.

\subsection{Ten-dimensional $N=1$ supergravity}
\label{sec-2-4-2}

The next case to be considered is $D=10$, which contains all supergravities arising as low-energy limits of critical superstring theories. In this dimension, there exist two theories with $N=2$ supersymmetry (Type IIA and Type IIB supergravity) and one theory with $N=1$ supersymmetry. The latter theory can be coupled to Yang-Mills fields and arises as a low-energy limit of heterotic and Type I string theory.

The minimal supersymmetry algebra in ten dimensions is the chiral $N=1$ algebra, which has the following representations 
\bea
\label{e-2-4-8}
\text{Gravity multiplet} \quad&:&\quad ( g_{\mu\nu} , B_{\mu\nu} , \phi , \psi^+_\mu , \chi^-),\nn\\
\text{Vector multiplet} \quad&:&\quad ( A_\mu, \lambda^+ ).
\eea
The $D=10$, $N=1$ supergravity coupled to super Yang-Mills theory is constructed by combining the gravity multiplet with $n$ vector multiplets transforming in the adjoint representation of a gauge group. The theory was first constructed in its 6--form version \cite{Chamseddine:1981ez} and later in the 2--form version \cite{Bergshoeff:1982um,Chapline:1983ww}. For the latter case, the Lagrangian is given by
\bea
\label{e-2-4-9}
e^{-1} \mathcal{L} &=& \frac{1}{4} R - \frac{1}{12} e^{-2 \phi} G_{\mu\nu\rho} G^{\mu\nu\rho} - \frac{1}{4} e^{-\phi} F^I_{\mu\nu} F_I^{\mu\nu} - \frac{1}{4} \partial_\mu \phi \partial^\mu \phi 
\nn\\ 
&&- \frac{1}{2} \bar{\psi}_\mu \Gamma^{\mu\nu\rho} D_\nu \psi_\rho - \frac{1}{2} \bar{\chi} \Gamma^\mu D_\mu \chi + \frac{1}{2} \tr (\bar{\lambda} \Gamma^\mu D_\mu \lambda )
\nn\\
&&- \frac{1}{24} e^{-\phi} \left(  \bar{\psi}_\lambda \Gamma^{\lambda\mu\nu\rho\sigma} \psi_\sigma + 6 \bar{\psi}^\mu \Gamma^\nu \psi^\rho - \sqrt{2} \bar{\psi}_\lambda \Gamma^{\mu\nu\rho} \Gamma^\lambda \chi + \bar{\lambda}^I \Gamma^{\mu\nu\rho} \lambda_I \right) G_{\mu\nu\rho} 
\nn\\  
&&+ \frac{1}{2 \sqrt{2}} e^{-\phi/2 } \tr \left[  \left( \bar{\lambda} \Gamma^\lambda \Gamma^{\mu\nu} \psi_\lambda + \frac{1}{\sqrt{2}} \bar{\lambda} \Gamma^{\mu\nu} \chi \right) F_{\mu\nu} \right] \nn\\ 
&&- \frac{1}{2} \bar{\psi}_\mu \Gamma^\nu \Gamma^\mu \chi \partial_\nu \phi + (\mathrm{Fermi})^4,
\eea
and is invariant under the supersymmetry transformations
\bea
\label{e-2-4-10}
\delta e^a_\mu &=& \bar{\epsilon} \Gamma^a \psi_\mu, 
\nn\\
\delta \phi &=& - \frac{1}{2} \bar{\epsilon} \chi, 
\nn\\
\delta B_{\mu\nu} &=& - e^{\phi} \left( \bar{\epsilon} \Gamma_{[\mu} \psi_{\nu]} - \frac{1}{2\sqrt{2}} \bar{\epsilon} \Gamma_{\mu\nu} \chi \right) + A^I_{[\mu} \delta A_{I \nu]}
\nn\\
\delta A_\mu &=& - \frac{1}{\sqrt{2}} e^{\phi/2} \bar{\epsilon} \Gamma_\mu \lambda,
\nn\\
\delta \psi_\mu &=& D_\mu \epsilon + \frac{1}{24} e^{-\phi} \Gamma^{\nu\rho\sigma} \Gamma_\mu G_{\nu\rho\sigma} \epsilon,
\nn\\
\delta \chi &=& \frac{1}{2} \Gamma^\mu \partial_\mu \phi \epsilon - \frac{1}{12} e^{-\phi} \Gamma^{\mu\nu\rho} G_{\mu\nu\rho} \epsilon,
\nn\\
\delta \lambda &=& \frac{1}{2 \sqrt{2}} e^{-\phi / 2} \Gamma^{\mu\nu} F_{\mu\nu} \epsilon. 
\eea
In the above we have ignored $(\mathrm{Fermi})^4$ terms in the Lagrangian and $(\mathrm{Fermi})^3$ terms in the supersymmetry transformation laws. Also, $G_3$ stands for the modified field strength
\be
\label{e-2-4-11}
G_3 = \dd B_2 - \omega_{Y,\mu\nu\rho}; \qquad \omega_{3Y} = \tr \left[  A \left( \dd A + \frac{2}{3} A^2 \right) \right],
\ee
involving the Yang-Mills field. Invariance of this field strength under Yang-Mills gauge transformations requires that $B_{\mu\nu}$ transform as
\be
\label{e-2-4-12}
\delta B_2 = \omega^1_{2Y}; \qquad \omega^1_{2Y} = \tr ( v \dd A ),
\ee
where $v$ is the Yang-Mills gauge transformation parameter.

The $D=10$, $N=1$ supergravity just described arises as a low-energy limit of Type I and heterotic string theories. In the first case the allowed gauge group can be $\mathrm{SO}(32)$ while in the second case it can be either $\mathrm{SO}(32)$ or $E_8 \times E_8$. These restrictions of the gauge group were first discovered through anomaly considerations. We will have a lot to say about that in Chapter 4.
\chapter{Anomalies}
\label{chap-3} 

Since their discovery, anomalies have played a very important role in field theory, both in Standard Model physics and in unified theories beyond the Standard Model. Their importance is even more profound in the context of higher-dimensional theories involving gravity, namely string theories, their effective supergravities and related models. In such cases, the requirement of anomaly cancellation seriously constrains the possible particle content of these theories and proves to be a crucial guiding principle for the construction of consistent models. Since anomalies are a central part of this thesis, in this chapter we aim to present a fairly detailed and self-contained review of this subject.

\section{Introduction to Anomalies}
\label{sec-3-1}

It is a well-known fact of classical field theory that a symmetry of the action of a field system under a continuous transformation gives rise to a conserved current. However, in certain cases, a current that is conserved in the classical theory fails to be so in the quantum theory. This phenomenon is called \emph{anomaly}.

Although anomalies arise in various occasions, we will here consider two specific types which, in standard terminology, are called \emph{singlet anomalies} and \emph{nonabelian anomalies}. The first type corresponds to global symmetries whose associated currents are not coupled to external fields; in our case, the symmetry to be considered is the chiral symmetry of massless fermions. The second type corresponds to local symmetries whose current couples to gauge and/or gravitational fields; in our case the symmetries to be considered are the gauge and GCT/LLT symmetries of chiral fermions and the associated anomalies are called \emph{gauge, gravitational and mixed anomalies}. 

The story of anomalies dates back to 1949 when Fukuda and Miyamoto \cite{Fukuda:1949} and Steinberger \cite{Steinberger:1949wx} calculated the triangle diagram for the electromagnetic decay $\pi^0 \to 2 \gamma$, essentially to find that gauge invariance of the amplitude is incompatible with the naive conservation law of the axial current. In 1951, Schwinger \cite{Schwinger:1951nm} also noted that regulating the axial current of QED in a gauge-invariant manner yields an effective Lagrangian giving a nonzero $\pi^0 \to 2 \gamma$ decay rate. The significance of these results was not however recognized at the time, the usual viewpoint being that they reflect ambiguities of perturbation theory. The $\pi^0 \to 2 \gamma$ decay problem was revisited during the 1960's in the context of PCAC where Sutherland \cite{Sutherland:1967vf} and Veltman proved that this decay cannot occur, in sharp disagreement with experiment; this paradox motivated further study of the problem. The breakthrough came in 1969 when Adler \cite{Adler:1969gk} and Bell-Jackiw \cite{Bell:1969ts} independently found that imposing gauge invariance on the vector vertices inevitably gives an anomaly in the conservation law of the axial current of QED, which results in the correct value for the $\pi^0 \to 2 \gamma$ decay rate. Subsequently, Bardeen \cite{Bardeen:1969md} extended this result to nonabelian gauge anomalies in theories with chiral fermions, while Adler and Bardeen \cite{Adler:1969er} showed that the anomaly is a strictly one-loop effect receiving no higher-order perturbative corrections. During the seventies it became clear that anomalies have a topological meaning and, in particular, singlet anomalies are related \cite{Jackiw:1977pu,Nielsen:1977aw} to the Atiyah-Singer index theorems \cite{Atiyah:1968mp,Atiyah:1968rj,Atiyah:1968ih,Atiyah:1971ws,Atiyah:1971rm}. Another key result was due to Fujikawa \cite{Fujikawa:1979ay,Fujikawa:1980eg} who related the anomaly with the variation of the fermion integration measure in the functional integral. However, it was during the eighties that anomalies was studied in a really systematic and unified manner. In particular, all types of anomalies were given a topological interpretation \cite{Alvarez:1984yi} in terms of index theorems, while Stora \cite{Stora:1983ct} and Zumino \cite{Zumino:1983rz,Zumino:1983ew} developed the descent formalism relating nonabelian anomalies in $2n$ dimensions to chiral anomalies in $2n+2$ dimensions; similar relations were obtained by Alvarez-Gaum\'e and Ginsparg \cite{Alvarez-Gaume:1984cs}. Further developments include the relations of anomalies to supersymmetric quantum mechanics \cite{Alvarez-Gaume:1983at,Friedan:1983xr} and the equivalence of Lorentz and Einstein gravitational anomalies \cite{Bardeen:1984pm}. The above culminated in a powerful formalism by means of which one can calculate all types of anomalies in any dimension and express the anomaly structure of a theory in a particularly elegant way. On the physical side, the implications of these results were striking, particularly in the context of 10D superstring theory. The most famous results are the calculation of gravitational anomalies in all dimensions by Alvarez-Gaum\'e and Witten \cite{Alvarez-Gaume:1984ig} which revealed that Type IIB supergravity is anomaly-free and the discovery of Green and Schwarz \cite{Green:1984sg} who found a mechanism for anomaly cancellation in $N=1$ supergravity. Since then, the study of anomalies continues to provide valuable insights in model building and still remains an active field of research.

In the rest of this section, we will give a very general description of singlet and nonabelian anomalies and we will discuss their most important physical aspects.

\subsection{Singlet anomalies}
\label{sec-3-1-1}

The first case we will consider are singlet anomalies, i.e. anomalies associated with global symmetry currents. The best-known such case is the chiral anomaly associated with spin $1/2$ fermions. To examine it, we consider a theory of massless \emph{Dirac} fermions in $2n$--dimensional Euclidean\footnote{In Sections 3.1--3.3, we will follow the standard convention in the literature and work in the Euclidean. From Section 3.4 on, we will switch back to Minkowski spacetime by analytic continuation.} spacetime, transforming in a representation $\MR$ of a gauge group $\MG$ and coupled to gravity. The classical action of the theory is
\be
\label{e-3-1-1}
S = \int \dd^{2n}  x \sqrt{g} \bar{\psi} \ii \fslash D \psi.
\ee 
This action is classically invariant under the chiral transformation
\be
\label{e-3-1-2}
\psi' = ( 1 + \ii \alpha \Gamma_{2n+1} ) \psi ,\qquad \bar{\psi}' = \bar{\psi} (1 + \ii \alpha \Gamma_{2n+1}).
\ee
To find the associated conserved current, we may employ the Noether procedure. We first take $\alpha$ to be position-dependent so that the action is no longer invariant but is modified by terms that contain the derivatives of $\alpha$. So, at first order, the variation of the action is
\be
\label{e-3-1-3}
\delta S = - \int \dd^{2n}x \sqrt{g} J^\mu_{2n+1} \partial_\mu \alpha = \int \dd^{2n} x \sqrt{g} \alpha D_\mu J^\mu_{2n+1},
\ee
where $J^\mu_{2n+1}$ is the axial current
\be
\label{e-3-1-4}
J^\mu_{2n+1} = \bar{\psi} \Gamma^\mu \Gamma_{2n+1} \psi.
\ee
Taking $\alpha$ back to a constant value, invariance of the action leads to the conservation law
\be
\label{e-3-1-5}
D_\mu J^\mu = 0.
\ee

In the quantum case, the object of interest is not the classical action, but the \emph{effective action}
\be
\label{e-3-1-6}
\varGamma = - \ln Z = - \ln \int \mathcal{D} \psi \mathcal{D} \bar{\psi} \exp \left( - S[\psi,\bar{\psi}] \right)
\ee
The effective action thus defined is invariant under any transformation of $(\psi,\bar{\psi})$ since the latter are just integration variables. However, this invariance does no longer guarantee that a classical current will be conserved in the quantum theory (in the expectation-value sense). The reason is that the variation of $\varGamma$ does not only receive the contribution from the variation of $S$, but may also receive an extra contribution from the {\it variation of the integration measure} $\mathcal{D} \psi \mathcal{D} \bar{\psi}$. Taking this into account, we write the transformation of the partition function $Z$ under (\ref{e-3-1-2}) as
\bea
\label{e-3-1-7}
\delta Z &=& \int \mathcal{D} \psi' \mathcal{D} \bar{\psi}' e^{-S[\psi',\bar{\psi}']} - \int \mathcal{D} \psi \mathcal{D} \bar{\psi} e^{-S[\psi,\bar{\psi}]}
= \int \mathcal{D} \psi \mathcal{D} \bar{\psi} ( J e^{-\delta S} - 1) e^{-S[\psi,\bar{\psi}]}
\eea
where $J$ is the Jacobian of the integration measure. To first order in $\alpha$, we can take 
\be
\label{e-3-1-8}
J = 1 - \ii \int \dd^{2n} x \sqrt{g} \alpha G ,\qquad e^{-\delta S} = 1 - \int \dd^{2n} x \sqrt{g} \alpha D_\mu J^\mu_{2n+1}.
\ee
Then, the variation of the effective action is given by   
\be
\label{e-3-1-9}
\delta \varGamma = - Z^{-1} \delta Z = \int \dd^{2n} x \alpha \left( D_\mu \langle J^\mu_{2n+1} \rangle  + \ii G \right) ,
\ee
and invariance of $\varGamma$ for constant $\alpha$ requires that
\be
\label{e-3-1-10}
D_\mu \langle J^\mu_{2n+1} \rangle = - \ii G.
\ee
which expresses a violation of the Ward identity of $J^\mu_{2n+1}$. The quantity $G$ is called the \emph{singlet anomaly}. The integrated expression
\be
\label{e-3-1-11}
G (\alpha) \equiv \alpha \int \dd^{2n}x \sqrt{g} G = \ii \alpha \int \dd^{2n}x \sqrt{g}  D_\mu \langle J^\mu_{2n+1} \rangle
\ee
will also be referred to as the singlet anomaly.

\subsubsection{Fujikawa's method}

Let us now calculate the anomaly according to the above guidelines. To find the variation of the integration measure, we will use \emph{Fujikawa's method} \cite{Fujikawa:1979ay,Fujikawa:1980eg}. According to that method, we introduce a complete set $\{ \phi_n(x) \}$ of (commuting) eigenvectors of $ \ii \fslash D $ satisfying the orthonormality condition
\be
\label{e-3-1-12}
\langle \phi_n | \phi_m \rangle \equiv \int \dd^{2n}x \sqrt{g} \phi_n^\dag(x) \phi_m(x) = \delta_{nm}.
\ee
and we expand $\psi(x)$ and $\bar{\psi}(x)$ as
\be
\label{e-3-1-13}
\psi(x) = \sum_n a_n \phi_n(x) ,\qquad \bar{\psi}(x) = \sum_n \bar{a}_n \phi_n^\dag(x),
\ee
where $a_n$ and $\bar{a}_n$ are constant Grassmann-valued coefficients, given by
\be
\label{e-3-1-14}
a_n = \langle \phi_n | \psi \rangle ,\qquad \bar{a}_n = \langle \bar{\psi} | \phi_n \rangle .
\ee
The functional integration measure is then defined by
\be
\label{e-3-1-15}
\mathcal{D} \psi \mathcal{D} \bar{\psi} = \prod_n \dd a_n \dd \bar{a}_n. 
\ee
Let us determine how the measure changes under the transformation (\ref{e-3-1-2}). The new coefficients $a'_n$ and $\bar{a}'_n$ of $\psi'$ and $\bar{\psi}'$ are given by
\bea
\label{e-3-1-16}
a'_n &=& \langle \phi_n | \psi' \rangle = \sum_m \langle \phi_n | 1 + \ii \alpha \Gamma_{2n+1} | \phi_m \rangle a_m \equiv \sum_m C_{nm} a_m, \nn\\ 
\bar{a}'_n &=& \langle \bar{\psi}' | \phi_n \rangle = \sum_m \bar{a}_m \langle \phi_m | 1 + \ii \alpha \Gamma_{2n+1} | \phi_n \rangle \equiv \sum_m \bar{a}_m C_{mn},
\eea
and so we arrive at the transformation law
\be
\label{e-3-1-17}
\mathcal{D} \psi' \mathcal{D} \bar{\psi}' = \prod_n \dd a'_n \dd \bar{a}'_n = (\det C)^{-2} \prod_n \dd a_n \dd \bar{a}_n = \exp (-2 \tr \ln C) \mathcal{D} \psi \mathcal{D} \bar{\psi},
\ee
where the minus signs arise from the fact that $a_n$ and $\bar{a}_n$ are Grassmann variables. It follows that the Jacobian of the transformation, to linear order in $\alpha$, is given by
\be
\label{e-3-1-18}
J = 1 - 2 \Tr \ln \left( 1 + \ii \alpha \Gamma_{2n+1}  \right) = 1 - 2 \ii  \Tr ( \alpha \Gamma_{2n+1} ) = 1 - 2 \ii \tr \sum_n \langle \phi_n | \alpha \Gamma_{2n+1} | \phi_n \rangle.
\ee
where ``$\Tr$'' denotes a functional trace and ``$\tr$'' denotes a trace over gamma-matrix and group indices. Comparing with (\ref{e-3-1-8}), we read off the integrated anomaly
\be
\label{e-3-1-19}
G (\alpha) = 2 \alpha \Tr \Gamma_{2n+1} = 2 \alpha \tr \sum_n \langle \phi_n | \Gamma_{2n+1} | \phi_n \rangle.
\ee 
As it stands, this expression is ill-defined since it involves the sum of a vanishing quantity ($\tr \Gamma_{2n+1}$) over an infinity of states. One must thus regularize this quantity in a manner that respects the symmetries of the problem. The standard (and least rigorous) method introduces a Gaussian UV cutoff by inserting the convergence factor $e^{- \beta (\ii \fslash D)^2 / 2}$ where $\beta$ should be taken to zero at the end of the day. This way, one obtains the regularized anomaly
\be
\label{e-3-1-20}
G (\alpha) = 2 \alpha \lim_{\beta \to 0} \Tr \{ \Gamma_{2n+1} \exp [ - \beta (\ii \fslash D)^2 / 2 ]\}.
\ee 
The usual way to compute the chiral anomaly is to write the trace in (\ref{e-3-1-20}) in terms of a plane-wave basis, expand the exponential and extract the term surviving at the $\beta \to 0$ limit. In four dimensions, the calculation is straightforward and yields the expression
\be
\label{e-3-1-21}
G^{(4)} (\alpha) = \alpha \int \dd^4 x \epsilon^{\mu\nu\rho\sigma} \left[ \frac{1}{16\pi^2} \tr ( F_{\mu\nu} F_{\rho\sigma} ) + \frac{\dim \MR}{384\pi^2} R_{\mu\nu\alpha\beta} R_{\rho\sigma}^{\phantom{\rho\sigma} \alpha\beta} \right],
\ee
where the two terms correspond to the gauge and gravitational part. However, the generalization of the plane-wave method to higher dimensions is quite messy, especially in theories involving gravity. Yet, it is possible to develop a powerful formalism that allows us to treat all cases simultaneously and derive results valid for any spacetime dimension. This will be the objective of Section 3.2.

\subsection{Nonabelian anomalies}
\label{sec-3-1-2}

We now turn to the anomalies associated with local symmetry currents, whose best-known examples are the spin $1/2$ gauge, gravitational and mixed anomalies. These arise in $2n$--dimensional theories of \emph{Weyl} fermions interacting with gauge and/or gravitational fields. The classical action is given by
\be
\label{e-3-1-22}
S = \int \dd^{2n}  x \sqrt{g} \bar{\psi} \ii \fslash D^+ \psi,
\ee
where $\ii \fslash D^+$ is the positive-chirality projection of the Dirac operator
\be
\label{e-3-1-23}
\ii \fslash D^+ = \ii \fslash D P^+ ,\qquad P^\pm \equiv \frac{1 \pm \Gamma_{2n+1}}{2}.
\ee
The classical action (\ref{e-3-1-22}) is invariant under local gauge transformations 
\be
\label{e-3-1-24}
\delta \psi = - v P^+ \psi ,\qquad \delta \bar{\psi} = \bar{\psi} v P^- ,\qquad \delta A_\mu = D_\mu v,
\ee
general coordinate transformations (GCT) 
\be 
\label{e-3-1-25}
\delta \psi = - \xi^\mu \partial_\mu \psi ,\qquad 
\delta \bar{\psi} = \xi^\mu \partial_\mu \bar{\psi} ,\qquad 
\delta e^a_\mu = e^a_\nu D_\mu \xi^\nu - \xi^\nu \omega_{\nu \phantom{a} b}^{\phantom{\nu} a} e^b_\mu ,\qquad 
\delta A_\mu = (\mathcal{L}_\xi A)_\mu,
\ee 
and local Lorentz transformations (LLT)
\be 
\label{e-3-1-26}
\delta \psi = - \frac{1}{4} \lambda_{ab} \Gamma^{ab} \psi ,\qquad 
\delta \bar{\psi} = - \frac{1}{4} \lambda_{ab} \bar{\psi} \Gamma^{ab} ,\qquad
\delta e^a_\mu = - \lambda^a_{\phantom a b} e^b_\mu.
\ee 

To find the associated conservation laws, it is advantageous in this case to reformulate them in terms of the external gauge and gravitational fields. To this end, we first define $\delta_v$, $\delta_\xi$ and $\delta_\lambda$ as the operators that generate a gauge transformation of $A_\mu$ with parameter $v$, a GCT of $g_{\mu\nu}$ with parameter $\xi$ and a LLT of $e^a_\mu$ with parameter $\lambda$ respectively, leaving the other fields invariant. The combined action of these operators on a functional $F[A,e]$ of $A_\mu$ and $e^a_\mu$ gives
\bea
\label{e-3-1-27}
\delta_{v,\xi,\lambda} F &=& \int \dd^{2n}x \left[ \tr \left( \frac{\delta F}{\delta A_\mu} D_\mu v \right) + \frac{\delta F}{\delta e^a_\mu} \left( e^a_\nu D_\mu \xi^\nu - \xi^\nu \omega_{\nu \phantom{a} b}^{\phantom{\nu} a} e^b_\mu - \lambda^a_{\phantom{a}b} e^b_\mu \right) \right] \nn\\
&=& - \int \dd^{2n}x \left[ \tr \left( v D_\mu \frac{\delta}{\delta A_\mu} \right) + \xi^\nu \left( D_\mu  e^a_\nu + \omega_{\nu \phantom{a} b}^{\phantom{\nu} a} e^b_\mu \right) \frac{\delta}{\delta e^a_\mu}  + \lambda^a_{\phantom{a}b} e^b_\mu \frac{\delta}{\delta e^a_\mu} \right] F,
\eea
and thus, we can write
\bea
\label{e-3-1-28}
\delta_v &=& - \int \dd^{2n}x \tr \left( v D_\mu \frac{\delta}{\delta A_\mu} \right) , \nn\\
\delta_\xi &=& - \int \dd^{2n} x \xi^\nu \left( D_\mu e^a_\nu + \omega_{\nu \phantom{a} b}^{\phantom{\nu} a} e^b_\mu \right) \frac{\delta}{\delta e^a_\mu} ,\nn\\
\delta_\lambda &=& - \int \dd^{2n} x \lambda^a_{\phantom{a}b} e^b_\mu \frac{\delta}{\delta e^a_\mu},
\eea
noting that the covariant derivative on the second line acts on all terms to its right.

In the classical case, we take $F$ to be the action $S$ with the understanding that the fermions are integrated out by their equations of motion. Noting that the variational derivatives of the action with respect to $A_\mu$ and $e^a_\mu$ give the gauge current and the energy-momentum tensor
\be
\label{e-3-1-29}
\frac{1}{\sqrt{g}} \frac{\delta S}{\delta A^I_\mu} = \ii \bar{\psi} \Gamma^\mu T_I P^+ \psi = J^\mu_I ,\qquad \frac{1}{\sqrt{g}} \frac{\delta S}{\delta e^a_\mu} = T_a^{\phantom{a} \mu},
\ee
we readily compute
\be
\label{e-3-1-30}
\delta_{v,\xi,\lambda} S = - \int \dd^{2n}x \sqrt{g} \left[ \tr \left( v D_\mu J^\mu \right) + \xi^\nu \left( D_\mu T^\mu_{\phantom{\mu} \nu} + \omega_{\nu a b} T^{[ab]} \right) + \lambda_{ab} T^{[ab]} \right].
\ee
Invariance of the action leads then to the laws
\be
\label{e-3-1-31}
D_\mu J^\mu = 0 ,\qquad D_\mu T^\mu_{\phantom{\mu} \nu} = 0 ,\qquad T^{[ab]} = 0.
\ee
expressing covariant conservation of $J^\mu$ and $T^{\mu\nu}$ and symmetry of $T^{ab}$.

Turning to the quantum case, we are interested in the effective action
\be
\label{e-3-1-32}
\varGamma = - \ln \int \mathcal{D} \psi \mathcal{D} \bar{\psi} \exp (-S[\psi,\bar{\psi}]),
\ee
whose variational derivatives with respect to $A_\mu$ and $e^a_\mu$ yield the induced gauge current and the induced energy-momentum tensor
\be
\label{e-3-1-33}
\frac{1}{\sqrt{g}} \frac{\delta \varGamma}{\delta A^A_\mu} = \langle J^{\mu A} \rangle ,\qquad \frac{\delta \varGamma}{\delta e^a_\mu} = \langle T_a^{\phantom{a} \mu} \rangle.
\ee
Then, in a similar manner as before, we have
\be
\label{e-3-1-34}
\delta_{v,\xi,\lambda} \varGamma = - \int \dd^{2n}x \sqrt{g} \left[ \tr \left( v D_\mu \langle J^\mu \rangle \right) + \xi^\nu \left( D_\mu \langle T^\mu_{\phantom{\mu} \nu} \rangle + \omega_{\nu a b} \langle T^{[ab]} \rangle \right) + \lambda_{ab} \langle T^{[ab]} \rangle \right].
\ee
If that were the total variation of $\varGamma$, its vanishing would lead to Eqs. (\ref{e-3-1-31}) for the expectation values of $J^\mu$, $T^{\mu\nu}$ and $T^{[ab]}$. However, now the fermions are not integrated out by their equations of motion but through the functional integration in (\ref{e-3-1-32}). So, the variation of $\varGamma$ may receive additional terms from the Jacobian of the integration measure. Parameterizing the Jacobian as
\be
\label{e-3-1-35}
J = 1 + \int \dd^{2n}x \sqrt{g} \left[ \tr \left(v G \right) + \xi_\mu G^\mu + \lambda_{ab} G^{ab} \right],
\ee 
we obtain, instead of (\ref{e-3-1-31}), the equations
\be
\label{e-3-1-36}
D_\mu \langle J^\mu \rangle = G ,\qquad D_\mu \langle T^{\mu}_{\phantom{\mu} \nu} \rangle = G_\nu - \omega_{\nu a b} G^{ab},\qquad T^{[ab]} = G^{ab}.
\ee
which express the violation of the Ward identities of $J^\mu$ and $T^{\mu}_{\phantom{\mu} \nu}$ and the failure of $T^{ab}$ to be symmetric. The quantities $G$, $G^\mu$ and $G^{ab}$ are called the gauge, Einstein and Lorentz contributions to the anomaly respectively. The integrated expressions
\bea
\label{e-3-1-37}
G (v) &=& \int \dd^{2n}x \sqrt{g} \tr \left( v D_\mu \langle J^\mu \rangle \right) = - \delta_v \varGamma ,\nn\\
G(\xi) &=& \int \dd^{2n}x \sqrt{g} \xi^\nu \left( D_\mu \langle T^\mu_{\phantom{\mu} \nu} \rangle + \omega_{\nu a b} \langle T^{[ab]} \rangle \right) =  - \delta_\xi \varGamma ,\nn\\
G(\lambda) &=& \int \dd^{2n}x \sqrt{g} \lambda_{ab} \langle T^{[ab]} \rangle =  - \delta_\lambda \varGamma,
\eea
will be called by the same names.

However, the above description is actually redundant \cite{Bardeen:1984pm} due to the fact that the Einstein and Lorentz anomalies are not two independent objects. Indeed, one can exploit the freedom of LLT's of the vielbein so as to fix a gauge where only one type of anomaly appears. The most convenient choice for calculations is to pick a gauge where $e_{a\mu}$ is symmetric so that the second of (\ref{e-3-1-33}) ensures that $T^{ab}$ is symmetric as well. In that case, only the Einstein anomaly appears and Eqs. (\ref{e-3-1-37}) simplify to
\be
\label{e-3-1-38}
G (v) = \int \dd^{2n}x \sqrt{g} \tr \left( v D_\mu \langle J^\mu \rangle \right) = - \delta_v \varGamma ,\qquad G(\xi) = \int \dd^{2n}x \sqrt{g} \xi^\nu D_\mu \langle T^\mu_{\phantom{\mu} \nu} \rangle =  - \delta_\xi \varGamma .
\ee
One can then switch to the Lorentz form of the anomaly by adding the so-called Bardeen-Zumino \cite{Bardeen:1984pm} counterterm to the action. The recipe is as follows. As we will see later, a pure Einstein anomaly has the form
\be
\label{e-3-1-39}
G(\xi) = c \int_{M_{2n}} \omega^1_{2n} \left( D_{[b} \xi_{a]} , \Gamma , R \right) ,
\ee
where $\omega^1_{2n}$ is a $2n$--form depending on $D_{[b} \xi_{a]}$, the affine connection $\Gamma$ and the curvature $R$ and $c$ is a numerical coefficient. The corresponding pure Lorentz anomaly is given by
\be
\label{e-3-1-40}
G(\lambda) = c \int_{M_{2n}} \omega^1_{2n} \left( \lambda_{ab} , \omega , R \right) 
\ee
that is, by the very same expression with the replacement $D_{[b} \xi_{a]} \to \lambda_{ab}$ and with the affine connection replaced by the spin connection.

The \emph{gauge}, \emph{gravitational} and \emph{mixed} anomalies are then defined as the terms of $G(v)$ depending only on $(F,v)$, the terms of $G(\xi)$ depending only on $(R,\xi)$ and the terms of $G(v) + G(\xi)$ with mixed dependence respectively.

\subsubsection{Consistency and covariance}

Nonabelian anomalies are constrained by a condition arising from the symmetries to which they correspond, called the \emph{Wess-Zumino (WZ) consistency condition} \cite{Wess:1971yu}. For the case of gauge transformations, this condition is derived by observing that, the commutator of two gauge transformations with parameters $v_1$ and $v_2$ is a gauge transformation with parameter $[v_1,v_2]$. Indeed, acting on $A_\mu$ with the commutator and using the Jacobi identity, we easily find
\be
\label{e-3-1-41}
[\delta_{v_1},\delta_{v_2}] A_\mu = \partial_\mu [v_1 , v_2] + \left[ A_\mu, [v_1 , v_2] \right] = \delta_{[v_1 , v_2]} A_\mu.
\ee
Eq. (\ref{e-3-1-41}) must hold for any functional of $A_\mu$ and, in particular, for $\varGamma$. Using (\ref{e-3-1-38}), we arrive at the WZ consistency condition
\be
\label{e-3-1-42}
\int \dd^{2n}x \sqrt{g} \tr \left[ \left( v_2 \delta_{v_1} - v_1 \delta_{v_2} \right) G \right] = \int \dd^{2n}x \sqrt{g} \tr \left( [v_1,v_2] G \right).
\ee
Since this is by definition obeyed by the anomaly $G(v)$, the latter is called the \emph{consistent anomaly}. Although the usefulness of the WZ consistency condition is not at all obvious at the present stage, it turns out that this condition is so strong that may even enable us to reconstruct the full expression for an anomaly given its leading term.

A property of consistent anomalies is that they violate the covariance of the associated current. To see this, we define the operator $\tilde{\delta}$ as the one implementing an arbitrary variation of $A_\mu$, $\tilde{\delta}_B A_\mu = B_\mu$. Then, considering the action of the commutator $[ \tilde{\delta}_B , \delta_{v} ]$ on $A_\mu$, we easily find
\be
\label{e-3-1-43}
[ \tilde{\delta}_B , \delta_{v} ] A_\mu = \tilde{\delta}_B [A_\mu,v]  = [B_\mu,v] = \tilde{\delta}_{[B,v]} A_\mu.
\ee
Again, the same relation must hold for $\varGamma[A]$. Using the first of (\ref{e-3-1-38}) and () and the Jacobi identity, we find
\be
\label{e-3-1-44}
\int \dd^{2n} x \sqrt{g} \tr \left( B_\mu \delta_v J^{\mu} \right) = \int \dd^{2n}x \sqrt{g} \tr \left( B_\mu [J^{\mu}, v ] \right) - \tilde{\delta}_B G(v).
\ee
So, in the presence of an anomaly, the covariant gauge transformation of $J^\mu$ is violated by an extra term. If we insist that the theory be phrased in terms of a covariant current, then we must search for a current of the form $\tilde{J}^\mu = J^\mu + K^\mu$, where $K^\mu$ should transform so that the non-covariant term in (\ref{e-3-1-44}) cancels,
\be
\label{e-3-1-45}
\int \dd^{2n}x \sqrt{g} \tr \left( B_\mu \delta_v K^{\mu} \right) = \int \dd^{2n}x \sqrt{g} \tr \left( B_\mu [ K^{\mu}, v ] \right) + \tilde{\delta}_B G(v).
\ee
If such a quantity exists, $\tilde{J}^\mu$ is the covariant current of the theory. Its divergence defines the {\it covariant anomaly} $\tilde{G} (v)$ \cite{Bardeen:1984pm} as follows
\be
\label{e-3-1-46}
\tilde{G} (v) \equiv \int \dd^{2n}x \sqrt{g} \tr ( v D_\mu \tilde{J}^\mu ).
\ee
Exactly analogous results hold for the GCT/LLT cases.

Hence, nonabelian anomalies admit two different equivalent descriptions. The consistent anomaly, specified in the usual way, is a fundamental quantity arising from the variation of the effective action under gauge transformations. On the other hand, the covariant anomaly is the one that is most easily computed. For that reason, in Section \ref{sec-3-3} we will compute the covariant anomaly and, later on, we will use the WZ condition to switch to the consistent form.

\subsubsection{Fujikawa's method}

To proceed, let us try to extend Fujikawa's method to deal with nonabelian anomalies. In doing so, we are immediately faced with the problem of defining the effective action for Weyl fermions. The naive definition
\be
\label{e-3-1-47}
\varGamma = - \ln \int \mathcal{D} \psi \mathcal{D} \bar{\psi} \exp \left( - \int \dd^{2n}  x \sqrt{g} \bar{\psi} \ii \fslash D^+ \psi \right) \stackrel{?}{=} - \ln \Det (\ii \fslash D^+),
\ee
is not valid because the integral does not correspond to a functional determinant of a differential operator. Indeed, for a positive-chirality spinor $\psi^+$ satisfying $P^+ \psi^+ = \psi^+$, one sees that $\ii \fslash D^+ \psi^+$ is a negative-chirality spinor satisfying $P^- (\ii \fslash D^+ \psi^+) = \ii \fslash D^+ \psi^+$ so that $\ii \fslash D^+$ is a map from the space of positive-chirality spinors to the space of negative-chirality spinors. Therefore, the eigenvalue problem of $\ii \fslash D^+$ is ill-defined and it does not make sense to define $\varGamma$ through its functional determinant nor to regulate the anomaly using its eigenvalues. To address the problem, one may either (i) replace $\ii \fslash D^+$ by some new operator with a well-posed eigenvalue problem, define $\varGamma$ in terms of its determinant and regulate the anomaly using its eigenvalues or (ii) keep the original definition of $\varGamma$ absorbing the projection operator in the definitions of the spinors and regulate the anomaly in terms of the eigenvalues of the ordinary Dirac operator $\ii \fslash D$. It turns out that the two methods lead to the consistent and the covariant anomaly respectively. Although we are ultimately interested in consistent anomalies, the calculation methods to be introduced in Sections 3.2 and 3.3 require that we use the second method which gives the covariant anomalies. The consistent form of the anomalies can then be deduced using the WZ condition.

To develop the method, we first rewrite the effective action (\ref{e-3-1-47}) in the equivalent form
\be
\label{e-3-1-48}
\varGamma = - \ln \int \mathcal{D} \psi^+ \mathcal{D} \bar{\psi}^- \exp \left( - \int \dd^{2n}  x \bar{\psi}^- \ii \fslash D \psi^+ \right),
\ee
where we defined the projections $\psi^+ = P^+ \psi$ and $\bar{\psi}^- = \bar{\psi} P^-$. These may be expanded as 
\be
\label{e-3-1-49}
\psi^+(x) = \sum_n a_n \phi^+_n (x) ,\qquad \bar{\psi}^- (x) = \sum_n \bar{b}_n \left[ \phi^-_n (x) \right]^\dag,
\ee
where $\phi^+_{n} = P^+ \phi_{n}$ and $\phi^-_{n} = P^- \phi_{n}$ denote the corresponding projections of the eigenvectors of $\ii \fslash D$. These satisfy the orthogonality relations $\langle \phi^\pm_{ n} | \phi^\pm_{ m} \rangle =\delta_{nm}$ and so Eqs. (\ref{e-3-1-49}) can be inverted leading to
\be
\label{e-3-1-50}
a_n = \langle \phi^+_{n} | \psi^+ \rangle ,\qquad \bar{b}_n = \langle \bar{\psi}^- | \phi^-_{n} \rangle.
\ee
The functional integration measure is then defined as
\be
\label{e-3-1-51}
\mathcal{D} \psi^+ \mathcal{D} \bar{\psi}^- = \prod_n a_n \bar{b}_n.
\ee

From these expressions, we can derive the required Fujikawa expressions for gauge and gravitational anomalies. 
Starting from the gauge anomaly, we consider the gauge transformation
\be 
\label{e-3-1-52}
\psi'^+ = ( 1 - v ) \psi^+ ,\qquad \bar{\psi}'^- = \bar{\psi}^- ( 1 + v ),
\ee 
under which the coefficients in (\ref{e-3-1-50}) transform as
\be
\label{e-3-1-53}
a'_n = C^+_{nm} a_m ,\qquad \bar{b}'_n = \sum_m \bar{b}_m C^-_{mn} ; \qquad C^\pm_{nm} \equiv \langle \phi^\pm_{n} | 1 \mp v | \phi^\pm_{m} \rangle,
\ee
resulting in the following transformation of the integration measure
\be
\label{e-3-1-54}
\mathcal{D} \psi'^+ \mathcal{D} \bar{\psi}'^- = \exp ( - \Tr \ln C^+ - \Tr \ln C^- ) \mathcal{D} \psi^+ \mathcal{D} \bar{\psi}^-.
\ee
So, the Jacobian is now given by
\be
\label{e-3-1-55}
J = 1 - \Tr \ln (1-v) - \Tr \ln (1+v)
= 1 + \tr  \sum_n \left( \langle \phi^+_{n} | v |\phi^+_{n} \rangle - \langle \phi^-_{n} | v |\phi^-_{n} \rangle \right) = 1 + \Tr ( \Gamma_{2n+1} v  ) 
\ee
and the last term gives the integrated anomaly. Since this expression involves the usual eigenvectors $\{ \phi_n \}$ of $\ii \fslash D$, we can regulate it exactly as in the singlet case, with the result
\be
\label{e-3-1-56a}
\tilde{G} (v) =  \lim_{\beta \to 0} \tr \{ \Gamma_{2n+1} v \exp [ - \beta (\ii \fslash D)^2 / 2 ] \}
\ee

The corresponding expression for the anomaly under GCT's is found in an analogous manner. Here, we consider a GCT on the fermion field which, modulo a LLT, is given by
\be 
\label{e-3-1-57}
\delta_\xi \psi^+ = - \xi^\mu D_\mu \psi^+ ,\qquad \delta_\xi \bar{\psi}^- = \xi^\mu D_\mu \bar{\psi}^-.
\ee 
Then, following exactly the same steps, we find the regularized expression
\be
\label{e-3-1-58}
\tilde{G} (\xi) =  \lim_{\beta \to 0} \tr \{ \Gamma_{2n+1} \xi^\mu D_\mu \exp [ - \beta (\ii \fslash D)^2 / 2 ] \}.
\ee

Eqs. (\ref{e-3-1-56a}) and (\ref{e-3-1-58}) have a similar form to (\ref{e-3-1-20}) and, for that reason, they may be computed by the same methods. E.g. using the plane-wave method in four dimensions gives
\be
\label{e-3-1-59}
\tilde{G}^{(4)} (v) = \int \dd^4 x \epsilon^{\mu\nu\rho\sigma} \left[ \frac{1}{32\pi^2}\tr ( v F_{\mu\nu} F_{\rho\sigma} ) + \frac{1}{768\pi^2} \tr v R_{\mu\nu\alpha\beta} R_{\rho\sigma}^{\phantom{\rho\sigma} \alpha\beta} \right]
\ee 
where the two terms correspond to the gauge and mixed anomaly (the latter is nonzero only for a $\mathrm{U}(1)$ gauge group in which case $\tr v \to \ii q v$). Again, the generalization to higher dimensions requires more efficient methods.

\subsection{Physical aspects of anomalies}
\label{sec-3-1-3}

After this introduction to anomalies, and before proceeding to their explicit computation, let us pause to discuss some of their physical aspects with the objective of clarifying their interpretation and their importance for the physical theories of interest.

To clarify the origin of anomalies, it is instructive to discuss the various alternative methods available for their computation. The main approaches are discussed below.
\begin{enumerate}
\item \emph{Fujikawa's method.} Using this method, anomalies are expressed in terms of the variation of the fermion integration measure in the functional integral defining the effective action, as in Eq. (\ref{e-3-1-19}). The resulting expression is ill-defined. To regulate it in a gauge-invariant manner, one can use (i) a Gaussian cutoff, (ii) heat-kernel regularization or (iii) $\zeta$--function regularization. In case (i) the anomaly itself can be computed using either plane-wave expansions or supersymmetric path integrals while in cases (ii,iii) it can be computed using Seeley coefficients. In all cases, the anomaly manifests itself as a finite term arising from the regulator.

\item \emph{Feynman-diagram method.} Using this method, anomalies in $2n$ dimensions are calculated by computing the divergence of the current through one-loop UV-divergent diagrams with one chiral fermion running in the loop and $n+1$ external legs; in particular, the singlet anomaly comes from diagrams with one axial current insertion and $n$ external gauge bosons and/or gravitons while gauge, gravitational and mixed anomalies correspond, respectively, to diagrams with $n+1$ external gauge bosons, gravitons and both, coupled to the loop fermion through $V-A$ vertices. Gauge-invariant regularization may be achieved using (i) the Pauli-Villars method involving extra regulator fields of large mass $M$ or (ii) dimensional regularization, where we work in $D = 2n + \varepsilon$ dimensions and use a careful definition of $\Gamma_{2n+1}$. In both cases, there is a finite term surviving the $M \to \infty$ and $\varepsilon \to 0$ limits respectively, giving the anomaly.

\item \emph{Adler's operator method.} Using this method, anomalies are calculated by directly computing the divergence of the current using its operator definition. Being a product of two local field operators at the same point, this current is singular. To regulate it, one uses the point-splitting method, i.e. separates the two local operators by a distance $\epsilon_\mu$ and inserts a Wilson line to maintain gauge invariance. Due to this insertion, there is a finite term surviving the $\epsilon \to 0$ limit, equal to the anomaly.
\end{enumerate}
So, whatever method one might choose to use, the origin of anomalies is exactly the same: \emph{it is impossible to regulate the current in a consistent way without violating its conservation law by new terms involving the external fields}. Put otherwise, the operators $\ii \fslash D$ and $\Gamma_{2n+1}$ ``do not commute'' and so it is impossible to maintain gauge invariance without spoiling invariance under chiral transformations. For details on these methods, the reader is referred to the original papers, the modern reviews \cite{Scrucca:2004jn,Harvey:2005it} and the book \cite{Bertlmann:1996xk}. 

The common aspect of all calculation methods outlined above is that anomalies appear during the regularization of ultraviolet-divergent quantities. Based on this, one would be tempted to interpret an anomaly is an ultraviolet effect. However, unlike the usual low-energy artifacts appearing during regularization of UV divergences, anomalies are \emph{finite} and \emph{regulator-independent} and cannot be removed by adding local counterterms to the action. It was first recognized by 't Hooft in \cite{'tHooft:1979bh} and further elucidated in \cite{Frishman:1981dq,Coleman:1982yg} that an anomaly is most appropriately interpreted as an \emph{infrared} effect which, if present in a fundamental theory, is also present in an effective theory and vice versa. 

Let us finally discuss the physical implications of anomalies. The existence of singlet anomalies in a theory alters its physical content but, nevertheless, it does not threaten its consistency as long as the symmetry current is not coupled to gauge fields\footnote{However, if the anomalous chiral current of a gravitational theory is coupled to a gauge field, the situation is problematic since the $\int \dd^{2n} x \sqrt{g} J^\mu A_\mu$ interaction acquires an anomalous variation; moreover, if one tries to absorb the anomaly into a redefinition of the current then GCT invariance is lost.}. In fact, the existence of the chiral anomaly was actually required for the understanding of many long-standing problems of high-energy physics, namely the $\pi^0 \to 2\gamma$ decay, the $\mathrm{U}(1)$ problem \cite{'tHooft:1976up,'tHooft:1976fv}, the mass of the $\eta^\prime$ meson and several types of hadronic and semileptonic decays \cite{Wess:1971yu}. On the other hand, the existence of nonabelian anomalies does pose serious threats to the consistency of a theory. To discuss them, we must distinguish between two cases. The first case refers to renormalizable theories, where Ward identities are needed to ensure that the unphysical polarizations of gauge fields decouple and that the S-matrix is unitary; in the presence of anomalies these identities are violated and such states may appear as poles in the S-matrix thus violating unitarity \cite{KorthalsAltes:1972aq} and renormalizability \cite{Gross:1972pv}. The loss of renormalizability is also easily seen by considering an $n$--point function and noting that the contributions involving anomalous diagrams scale in a different way from the other contributions. The second case refers to non-renormalizable effective theories. Here, loss of unitarity would not be fatal if there was a way to cancel it in the UV-complete theory; however, the infrared nature of anomalies rules out this possibility and so an anomalous theory is again inconsistent. For that reason, the study of anomalies is of utmost importance in low-energy effective theories such as the supergravities considered in this thesis.

\section{Calculation of Singlet Anomalies}
\label{sec-3-2}

In this section, we will present the actual calculation of all possible types of singlet anomalies, as first done in the seminal paper \cite{Alvarez-Gaume:1984ig} of Alvarez-Gaum\'e and Witten. In particular, we will illustrate the second approach of employed in that paper (see also \cite{Alvarez-Gaume:1983at,Friedan:1983xr}), which is based on supersymmetric quantum-mechanical path integrals, following the more modern treatment of \cite{Scrucca:2004jn}. Using this method, we will present the detailed calculation of the spin $1/2$ chiral anomaly which also serves as a proof of the Atiyah-Singer index theorem. By extrapolating this result to the cases of spin $3/2$ fermions and self-dual $(n-1)$--forms, we will then deduce the anomaly for these fields as well.

\subsection{The spin--$1/2$ anomaly}
\label{sec-3-2-1} 

To calculate spin--$1/2$ singlet anomalies, we will proceed as follows. As a preliminary step, we will start by the Fujikawa expression (\ref{e-3-1-20}) and relate it to the Dirac index and to the Witten index of a suitable supersymmetric theory. Then, using the latter description, we will compute the anomalies by means of supersymmetric path integrals.

\subsubsection{The singlet anomaly as a Dirac index}

To make a connection between anomalies and index theory, consider Eq. (\ref{e-3-1-20}) for the spin--$1/2$ chiral anomaly, which involves the trace of $\Gamma_{2n+1}$ over the eigenvectors of $\ii \fslash D$. It is easy to see that, given an eigenvector $| \phi_n \rangle$ with eigenvalue $\lambda_n$, $\Gamma_{2n+1} | \phi_n \rangle$ is also an eigenvector with eigenvalue $- \lambda_n$. Since $\ii \fslash D$ is Hermitian, its eigenvectors that correspond to different eigenvalues are orthogonal, i.e. $\langle \phi_n | \Gamma_{2n+1} | \phi_n \rangle = 0$ if $\lambda_n \ne 0$. So, the only states that give nonzero contributions to the sum (\ref{e-3-1-20}) are the \emph{zero modes} of $\ii \fslash D$. Denoting these states by $| \phi_0^i \rangle$, $i=1,\ldots,N$, we can split them into two irreducible representations $| \phi_0^{i_\pm} \rangle$ of the Clifford algebra, classified by the $\Gamma_{2n+1}$ eigenvalue according to
\be
\label{e-3-2-1}
\Gamma_{2n+1} | \phi_0^{i_\pm} \rangle = \pm | \phi_0^{i_\pm} \rangle .
\ee
Then, (\ref{e-3-1-20}) can be written as follows
\be
\label{e-3-2-2}
G_{1/2} (\alpha) = 2 \alpha \sum_i \langle \phi_0^i  | \Gamma_{2n+1} | \phi_0^i \rangle = 2 \alpha \Bigl( \sum_{i_+} \langle \phi_0^{i^+}  | \phi_0^{i^+} \rangle - \sum_{i_-} \langle \phi_0^{i^-}  | \phi_0^{i^-} \rangle \Bigr) = 2 \alpha n^{1/2}
\ee
where $n^{1/2}$ is the difference between the number of zero modes with positive and negative chirality. Considering the positive- and negative-chirality projections $\ii \fslash D^+ = \ii \fslash D P^+$ and $\ii \fslash D^- = \ii \fslash D P^-$ and noting that $\ii \fslash D^- = ( \ii \fslash D^+ ) ^\dag$, we can write (\ref{e-3-2-2}) in the form 
\be
\label{e-3-2-3}
G_{1/2} (\alpha) = 2 \alpha \bigl[ \dim \ker \ii \fslash D^+ - \dim \ker ( \ii \fslash D^+)^\dag \bigr] = 2 \alpha \ind ( \ii \fslash D )
\ee 
This establishes a relation between anomalies and index theory: the spin--$1/2$ chiral anomaly is equal to $2 \alpha$ times the index of the operator $\ii \fslash D$.

\subsubsection{The singlet anomaly as a Witten index}

The chiral anomaly admits yet another interpretation as the Witten index of a certain supersymmetric quantum-mechanical theory. To see how this comes about, let us first recall some elementary facts about $N=1/2$ supersymmetric quantum mechanics. A quantum-mechanical system with $N=1/2$ supersymmetry is characterized by (i) a Hilbert space $\mathcal{H}$ consisting of a bosonic and a fermionic subspace (ii) an operator $(-1)^F$ that anticommutes with all fermion operators and has eigenvalues $+1$ ($-1$) in a bosonic (fermionic) state and (iii) one hermitian supercharge $Q$ mapping bosonic to fermionic states and vice versa. The Hamiltonian of such a system is given by
\be
\label{e-3-2-4}
H = Q^2.
\ee 
To see an interesting property of such a theory, consider a bosonic energy eigenstate $| \phi_{\mathrm{b}} \rangle$ with $H | \phi_{\mathrm{b}} \rangle = E | \phi_{\mathrm{b}} \rangle$. From this, we can construct the fermionic state $| \phi_{\mathrm{f}} \rangle = Q | \phi_{\mathrm{b}} \rangle$, which, since $[H,Q]=0$, is also an energy eigenstate of eigenvalue $E$. Hence, if $E > 0$, for every bosonic energy eigenstate there exists a fermionic energy eigenstate with the same eigenvalue and vice versa. On the other hand, if $E=0$, the corresponding energy eigenstates can be non-degenerate.

The {\it Witten index} \cite{Witten:1982df} of the theory is defined as the difference between the numbers of bosonic and fermionic states of the theory,
\be
\label{e-3-2-5}
\ind_\mathrm{W} (H) \equiv \sum_i \langle \phi_{\mathrm{b},i} | \phi_{\mathrm{b},i} \rangle - \sum_i \langle \phi_{\mathrm{f},i} | \phi_{\mathrm{f},i} \rangle = \sum_i \langle \phi_i | (-1)^F | \phi_i \rangle = \Tr  (-1)^F .
\ee
and, by the arguments of the previous paragraph, it only receives contributions from the zero-energy sector of $\MH$. Hence, it does no harm to insert a factor of $e^{-\beta H}$ and write 
\be 
\label{e-3-2-6}
\ind_\mathrm{W} (H) = \Tr \left[ (-1)^F e^{-\beta H} \right].
\ee 
Apart from the insertion of $(-1)^F$, this looks much like the partition function of a statistical-mechanical system at inverse temperature $\beta$ and may be calculated by standard methods, one of which is based on the Euclidean path integral. As for the insertion of $(-1)^F$, its only effect is to switch the boundary conditions of the fermions from antiperiodic to periodic\footnote{This is easily verified by noticing that fermion correlators with one insertion of $(-1)^F$ satisfy relations like e.g. $\langle (-1)^F \psi(\tau) \psi(0) \rangle = \langle (-1)^F \psi(\tau) \psi(\beta) \rangle$, which are consistent only with periodic boundary conditions.}.

The relevance of the above to anomalies is evident once we compare (\ref{e-3-2-6}) and (\ref{e-3-1-20}). The comparison reveals that, given a supersymmetric quantum-mechanical theory which realizes the relations
\be
\label{e-3-2-8}
(-1)^F = \Gamma_{2n+1} ,\qquad Q = \frac{\ii \fslash D}{\sqrt{2}} 
\ee
on its Hilbert space, the singlet anomaly is proportional to the Witten index of the theory, 
\be
\label{e-3-2-7}
G_{1/2} (\alpha) = 2 \alpha \lim_{\beta \to 0} \ind_\mathrm{W} ( H ).
\ee 

Such a theory does indeed exist. It involves $D=2n$ bosonic fields $x^\mu$ that act as coordinates on $M_{2n}$, their fermionic superpartners $\psi^\mu$, a set of fermions $\bar{c}_A$ and $c^A$ transforming in the representation $\MR$ of the gauge group and its conjugate $\bar{\MR}$ respectively and a set of external fields given by the vielbein $e^a_\mu$, the spin connection $\omega_{\mu a b}$ and the gauge field $A_\mu = A_\mu^I T_I$. Its Lagrangian is
\be
\label{e-3-2-9}
L = \frac{1}{2} g_{\mu\nu} \dot{x}^\mu \dot{x}^\nu + \frac{\ii}{2} \psi_a \dot{\psi}^a + \frac{\ii}{4} \omega_{\mu a b} [ \psi^a, \psi^b] \dot{x}^\mu + \ii \bar{c}_A \left( \dot{c}^A + A^{\phantom{\mu} A}_{\mu \phantom{A} B} c^B \dot{x}^\mu \right) + \frac{1}{2} \bar{c}_A c^B \psi^a \psi^b F_{ab \phantom{A} B}^{\phantom{ab} A}.
\ee 
and is invariant under the supersymmetry transformations
\bea
\label{e-3-2-10}
&\delta x^\mu = \ii \epsilon \psi^\mu,\qquad \delta \psi^a = - \epsilon \left( \dot{x}^a + \frac{\ii}{2} \omega_{\mu a b} [\psi^\mu,\psi^b] \right), \nn\\
&\delta c^A = \ii \epsilon c^B A^{\phantom{\mu} A}_{\mu \phantom{A} B} \psi^\mu ,\qquad \delta \bar{c}_A = - \ii \epsilon \bar{c}_B A^{\phantom{\mu} B}_{\mu \phantom{B} A} \psi^\mu.
\eea
This theory may be constructed as a dimensional reduction of the 2D sigma model \cite{Hull:1985jv} used to describe the heterotic string or in the context of 1D representations of gauge field theories \cite{D'Hoker:1995bj}. The Witten index of this theory is easily shown to admit the path-integral expression
\be
\label{e-3-2-11}
\ind_\mathrm{W} (H) = \Tr \left[ (-1)^F e^{-\beta H} \right] = \tr_{c,\bar{c}} \int_\mathrm{P} \mathcal{D} x \mathcal{D} \psi e^{- \int_0^\beta \dd \tau R_\mathrm{E}}
\ee 
where $\tr_{c,\bar{c}}$ denotes the trace over the \emph{one-particle states} created by $c^i$ and $\bar{c}_i$ and $R_\mathrm{E}$ is the Euclidean continuation of the Routhian $R = L - \ii \bar{c}_A \dot{c}^A$, defined as a Lagrangian with respect to $x^\mu$ and $\psi^a$ and as a Hamiltonian with respect to $c^A$ and $\bar{c}_A$. To prove that the Lagrangian (\ref{e-3-2-9}) leads to the identifications (\ref{e-3-2-8}) for $Q$ and $(-1)^F$, we first compute the conjugate momenta
\be 
\label{e-3-2-13}
\pi^x_\mu = \dot{x}_\mu + \frac{\ii}{4} \omega_{\mu a b} [\psi^a,\psi^b] + \ii \bar{c}_A  A^{\phantom{\mu} A}_{\mu \phantom{A} B} c^B ,\qquad \pi^c_A = \ii \bar{c}_A ,\qquad \pi^\psi_a = \frac{\ii}{2} \psi_a,
\ee
and we find the supercharge
\be
\label{e-3-2-14}
Q = - \psi^\mu \dot{x}_\mu.
\ee
Next, we must quantize the theory. For $x^\mu$ and $c_A$ this is carried out as usual by imposing the (anti)commutation relations $[ \phi^i,\pi^\phi_j ]_\pm = \ii \delta^i_j$ while, for $\psi^a$, the fact that the canonical momentum is a multiple of the field itself requires that we employ Dirac's quantization method treating the third of (\ref{e-3-2-13}) as a second-class constraint. This leads to the relations
\be
\label{e-3-2-15}
[ x^\mu , \pi^x_\nu ] = \ii \delta^\mu_\nu ,\qquad \{ c^A , \bar{c}_B \} = \delta^A_B, \qquad \{ \psi^a , \psi^b \}_D = \delta^{ab},
\ee
the last one being a Dirac bracket. These relations can be realized through the assignments
\be 
\label{e-3-2-17}
\pi^x_\mu \to - \ii \partial_\mu ,\qquad \psi^a \to \frac{1}{\sqrt{2}} \Gamma^a ,\qquad \bar{c}_A T_{I \phantom{A} B}^{\phantom{I} A} c^B \to T_I,
\ee 
which, upon insertion in (\ref{e-3-2-14}), lead to the first identification
\be
\label{e-3-2-18}
Q = \frac{\ii}{\sqrt{2}} \Gamma^\mu \left( \partial_\mu + \frac{1}{4} \omega_{\mu a b} \Gamma^{ab} + A_\mu \right) = \frac{\ii \fslash D}{\sqrt{2}}.
\ee
As for $(-1)^F$, a consistent choice is given by the product $(2\ii)^n \psi^0 \ldots \psi^{2n-1}$, which anticommutes with all $\psi^a$'s and has eigenvalues $\pm 1$. Then, the second of (\ref{e-3-2-17}) leads to the second identification
\be 
\label{e-3-2-19}
(-1)^F = \Gamma_{2n+1}.
\ee 

\subsubsection{The calculation}

Let us turn to the actual calculation. According to the above, the chiral anomaly is written as
\be
\label{e-3-2-20}
G_{1/2} (\alpha) = 2 \alpha \lim_{\beta \to 0} \tr_{c,\bar{c}} \int_\mathrm{P} \mathcal{D} x \mathcal{D} \psi e^{- \int_0^1 \dd \tau R_\mathrm{E}}.
\ee
where we employed the change of variable $\tau \to \beta \tau$ and $R_\mathrm{E}$ is the Euclidean Routhian 
\be 
\label{e-3-2-21}
R_\mathrm{E} = \frac{1}{2 \beta} g_{\mu\nu} \dot{x}^\mu \dot{x}^\nu + \frac{1}{2} g_{\mu\nu} \psi^\mu \left( \dot{\psi}^\nu + \Gamma^\nu_{\rho\sigma} \dot{x}^\rho \psi^\sigma \right) + \bar{c}_A A^{\phantom{\mu} A}_{\mu \phantom{A} B} c^B \dot{x}^\mu - \frac{\beta}{2} \bar{c}_A c^B \psi^\mu \psi^\nu F_{\mu\nu \phantom{A} B}^{\phantom{ab} A}.
\ee 
In the limit $\beta \to 0$, the leading contributions to the path integral come from constant solutions. To evaluate the integral, we expand about these solutions according to
\be
\label{e-3-2-22}
x^\mu(\tau) = x^\mu_0 + \sqrt{\beta} y^\mu(\tau) ,\qquad \psi^\mu(\tau)= \frac{1}{\sqrt{\beta}} \psi^\mu_0 + \eta^\mu(\tau)
\ee
and compute the terms of (\ref{e-3-2-21}) quadratic in the fluctuating fields. The expansion simplifies if we use Riemann normal coordinates around $x_0$. In this particular coordinate system, we have
\be
\label{e-3-2-23}
\partial_\rho g_{\mu\nu}(x_0) = 0 ,\qquad \Gamma^\mu_{\nu \rho} (x_0) = 0 ,\qquad \partial_\sigma \Gamma^\mu_{\nu \rho} (x_0) = - \frac{1}{3} ( R^\mu_{\phantom \mu \nu \rho \sigma} + R^\mu_{\phantom \mu \rho \nu \sigma}),
\ee 
so that
\be
\label{e-3-2-24}
\partial_\mu \Gamma_{\rho \nu \sigma} \psi^\rho_0 \psi^\sigma_0 = \frac{1}{2} ( R_{\mu \nu \rho \sigma} - R_{\mu [ \nu \rho \sigma ]}) \psi^\rho_0 \psi^\sigma_0 = \frac{1}{2} R_{\mu \nu \rho \sigma} \psi^\rho_0 \psi^\sigma_0 
\ee
Then, the leading high-temperature contribution from the fluctuations of the fields is written as
\be
\label{e-3-2-25}
R^{(2)}_\mathrm{E} = \frac{1}{2} g_{\mu\nu} \dot{y}^\mu \dot{y}^\nu + \frac{1}{4} R_{\mu \nu \rho \sigma} \psi^\rho_0 \psi^\sigma_0 y^\mu \dot{y}^\nu + \frac{1}{2} g_{\mu\nu} \eta^\mu \dot{\eta}^\nu - \frac{1}{2} \bar{c}_A F_{\mu\nu \phantom{A} B}^{\phantom{\mu\nu} A} c^B \psi_0^\mu \psi_0^\nu.
\ee
and is $\beta$--independent. Passing back to the vielbein basis, integrating by parts and introducing the ``2--forms''
\be
\label{e-3-2-26}
R_{ab} \equiv \frac{1}{2} R_{abcd} \psi^c_0 \psi^d_0 ,\qquad F^A_{\phantom{A} B} \equiv \frac{1}{2} F_{\mu\nu \phantom{A} B}^{\phantom{\mu\nu} A} \psi_0^\mu \psi_0^\nu
\ee 
we write
\be
\label{e-3-2-27}
R^{(2)}_\mathrm{E} = \frac{1}{2} y^a \left( - \delta_{ab} \partial_\tau^2 + R_{ab} \partial_\tau \right) y^b + \frac{1}{2} \eta^a \left( \delta_{ab} \partial_\tau \right) \eta^b - \bar{c}_A F^A_{\phantom{A} B} c^B.
\ee
Therefore, the anomaly is given by the expression
\be
\label{e-3-2-28}
G_{1/2} (\alpha) = 2 \alpha \ii^n \int \dd^{2n} x_0 \dd^{2n} \psi_0 I_\mathrm{c} I_\mathrm{\eta} I_\mathrm{y},
\ee
where $I_\mathrm{c}$, $I_\mathrm{\eta}$ and $I_\mathrm{y}$ denote the one-particle trace over $(c^A,\bar{c}_A)$ and the path integrals over the fluctuations $\eta^a$ and $y^a$ respectively. The first is easily found to be
\be
\label{e-3-2-29}
I_\mathrm{c} = \tr_{c,\bar{c}} e^{\bar{c}_A F^A_{\phantom{A} B} c^B} = \tr e^F.
\ee
As for $I_\eta$, it is equal to the free (periodic) fermion determinant
\be
\label{e-3-2-30}
I_\eta =  \int \mathcal{D} \eta e^{ - \frac{1}{2} \int_0^1 \dd \tau \eta^a \left( \delta_{ab} \partial_\tau \right) \eta^b } = (- \ii)^n.
\ee
Finally, $I_y$ is given by the expression
\be
\label{e-3-2-31}
I_y = \int \mathcal{D} y e^{ -  \frac{1}{2}  \int_0^1 \dd \tau y^a \left( - \delta_{ab} \partial_\tau^2 + R_{ab} \partial_\tau \right) y^b } = \frac{1}{(2\pi)^n} \left[ \frac{ \Det ' ( - \delta_{ab} \partial_\tau^2 + R_{ab} \partial_\tau ) }{\Det ' ( - \delta_{ab} \partial_\tau^2 )} \right]^{-1/2},
\ee 
where the prime implies that the zero mode is to be excluded and where we multiplied and divided by the free boson determinant. To evaluate the determinant in the numerator, we first write it as an infinite product of ordinary determinants as follows
\be
\label{e-3-2-32}
\Det' ( - \delta_{ab} \partial_\tau^2 + R_{ab} \partial_\tau ) = \prod_{m=1}^\infty \det ( - \delta_{ab} \lambda_m^2 + R_{ab} \lambda_m ),
\ee
where $\lambda_m = 2 \pi \ii m$ are the eigenvalues of $\partial_\tau$ on the circle. Bringing $R_{ab}$ to a skew-diagonal form, we see that each determinant in (\ref{e-3-2-32}) breaks into a product of $2 \times 2$ determinants so that
\be
\label{e-3-2-33}
\det ( - \delta_{ab} \lambda_m^2 + R_{ab} \lambda_m ) = \prod_{i=1}^n \lambda_m^4 \left( 1 + \frac{4 \pi^2 x_i^2}{\lambda_m^2} \right).
\ee 
Substituting (\ref{e-3-2-32}) and (\ref{e-3-2-33}) in (\ref{e-3-2-31}), we find 
\be
\label{e-3-2-34}
I_y = \frac{1}{(4\pi)^n} \prod_{i=1}^n \prod_{m=1}^\infty \left[ 1 + \frac{( \ii \pi x_i)^2}{\pi^2 m^2} \right]^{-1/2} = \frac{1}{(4\pi)^n} \prod_{i=1}^n \frac{ \ii \pi  x_i}{\sinh(  \ii \pi x_i)}
\ee
Putting everything together, we write the anomaly as
\be
\label{e-3-2-35}
G_{1/2} (\alpha) = 2 \alpha \left( \frac{\ii}{2\pi} \right)^n \int \dd^{2n}  x_0 \dd^{2n}  \psi_0 \prod_{i=1}^n \frac{ \ii \pi x_i}{\sinh(  \ii \pi x_i )} \tr e^F
\ee
To saturate the $\psi_0$ integral, we need the term that contains exactly $2n$ occurrences of $\psi_0$. Then, the prefactor $(\ii / 2\pi)^n$ can be absorbed in the $x_i$'s and $F$. Making use of the identity
\be
\label{e-3-2-36}
\int \dd^{2n}  x_0 \dd^{2n}  \psi_0 \psi_0^{a_1} \ldots \psi_0^{a_{2n}} = \int_{M_{2n}} dx^{a_1} \wedge \ldots \wedge dx^{a_{2n}},
\ee
we arrive at our final expression for the anomaly,
\be
\label{e-3-2-37}
G_{1/2} (\alpha) = 2 \alpha \int_{M_{2n}} \left\{ \left[ \prod_{i=1}^n \frac{x_i/2}{\sinh( x_i/2) } \right] \tr e^{\ii F / 2 \pi} \right\}_{2n} ,
\ee
where the $x_i$'s and $F$ are from now on understood as 2-forms in the usual sense and where the subscript ``${2n}$'' indicates that only terms proportional to the volume form on $M_{2n}$ should be kept.

The objects in (\ref{e-3-2-37}) are quite familiar from the theory of characteristic classes. The product in square brackets is just the Dirac genus $\widehat{A}(R)$, while $\tr e^{\ii F / 2 \pi}$ is the Chern character $\cch_{\MR} (F)$ of the bundle associated with the gauge field. Therefore, the anomaly may be written as 
\be
\label{e-3-2-38}
G_{1/2} (\alpha) = 2 \alpha \int_{M_{2n}} \left[ \widehat{A}(R) \cch_{\MR}(F) \right]_{2n}.
\ee 
Recalling that the anomaly is also $2 \alpha$ times the index of $\ii \fslash D$, we have actually proven the Atiyah-Singer index theorem for the curved-space Dirac operator. In the literature, this is known as the supersymmetric proof \cite{Alvarez-Gaume:1983at,Friedan:1983xr}. 

\subsection{The spin--$3/2$ anomaly}
\label{sec-3-2-2}

The above method can be extended in a straightforward way to the spin--$3/2$ chiral anomaly. The trick there is to enlarge the gauge group by adding an extra $\mathrm{SO}(2n)$ factor with the gauge field being the spin connection on $M_{2n}$ so as to yield an extra vector index for the fermion; the relevant $c$--fields must be taken to transform in the vector representation. Therefore, we may suspect that the spin--$3/2$ chiral anomaly is obtained by the spin--$1/2$ mixed chiral anomaly by inserting a factor of $\tr e^{\ii R / 2\pi}$ to account for the extra $\mathrm{SO}(2n)$ factor, i.e. is given by
\be
\label{e-3-2-39}
G_{3/2} (\alpha) \stackrel{?}{=} 2 \alpha \int \left[ \widehat{A}(R) \tr e^{\ii R / 2\pi} \cch_{\MR}(F) \right]_{2n} .
\ee
However, Eq. (\ref{e-3-2-39}) would represent the spin $3/2$ anomaly if all degrees of freedom of the gravitino $\psi_\mu$ were unconstrained. However, the chiral gravitino $\psi_\mu$ is actually subject to the constraints $\partial_\mu \psi^\mu = 0$, $\psi_\mu \cong \psi_\mu + \partial_\mu \chi$ and $\Gamma^\mu \psi_\mu = 0$, whose net effect is the removal of one spin $1/2$ degree of freedom of the same chirality as $\psi_\mu$. This can be taken into account by subtracting the spin--$1/2$ chiral anomaly from (\ref{e-3-2-39}), leading to
\be
\label{e-3-2-40}
G_{3/2} (\alpha) = 2 \alpha \int \left[ \widehat{A}(R) \left( \tr e^{\ii R / 2\pi} -1 \right) \cch_{\MR}(F) \right]_{2n} .
\ee 
By now it should be no surprise that the integral on the RHS of (\ref{e-3-2-40}) is actually the index of the Rarita-Schwinger operator. 

\subsection{The self-dual $(n-1)$--form anomaly}
\label{sec-3-2-3}

Another field that gives rise to a singlet gravitational anomaly in $2n=4k$ dimensions is an $(n-1)$--form potential $A_{n-1}$ with a (anti-)self-dual $n$--form field strength $F_n$. Although such a field is {\it bosonic} rather than fermionic, it gives rise to anomalies due to the fact that the antisymmetric tensor representations of the Lorentz group with (anti-)self-dual field strengths are constructed by taking the tensor product of two Weyl spinors with equal chirality. The self-dual $(n-1)$--form anomaly may be calculated as before. However, we have established relations between anomalies and index theorems that enable us to deduce its form rather easily. In particular, we expect that the anomaly is related to the index of the operator $\ii \fslash D_\phi$ associated with a bispinor $\phi_{\alpha\beta}$, which is equal to the integral $\int \left[ L(R) \right]_{2n}$ of the Hirzebruch polynomial. To find the correct relation, we must take into account the various constraints satisfied by $A_{n-1}$. In particular, the second spinor index of $\phi_{\alpha\beta}$ should be projected to the same chirality as the first index, while the (Minkowski-space) field strength of $A_{n-1}$ is required to be real. Moreover, there is an extra minus sign since the $(n-1)$--form obeys Bose rather than Fermi statistics. This leads to an overall factor of $- \frac{1}{2} \cdot \frac{1}{2} = -\frac{1}{4}$ so that the anomaly reads
\be
\label{e-3-2-41}
G_A (\alpha) = 2 \alpha \left( - \frac{1}{4} \right) \ind ( \ii \fslash D_\phi ) = 2 \alpha \int_{M_{2n}} \left( - \frac{1}{4} \right) \left[ L(R) \right]_{2n} .
\ee
This result coincides with that of the explicit calculation in \cite{Alvarez-Gaume:1984ig}. 

\section{Calculation of Nonabelian Anomalies} 
\label{sec-3-3}

The formalism of the previous section can be extended in a straightforward way to compute nonabelian anomalies. As before, we will present a detailed calculation for the spin $1/2$ case and deduce the results for the spin $3/2$ and self-dual $(n-1)$--form case. 

\subsection{The spin--$1/2$ anomaly} 
\label{sec-3-3-1}

The calculation of the spin $1/2$ nonabelian anomalies proceeds in much the same way as for the chiral anomalies. As before, we will use Fujikawa's method to transcribe the anomalies to functional traces and then we will compute the latter using supersymmetric path integrals.

\subsubsection{The calculation}

Starting from the gauge anomaly (\ref{e-3-1-56a}), we can again use the supersymmetric sigma model, the only difference being the insertion of $v$. So, the gauge anomaly is given by the path integral
\be
\label{e-3-3-1}
\tilde{G}_{1/2} (v) = \lim_{\beta \to 0} \tr_{c,\bar{c}} \int_\mathrm{P} \mathcal{D} x \mathcal{D} \psi v e^{- \int_0^1 \dd \tau R^{(2)}_\mathrm{E}}.
\ee
where $R_\mathrm{E}$ is given by (\ref{e-3-2-25}). To proceed, it is convenient to exponentiate this term according to 
\be
\label{e-3-3-2}
v \to \tr_{c,\bar{c}} e^{ \bar{c}_A v^A_{\phantom{A} B} c^B },
\ee
keeping in mind that only terms linear in $v$ must be retained. This way, we find the expression
\be
\label{e-3-3-3}
\tilde{G}_{1/2}(v) = - \lim_{\beta \to 0} \tr_{c,\bar{c}} \left[ \int_\mathrm{P} \mathcal{D} x \mathcal{D} \psi e^{-\int_0^1 \dd\tau R^{(2)}_\mathrm{E} (F')} \right]_{\mathcal{O}(v)}. 
\ee 
which has a form similar to (\ref{e-3-2-20}), but with $F$ replaced by
\be
\label{e-3-3-4}
F' = F + v.
\ee
So, we immediately read off the gauge anomaly
\be
\label{e-3-3-5}
\tilde{G}_{1/2}(v) = - \left( \frac{\ii}{2\pi} \right)^n \int \dd^{2n}  x_0 \dd^{2n} \psi_0 \prod_{i=1}^n \frac{ \ii \pi x_i}{\sinh( \ii \pi x_i )} \left[ \tr e^{F'} \right]_{\mathcal{O}(v)}.
\ee
To isolate the terms that saturate the $\psi_0$ integral and are linear in $v$, we have to pick the monomials containing $n+1-k$ $x_i$'s and $k$ $F$'s. Absorbing $n+1$ factors of $(\ii / 2\pi)$ in the $x_i$'s and $F'$, we write
\be
\label{e-3-3-6}
\tilde{G}_{1/2}(v) = 2 \pi \ii \int_{M_{2n}} \left\{ \prod_{i=1}^n \frac{x_i/2}{\sinh( x_i/2) }  \left[ \tr e^{\ii F' / 2 \pi} \right]_{\mathcal{O}(v)} \right\}_{2n+2} .
\ee 

Turning to the gravitational anomaly (\ref{e-3-1-58}), we must evaluate the supersymmetric path integral with the insertion of the sigma-model quantity corresponding to $\xi^\mu D_\mu$. As seen from (\ref{e-3-2-17}), this quantity is given by $\ii \xi_\mu \dot{x}^\mu$ so that the appropriate path integral is
\be
\label{e-3-3-7}
\tilde{G}_{1/2}(\xi) = \lim_{\beta \to 0} \tr_{c,\bar{c}} \int_\mathrm{P} \mathcal{D} x \mathcal{D} \psi \xi_\mu \dot{x}^\mu e^{- \int_0^1 \dd \tau R_\mathrm{E}}.
\ee
where we took account of the Euclidean continuation. Recalling that at the high-temperature limit we expand about constant solutions, we must also consider the expansion of $\xi_\mu \dot{x}^\mu$, given by
\be
\label{e-3-3-8}
\xi_\mu \dot{x}^\mu = [ \xi_\mu (x_0) + y^\nu \partial_\nu \xi_\mu (x_0) ] \dot{y}^\mu \to D_a \xi_b(x_0) y^a\dot{y}^b,
\ee 
where the last equation follows from the fact that $\xi_\mu \dot{y}^\mu$ does not contribute and from the use of Riemann normal coordinates. To proceed, we use the exponentiation
\be
\label{e-3-3-9}
- D_a \xi_b y^a \dot{y}^b \to e^{ - \int_0^1 \dd \tau D_a \xi_b y^a \dot{y}^b } = e^{ - \int_0^1 \dd \tau D_{[a} \xi_{b]} y^a \dot{y}^b },
\ee  
with the understanding that only terms linear in $\xi$ must be kept. This way, we find the expression
\be
\label{e-3-3-10}
\tilde{G}_{1/2}(\xi) = - \lim_{\beta \to 0} \tr_{c,\bar{c}} \left[ \int_\mathrm{P} \mathcal{D} x \mathcal{D} \psi e^{-\int_0^1 \dd\tau R^{(2)}_\mathrm{E} (R'_{ab}) } \right]_{\mathcal{O}(\xi)}. 
\ee 
which has a form similar to (\ref{e-3-2-20}), but with $R_{ab}$ replaced by
\be
\label{e-3-3-11}
R'_{ab} = R_{ab} + D_a \xi_b - D_b \xi_a.
\ee
So, the gravitational anomaly is given by
\be
\label{e-3-3-12}
\tilde{G}_{1/2}(\xi) = - \left( \frac{\ii}{2\pi} \right)^n \int \dd^{2n}  x_0 \dd^{2n}  \psi_0 \left[ \prod_{i=1}^n \frac{ \ii \pi x'_i}{\sinh(  \ii \pi x'_i )} \right]_{\mathcal{O}(\xi)} \tr e^F .
\ee
where $x'_i$ are the skew-eigenvalues of $R'_{ab}/2\pi$. Picking up the terms that saturate the $\psi_0$ integral and are linear in $\xi$, we write
\be
\label{e-3-3-13}
\tilde{G}_{1/2}(\xi) = 2 \pi \ii \int_{M_{2n}} \left\{ \left[ \prod_{i=1}^n \frac{x'_i/2}{\sinh( x'_i/2) } \right]_{\mathcal{O}(\xi)} \tr e^{\ii F / 2 \pi} \right\}_{2n} .
\ee 
The purely gravitational anomaly corresponds to the case $F=0$, in which case the terms of interest contain $n+1$ $x'_i$'s. Since the integrand in (\ref{e-3-3-13}) is even in the $x'_i$'s, $n+1$ must also be even. Thus, we establish that spin $1/2$ gravitational anomalies can only occur in $D=4k+2$ dimensions. 

The above results can also be represented by the use of characteristic classes. Indeed, the anomaly under gauge transformations can be written in the form
\be
\label{e-3-3-14}
\tilde{G}_{1/2}(v) = 2 \pi \ii \int_{M_{2n}} \left\{ \widehat{A}(R) \left[ \cch_{\MR}(F') \right]_{\mathcal{O}(v)} \right\}_{2n},
\ee 
while the anomaly under GCT's is given by
\be
\label{e-3-3-15}
\tilde{G}_{1/2}(\xi) = 2 \pi \ii \int_{M_{2n}} \left\{ \left[ \widehat{A}(R') \right]_{\mathcal{O}(\xi)}  \cch_{\MR}(F) \right\}_{2n} .
\ee 
To apply the above expressions, one has to use the characteristic class formulas of () keeping the terms that are proportional to the volume form on $M_{2n}$ and linear in the transformation parameters. The gauge, gravitational and mixed contributions are then obtained in the way specified in \S\ref{sec-3-1-2}.

Finally, we are now in a position to understand why the anomalies computed above are covariant. Considering the gauge case, Eq. (\ref{e-3-3-14}) is an expression involving only the curvature $F$ and the parameter $v$ and thus the divergence of the associated current is guaranteed to transform covariantly under gauge transformations. The reason for this can be traced back to Eq. (\ref{e-3-1-56a}): since the regulator involves the ordinary Dirac operator $\ii \fslash D$, there is just one occurrence of $\Gamma_{2n+1}$ and the relevant contributions correspond to Feynman diagrams with one axial and $n$ vector currents. In contrast, the diagrams required for the computation of the consistent anomaly involve $n+1$ $V-A$ currents, leading to a different expression involving the connection $A$ explicitly. Moreover, the extra Bose symmetry present in the latter case yields an extra $1/(n+1)$ factor for the leading $v(\dd A)^n$ terms. This information, along with the WZ condition, is all we need in order to obtain the consistent anomalies from the covariant anomalies found here.

\subsection{The spin--$3/2$ anomaly}
\label{sec-3-3-2}

The spin $3/2$ anomalies are obtained in a straightforward way by applying the modifications of the previous subsection to the results of \S\ref{sec-3-2-2}. So, the anomaly under gauge transformations is
\be
\label{e-3-3-16}
\tilde{G}_{3/2}(v) = 2 \pi \ii \int_{M_{2n}} \left\{ \widehat{A}(R) \left( \tr e^{\ii R / 2\pi} -1 \right) \left[ \cch_{\MR}(F') \right]_{\mathcal{O}(v)} \right\}_{2n},
\ee 
the anomaly under GCT's is
\be
\label{e-3-3-17}
\tilde{G}_{3/2}(\xi) = 2 \pi \ii \int_{M_{2n}} \left\{ \left[ \widehat{A}(R') \left( \tr e^{ \ii R' / 2\pi} - 1 \right) \right]_{\mathcal{O}(\xi)} \cch_{\MR}(F) \right\}_{2n} ,
\ee 
and the gauge, gravitational and mixed contributions are collected as before. Again, one easily sees that spin $3/2$ gravitational anomalies can only occur in $D=4k+2$ spacetime dimensions.

\subsection{The self-dual $(n-1)$--form anomaly}
\label{sec-3-3-3}

Let us finally turn to the case of the self-dual $(n-1)$--form anomaly, which is purely gravitational. By the same arguments as before, one finds the expression
\be
\label{e-3-3-18}
\tilde{G}_A (\xi) = 2 \pi \ii \int_{M_{2n}} \left( - \frac{1}{4} \right) \left\{ \left[ L(R') \right]_{\mathcal{O}(\xi)}  \right\}_{2n},
\ee
and, again, the anomaly is nonvanishing only in the case $D=4k+2$.

\subsection{The consistent anomaly from the descent equations}
\label{sec-3-3-4}

The expressions obtained in the previous section for nonabelian anomalies have a strong resemblance to those for chiral anomalies, suggesting that there also exists an interpretation of the former in terms of index theorems. On the other hand, a close look at these equations reveals that the contributions to the anomaly in $2n$ dimensions come from monomials with $n+1$ powers of the curvatures so that the ``index theorems'' at hand, if any, must actually refer to $2n+2$ dimensions. This could be regarded as a hint that gauge/gravitational anomalies in $2n$ dimensions are somehow related to chiral anomalies in $2n+2$ dimensions. To investigate this relation, there are several approaches (see for example \cite{Alvarez:1984yi} and \cite{Alvarez-Gaume:1984cs}). Here, will use the simplest of these methods, which consists in applying the WZ consistency condition and the descent equations discussed in Appendix \ref{app-d} to deduce the form of the consistent anomaly and its relation to the chiral one. 

Let us for simplicity consider the spin--$1/2$ gauge anomaly in $2n$ dimensions. Following the formalism of \S\ref{sec-c-2}, we introduce extra coordinates $\{\theta^\alpha\}$, we apply a gauge transformation $g(x,\theta)$ to transform the gauge field $A$ to the $\theta$--dependent gauge field $\widehat{A} = g^{-1} A g + g^{-1} \dd g$ and we introduce the BRS operator $s$ whose action on $\hat{A}$ is a gauge transformation with parameter $\hat{v} = g^{-1} s g$. Let us now specialize to the case where the 1--form $\hat{v}$ takes the infinitesimal value
\be
\label{e-3-3-19}
\hat{v} = v_\alpha \dd\theta^\alpha ;\qquad v_\alpha = g^{-1} \frac{\partial}{\partial \theta^\alpha} g,
\ee 
and set $\hat{A} = A$, i.e. evaluate all quantities at $\theta=0$. Then, the action of $\sss$ on $A$ yields
\be
\label{e-3-3-20}
\sss A = - ( \dd \hat{v} + [\hat{A},\hat{v}]  ) \bigr|_{\theta=0} = \dd \theta^\alpha ( \dd v_\alpha + [A,v_\alpha] ) = \dd \theta^\alpha \delta_{v_\alpha} A,
\ee 
that is, amounts to a gauge transformation with parameter $\hat{v}$. It follows that the action of $\sss$ on the effective action yields the gauge anomaly,
\be
\label{e-3-3-21}
G (\hat{v}) = - \delta_{\hat{v}} \varGamma = - \sss \varGamma.
\ee 
Then, due to the nilpotency of $\sss$, $G (\hat{v})$ is $\sss$--closed,
\be
\label{e-3-3-22}
\sss G (\hat{v}) = 0.
\ee
To write this condition in a more familiar form, we consider the case where there are just two extra coordinates $\theta^\alpha$ so that $\hat{v} = v_1 \dd \theta^1 + v_2 \dd \theta^2$. Then, we can write the condition (\ref{e-3-3-22}) in the form
\bea
\label{e-3-3-23}
0 &=& \sss \int \dd^{2n}x \sqrt{g} \tr( \hat{v} G) = \int \dd^{2n}x \sqrt{g} \tr [ \sss (\hat{v} G) - (\sss \hat{v} + \hat{v}^2) G ]
= - \int \dd^{2n}x \sqrt{g} \tr ( \hat{v} \sss G + \hat{v}^2 G )
\nonumber\\ &=& \dd\theta^1 \dd\theta^2 \int \dd^{2n}x \sqrt{g} \tr \{ ( v_2 \delta_{v_1} - v_1 \delta_{v_2} ) G - [v_1,v_2] G \},
\eea
which is just the WZ consistency condition! So, an anomaly that is $\sss$--closed is automatically consistent.

Next, consider the \emph{singlet} anomaly in $2n+2$ dimensions. For later convenience, we write a general singlet anomaly in the form
\be
\label{e-3-3-24}
G^{(2n+2)} (\alpha) = \frac{1}{\pi} \alpha \int_{M_{2n+2}} \hat{I}_{2n+2}
\ee 
where $\hat{I}_{2n+2}$ equals $2\pi$ times the integrand in (\ref{e-3-2-38}), (\ref{e-3-2-40}) or (\ref{e-3-2-41}). Since $\hat{I}_{2n+2}$ is a closed form, it can be expressed in terms of a Chern-Simons form as 
\be
\label{e-3-3-25}
\hat{I}_{2n+2} = \dd \hat{I}_{2n+1}.
\ee
Acting with the BRS operator $\sss$ on $\hat{I}_{2n+1}$ and using the first descent equation, we find
\be 
\label{e-3-3-26}
\sss \hat{I}_{2n+1} = - \dd \hat{I}^1_{2n},
\ee
Finally, acting with $\sss$ on $\hat{I}^1_{2n}$, using the second of the descent equations and integrating on a $2n$--dimensional boundary-free submanifold $M_{2n}$, we find
\be
\label{e-3-3-27}
\sss \int_{M_{2n}} \hat{I}^1_{2n} = - \int_{M_{2n}} \dd \hat{I}^2_{2n-1} = 0.
\ee
that is, $\int_{M_{2n}} \hat{I}^1_{2n}$ is $\sss$--closed.

According to the previous paragraphs, an obvious candidate for the consistent anomaly is 
\be
\label{e-3-3-28}
G_{1/2}^{(2n)} (v) = c \int_{M_{2n}} \hat{I}^1_{2n},
\ee
and, in fact it is the only choice since the solution of the WZ consistency condition is uniquely defined up to normalization. To determine the coefficient $c$, we consider the leading term (i.e. the term containing the largest number of powers of $\dd A$) of the above expression. It is given by
\be
\label{e-3-3-29}
G_{1/2}^{(2n)} (v) \bigr|_\mathrm{leading} = c \frac{\ii}{(n+1)!} \left( \frac{\ii}{2\pi} \right)^n \int_{M_{2n}} \tr \left[ v ( \dd A )^n \right].
\ee
On the other hand, the leading term of the \emph{covariant} gauge anomaly (\ref{e-3-3-14}) is
\be
\label{e-3-3-30}
\tilde{G}_{1/2}^{(2n)} (v) \bigr|_\mathrm{leading}  = - \frac{1}{n!} \left( \frac{\ii}{2\pi} \right)^n \int_{M_{2n}} \tr \left[ v (\dd A)^n \right].
\ee 
and the leading term of the \emph{consistent} gauge anomaly must be given by the same expression, divided by $n+1$ to account for Bose symmetrization. So, we have
\be
\label{e-3-3-31}
G_{1/2}^{(2n)} (v) \bigr|_\mathrm{leading}  = - \frac{1}{(n+1)!} \left( \frac{\ii}{2\pi} \right)^n \int_{M_{2n}} \tr \left[ v (\dd A)^n \right].
\ee 
This fixes the normalization factor to $c=\ii$ and so the consistent anomaly is given by
\be
\label{e-3-3-32}
G_{1/2}^{(2n)} (v) = \ii \int_{M_{2n}} \hat{I}^1_{2n},
\ee
as can be verified by explicit computation. The same arguments hold for all other anomalies. 

\section{The Anomaly Polynomials}
\label{sec-3-4}

The end result of the previous section is that the consistent nonabelian anomalies in $2n$ dimensions follow from singlet anomalies in $2n+2$ dimensions through the descent equations. We are now in a position to summarize these results and express the anomaly structure of a theory in a compact and elegant way through the so-called anomaly polynomials.

From this section on, we will switch back to Minkowski spacetime through the Wick rotation $G_{\mathrm{M}} = -G_{\mathrm{E}}$ and $\varGamma_{\mathrm{M}} = \ii \varGamma_{\mathrm{E}}$. Our starting point is Eq. (\ref{e-3-3-24}) for singlet anomalies in $2n+2$ dimensions, whose Minkowski continuation reads
\be
\label{e-3-4-1}
G^{(2n+2)} (\alpha) = - \frac{1}{\pi} \alpha \int_{M_{2n+2}} \hat{I}_{2n+2},
\ee
The explicit expressions for $\hat{I}_{2n+2}$ for the various cases are given by
\bea
\label{e-3-4-2}
\hat{I}^{1/2}_{2n+2} &=& 2 \pi \left[ \widehat{A}(R) \cch_{\MR} (F) \right]_{2n+2}, \nn\\
\hat{I}^{3/2}_{2n+2} &=& 2 \pi \left[ \widehat{A}(R) \left( \tr e^{ \ii R / 2\pi } -1 \right) \cch_{\MR}(F) \right]_{2n+2}, \nn\\
\hat{I}^{A}_{2n+2} &=& 2\pi \left[ \frac{1}{2} \left( \frac{1}{4} \right) L(R) \right]_{2n+2} .
\eea
where the superscripts ``$1/2$'', ``$1/2$'' and ``$A$'' refer respectively to spin--$1/2$, spin--$3/2$ and antisymmetric $n$--form fields. Using the descent equations
\be
\label{e-3-4-3}
\dd \hat{I}_{2n+1} = \hat{I}_{2n+2} ,\qquad \delta_{v,\lambda} \hat{I}_{2n+1} = - \dd \hat{I}^1_{2n},
\ee
one then obtains the consistent Minkowski anomalies according to
\be
\label{e-3-4-4}
\delta_{v,\lambda} \varGamma = - G^{(2n)} (v,\lambda) = - \int_{M_{2n}} \hat{I}^1_{2n},
\ee
The explicit expressions for $\hat{I}^1_{2n}$ are given by
\bea
\label{e-3-4-5}
\hat{I}_{2n}^{1,1/2}(v) &=& 2 \pi \left[ \widehat{A}(R) \cch^1_{\MR}(v,A,F) \right]_{2n}, \nn\\
\hat{I}_{2n}^{1,1/2}(\lambda) &=& 2 \pi \left[ \widehat{A}^1(\lambda,\omega,R) \cch_{\MR}(F) \right]_{2n}, \nn\\
\hat{I}_{2n}^{1,3/2}(v) &=& 2 \pi \left[ \widehat{A}_{3/2}(R) \cch^1_{\MR}(v,A,F) \right]_{2n} ,\nn\\ 
\hat{I}_{2n}^{1,3/2}(\lambda) &=& 2 \pi \left[ \widehat{A}_{3/2}^1(\lambda,\omega,R) \cch_{\MR}(F) \right]_{2n} ,\nn\\
\hat{I}_{2n}^{1,A} (\lambda) &=& 2\pi \left[ \frac{1}{2} \left( \frac{1}{4} \right) L^1(\lambda,\omega,R) \right]_{2n}.
\eea
Here, we have used the shorthand $\widehat{A}_{3/2}(R) \equiv \widehat{A}(R) \left( \tr e^{\ii R / 2\pi} -1 \right)$ and the superscript ``$1$'' over a characteristic class to denote its descent. Also, the last equations in (\ref{e-3-4-2}) and (\ref{e-3-4-2}) include an extra factor of $1/2$: this is a consequence of the fact that, for the case of $2n$--dimensional anomalies, the trace must be taken over $\mathrm{SO}(2n-1,1)$ rather than $\mathrm{SO}(2n+1,1)$ spinor indices.

So, in the end, the complete anomaly structure of a $2n$--dimensional theory is completely determined by the formal $(2n+2)$--forms $\hat{I}_{2n+2}$. Expanding these forms according to the formulas of Appendix \ref{app-d}, one easily sees that they always contain prefactors proportional to $1/ (2\pi)^n$. To further simplify the notation, it is useful to absorb such factors in a suitable redefinition of the anomaly polynomials. For the cases of interest, $n=5$ and $n=3$, we find it convenient to define 
\be
\label{e-3-4-6}
I_{12} \equiv 720 (2\pi)^5 \hat{I}_{12} ,\qquad I_8 \equiv -16 (2\pi)^3 \hat{I}_8
\ee
The quantities $I_{2n+2}$ thus defined are called the \emph{anomaly polynomials}. The gauge, gravitational and mixed contributions to these polynomials are extracted by keeping the terms involving only $F$, the terms involving only $R$ and the terms involving both $F$ and $R$ respectively and they are denoted by $I_{2n+2}(F)$, $I_{2n+2}(R)$ and $I_{2n+2}(F,R)$. The results for the spacetime dimensions of interest, $D=10$ and $D=6$ are as follows.
\begin{itemize}
\item $D=10$. The anomaly polynomials in 10 dimensions are given by
\bea
\label{e-3-4-7}
I^{1/2}_{12} (R)   &=& - \frac{1}{504} \tr R^6 - \frac{1}{384} \tr R^4 \tr R^2 - \frac{5}{4608} (\tr R^2)^3, \nn\\
I^{1/2}_{12} (F)   &=& \tr F^6, \nn\\
I^{1/2}_{12} (F,R) &=& \frac{1}{16} \tr R^4 \tr F^2 + \frac{5}{64} (\tr R^2)^2 \tr F^2 - \frac{5}{8} \tr R^2 \tr F^4, \nn\\
I^{3/2}_{12} (R)   &=& \frac{55}{56} \tr R^6 - \frac{75}{128} \tr R^4 \tr R^2 + \frac{35}{512} (\tr R^2)^3 ,\nn\\
I^A_{12} (R) 	   &=& - \frac{496}{504} \tr R^6 + \frac{7}{12} \tr R^4 \tr R^2 - \frac{5}{72} (\tr R^2)^3. 
\eea
In the above, we have neglected the spin--$3/2$ gauge and mixed anomalies due to the fact that the gravitino does not couple to gauge fields in the theories of interest.
\item $D=6$. The anomaly polynomials in 6 dimensions are given by
\bea
\label{e-3-4-8}
I^{1/2}_{8} (R) 	  &=& \frac{1}{360} \tr R^4 + \frac{1}{288} (\tr R^2)^2, \nn\\
I^{1/2}_{8} (F) 	  &=& \frac{2}{3} \tr F^4, \nn\\
I^{1/2}_{8} (F_X,F_Y) &=& 4 \tr F_X^2 \tr F_Y^2, \nn\\
I^{1/2}_{8} (F,R) 	  &=& - \frac{1}{6} \tr R^2 \tr F^2, \nn\\
I^{3/2}_{8} (R) 	  &=& \frac{49}{72} \tr R^4 - \frac{43}{288} (\tr R^2)^2 ,\nn\\
I^{3/2}_{8} (F) 	  &=& \frac{10}{3} \tr F^4, \nn\\
I^{3/2}_{8} (F,R) 	  &=& \frac{19}{6} \tr R^2 \tr F^2, \nn\\
I^A_{8} (R) 		  &=& - \frac{7}{90} \tr R^4 + \frac{1}{36} (\tr R^2)^2.
\eea
The third polynomial in the list is relevant to the case where there are spin-1/2 fermions charged under two gauge group factors, $\MG_X$ and $\MG_Y$, and can be derived by a straightforward generalization of the above methods.	
\end{itemize} 

\section{Anomaly Cancellation}
\label{sec-3-5}

As stressed in \S\ref{sec-3-1-3}, gauge, gravitational and mixed anomalies threaten the consistency of physical theories even if the latter are regarded as effective low-energy models. The only way out is that these anomalies cancel by some mechanism. The simplest case is when the anomalies of a theory cancel out when summed over all channels, as happens in the Standard Model. However, in effective theories, anomaly cancellation may be achieved even if the net anomaly is nonzero; this may happen if the effective action contains terms involving $p$--form fields whose classical variation is equal and opposite to the quantum anomaly. The first and most known example is the celebrated {\it Green-Schwarz mechanism} \cite{Green:1984sg} (see also \cite{Green:1987mn}), which we will describe in detail here.

\subsection{The Green-Schwarz mechanism} 
\label{sec-3-5-1}

The basic idea behind the Green-Schwarz mechanism is that, under certain conditions, the anomalies of a theory (a one-loop effect) may be cancelled by the anomalous variation of certain classical terms in the effective action (a tree-level effect). To explain this rather counterintuitive idea, it is useful to recall what a low-energy effective action really is. In the context of effective theories (e.g. the low-energy supergravities of string theory), the low-energy effective action is usually said to be the action obtained by \emph{truncating} the massive modes. However, according to its proper definition, the effective action is the action obtained by \emph{integrating out} the massive modes. This would result in additional terms involving irrelevant higher-derivative operators. Since these terms have no a priori reason to respect the gauge symmetries of the theory, they can generally have anomalous variations. It may be then the case that these variations cancel the anomalies of the naive low-energy theory. 

Let us consider a $2n$--dimensional theory containing nonabelian anomalies and let $I_{2n+2}$ be the total anomaly polynomial. Then the variation of the effective action under a gauge/Lorentz transformation is given by
\be 
\label{e-3-5-1}
\delta_{v,\lambda} \varGamma = - \int_{M_{2n}} I^1_{2n} (v,\lambda,F,R), 
\ee 
where $I^1_{2n}$ is obtained from $I_{2n+2}$ through the descent equations 
\be
\label{e-3-5-2}
I_{2n+2} = \dd I_{2n+1} ,\qquad \delta_{v,\lambda} I_{2n+1} = - \dd I^1_{2n}. 
\ee 
Now, {\it suppose} that (i) the theory under consideration contains a $2p$--form potential $B_{2p}$ and (ii) the anomaly polynomial $I_{2n+2}$ can be written in the {\it factorized} form
\be 
\label{e-3-5-3}
I_{2n+2} = \Omega_{2p+2} \Omega_{2n-2p},
\ee
where $\Omega_{2p+2}$ and $\Omega_{2n-2p}$ are two polynomials constructed out of the same curvature invariants as $I_{2n+2}$. Then the anomalies of the theory can be cancelled by adding to the action suitable terms involving $B_{2p}$, $\Omega_{2p+2}$ and $\Omega_{2n-2p}$. We must stress here that the factorization condition (\ref{e-3-5-3}) is not trivially satisfied: in general, it imposes very stringent constraints on the field content of the theory.

Let us calculate the variation of the effective action corresponding to the anomaly polynomial (\ref{e-3-5-3}). By the first of (\ref{e-3-5-2}), $I_{2n+2}$ is exact and thus the same must hold for $\Omega_{2p+2}$ and $\Omega_{2n-2p}$. Therefore, the latter can be expressed in terms of Chern-Simons forms as follows
\be 
\label{e-3-5-4}
\Omega_{2p+2} = \dd \Omega_{2p+1} ,\qquad \Omega_{2n-2p} = \dd \Omega_{2n-2p-1}.
\ee  
Then, since $\Omega_{2p+2}$ and $\Omega_{2n-2p}$ are closed, we see that the $(2n+1)$--forms
\be
\label{e-3-5-5}
I^{(a)}_{2n+1} = \Omega_{2p+2} \Omega_{2n-2p-1} ,\qquad I^{(b)}_{2n+1} = \Omega_{2p+1} \Omega_{2n-2p}.
\ee
are candidates for the Chern-Simons form $I_{2n+1}$ of $I_{2n+2}$. Thus, the most general form of $I_{2n+1}$ is given by
\be
\label{e-3-5-6}
I_{2n+1} = a \Omega_{2p+2} \Omega_{2n-2p-1} + (1-a) \Omega_{2p+1} \Omega_{2n-2p},
\ee
where the coefficient $a$ reflects the ambiguity in the definition of $I_{2n+1}$ up to an exact form and may be deduced by the WZ consistency condition. Now, considering the variations of $\Omega_{2p+1}$ and $\Omega_{2n-2p-1}$ under a local gauge/Lorentz transformation,
\be
\label{e-3-5-7}
\delta_{v,\lambda} \Omega_{2p+1} = \dd \Omega^1_{2p} ,\qquad \delta_{v,\lambda} \Omega_{2n-2p-1} = \dd \Omega^1_{2n-2p-2},
\ee
we find that the corresponding variation of $I_{2n+1}$ is given by the second of (\ref{e-3-5-2}) with
\be
\label{e-3-5-8}
I^1_{2n} = - a \Omega_{2p+2} \Omega^1_{2n-2p-2} - ( 1 - a ) \Omega^1_{2p} \Omega_{2n-2p},
\ee
Therefore, the anomalous variation of the effective action under the transformation is given by
\be
\label{e-3-5-9}
\delta_{v,\lambda} \varGamma = \int_{M_{2n}} \left[ a \Omega_{2p+1} \dd \Omega^1_{2n-2p-2} + ( 1 - a ) \Omega^1_{2p} \Omega_{2n-2p} \right].
\ee

Next, consider the $2p$--form field $B_{2p}$. Normally, this field does not transform under {\it Yang-Mills} gauge transformations and LLT's, while its field-strength is defined in the usual way, $H_{2p+1} = \dd B_{2p}$. Here, we endow $B_{2p}$ with the gauge/Lorentz transformation property
\be
\label{e-3-5-10}
\delta_{v,\lambda} B_{2p} = \Omega^1_{2p},
\ee
and we replace $H_{2p+1}$ by the modified field-strength
\be
\label{e-3-5-11}
\tilde{H}_{2p+1} = \dd B_{2p} - \Omega_{2p+1},
\ee
which is gauge and Lorentz invariant. Then, we add to the action the {\it Green-Schwarz term}
\be
\label{e-3-5-12}
S_{GS} = - \int_{M_{2n}} \left( B_{2p} \Omega_{2n-2p} + a \Omega_{2p+1} \Omega_{2n-2p-1}  \right),
\ee 
which contains the interaction $B_{2p} \Omega_{2n-2p}$ plus an irrelevant counterterm. The classical variation of this term is given by
\bea
\label{e-3-5-13}
\delta S_{GS} &=& - \int_{M_{2n}} \left( \Omega^1_{2p} \Omega_{2n-2p} + a \dd \Omega^1_{2p} \Omega_{2n-2p-1} + a \Omega_{2p+1} \dd \Omega^1_{2n-2p-2} \right) \nonumber\\ &=& - \int_{M_{2n}} \left[ a \Omega_{2p+1} \dd \Omega^1_{2n-2p-2} + (1-a) \Omega^1_{2p} \Omega_{2n-2p} \right],
\eea
that is, it is equal and opposite to the variation of $\varGamma$. Hence, the inclusion of this term renders the theory anomaly-free since
\be
\label{e-3-5-14}
\delta ( \varGamma + S_{GS} ) = 0.
\ee  
This is the Green-Schwarz anomaly cancellation mechanism.

From the above, the inclusion of the Green-Schwarz term may seem a little {\it ad hoc}. In the context of model building, this is not a serious problem: one simply regards this term as a necessary addition to the action of the theory in order to restore its consistency at the quantum level. However, in the context of a low-energy effective theory, this term must be present in the underlying theory because the latter would be otherwise inconsistent. Fortunately, it turns out that, in all theories of the second type that are of interest, such a term exists. The reason why it may not be present in the effective action in the first place is because it may correspond to a higher-derivative correction.

\section{Global anomalies}
\label{sec-3-6}

Apart from the local anomalies analyzed in the preceding sections, there is also another type of anomalies that can arise in $2n$--dimensional theories with chiral fermions. These are called global anomalies and they may arise when the $2n$--th homotopy group of the gauge is non-trivial. In such cases, there exist ``large'' gauge transformations, under which the fermionic determinant can acquire a phase factor that renders the quantum theory ill-defined. Below, we review the essential facts about global anomalies and we state the necessary and sufficient conditions for their absence in six dimensions.

\subsection{Witten's $\mathrm{SU}(2)$ anomaly}

To describe global anomalies in gauge theories, we will follow Witten's \cite{Witten:1982fp} original approach, referring to an $\mathrm{SU}(2)$ gauge theory with one doublet of Weyl fermions in four-dimensional Euclidean spacetime. Taking for the moment the gauge fields to be classical and ignoring their action, we write the partition function of the theory as (cf. Eq. (\ref{e-3-1-48})), 
\be
\label{e-3-6-1}
Z [A] = \int \mathcal{D} \psi^+ \mathcal{D} \bar{\psi}^- \exp \left( - \int \dd^4  x \bar{\psi}^- \ii \fslash D \psi^+ \right) \equiv [\Det (\ii \fslash D(A)) ]^{1/2},
\ee
where the square root arises because only half of the chiral components of $\psi$ contribute. However, this square root is a source of ambiguity since it is not clear which sign we should take. To be specific, we saw in \S\ref{sec-3-2-1} that the eigenvalues of $\ii \fslash D(A)$ come in pairs of opposite sign. Hence, the square root of the determinant in (\ref{e-3-6-1}) is obtained by picking only one eigenvalue from each such pair and, by convention, we can define $[ \det(\ii \fslash D(A)) ]^{1/2}$ to be the product of all positive eigenvalues. Now, consider a gauge transformation of $A$ to $A^g = g^{-1} A g + g^{-1} \dd g$ where the $\mathrm{SU}(2)$--valued gauge function $g(x)$ satisfies $\lim_{|x| \to \infty} g(x) = 1$, i.e. corresponds to a map from $\MBFS^4$ to $\mathrm{SU}(2)$. If $g(x)$ is continuously connected to the identity then the above definition of $[ \det(\ii \fslash D(A)) ]^{1/2}$ is invariant under the gauge transformation; the sign of an eigenvalue of $\ii \fslash D(A)$ cannot change under an infinitesimal transformation. However, the peculiar feature of the gauge group $\mathrm{SU}(2)$ in four dimensions is that its fourth homotopy group is non-trivial,
\be
\label{e-3-6-2}
\pi_4 (\mathrm{SU}(2)) = \MBBZ_2,
\ee
which means that there exist ``large'' gauge transformations where $g(x)$ ``wraps'' twice around $\mathrm{SU}(2)$ and thus cannot be continuously deformed to the identity. Under such a transformation, the sign of an eigenvalue of $\ii \fslash D(A)$ \emph{can} change. If there occurs an odd number of such changes, then $[ \det(\ii \fslash D(A)) ]^{1/2}$ picks up an overall minus sign, i.e.
\be
\label{e-3-6-3}
Z[A^g] = - Z[A].
\ee
This phenomenon is called a \emph{global anomaly}.

The existence of global anomalies in a theory implies that the theory is not self-consistent. The reason is that, in their presence, the partition function (\ref{e-3-6-1}) cannot be integrated any further. Indeed, treating $A$ as a quantum field and considering the full partition function 
\be
\label{e-3-6-4}
\MZ = \int \MD \!A Z[A] \exp \left( - \frac{1}{4} \int \dd^4 x F^i_{\mu\nu} F_i^{\mu\nu} \right)
\ee
we see that, due to Eq. (\ref{e-3-6-3}), the contribution from each $A$ is cancelled by an equal and opposite contribution from the corresponding $A^g$. Thus, the partition function $\MZ$ equals zero and the observables of the theory are ill-defined.

We now proceed to show that, in the theory under consideration, such an anomaly does exist. To see this, we introduce a fifth coordinate $t$ ranging from $-\infty$ to $\infty$ and we let
\be
\label{e-3-6-5}
\ii \fslash D^{(5)} (A^{(5)}) = \ii \Gamma^t D_t(A^{(5)}) + \ii \fslash D(A^{(5)})
\ee
be the corresponding 5D Dirac operator. The 5D gauge field $A^{(5)}(x,t)$ is chosen as
\be
\label{e-3-6-6}
A^{(5)}_t(x,t) = 0 ,\qquad \lim_{t \to - \infty} A^{(5)}_\mu (x,t) = A_\mu(x) ,\qquad \lim_{t \to \infty} A^{(5)}_\mu (x,t) = A^g_\mu(x),
\ee
so that $A^{(5)}_\mu (x,t)$ evolves \emph{adiabatically} from $A_\mu(x)$ to $A^g_\mu(x)$ as $t$ varies from $-\infty$ to $\infty$. For the given configuration, there exists an index theorem, called the Atiyah-Singer ``$\textrm{mod } 2$'' index theorem \cite{Atiyah:1971ws}, which states that the number of zero modes of $\ii \fslash D^{(5)}$ is odd. To see the relevance of this to our problem, we let $\Phi(x,t)$ be such a zero mode. By (\ref{e-3-6-5}) and (\ref{e-3-6-6}), $\Phi(x,t)$ satisfies
\be
\label{e-3-6-7}
\frac{\dd \Phi(x,t)}{\dd t} = - \Gamma^t \fslash D (A^{(5)}) \Phi(x,t).
\ee
In the adiabatic approximation, one may use the separation of variables $\Phi(x,t) = f(t) \phi_t(x)$, where $\phi_t(x)$ is chosen to be an eigenvector of $-\Gamma^t \fslash D$ with eigenvalue $\lambda(t)$, 
\be
\label{e-3-6-8}
-\Gamma^t \fslash D(A^{(5)}) \phi_t(x) = \lambda(t) \phi_t(x),
\ee
noting that an eigenvalue of $-\Gamma^t \fslash D$ is the same thing as an eigenvalue of $\ii \fslash D$ (as $\ii \Gamma^t \Gamma^\mu$ is also a basis for the Clifford algebra). Then, Eq. (\ref{e-3-6-7}) leads to the following equation for $f(t)$
\be
\label{e-3-6-9}
\frac{\dd f(t)}{\dd t} = \lambda(t) f(t),
\ee
which is solved by
\be
\label{e-3-6-10}
f(t) = f(0) \exp \left[ \int_0^t \dd t' \lambda(t') \right].
\ee
In order for this solution to be normalizable, $\lambda(t)$ must be positive as $t \to -\infty$ and negative as $t \to \infty$. Hence, in the adiabatic approximation, each zero mode of $\ii \fslash D^{(5)}(A^{(5)})$ signifies a change of sign in an eigenvalue of $\ii \fslash D(A^{(5)})$ as $t$ varies from $-\infty$ to $\infty$, i.e. a change of sign in an eigenvalue of $\ii \fslash D(A)$ as $A$ goes to $A^g$. Going beyond the adiabatic approximation only corrects this exact correspondence to a $\textrm{mod } 2$ correspondence. Hence, since the number of zero modes of $\ii \fslash D^{(5)}$ is odd, so is the number of sign changes in the eigenvalues of $\ii \fslash D(A)$ as $A$ goes to $A^g$ and Eq. (\ref{e-3-6-3}) does indeed hold.

To generalize the above problem, we first consider the case where the theory contains a net number $n_\mathbf{2}$ of positive-chirality $\mathrm{SU}(2)$ doublets (since two Weyl spinors of opposite chirality make up a Dirac spinor, only the net number counts). The effect of this change is that the RHS of (\ref{e-3-6-3}) now equals $[\Det (\ii \fslash D(A)) ]^{n_\mathbf{2}/2}$. Hence, if $n_\mathbf{2}$ is odd then $Z[A]$ changes its sign while, if $n_\mathbf{2}$ is even, then $Z[A]$ is equal to an integer power of $\Det (\ii \fslash D(A))$ and we need not worry about the definition of the square root. Even more generally, we may consider a theory with a net number $n_{\MR}$ of positive-chirality multiplets in an arbitrary representation $\MR$ of $SU(2)$. Writing $\tr_{\MR} F^2 = c_{\MR} \tr F^2$, the ``$\textrm{mod } 2$'' index theorem for the case at hand implies that $Z[A]$ changes its sign if the quantity $n_{\MR} c_{\MR}$ (which is always an integer) is odd. Therefore, in the most general case, the condition for the absence of global anomalies is
\be
\label{e-3-6-11}
\sum_{\MR} n_{\MR} c_{\MR} = 0 \mod 2.
\ee

\subsection{Generalizations}

The above considerations can be generalized to any $2n$--dimensional theory with Weyl fermions in representations of a gauge group $\MG$ with non-trivial $2n$--th homotopy group,
\be
\label{e-3-6-12}
\pi_{2n} ( \MG ) \ne 0.
\ee
In such a case, under a non-trivial gauge transformation, the partition function changes as
\be
\label{e-3-6-13}
Z[A^g] = \exp(i \gamma) Z[A].
\ee
where $\gamma$ is a phase factor which, unless it equals a multiple of $2 \pi$, signifies the presence of a global anomaly (for the case of $\mathrm{SU}(2)$ in 4D, $\gamma=\pi$). As before, such an anomaly is dangerous, since the full partition function receives which add up to zero, rendering the theory ill-defined.

The value of the phase factor can be computed according to two alternative methods (i) generalize the original approach of \cite{Witten:1982fp} and study the evolution of the eigenvalues of $\ii \fslash D$ by invoking suitable index theorems or (ii) embed $\MG$ in a larger group $\hat{\MG}$ with trivial  $\pi_{2n}(\hat{\MG})$ and calculate the global anomaly in terms of the local gauge anomaly for $\hat{\MG}$ \cite{Elitzur:1984kr,Kiritsis:1986mf,Bershadsky:1997sb}. Since the meaning of global anomalies is by now quite clear, we leave technical details aside and we just present the results.

In the six-dimensional case of interest, the possible gauge groups that may lead to global anomalies are $\mathrm{SU}(2)$, $\mathrm{SU}(3)$ and $G_2$, for which
\be
\label{e-3-6-14}
\pi_6 (\mathrm{SU}(2) ) = \MBBZ_{12} ,\qquad \pi_6 ( \mathrm{SU}(3) ) = \MBBZ_6 ,\qquad \pi_6 (G_2) = \MBBZ_3.
\ee
The conditions for the absence of global anomalies can be computed as indicated above (actually, in the 6D case, there is another \cite{Bershadsky:1997sb} possible method based on geometrical engineering). They are expressed in terms of the coefficients $b_{\MR}$, defined by $\tr_{\MR} F^4 = b_{\MR} (\tr F^2)^2$, and they are given by
\begin{align}
\label{e-3-6-15}
&\MG=\mathrm{SU}(2) : &&- 2 \sum_{\MR} n_{\MR} b_{\MR} = 0 \mod 6,\nn\\
&\MG=\mathrm{SU}(3) : &&- 2 \sum_{\MR} n_{\MR} b_{\MR} = 0 \mod 6, \nn\\
&\MG=G_2 :            &&- 4 \sum_{\MR} n_{\MR} b_{\MR} = 0 \mod 3.
\end{align}
These conditions, when applicable, supplement the local anomaly cancellation conditions in six-dimensional theories.
\chapter{Anomaly Cancellation in Superstring Theory and M-Theory}
\label{chap-4}

In this chapter, we prepare the setting for the main part of the thesis by giving a detailed account of anomaly cancellation in superstring and M-theory models. We begin by reviewing the classic examples in ten dimensions, namely the automatic cancellation of anomalies in Type IIB supergravity and the Green-Schwarz anomaly cancellation in heterotic and Type I theories. We next review the construction of Ho\v rava and Witten which considered eleven-dimensional supergravity on a $\mathbf{S}^1 / \mathbb{Z}_2$ orbifold, regarded as a manifold with boundary on the orbifold fixed points. In that case, the symmetries of the theory impose chiral boundary conditions for the fermions on the fixed points, yielding a chiral theory with an anomalous spectrum. Anomalies are cancelled by introducing extra fields living on the fixed points and applying a special version of the Green-Schwarz mechanism.

\section{Anomaly Cancellation in Superstring Theory}
\label{sec-4-1}

Here, we give an account of anomaly cancellation in superstring theory, based the developments made during the first superstring revolution of 1984-5. In particular, we demonstrate that $N=2$ Type IIB theory is manifestly anomaly-free while $N=1$ heterotic/Type I theory can be made anomaly-free, for certain choices of the gauge group, through the Green-Schwarz mechanism. 

\subsection{Anomaly cancellation in Type IIB supergravity} 
\label{sec-4-1-1}

Let us start with Type IIB supergravity, whose field content is given by the gravity multiplet $( g_{\mu\nu} , B_{\mu\nu} , \phi, C , C_{\mu\nu} , C^+_{\mu\nu\rho\sigma}, \psi^{I+}_\mu, \chi^{I-} )$, where $I=1,2$. The anomalies of this theory are purely gravitational and arise from $\psi^{I+}_\mu$, $\chi^{I-}$ and $C^+_4$. Noting that the fermions are Majorana-Weyl, the fermionic part of the anomaly equals the sum of the contributions from one positive-chirality Weyl gravitino and a negative-chirality Weyl spinor. Including the contribution from the antisymmetric tensor field, we find the total anomaly polynomial
\be 
\label{e-4-1-1}
I_{12} = I^{3/2}_{12} (R) - I^{1/2}_{12} (R) + I^{A}_{12} (R),
\ee  
Using the explicit formulas from Eq. (\ref{e-3-4-7}), we immediately find that the total anomaly {\it vanishes},
\be
\label{e-4-1-2}
I_{12} = 0.
\ee
This is the ``miraculous" anomaly cancellation of Type IIB superstring theory, first proven in \cite{Alvarez-Gaume:1984ig}. This is due to the fact that the three anomaly polynomials in (\ref{e-4-1-1}) are linearly dependent and so cancellation may occur if the spectrum of the theory is chosen appropriately. Note that in the absence of the gravitino, we would only be left with the polynomials $I^{1/2} (R)$ and $I^{A} (R)$ which are linearly independent; this is an indication that the only consistent 10D gravitational theories with chiral fields are supergravity theories.

\subsection{Green-Schwarz anomaly cancellation in heterotic/Type I theory}
\label{sec-4-1-2}

Next, let us apply the above to $D=10$, $N=1$ supergravity coupled to super Yang-Mills theory, which is the low-energy limit of heterotic and Type I superstring theories, following the original work \cite{Green:1984sg}. The field content of the theory is given by the gravity multiplet $( g_{\mu\nu} , B_{\mu\nu} , \phi , \psi^+_\mu , \chi^-)$ and $n$ vector multiplets $( A_\mu, \lambda^+ )$, transforming in the adjoint representation of some gauge group of dimension $n$. Now the theory contains anomalies of all types, arising from $\psi^+_\mu$, $\chi^-$ and $\lambda^+$. The gravitational anomaly receives contributions from all three types of fields and is given by
\bea
\label{e-4-1-3}
I_{12} (R) &=& \frac{1}{2} \left[ I^{3/2}_{12} (R) + (n-1) I^{1/2}_{12} (R) \right] \nn\\
&=& \frac{1}{2} \left[ \frac{496-n}{504} \tr R^6 - \frac{224+n}{384} \tr R^2 \tr R^4 + \frac{320 - 5 n}{4608} (\tr R^2)^3 \right]
\eea
The gauge and mixed anomalies receive contributions only from the gauginos. The gauge anomaly is simply given by
\be
\label{e-4-1-4}
I_{12} (F) = \frac{1}{2} I^{1/2}_{12} (F) = \frac{1}{2} \Tr F^6,
\ee
while the mixed anomaly reads
\be
\label{e-4-1-5}
I_{12} (F,R) = \frac{1}{2} I^{1/2}_{12} (F,R) = \frac{1}{2} \left[ \frac{1}{16} \Tr F^2 \tr R^4 + \frac{5}{64} \Tr F^2 (\tr R^2)^2 - \frac{5}{8} \Tr F^4 \tr R^2 \right].
\ee
In the above, ``$\Tr$'' stands for the trace in the adjoint representation; the trace in the fundamental representation will be denoted as ``$\tr$''. Putting everything together, we find the total anomaly polynomial
\bea
\label{e-4-1-6}
I_{12} = \frac{1}{2} &\biggl[& \frac{496-n}{504} \tr R^6 - \frac{224+n}{384} \tr R^2 \tr R^4 + \frac{320 - 5 n}{4608} (\tr R^2)^3 + \Tr F^6 \nonumber\\ &&+ \frac{1}{16} \Tr F^2 \tr R^4 + \frac{5}{64} \Tr F^2 (\tr R^2)^2 - \frac{5}{8} \Tr F^4 \tr R^2 \biggr], 
\eea

Since the above anomaly is nonvanishing, the only way to achieve anomaly cancellation is via the Green-Schwarz mechanism of \S\ref{sec-3-5-1}. Since the theory contains a 2--form $B_2$, the factorization condition (\ref{e-3-5-3}) takes the specific form
\be
\label{e-4-1-7}
I_{12} = \Omega_4 \Omega_8,
\ee
and, noting that the only curvature invariants available for the construction of $\Omega_4$ are $\tr R^2$ and $\Tr F^2$, we may choose our normalization so that
\be
\label{e-4-1-8}
\Omega_4 = k \Tr F^2 - \tr R^2,
\ee 
where $k$ is a yet undetermined constant. A first condition for the factorization (\ref{e-4-1-7}) to be possible is that the irreducible parts of $\tr R^6$ and $\Tr F^6$ vanish. Regarding the former, it is known that $\mathrm{SO}(9,1)$ has a sixth-order Casimir, which implies that $\tr R^6$ cannot be expressed in terms of lower-order traces. Hence, in order for factorization to be possible, the coefficient of $\tr R^6$ must vanish, i.e. we must have
\be
\label{e-4-1-9}
n = 496.
\ee
This is a first dramatic demonstration of the restrictions imposed by anomaly cancellation: the factorization condition tells us that anomaly cancellation requires the presence of vector multiplets and also uniquely fixes the dimension of the gauge group. Turning to the $\Tr F^6$ term and noting that its coefficient is fixed, factorization requires that the adjoint of the gauge group must have no sixth-order Casimirs, i.e. that $\Tr F^6$ must be reducible according to
\be
\label{e-4-1-10}
\Tr F^6 = \alpha \Tr F^2 \Tr F^4 + \beta (\Tr F^2)^3.
\ee
where $\alpha$ and $\beta$ are two constants. Provided that (\ref{e-4-1-9}) and (\ref{e-4-1-10}) hold, the anomaly polynomial (\ref{e-4-1-6}) is written as
\bea
\label{e-4-1-11}
I_{12} = \frac{1}{2} &\biggl[& - \frac{15}{8} \tr R^2 \tr R^4 - \frac{15}{32} (\tr R^2)^3 + \alpha \Tr F^2 \Tr F^4 + \beta (\Tr F^2)^3 \nn\\
&& + \frac{1}{16} \Tr F^2 \tr R^4 + \frac{5}{64} \Tr F^2 (\tr R^2)^2 - \frac{5}{8} \Tr F^4 \tr R^2 \biggr].
\eea
From this expression, we can fix the value of the coefficient $k$ appearing in (\ref{e-4-1-8}). Indeed, from the fixed values of the coefficients of $\Tr F^2 \tr R^4$ and $\tr R^2 \tr R^4$, we immediately see that we must have $k = 1 / 30$. Therefore, the factorized expression (\ref{e-4-1-7}) should read
\be
\label{e-4-1-12}
I_{12} = \left( \frac{1}{30} \Tr F^2 - \tr R^2 \right) \left[ \frac{15}{16} \tr R^4 + a (\tr R^2)^2 + b \Tr F^4 + c (\Tr F^2)^2 + d \tr R^2 \Tr F^2 \right].
\ee 
where $a$, $b$, $c$ and $d$ are four more constants. The six constants in Eqs. (\ref{e-4-1-11}) and (\ref{e-4-1-12}) can now be determined by comparing the two expressions term-by-term. The comparison yields
\be
\label{e-4-1-13}
a = \frac{15}{64} ,\quad
b = \frac{5}{16} ,\quad
c = - \frac{1}{960} ,\quad
d = - \frac{1}{32} ,\quad
\alpha = \frac{1}{48} ,\quad
\beta = - \frac{1}{14400}.
\ee

To summarize, the factorization condition requires that (i) the gauge group of the theory has dimension $n=496$ and (ii) the adjoint representation of the gauge group has no sixth-order Casimir, with $\Tr F^6$ being reducible according to
\be
\label{e-4-1-14}
\Tr F^6 = \frac{1}{48} \Tr F^2 \Tr F^4 - \frac{1}{14400} (\Tr F^2)^3.
\ee
If this is the case, the anomaly polynomial takes the factorized form (\ref{e-4-1-7}) with 
\be
\label{e-4-1-15}
\Omega_4 = \frac{1}{30} \Tr F^2 - \tr R^2,
\ee
and
\be
\label{e-4-1-16}
\Omega_8 = \frac{15}{16} \tr R^4 + \frac{15}{64} (\tr R^2)^2 + \frac{5}{16} \Tr F^4 - \frac{1}{960} (\Tr F^2)^2 - \frac{1}{32} \tr R^2 \Tr F^2 .
\ee
and the Green-Schwarz mechanism may operate.

To determine the possible gauge groups for which anomaly cancellation may occur, one is thus instructed to search of gauge groups of dimension $496$ satisfying Eq. (\ref{e-4-1-14}). Originally, Green and Schwarz started from the second requirement. In searching for groups with no sixth-order Casimir, they encountered the $\mathrm{SO}(N)$ identity
\be
\label{e-4-1-17}
\Tr F^6 = (N-32) \tr F^6 +  15 \tr F^2 \tr F^4,
\ee
expressing the sixth-order trace in the adjoint in terms of traces in the fundamental. For the case of $\mathrm{SO}(32)$, the first term on the RHS vanishes and $\Tr F^6$ is reducible. The remarkable facts are that (i) the reduction of $\Tr F^6$ is \emph{exactly} given by (\ref{e-4-1-14}) and that (ii) the dimension of $\mathrm{SO}(32)$ is $\frac{32 \cdot 31}{2} = 496$, exactly that required by (\ref{e-4-1-9})! Searching for other groups with the same properties, they also learned that the exceptional group $E_8$ also has no sixth-order Casimirs. Since the dimension of $E_8$ is $248 = \frac{496}{2}$, they were led to consider the 496--dimensional gauge $E_8 \times E_8$ which again turned out that this group satisfied all anomaly cancellation conditions. Let us briefly review these two examples.
\begin{itemize}
\item $\mathrm{SO}(32)$. For $\mathrm{SO}(32)$, the decompositions of second-, fourth- and sixth-order adjoint traces in terms of fundamental ones are given by
\be
\label{e-4-1-18}
\Tr F^2 = 30 \tr F^2 ,\qquad \Tr F^4 = 24 \tr F^4 + 3 ( \tr F^2 )^2 ,\qquad \Tr F^6 = 15 \tr F^2 \tr F^4.
\ee
Combining these relations, we easily find the required relation
\bea
\label{e-4-1-18a}
\Tr F^6 &=& 15 \left( \frac{1}{30} \Tr F^2 \right) \left[ \frac{1}{24} \Tr F^4 - \frac{1}{8} \left( \frac{1}{30} \Tr F^2 \right)^2 \right] \nn\\
&=& \frac{1}{48} \Tr F^2 \Tr F^4 - \frac{1}{14400} (\Tr F^2)^3.
\eea

\item $E_8 \times E_8$. For each of the $E_8$ factors, labelled by $i=1,2$, the decompositions of fourth- and sixth-order traces are given by
\be
\label{e-4-1-19}
\Tr F_i^4 = \frac{1}{100} ( \Tr F_i^2 )^2 ,\qquad \Tr F_i^6 = \frac{1}{7200} ( \Tr F_i^2 )^3 .
\ee
For the full $E_8 \times E_8$ group, we write 
\be
\label{e-4-1-20}
\Tr F^{2m} = \Tr F_1^{2m} + \Tr F_2^{2m}.
\ee
Applying this to $\Tr F^6$ and using (\ref{e-4-1-19}) and the identity $2(a^3+b^3) = 3(a+b)(a^2+b^2) - (a+b)^3$, we find again the desired relation
\be
\label{e-4-1-21}
\Tr F^6 = \frac{1}{7200} [( \Tr F_1^2 )^3 + ( \Tr F_2^2 )^3] =  \frac{1}{48} \Tr F^2 \Tr F^4 - \frac{1}{14400} (\Tr F^2)^3.
\ee
In what follows, it will be convenient to \emph{define} the ``fundamental'' trace for $E_8 \times E_8$ as
\be
\label{e-4-1-22}
\tr F^2 \equiv \frac{1}{30} \Tr F^2,
\ee
so that its normalization coincides with that of the $\mathrm{SO}(16) \times \mathrm{SO}(16)$ subgroup. 
\end{itemize} 
Today, it is known that the \emph{only} groups for which anomaly cancellation is possible are $\mathrm{SO}(32)$, $E_8 \times E_8$, $E_8 \times U(1)^{248}$ and $U(1)^{496}$.

The application of the Green-Schwarz anomaly cancellation mechanism is now straightforward. Applying the first descent equation, we write the 4--form $\Omega_4$ in (\ref{e-4-1-15}) as
\be
\label{e-4-1-23}
\Omega_4 = \dd \Omega_3  ;\qquad \Omega_3 = \omega_{3Y} - \omega_{3L},
\ee
where $\omega_{3Y}$ and $\omega_{3L}$ are the Chern-Simons forms of $\tr F^2$ and $\tr R^2$ respectively. By the descent equations, the variation of $\Omega_3$ is
\be
\label{e-4-1-24}
\delta_{v,\lambda} \Omega_3 = \dd \Omega^1_2 ;\qquad \Omega^1_2 = \omega^1_{2Y} - \omega^1_{2L}.
\ee
Similarly, we consider the 8--form $\Omega_8$ and we let $\Omega^1_6$ denote its descent. Then, we find that the anomalous gauge/Lorentz variation of the effective action is
\be
\label{e-4-1-25}
\delta_{v,\lambda} \varGamma = \int_{M_{10}} \left[ a \Omega_3 \dd \Omega^1_6 + (1-a) \Omega^1_2 \Omega_8 \right],
\ee 
Replacing the ordinary field strength (\ref{e-2-4-11}) by
\be
\label{e-4-1-26}
\tilde{G}_3 = \dd B_2 - \Omega_3 = \dd B_2 - \omega_{3Y} + \omega_{3L},
\ee
and changing the transformation law (\ref{e-2-4-12}) of $B_2$ to
\be
\label{e-4-1-27}
\delta_{v,\lambda} B_2 = \Omega^1_2 = \omega^1_{2Y} - \omega^1_{2L},
\ee
we see that the anomaly is cancelled by the Green-Schwarz term
\be
\label{e-4-1-28}
S_{GS} = - \int_{M_{10}} \left( B_2 \Omega_8 + a \Omega_3 \Omega_7 \right).
\ee 

The above discoveries had a striking impact on the development of string theory. Regarding the $\mathrm{SO}(32)$ theory, Green and Schwarz showed that the corresponding Type I superstring theory also satisfies the RR tadpole cancellation condition \cite{Green:1984ed} and proved anomaly cancellation using string-theory arguments \cite{Green:1984qs}. As for the $E_8 \times E_8$ theory, the fact that there was no known superstring model realizing its gauge group motivated the search for such a model and eventually led to the discovery of the heterotic string \cite{Gross:1985dd}, which admits both $\mathrm{SO}(32)$ and $E_8 \times E_8$ as possible gauge groups. Moreover, it was found that the $B_2 \Omega_8$ Green-Schwarz interaction in (\ref{e-4-1-28}) is present in superstring theory, in the form of a one-loop string correction to the supergravity effective action, with exactly the right coefficient to cancel the variation (\ref{e-4-1-25}). These discoveries marked the onset of the first superstring revolution of 1984-1986, which firmly established string theory as an active area of theoretical physics.

\section{Anomaly Cancellation in Heterotic M-theory}
\label{sec-4-2}

An archetypal model for orbifold compactification which provides an impressive exhibition of the role of anomaly cancellation in model building is provided by \emph{heterotic M-theory} or \emph{Ho\v rava-Witten theory} \cite{Horava:1996qa,Horava:1996ma}, that is, 11D supergravity compactified on the $\mathbf{S}^1/\mathbb{Z}_2$ orbifold. Compactification on this orbifold results in a chiral spectrum on the orbifold fixed planes, but the price to pay is the appearance of anomalies. The only way to cancel these anomalies is through the introduction of 10D vector multiplets on the orbifold fixed planes and the anomaly cancellation conditions uniquely fix the gauge group on each plane to be $E_8$. The important fact is that, after a modification of the Green-Schwarz mechanism and a careful examination of the role played by the supergravity Chern-Simons term, the theory turns out to be anomaly-free and, moreover, the gauge coupling is uniquely fixed. The dynamics of the resulting theory are believed to correspond to a strongly-coupled version of the $E_8 \times E_8$ heterotic string.

\subsection{11--dimensional supergravity on $\mathbf{S}^1/\mathbb{Z}_2$}
\label{sec-4-2-1}

The Ho\v rava-Witten construction begins by considering 11D supergravity on the orbifold $\mathbf{S}^1/\mathbb{Z}_2$. The starting point is thus the 11D supergravity Lagrangian and transformation rules of \S\ref{sec-2-4-1}. Reinstating the gravitational coupling $\kappa$ and ignoring $(\mathrm{Fermi})^4$ terms, we have the Lagrangian
\bea
\label{e-4-2-1}
\kappa^2 E^{-1} \ML_{11} &=& \frac{1}{4} R - \frac{1}{48} F_{MNPQ} F^{MNPQ} - \frac{1}{2} \bar{\psi}_M \Gamma^{MNP} D_N \psi_P \nn\\
&& - \frac{1}{96} \left( \bar{\psi}_L \Gamma^{LMNPQR} \psi_R + 12 \bar{\psi}^M \Gamma^{NP} \psi^Q \right) F_{MNPQ} \nn\\
&& + \frac{1}{3 \cdot 3456} E^{-1} \epsilon^{M_1 \ldots M_{11}} A_{M_1 M_2 M_3} F_{M_4 M_5 M_6 M_7} F_{M_8 M_9 M_{10} M_{11}} .
\eea
and the supersymmetry transformations
\bea
\label{e-4-2-2}
\delta E^A_M &=& \bar{\epsilon} \Gamma^A \psi_M,
\nn\\
\delta A_{MNP} &=& - \frac{3}{2} \bar{\epsilon} \Gamma_{[MN} \psi_{P]},
\nn\\
\delta \psi_M &=& D_M \epsilon + \frac{1}{144} \left( \Gamma_M^{\phantom{M} NPQR} - 8 \delta_M^N \Gamma^{PQR} \right) F_{NPQR} \epsilon.
\eea

From our discussion of the symmetry properties of 11D supergravity in \S\ref{sec-2-4-1}, we know that it is invariant under a $\mathbb{Z}_2$ symmetry corresponding to parity transformations and certain transformations of the fields. Let us examine this symmetry more carefully. We choose $\mathbb{Z}_2$ to act as a parity transformation along the eleventh dimension $x_{11}$,
\be
\label{e-4-2-3}
\mathbb{Z}_2 \qquad:\qquad x_{11} \to - x_{11}.
\ee
The $\mathbb{Z}_2$ parity assignments to the fields follow from invariance of the action and supersymmetry transformation laws. For the metric, we obviously have
\be
\label{e-4-2-4}
\mathbb{Z}_2 \qquad:\qquad g_{\mu\nu} \to g_{\mu\nu} ,\quad g_{\mu,11} \to - g_{\mu,11} ,\quad g_{11,11} \to g_{11,11}.
\ee
For $A_3$, we consider the Chern-Simons term in (\ref{e-4-2-1}) whose invariance requires that all combinations of the type $A_3 F_4 F_4$ must be $\mathbb{Z}_2$--even. Using $F_4 = \dd A_3$ and the fact that $\partial_\mu$ is $\mathbb{Z}_2$--even while $\partial_{11}$ is $\mathbb{Z}_2$--odd, we see that the only consistent way to achieve invariance is through the assignment
\be
\label{e-4-2-5}
\mathbb{Z}_2 \qquad:\qquad A_{\mu\nu\rho} \to - A_{\mu\nu\rho} ,\quad A_{\mu\nu11} \to A_{\mu\nu,11}.
\ee
Finally, for $\epsilon$ and $\psi_M$, we can consider the $\delta \psi_M$ transformation law. We first parameterize the parity assignments as $\epsilon \to P \epsilon$, $\psi_\mu \to Q \psi_\mu$ and $\psi_{11} \to R \psi_{11}$. Invariance of the $\delta \psi_M = D_M \epsilon$ part immediately implies that $P=Q=-R$. To determine $P$, we consider the $F_4$--dependent term of $\delta \psi_\mu$, which we write schematically as $\delta \psi_\mu \sim \gamma_\mu^{\phantom{\mu} \nu\rho\sigma\tau} F_{\nu\rho\sigma\tau} \epsilon + \gamma_\mu^{\phantom{\mu} \nu\rho\sigma} \Gamma_{11} F_{\nu\rho\sigma 11} \epsilon$. Under $\mathbb{Z}_2$, this transforms to $P \delta \psi_\mu \sim - \gamma_\mu^{\phantom{\mu} \nu\rho\sigma\tau} F_{\nu\rho\sigma\tau} P \epsilon + \gamma_\mu^{\phantom{\mu} \nu\rho\sigma} \Gamma_{11} F_{\nu\rho\sigma 11} P \epsilon$. Consistency of the two expressions demands that $P$ anticommutes with $\Gamma_\mu^{\phantom{\mu} \nu\rho\sigma\tau}$ and commutes with $\Gamma_\mu^{\phantom{\mu} \nu\rho\sigma}$. The obvious choice is $P = \pm \Gamma_{11}$, where the normalization follows from invariance of the action. Finally, invariance of $\delta E^A_M$ and $\delta A_{MNP}$ rules out the minus sign and we find $P=Q=-R=\Gamma_{11}$, that is,
\be
\label{e-4-2-6}
\mathbb{Z}_2 \qquad:\qquad \epsilon \to \Gamma_{11} \epsilon ,\quad \psi_\mu \to \Gamma_{11} \psi_\mu ,\quad \psi_{11} \to - \Gamma_{11} \psi_{11}.
\ee

Now, consider compactification of $x_{11}$ on the $\mathbf{S}^1/\mathbb{Z}_2$ orbifold. This orbifold is constructed by taking $\mathbf{S}^1$ and modding out the $\mathbb{Z}_2$ symmetry, that is, it corresponds to a fundamental domain of the  $\mathbb{Z}_2$ action on $\mathbf{S}^1$. Parameterizing $\mathbf{S}^1$ by the interval $[-\pi R,\pi R]$ with the ends identified, we see that the orbifold can be represented by the interval $I = [0,\pi R]$ whose endpoints at $x_{11} = 0$ and $x_{11} = \pi R$ are fixed points of the $\mathbb{Z}_2$ action. To find the particle spectrum of the theory on the 10D fixed planes, we first consider the usual Kaluza-Klein decomposition of the 11D fields on $\mathbf{S}^1$,
\bea
\label{e-4-2-7}
g_{MN} \to ( g_{\mu\nu} , A_\mu , \phi ) , \qquad A_{MNP} \to ( A_{\mu\nu\rho} , B_{\mu\nu} ) , \qquad \psi_M \to (\psi_\mu, \chi) ,
\eea
with $A_{\mu} = g_{\mu, 11}$, $\phi = g_{11,11}$, $B_{\mu\nu} = A_{\mu\nu 11}$ and $\chi = \psi_{11}$
and we project out the fields that are odd under $\mathbb{Z}_2$. The spectrum on the fixed planes is then obtained from (\ref{e-4-2-5}) by discarding the $\mathbb{Z}_2$--odd fields . Using (\ref{e-4-2-4})-\ref{e-4-2-6}, we easily see that the surviving fields are 
\be
\label{e-4-2-8}
(g_{\mu\nu} , B_{\mu\nu} , \phi , \psi^+_\mu, \chi^-).
\ee
where the $\pm$ subscripts on spinors indicate 10D chirality. This spectrum is nothing but the \emph{chiral} $D=10$, $N=1$ gravity multiplet of (\ref{e-2-1-30}). This is the first success of Ho\v rava-Witten theory: compactification of 11D supergravity on the $\mathbf{S}^1/\mathbb{Z}_2$ orbifold gives rise to 10D chiral fermions, evading Witten's \cite{Witten:1983ux} no-go theorem about compactifications of non-chiral theories on ordinary manifolds. 

On the other hand, the appearance of chiral fermions in the theory inevitably gives rise to anomalies, which imply that the theory is inconsistent at this stage. The idea of Ho\v rava and Witten was to introduce extra \emph{ten-dimensional} fields living only on the orbifold fixed planes with the purpose of cancelling these anomalies. Since the theory already contains gravity, the only available multiplet is the $D=10$, $N=1$ vector multiplet with field content
\be
\label{e-4-2-9}
\text{Vector multiplet} \quad:\quad ( A_\mu, \lambda^+ ).
\ee
The remaining tasks are to construct the action describing the boundary vector multiplets, state the anomaly cancellation conditions and identify the mechanism by which anomalies may cancel.

Before we proceed, we note that the theory can be formulated according to two different, but equivalent, formalisms. In the ``upstairs'' formalism, one regards the 11D spacetime as $M_{11,U} = M_{10} \times \mathbf{S}^1$ subject to $\mathbb{Z}_2$ invariance. This \emph{orbifold} has no boundary and partial integrations with respect to $x_{11}$ do not give rise to surface terms. In the ``downstairs'' formalism, the spacetime manifold is thought of to be $M_{11,D} = M_{11,U}/ \mathbb{Z}_2 = M_{10} \times I$ with $I = [0,\pi R]$. This is a manifold with two boundaries located on the two $\mathbb{Z}_2$ fixed planes and partial integrations with respect to $x_{11}$ give rise to surface terms. Since $M_{11,U}$ has twice the volume of $M_{11,D}$, the requirement that the action be the same in both approaches implies that the associated gravitational couplings must be related according to $\kappa_U^2 = 2 \kappa_D^2$. Also, by arguments related to M2--branes and $F_4$--flux quantization conditions, it has been proven \cite{Harmark:1998bs} that $\kappa_D$ is the true 11D gravitational coupling. In what follows, we will mainly use the downstairs approach and, for definiteness, we will restrict our considerations to the fixed plane at $x_{11}=0$, which we will call $M_{10}$.

\subsection{Anomaly analysis}
\label{sec-4-2-2}

For the problem at hand, it is advantageous to begin the analysis by examining the anomaly structure of the theory.  To determine the gravitational anomalies resulting from the bulk fields, we note that, in the limit of a small compactification radius where the two planes effectively coincide, the total anomaly must be equal to that of a 10D positive-chirality gravitino plus that of a 10D negative-chirality spinor. Given the fact that the two fixed planes are symmetric, we conclude that the anomaly contribution on each fixed plane is given by half the above sum. Including another factor of $1/2$ due to the Majorana-Weyl property of spinors, we find
\be
\label{e-4-2-10}
I^{bulk}_{12} (R) = \frac{1}{2} \left\{ \frac{1}{2} \left[ I^{3/2}_{12} (R) - I^{1/2}_{12} (R) \right] \right\} =  \frac{1}{2} \left[ \frac{248}{504} \tr R^6 - \frac{112}{384} \tr R^2 \tr R^4 + \frac{160}{4608} (\tr R^2)^3 \right].
\ee 
The inclusion of the boundary vector multiplets gives rise to extra gravitational, gauge and mixed anomalies, all arising from the positive-chirality gaugino. The gravitational contribution reads
\be
\label{e-4-2-11}
I^{bdy}_{12} (R) = \frac{1}{2} n I^{1/2}_{12} (R) = \frac{1}{2} \left[ - \frac{n}{504} \tr R^6 - \frac{n}{384} \tr R^2 \tr R^4 - \frac{5 n}{4608} (\tr R^2)^3 \right],
\ee
while the gauge and mixed contributions are read off from (\ref{e-4-1-4}) and (\ref{e-4-1-5}) respectively. Adding all contributions, we find the total anomaly
\bea
\label{e-4-2-12}
I_{12} = \frac{1}{2} &\biggl[& \frac{248-n}{504} \tr R^6 - \frac{112+n}{384} \tr R^2 \tr R^4 + \frac{160 - 5 n}{4608} (\tr R^2)^3 + \Tr F^6 \nn\\ 
&& + \frac{1}{16} \Tr F^2 \tr R^4 + \frac{5}{64} \Tr F^2 (\tr R^2)^2 - \frac{5}{8} \Tr F^4 \tr R^2 \biggr].
\eea

Now, we can proceed as in the case of $D=10$, $N=1$ supergravity. As will become clear later, we are again interested in factorizations of the form (\ref{e-4-1-7}) where now we set
\be
\label{e-4-2-13}
\Omega_4 = \frac{1}{2} ( k \Tr F^2 -  \tr R^2 ).
\ee
The requirement that the coefficient of $\tr R^6$ vanishes leads to
\be
\label{e-4-2-14}
n = 248,
\ee
while the $\Tr F^6$ term must factorize as
\be
\label{e-4-2-15}
\Tr F^6 = \alpha \Tr F^2 \Tr F^4 + \beta (\Tr F^2)^3.
\ee
Provided that these relations hold, Eq. (\ref{e-4-2-12}) reduces to 
\bea
\label{e-4-2-16}
I_{12} &=& \frac{1}{2} \biggl[ - \frac{15}{16} \tr R^2 \tr R^4 - \frac{15}{64} (\tr R^2)^3 + \alpha \Tr F^2 \Tr F^4 + \beta (\Tr F^2)^3 \nn\\
&& \quad + \frac{1}{16} \Tr F^2 \tr R^4 + \frac{5}{64} \Tr F^2 (\tr R^2)^2 - \frac{5}{8} \Tr F^4 \tr R^2 \biggr].
\eea
which implies that the coefficient $k$ in (\ref{e-4-2-13}) should be equal to $\frac{1}{15}$. Then, the factorized anomaly should have the form
\be
\label{e-4-2-17}
I_{12} = \left( \frac{1}{30} \Tr F^2 - \frac{1}{2} \tr R^2 \right) \left[ \frac{15}{16} \tr R^4 + a (\tr R^2)^2 + b \Tr F^4 + c (\Tr F^2)^2 + d \tr R^2 \Tr F^2 \right].
\ee
and comparison with (\ref{e-4-2-16}) yields the numerical coefficients
\be
\label{e-4-2-18}
a = \frac{15}{64} ,\quad
b = \frac{5}{8} ,\quad
c = - \frac{1}{240} ,\quad
d = - \frac{1}{16} ,\quad
\alpha = \frac{1}{24} ,\quad
\beta =  - \frac{1}{3600}.
\ee
Hence, the decomposition of $\Tr F^6$ should read
\be
\label{e-4-2-19}
\Tr F^6 = \frac{1}{24} \Tr F^2 \Tr F^4 - \frac{1}{3600} (\Tr F^2)^3.
\ee
Provided that this holds, the factorized form of the anomaly polynomial is given by (\ref{e-4-1-7}) with
\be
\label{e-4-2-20}
\Omega_{4} = \frac{1}{30} \Tr F^2 - \frac{1}{2} \tr R^2 = \tr F^2 - \frac{1}{2} \tr R^2,
\ee
and
\be
\label{e-4-2-21}
\Omega_8 = \frac{15}{16} \tr R^4 + \frac{15}{64} (\tr R^2)^2 + \frac{5}{8} \Tr F^4 - \frac{1}{240} (\Tr F^2)^2 - \frac{1}{16} \tr R^2 \Tr F^2 .
\ee
For reasons that will become clear later, it is convenient to express $\Omega_8$ in terms of $\Omega_4$ and another 8--form $\tilde{\Omega}_8$ that only depends on the Lorentz curvature $R$. Using (\ref{e-4-1-22}) and performing straightforward manipulations, we write
\be
\label{e-4-2-22}
\Omega_8 = \frac{15}{8} \left( \tr F^2  - \frac{1}{2} \tr R^2 \right)^2 + \left[ \frac{15}{16}\tr R^4 - \frac{15}{64} (\tr R^2)^2 \right] = \frac{15}{8} \Omega_4   \Omega_4 + \tilde{\Omega}_8.
\ee
where $\tilde{\Omega}_8$ is the $F$--independent form
\be
\label{e-4-2-23}
\tilde{\Omega}_8 \equiv \frac{15}{16}\tr R^4 - \frac{15}{64} (\tr R^2)^2 .
\ee
This way, our factorized anomaly takes the form
\be
\label{e-4-2-24}
I_{12} = \frac{15}{8} \Omega_4 \Omega_4 \Omega_4 + \Omega_4 \tilde{\Omega}_8.
\ee

To determine the possible gauge groups that may yield the desired cancellation, we may proceed as before. Now, the obvious candidate for the gauge group is the 248--dimensional $E_8$ group. Indeed, the $E_8$ identities (\ref{e-4-1-19}) 
\be
\label{e-4-2-25}
\Tr F^6 = \frac{1}{24} \Tr F^2 \frac{1}{100} (\Tr F^2 )^2  - \frac{1}{3600} ( \Tr F^2 )^3 = \frac{1}{24} \Tr F^2 \Tr F^4 - \frac{1}{3600} (\Tr F^2)^3,
\ee
lead indeed to the desired decomposition (\ref{e-4-2-19}). Moreover, it may be proven that only this group has this property. Therefore, anomaly cancellation on the fixed planes uniquely determines the gauge group on each fixed plane to be $E_8$.

A very important result that emerges from this anomaly analysis is that the particle spectrum of the full theory is constrained to be exactly that of the $E_8 \times E_8$ 10--dimensional heterotic string, this time however with one $E_8$ localized at each 10D boundary of the 11D world. This led Ho\v rava and Witten to conjecture that 11--dimensional supergravity compactified on $\mathbf{S}^1/\mathbb{Z}_2$ corresponds to some special limit of the $E_8 \times E_8$ heterotic string (hence the name ``heterotic M-theory'') and showed that this limit must be a strong-coupling one. Therefore, anomaly considerations have again taken us a long way towards discovering new, unexpected, duality relations between string theory and M-theory.

\subsection{The boundary action}
\label{sec-4-2-3}

Having introduced new boundary fields, necessary for anomaly cancellation, we must now determine the action describing those fields and their couplings to the bulk supergravity. This can be done by starting from the globally supersymmetric Yang-Mills action and apply the Noether method introducing additional interactions and modifications to the supersymmetry transformations until the theory becomes locally supersymmetric. However, at the present stage, there are several unresolved issues which should be taken into account.

First of all, the super Yang-Mills action introduces a gauge coupling $g$, which is \emph{a priori} undetermined. Moreover, one can combine it with the gravitational coupling $\kappa$ and form the dimensionless combination
\be
\label{e-4-2-26}
\eta = \frac{g^6}{\kappa^4}.
\ee
Given that the theory is to be dual to the heterotic string, the parameter $\eta$ must be fixed by some scalar expectation value of the string theory. However, the only scalar in heterotic string theory is the dilaton $\phi$ which, in the given compactification, is determined in terms of compactification radius $R$. Therefore, from the string viewpoint, the parameter $\eta$ is an extra parameter introduced into the theory. In order for the duality with the heterotic string to work, the parameter $\eta$ must be fixed within the present theory, i.e. the value of the gauge coupling $g$ must be fixed in terms of the gravitational coupling $\kappa$.

Second, the mechanism for anomaly cancellation is not obvious. Since the theory does not contain a $2p$--form field localized on the boundary, the usual Green-Schwarz mechanism is clearly not applicable. An important hint is provided by the calculations of \cite{Vafa:1995fj,Duff:1995wd} which show that string loop corrections to Type IIA superstring theory induce a $B_2 \tilde{\Omega}_8$ Green-Schwarz-type interaction where $\tilde{\Omega}_8$ is given in (\ref{e-4-2-23}). Since M-theory reduces to the Type IIA theory on $\mathbf{S}^1$, it must contain an $A_3 \tilde{\Omega}_8$ interaction. Then, if the boundary conditions of $A_3$ are modified so as to induce an anomalous variation on the boundary, this interaction might serve to cancel part of the anomaly. Moreover, such a modification to the boundary conditions also affects the Chern-Simons $A_3 F_4 F_4$ interaction and so we have a second candidate for an anomaly-cancelling term. However, this still leaves us with the question of what gives $A_3$ its anomalous variation.

The above issues are actually more closely related than it might seem at first and can be resolved in an elegant and consistent way. The additional interactions required for local supersymmetry require the modification of the boundary value (or the Bianchi identity) of $F_4$ by a term proportional to $\frac{\kappa^2}{g^2} \Omega_4$, where $\Omega_4$ is the anomaly 4--form in (\ref{e-4-2-20}). Solving the resulting equations for $A_3$, we find that $A_3$ has an anomalous variation of the boundary. Then, it turns out that the first part of the anomaly (\ref{e-4-2-24}) is cancelled by the variation of $A_3 F_4 F_4$ while the second part is cancelled by the variation of $A_3 \tilde{\Omega}_8$, with all coefficients working out exactly right if the gauge coupling $g$ is related to the gravitational coupling $\kappa$ in a certain way. The final result is an anomaly-free, self-consistent theory with no undetermined fundamental parameters.

\subsubsection{The vector multiplet action}

To construct the locally supersymmetric action for the boundary vector multiplets, we start from the globally supersymmetric $D=10$, $N=1$ super Yang-Mills action
\be
\label{e-4-2-27}
g^2 e^{-1} \ML^{(0)}_{10} =  - \frac{1}{4} F^I_{\mu\nu} F_I^{\mu\nu} - \frac{1}{2} \bar{\lambda}^I \Gamma^\mu D_\mu \lambda_I.
\ee
which is invariant under the supersymmetry transformations
\be
\label{e-4-2-28}
\delta A^I_\mu = \frac{1}{2} \bar{\epsilon} \Gamma_\mu \lambda^I ,\qquad \delta \lambda^I = - \frac{1}{4} \Gamma^{\mu\nu} F^I_{\mu\nu} \epsilon.
\ee
To couple the theory to the bulk supergravity, we introduce the usual Noether coupling of the supercurrent to the gravitino, given by
\be
\label{e-4-2-29}
\ML^{(1)}_{10} = - \frac{1}{4 g^2} e \bar{\psi}_\mu \Gamma^{\nu\rho} \Gamma^\mu F^I_{\nu\rho} \lambda_I.
\ee
However, this still leaves the residual variation
\be
\label{e-4-2-30}
\Delta = \frac{1}{16 g^2} e \bar{\psi}_\mu \Gamma^{\mu\nu\rho\sigma\tau} F^I_{\nu\rho} F_{I \sigma\tau} \epsilon,
\ee
which cannot be cancelled neither by introducing additional boundary interactions nor by modifying the supersymmetry transformation rules. However, a crucial observation is that the \emph{bulk} theory contains the interaction
\be
\label{e-4-2-31}
- \frac{1}{192 \kappa^2} \int_{M_{11}} \dd^{11} x E \bar{\psi}_L \Gamma^{LMNPQR} \psi_R F_{MNPQ}
\ee
whose variation under $\delta \psi_M = D_M \epsilon$ contains the term
\be
\label{e-4-2-32}
- \frac{1}{96 \kappa^2} \int_{M_{11}} \dd^{11} x E \bar{\psi}_\mu \Gamma^{\mu\nu\rho\sigma\tau} F_{\nu\rho\sigma\tau} \Gamma^{10} \partial_{11} \epsilon = - \frac{1}{96 \kappa^2} \int_{M_{11}} \dd^{11} x e \bar{\psi}_\mu \Gamma^{\mu\nu\rho\sigma\tau} F_{\nu\rho\sigma\tau} \partial_{11} \epsilon.
\ee
where we used $\Gamma^{10} \epsilon = \frac{e}{E} \Gamma_{11} \epsilon = \frac{e}{E} \epsilon$. This term has the right structure to cancel $\Delta$, \emph{provided that the boundary condition (or Bianchi identity) of} $F_4$ \emph{is appropriately modified on the boundary}.

In the downstairs approach where the theory lives on a manifold $M_{11,D}$ with boundary, this relation is formulated as follows. We first integrate (\ref{e-4-2-32}) by parts and obtain the surface term
\be
\label{e-4-2-33}
- \frac{1}{96 \kappa^2} \int_{M_{11,D}} \dd^{11} x e \partial_{11} \left( \bar{\psi}_\mu \Gamma^{\mu\nu\rho\sigma\tau} F_{\nu\rho\sigma\tau} \epsilon \right) = \frac{1}{96 \kappa^2} \int_{M_{10}} \dd^{10} x e \bar{\psi}_\mu \Gamma^{\mu\nu\rho\sigma\tau} F_{\nu\rho\sigma\tau} \epsilon,
\ee
where we restricted our considerations to the fixed plane at $x_{11}=0$. This surface term can cancel the variation $\Delta$, provided that the boundary value of $F_{\mu\nu\rho\sigma}$ (previously equal to zero by the $\mathbb{Z}_2$ projection) is taken to be
\be
\label{e-4-2-34}
F_{\mu\nu\rho\sigma} \bigr|_{x_{11}=0} = - \frac{6 \kappa^2}{g^2} F^I_{[\mu\nu} F_{I \rho\sigma]} = \frac{6 \kappa^2}{g^2} \tr \left( F_{[\mu\nu} F_{\rho\sigma]} \right),
\ee
or, in form notation,
\be
\label{e-4-2-35}
F_4 \bigr|_{x_{11}=0} = \frac{\kappa^2}{g^2} \tr F^2.
\ee
Actually, from the structure of the anomaly 4--form $\Omega_4$ in (\ref{e-4-2-20}), we deduce that the above term must be supplemented by a gravitational contribution. Thus, we write
\be
\label{e-4-2-36}
F_4 \bigr|_{x_{11}=0} = \frac{\kappa^2}{g^2} \left( \tr F^2 - \frac{1}{2} \tr R^2 \right) = \frac{\kappa^2}{g^2} \Omega_4.
\ee
Eq. (\ref{e-4-2-36}) must be solved for the gauge field $A_3$. To this end, we introduce the Chern-Simons form of $\Omega_4$ according to 
\be
\label{e-4-2-37}
\Omega_{4} = \dd \Omega_3 ;\qquad \Omega_3 = \omega_{3Y} - \frac{1}{2} \omega_{3L},
\ee
and we see that the boundary value (\ref{e-4-2-36}) for $F_4$ can be obtained by requiring that
\be
\label{e-4-2-38}
A_3 \bigr|_{x_{11}=0} = \frac{\kappa^2}{g^2} \Omega_3.
\ee
From the descent equation
\be
\label{e-4-2-39}
\delta_{v,\lambda} \Omega_3 = \dd \Omega^1_2 ;\qquad \Omega^1_2 =  \omega^1_{2Y} - \frac{1}{2} \omega^1_{2L}.
\ee
it follows that $A_3$ acquires an anomalous gauge/Lorentz variation, given by
\be
\label{e-4-2-40}
\delta_{v,\lambda} A_3 \bigr|_{x_{11}=0} = \frac{\kappa^2}{ g^2} \dd \Omega^1_2.
\ee

In the upstairs approach, originally used in \cite{Horava:1996ma}, the partial integration of the variation (\ref{e-4-2-32}) does not give rise to a surface term but instead leaves the 11D variation
\be
\label{e-4-2-41}
\frac{1}{192 \kappa^2} \int_{M_{11,\mathrm{U}}} \dd^{11} x E \bar{\psi}_\mu \Gamma^{\mu\nu\rho\sigma\tau} \epsilon \partial_{11} F_{\nu\rho\sigma\tau}
\ee
This can cancel the variation $\Delta$, provided that the Bianchi identity of $F_4$ is modified to
\be
\label{e-4-2-42}
dF_{4} = - \frac{2 \kappa^2}{g^2} \delta_1 \Omega_4,
\ee
where we introduced the distribution $\delta_1 = \delta(x_{11}) dx^{11}$. This modified Bianchi identity admits a 1--parameter family of solutions for $A_3$. This seems to bring an extra free parameter in the theory but it turns out that anomaly cancellation eventually fixes the value of this parameter. For a very careful treatment of anomaly cancellation and related issues in the upstairs approach, the reader is referred to \cite{Bilal:1999ig,Bilal:2003es}.

The construction of the rest of the action is straightforward. Without going into much detail, we note that a modification of the supersymmetry transformation law for $F_4$ induces an uncancelled variation of the $F_4$ kinetic term while the supercurrent interaction $\ML^{(2)}_{10}$ also has an uncancelled variation under $\delta \Psi_M \sim \Gamma F_4 \epsilon$, both variations being of the form $\bar{\lambda} \Gamma F_4 F_2 \epsilon$ \cite{Horava:1996ma}. Eventually both terms cancel if we introduce the extra interaction
\be
\label{e-4-2-43}
\ML^{(2)}_{10} = \frac{1}{24 g^2} e \lambda^I \Gamma^{\mu\nu\rho} \lambda_I F_{\mu\nu\rho 11} .
\ee
This completes the construction of the theory, up to $(\mathrm{Fermi})^4$ terms.

\subsection{Anomaly cancellation}
\label{sec-4-2-4}

Let us now study the anomaly cancellation mechanism of the theory. As remarked earlier on, the possible sources for the cancellation of anomalies are the $A_3 F_4 F_4$ Chern-Simons term and the $A_3 \tilde{\Omega}_8$ ``Green-Schwarz'' term, and we have just verified that the requirement for supersymmetry induces a modified boundary condition for $F_4$ and an anomalous variation for $A_3$. What remains is to see whether the coefficients work out right so that the anomalies cancel and how the value of the parameter $\eta$ is determined.

We start from the Chern-Simons term of (\ref{e-4-2-1}). Using form notation, we write it as
\be
\label{e-4-2-44}
S_{CS} = \frac{1}{3 \kappa^2} \int_{M_{11,D}} A_3 F_4 F_4.
\ee
Then, using the anomalous variation (\ref{e-4-2-40}) of $A_3$ and the boundary condition (\ref{e-4-2-36}) for $F_4$, we readily find that the gauge/Lorentz variation of $S_{CS}$ is
\be
\label{e-4-2-45}
\delta_{v,\lambda} S_{CS} = \frac{1}{3 g^2} \int_{M_{11,D}} \dd \Omega^1_2 F_4 F_4 = \frac{1}{3g^2} \int_{M_{11,D}} \dd \left( \Omega^1_2 F_4 F_4 \right) = - \frac{\kappa^4}{3 g^6} \int_{M_{10}} \Omega^1_2 \Omega_4 \Omega_4.
\ee
This variation corresponds to the anomaly 10--form
\be
\label{e-4-2-46}
\hat{I}^{1}_{10,CS} = \frac{\kappa^4}{3 g^6} \Omega^1_2 \Omega_4 \Omega_4.
\ee
The corresponding anomaly polynomial is easily computed by the descent equations (in reverse order). We have
\be
\label{e-4-2-47}
\dd \hat{I}^{1}_{10,CS} = - \delta \hat{I}^{CS}_{11} ;\qquad \hat{I}^{CS}_{11} = - \frac{\kappa^4}{3 g^6} \Omega_3 \Omega_4 \Omega_4,
\ee
and
\be
\label{e-4-2-48}
\hat{I}_{12,CS} = \dd \hat{I}_{11,CS} = - \frac{\kappa^4}{3 g^6} \Omega_4   \Omega_4   \Omega_4 .
\ee
In our ``unhatted'' normalization of \S\ref{sec-3-4}, we write
\be
\label{e-4-2-49}
I_{12,CS} = (2\pi)^5 6! \hat{I}_{12,CS} = - \frac{7680 \pi^5 \kappa^4}{g^6} \Omega_4 \Omega_4 \Omega_4.
\ee
This anomaly polynomial has exactly the right form to cancel the first part of the anomaly (\ref{e-4-2-24}), provided that the respective coefficients are equal and opposite. This leads to the relation
\be
\label{e-4-2-50a}
\eta = \frac{g^6}{\kappa^4} = 4 ( 4 \pi )^5.
\ee
which determines the numerical value of the dimensionless coupling $\eta$ of the theory\footnote{The difference from the standard value $\eta = ( 4 \pi )^5$ found e.g. in \cite{Harmark:1998bs,Bilal:2003es} is only due to our different conventions for the gravitational coupling which follow those of \cite{Cremmer:1978km}.}.

Let us pass to the $A_3 \tilde{\Omega}_8$ ``Green-Schwarz'' term of the theory. A computation through string loop corrections \cite{Vafa:1995fj} or fivebrane considerations \cite{Duff:1995wd} gives, in our present conventions,
\be
\label{e-4-2-51}
S_{GS} = \frac{1}{720 \pi^3 (8 \pi\kappa^2)^{1/3}} \int_{M_{11,D}} A_3 \tilde{\Omega}_8.
\ee
where the form $\tilde{\Omega}_8$ is exactly the one appearing in \ref{e-4-2-23}, hinting that this term can completely cancel the remaining anomaly. The application of the Green-Schwarz mechanism is then quite straightforward. Evaluating the gauge/Lorentz variation of $S_{GS}$ according to (\ref{e-4-2-40}) and using (\ref{e-4-2-50a}), we find
\bea
\label{e-4-2-52}
\delta_{v,\lambda} S_{GS} &=& \frac{1}{720 \pi^3 (8\pi\kappa^2)^{1/3}} \int_{M_{11,D}} \delta_{v,\lambda} A_3   \tilde{\Omega}_8 = \frac{1}{720 \pi^3 (8 \pi)^{1/3}} \left( \frac{g^6}{\kappa^4} \right)^{-1/3} \int_{M_{11,D}} \dd \Omega^1_2 \tilde{\Omega}_8 \nn\\
&=& \frac{1}{23040 \pi^5} \int_{M_{11,D}} \dd \left( \Omega^1_2   \tilde{\Omega}_8 \right) = - \frac{1}{23040 \pi^5} \int_{M_{10}} \Omega^1_2   \tilde{\Omega}_8.
\eea
This leads to the anomaly 10--form
\be
\label{e-4-2-53}
\hat{I}^1_{10,GS} = \frac{1}{23040 \pi^5} \Omega^1_2 \tilde{\Omega}_8,
\ee
and to the anomaly polynomial
\be
\label{e-4-2-54}
I_{12,GS} = - \Omega_4 \tilde{\Omega}_8.
\ee
This is equal and opposite to the second term of (\ref{e-4-2-24}) and thus this part of the anomaly does also cancel. It is remarkable that the Green-Schwarz term (\ref{e-4-2-51}), deduced through string loop calculations, involves exactly the form $\tilde{\Omega}_8$ appearing in our anomaly analysis and has exactly the right coefficient to cancel the anomaly when we fix $\eta$ to the value given in (\ref{e-4-2-50a}).

\chapter{Anomaly-Free Supergravities in Six Dimensions}
\label{chap-5} 

The issue of anomaly freedom in chiral supergravities can be examined not only in the usual context of superstring theories, but also in the context of lower-dimensional supergravities, many of which arise from superstring and M-theory compactifications. Among these latter theories, six-dimensional chiral supergravities play a prominent role due to the fact that the presence of gravitational and mixed nonabelian anomalies on top of gauge and mixed abelian ones in these theories implies that the requirement of anomaly cancellation may lead to powerful constraints singling out a relatively small number of consistent models. Moreover, there also exists the option of gauging the R-symmetry of the theory, which leads to supergravities that cannot be realized in terms of straightforward string compactifications. In this chapter, we will review $D=6$, $N=2$ minimal supergravities, we will thoroughly examine the issue of anomaly cancellation in these theories and we will present the results of a systematic search for anomaly-free theories, under a certain set of conditions on the allowed gauge groups and representations. We also discuss anomaly cancellation on the six-dimensional theories living on the fixed points of $D=7$, $N=2$ supergravity on $\mathbf{S}^1 / \mathbb{Z}_2$.

\section{$D=6$, $N=2$ Supergravity}
\label{sec-5-1} 

\subsection{A quick survey of $D=6$ supergravities}
\label{sec-5-0} 

Before we begin our study of $D=6$, $N=2$ minimal supergravities, let us first present a survey of the various six-dimensional supergravity theories. The available possibilities can be read off Table 2.1 and the main properties of the resulting theories are summarized as follows.

\begin{enumerate}
\item $N=8$. There exists a non-chiral $(N_+,N_-)=(4,4)$ supergravity, constructed in \cite{Tanii:1984zk}, as well as chiral $(N_+,N_-)=(8,0)$ and $(N_+,N_-)=(6,2)$ supergravities \cite{Townsend:1983xt} which have not been constructed up to date. These theories are considered of limited interest.

\item $N=4$. There exists a non-chiral $(N_+,N_-)=(2,2)$ supergravity, whose first version was constructed in \cite{Giani:1984dw}. This theory contains the $(N_+,N_-)=(2,2)$ gravity and vector multiplets, it has an $\mathrm{SU}(2)$ gauge group and admits no stable maximally symmetric vacua. The construction was generalized by Romans \cite{Romans:1985tw} to include a mass term for the 2--form of the supergravity multiplet and leads to distinct theories, determined by the values of the $\mathrm{SU}(2)$ coupling $g$ and the mass parameter $m$. For $g=0$, one has an ungauged supergravity with a 6D Minkowski vacuum. The theory with $g > 0$ and $m > 0$ is  a gauged theory known as Romans supergravity. It is symmetric under the anti-de Sitter superalgebra $F(4)$, it has a scalar potential with two AdS extrema (one fully supersymmetric and one non-supersymmetric) and it is the only pure supergravity that exhibits the Higgs mechanism in a supersymmetric ground state. This theory can also be obtained \cite{Cvetic:1999un} by a warped $\MBFS^4$ reduction of massive Type IIA supergravity. There also exists a chiral $(N_+,N_-)=(4,0)$ theory \cite{Townsend:1983xt,Romans:1986er} containing the $(N_+,N_-)=(4,0)$ gravity and tensor multiplets.

\item $N=2$. The $N=2$ supergravities are necessarily chiral and were first constructed in \cite{Nishino:1984gk,Nishino:1986dc}. They contain the $N=2$ supergravity multiplet plus arbitrary numbers of vector multiplets, tensor multiplets and hypermultiplets. The vector multiplets may belong to a gauge group under which the hypermultiplets may be charged. The case when the gauge group does not include the $\mathrm{USp}(2)$ R-symmetry group leads to ungauged theories with a 6D Minkowski vacuum. The case when the gauge group includes a subgroup of the R-symmetry group leads to gauged theories which have the remarkable property that they admit a vacuum consisting of 4D Minkowski space times an internal $\MBFS^2$. Many of the ungauged theories may be obtained from string theory via various mechanisms; on the contrary, the origin of the gauged theories still remains unclear.
\end{enumerate}

From the above, it is already obvious that the $N=2$ chiral supergravities are by far the most interesting among the six-dimensional supergravities, since they have the richest structure and they are the ones that can potentially be used to make contact with 4D phenomenology. We turn to a detailed study of these theories immediately on.

\subsection{General facts about $D=6$, $N=2$ supergravity}
\label{sec-5-00} 

The minimal supersymmetry algebra in six dimensions is the $N=2$ algebra\footnote{In the literature, one often encounters an alternative convention where the number of supersymmetries is counted in terms of Weyl spinors; in that convention, the algebra is called $N=1$.} which is chiral and has $\mathrm{USp}(2)$ as its R-symmetry group. As we saw in \S\ref{sec-2-1-2}, its massless representations are classified in terms of the $\mathrm{SO}(4) \cong \mathrm{SU}(2) \times \mathrm{SU}(2)$ little group and the $\mathrm{USp}(2)$ R-symmetry group and are given by
\bea
\label{e-5-1-1}
\text{Supergravity multiplet} \quad&:&\quad (\mathbf{3},\mathbf{3};\mathbf{1}) + (\mathbf{3},\mathbf{1};\mathbf{1}) + (\mathbf{3},\mathbf{2};\mathbf{2}) = ( g_{\mu\nu} , B^+_{\mu\nu} , \psi^{A-}_\mu ),\nn\\
\text{Tensor multiplet}       \quad&:&\quad (\mathbf{1},\mathbf{3};\mathbf{1}) + (\mathbf{1},\mathbf{1};\mathbf{1}) + (\mathbf{1},\mathbf{2};\mathbf{2}) = ( B^-_{\mu\nu} , \phi , \chi^{A+} ), \nn\\
\text{Vector multiplet}        \quad&:&\quad (\mathbf{2},\mathbf{2};\mathbf{1}) + (\mathbf{2},\mathbf{1};\mathbf{2}) = ( A_\mu, \lambda^{A-} ), \nn\\
\text{Hypermultiplet}          \quad&:&\quad 4 (\mathbf{1},\mathbf{1};\mathbf{2}) + 2 (\mathbf{1},\mathbf{2};\mathbf{1}) = ( 4 \varphi , 2 \psi^{+} ).
\eea
Here, the spinors are symplectic Majorana-Weyl, the index $A=1,2$ takes values in the fundamental representation of $\mathrm{USp}(2)$ and the $+$ ($-$) superscripts denote positive (negative) chirality for spinors and (anti-)self-duality for 2--forms.

A general $D=6$, $N=2$ supergravity theory coupled to matter is constructed by combining one supergravity multiplet with $n_T$ tensor multiplets, $n_V$ vector multiplets and $n_H$ hypermultiplets, where $n_T$, $n_V$ and $n_H$ are defined so as to include group multiplicities. The $n_T$ real scalars in the tensor multiplet parameterize the coset space $\mathrm{SO}(1,n_T) / \mathrm{SO}(n_T)$. The $4 n_H$ real hyperscalars parameterize a quaternionic manifold of the form
\be
\label{e-5-1-2}
\MM = \frac{G}{H \times \mathrm{USp}(2) },
\ee
where the $\mathrm{USp}(2)$ subgroup is identified with the R-symmetry group, and the hyperinos furnish an appropriate representation of $H$. The allowed choices for $(G,H)$ \cite{Nishino:1986dc} are the canonical case $( \mathrm{USp}(2n_H,2) , \mathrm{USp}(2 n_H) )$ plus the extra cases $( \mathrm{SU}(n_H,2) , \mathrm{SU}(n_H) \times \mathrm{U}(1) )$, $(\mathrm{SO}(n_H,4) , \mathrm{SO}(n_H) \times \mathrm{SO}(3) )$, $( E_8 , E_7 )$, $( E_7 , \mathrm{SO}(12) )$, $( E_6 , \mathrm{SU}(6) )$, $( F_4 , \mathrm{USp}(6) )$ and $( G_2, \mathrm{USp}(2) )$. The vector multiplets may belong to a gauge group $\MG$ which is the product of a subgroup of the isometry group $G$ and a possible ``shadow'' group $S$ under which all other multiplets are inert. In the first three cases, where $G$ is non-compact, this essentially means\footnote{For the remaining cases where $G$ is compact, the gauge group can be any subgroup of $G$ times $S$ but the hyperinos are restricted to transform only under $H$.} that $\MG$ is a subgroup of the $H \times \mathrm{USp}(2)$ holonomy group times $S$. We write this gauge group as $\MG = \MG_h \times \MG_r$, where $\MG_h$ contains the factors from $H$ and $S$ while $\MG_r$ is the R-symmetry factor arising in gauged theories and can be either $\mathrm{USp}(2)$ or a $\mathrm{U}(1)$ subgroup thereof. Introducing an extended index $x=1,\ldots,N$ that runs over all group factors in $\MG_h \times \MG_r$ (i.e. $x=\alpha$ for ungauged theories and $x =( \alpha,r )$ for gauged theories), we write the full group as $\MG = ( \prod_A \MG_A ) \times \MG_a$.

The transformation properties of the various fermions under the gauge group are as follows. Under $\MG_h$, the hyperinos may transform in arbitrary representations while the gravitino and tensorinos are inert. Under $\MG_r$, the hyperinos are inert (although the hyperscalars are charged) while the gravitino, tensorinos and gauginos transform non-trivially. In particular, in the case where the whole $\mathrm{USp}(2)$ is gauged, Eq. (\ref{e-5-1-1}) indicates that the gravitino, the tensorinos and the $\MG_s$ gauginos transform in the fundamental $\mathbf{2}$ while the $\mathrm{USp}(2)$ gauginos transform in the $\mathbf{3} \times \mathbf{2} = \mathbf{2} + \mathbf{4}$. In the case where only a $\mathrm{U}(1) \subset \mathrm{USp}(2)$ is gauged, the gravitino, the tensorinos and all gauginos have unit charge.

The $D=6$, $N=2$ supergravity coupled to matter was first constructed by Nishino and Sezgin in \cite{Nishino:1984gk,Nishino:1986dc} and further generalized in \cite{Nishino:1997ff,Ferrara:1998gh,Riccioni:2001bg} to include new couplings and more than one tensor multiplet. In the general case, the standard construction method is to start with a general ansatz for the supersymmetry variations and equations of motion involving a set of undetermined coefficients and then fix all coefficients through the requirement of closure of the supersymmetry algebra on the fields. This approach, although rather technical, has the merit that it is valid for arbitrary $n_T$ where a Lagrangian formulation of the theory is not possible due to the self-duality conditions; in the $n_T=1$ case, the invariant Lagrangian can be derived by integrating the equations of motion. On the other hand, since we are mostly interested in the case $n_T=1$, it is more natural to follow the somewhat simpler Noether procedure to derive the full supergravity Lagrangian, up to $(\mathrm{Fermi})^4$ terms.

\subsection{Gravity and tensor multiplets}
\label{sec-5-1-1} 

Our starting point is the the locally supersymmetric Lagrangian describing the gravity and tensor multiplets, i.e. the set of fields $(g_{\mu\nu},B_{\mu\nu},\psi_\mu,\phi,\chi)$. This can be derived quite easily by standard methods and we do not find it necessary to repeat the discussion here. The result is the Lagrangian
\bea
\label{e-5-1-3}
e^{-1} \mathcal{L}_{GT} &=& \frac{1}{4} R - \frac{1}{12} e^{2 \phi} G_{\mu\nu\rho} G^{\mu\nu\rho} - \frac{1}{4} \partial_\mu \phi \partial^\mu \phi - \frac{1}{2} \bar{\psi}_\mu \Gamma^{\mu\nu\rho} D_\nu \psi_\rho - \frac{1}{2} \bar{\chi} \Gamma^\mu D_\mu \chi
\nn\\
&-& \frac{1}{24} e^\phi \left( \bar{\psi}^\lambda \Gamma_{[\lambda} \Gamma^{\mu\nu\rho} \Gamma_{\sigma]} \bar{\psi}^\sigma - 2 \bar{\chi} \Gamma^\lambda \Gamma^{\mu\nu\rho} \psi_\lambda - \bar{\chi} \Gamma^{\mu\nu\rho} \chi \right) G_{\mu\nu\rho} 
\nn\\ 
&-& \frac{1}{2} \bar{\chi} \Gamma^\nu \Gamma^\mu \psi_\nu \partial_\mu \phi + (\mathrm{Fermi})^4,
\eea 
which is invariant under the set of supersymmetry transformations
\bea
\label{e-5-1-4}
\delta e^a_\mu &=& \bar{\epsilon} \Gamma^a \psi_\mu, 
\nn\\
\delta B_{\mu\nu} &=& e^{-\phi} \left( - \bar{\epsilon} \Gamma_{[\mu} \psi_{\nu]} + \frac{1}{2} \bar{\epsilon} \Gamma_{\mu\nu} \chi \right)
\nn\\
\delta \phi &=& \bar{\epsilon} \chi, 
\nn\\
\delta \psi_\mu &=& D_\mu \epsilon + \frac{1}{24} e^\phi \Gamma^{\nu\rho\sigma} \Gamma_\mu G_{\nu\rho\sigma} \epsilon,
\nn\\
\delta \chi &=& \frac{1}{2} \Gamma^\mu \partial_\mu \phi \epsilon - \frac{1}{12} e^\phi \Gamma^{\mu\nu\rho} G_{\mu\nu\rho} \epsilon.
\eea

\subsection{Vector multiplets}
\label{sec-5-1-2} 

The next step in our construction is to couple the theory to vector multiplets in a way consistent with local supersymmetry. For definiteness, we take the gauge group $\MG$ to be the full holonomy group of the scalar manifold, $\MG = H \times \mathrm{USp}(2)$, and we let $\hat{I}$ be the adjoint index for $\MG$ which is decomposed into the adjoint indices $I$ and $i$ of $H$ and $\mathrm{USp}(2)$ respectively (the restriction to a subgroup of $H \times \mathrm{USp}(2)$ as well as the addition of abelian factors is a trivial matter). Our starting point is the well-known globally supersymmetric Lagrangian for vector multiplets,
\be
\label{e-5-1-5} 
e^{-1} \mathcal{L}^{(0)}_V = - \frac{1}{4} e^{-\phi} v_X ( F^{\hat{I}}_{\mu\nu} F^{\mu\nu}_{\hat{I}} )_X - \frac{1}{2} v_X ( \bar{\lambda}^{\hat{I}} \Gamma^\mu D_\mu \lambda_{\hat{I}} )_X,
\ee
where the summation index $X$ labels the simple factors of the gauge group and $v_X$ are some numerical constants that will be determined by anomaly considerations in the next section. The Lagrangian (\ref{e-4-4}) is invariant under the rigid supersymmetry transformations
\be
\label{e-5-1-6}
\delta A^{\hat{I}}_\mu = \frac{1}{\sqrt{2}} e^{\phi/2} \bar{\epsilon} \Gamma_\mu \lambda^{\hat{I}} ,\qquad\qquad \delta \lambda^{\hat{I}} = - \frac{1}{2 \sqrt{2}} e^{-\phi/2} \Gamma^{\mu\nu} \epsilon F^{\hat{I}}_{\mu\nu}.
\ee
Our first step towards obtaining a locally supersymmetric theory is to introduce the usual Noether coupling of the gravitino to the supercurrent of the multiplet. The required term is
\be
\label{e-5-1-7}
\mathcal{L}^{(1)}_V = - \frac{1}{2 \sqrt{2}} e e^{-\phi/2} v_X ( \bar{\psi}_\mu \Gamma^{\nu\rho} \Gamma^\mu \lambda^{\hat{I}} F_{\hat{I} \nu\rho} )_X.
\ee
Next, we must cancel the $\bar{\lambda} \Gamma F \partial \phi \epsilon$ variation of $\mathcal{L}^{(0)}_V$. This variation is found to be
\be
\label{e-5-1-8}
\Delta^{(1)}_V = \frac{1}{4 \sqrt{2}} e e^{-\phi/2} v_X ( \bar{\lambda}^{\hat{I}} \Gamma^{\mu\nu} \Gamma^\rho \epsilon_i F_{\hat{I} \mu\nu} \partial_\rho \phi )_X,
\ee
and can be cancelled by the $\delta \chi \sim \Gamma \partial \phi \epsilon$ variation of the additional term
\be
\label{e-5-1-9}
\mathcal{L}^{(2)}_V = - \frac{1}{2 \sqrt{2}} e e^{-\phi/2} v_X ( \bar{\lambda}^{\hat{I}} \Gamma^{\mu\nu} \chi F_{\hat{I} \mu\nu} )_X.
\ee
The introduction of these new interactions results in additional uncancelled terms of the form $\bar{\lambda} \Gamma F G \epsilon$ coming from the $\delta \psi$ and $\delta \chi$ variations of $\mathcal{L}^{(1)}_V$ and
$\mathcal{L}^{(2)}_V$ respectively. The first one vanishes by the 6D identity $\Gamma^\mu \Gamma^{\nu\rho\sigma} \Gamma_\mu = 0$, while the second one is given by
\be
\label{e-5-1-10}
\Delta^{(2)}_V = \frac{1}{24 \sqrt{2}} e e^{\phi/2} v_X ( \bar{\lambda}^{\hat{I}} \Gamma^{\mu\nu} \Gamma^{\rho\sigma\tau} \epsilon G_{\rho\sigma\tau} F_{\hat{I} \mu\nu} )_X.
\ee
This can be cancelled by introducing the additional interaction
\be
\label{e-5-1-11}
\mathcal{L}^{(3)}_V = \frac{1}{24} e e^{\phi} v_X ( \bar{\lambda}^{\hat{I}} \Gamma^{\mu\nu\rho} \lambda_{\hat{I}}
G_{\mu\nu\rho} )_X.
\ee
What remains is to cancel the $\bar{\psi} \Gamma F^2 \epsilon$ and $\bar{\chi} \Gamma F^2 \epsilon$ terms coming from the $\delta \lambda$ variations of $\mathcal{L}^{(1)}_V$ and $\mathcal{L}^{(2)}_V$. These terms are given by
\be
\label{e-5-1-12}
\Delta^{(3)}_V = \frac{1}{8} e e^{-\phi} v_X ( \bar{\psi}_\mu \Gamma^{\mu\nu\rho\sigma\tau} \epsilon F^{\hat{I}}_{\nu\rho} F_{\hat{I} \sigma\tau} )_X,
\ee
and
\be
\label{e-5-1-13}
\Delta^{(4)}_V = \frac{1}{8} e e^{-\phi} v_X ( \bar{\chi} \Gamma^{\mu\nu\rho\sigma} \epsilon F^{\hat{I}}_{\mu\nu} F_{\hat{I} \rho\sigma} )_X.
\ee
To cancel these terms, we use standard spinor identities plus the gamma-matrix duality relation (\ref{e-b-1-7}) to write thir sum in the form
\be
\label{e-5-1-14}
\Delta^{(3)}_V + \Delta^{(4)}_V = - \frac{1}{8} \epsilon^{\mu\nu\rho\sigma\tau\upsilon} \left[ e^{-\phi} \left( - \bar{\epsilon} \Gamma_{[\mu} \psi_{\nu]} + \frac{1}{2} \bar{\epsilon} \Gamma_{\mu\nu} \chi \right) \right] v_X ( F^{\hat{I}}_{\rho\sigma} F_{\hat{I} \tau\upsilon} )_X .
\ee
Noting that the term inside brackets is exactly the supersymmetry variation of $B_{\mu\nu}$ in (\ref{e-5-1-4}), we see that one can cancel this variation by introducing the interaction
\be
\label{e-5-1-15}
\mathcal{L}^{(4)}_V = \frac{1}{8} \epsilon^{\mu\nu\rho\sigma\tau\upsilon} B_{\mu\nu} v_X ( F^{\hat{I}}_{\rho\sigma} F_{\hat{I} \tau\upsilon} )_X.
\ee
This interaction, immediately recognized as a Green-Schwarz term, plays an important role in the context of anomaly cancellation. Here, we note that, in contrast to the 10D case, the 6D Green-Schwarz term is not a higher-derivative correction to the action of the theory but it is present in the low-energy action in the first place.

Combining the above terms, we finally find that the locally supersymmetric Lagrangian describing the vector multiplets and their couplings to the gravity and tensor multiplet is
\bea
\label{e-5-1-16} 
e^{-1} \mathcal{L}_V &=& - \frac{1}{4} e^{-\phi} v_X ( F^{\hat{I}}_{\mu\nu} F^{\mu\nu}_{\hat{I}} )_X - \frac{1}{2} v_X ( \bar{\lambda}^{\hat{I}} \Gamma^\mu D_\mu \lambda_{\hat{I}} )_X - \frac{1}{2 \sqrt{2}} e e^{-\phi/2} v_X ( \bar{\psi}_\mu \Gamma^{\nu\rho} \Gamma^\mu \lambda^{\hat{I}} F_{\hat{I} \nu\rho} )_X
\nn\\
&& - \frac{1}{2 \sqrt{2}} e e^{-\phi/2} v_X ( \bar{\lambda}^{\hat{I}} \Gamma^{\mu\nu} \chi F_{\hat{I} \mu\nu} )_X + \frac{1}{24} e e^{\phi} v_X ( \bar{\lambda}^{\hat{I}} \Gamma^{\mu\nu\rho} \lambda_{\hat{I}} G_{\mu\nu\rho} )_X
\nn\\
&& + \frac{1}{8} \epsilon^{\mu\nu\rho\sigma\tau\upsilon} B_{\mu\nu} v_X ( F^{\hat{I}}_{\rho\sigma} F_{\hat{I} \tau\upsilon} )_X.
\eea

\subsection{Hypermultiplets}
\label{sec-5-1-3} 

Our final step in constructing the $D=6$, $N=2$ supergravity action is to couple the theory to hypermultiplets. In what follows, we shall derive the action describing the hypermultiplets and their couplings to the other multiplets of the theory in two steps. In the first step, we will consider the case where the hypermultiplets are inert under the gauge group and we construct the appropriate locally supersymmetric action. In the second step, we will gauge the theory by identifying the gauge group with a subgroup of the holonomy group of the scalar manifold, letting the hypermultiplets transform under this gauge group and introducing extra interactions so as to maintain local supersymmetry. In the course of the discussion, we will explain the fact that the scalar manifold must be quaternionic in the locally supersymmetric case. We will also see the appearance of a scalar potential, which plays a significant role when the $\mathrm{USp}(2)$ R-symmetry is gauged.

\subsubsection{The action for neutral hypermultiplets}

We begin by constructing the locally supersymmetric action for the case where the hypermultiplets are neutral under the gauge group. For the moment, we take the scalar manifold to have the general form $\MM = G / H \times \mathrm{USp}(2)$, with $G$ and $H$ chosen so that its dimension is $4 n_H$; the metric of that space is denoted by $g_{\alpha \beta} ( \varphi )$. The restrictions imposed by supersymmetry on the scalar manifold will arise during the discussion.

Let us first examine the parameterization of the scalar manifold. According to the guidelines of \S\ref{sec-2-3}, we consider a coset representative given by a matrix $L$ whose Maurer-Cartan form decomposes as
\be
\label{e-5-1-17}
L^{-1} \partial_\alpha L = \MA_\alpha^{\phantom{\alpha} I} T_I + \MA_\alpha^{\phantom{\alpha} i} T_i +
\MV_\alpha^{\phantom{\alpha} a A} T_{a A}, 
\ee
$T_I$, $T_i$ and $T_{a A}$ are the generators of $H$, $\mathrm{USp}(2)$ and the coset, $\MA_\alpha^{\phantom{\alpha} I}$ and $\MA_\alpha^{\phantom{\alpha} i}$ are the $\mathrm{USp}(2 n_H)$ and $\mathrm{USp}(2)$ connections and $\MV_\alpha^{\phantom{\alpha} a A}$ is the coset vielbein. The pullback of this Maurer-Cartan form is given by
\be
\label{e-5-1-17a}
L^{-1} \partial_\mu L = \MQ_\alpha^{\phantom{\mu} I} T_I + \MQ_\mu^{\phantom{\mu} i} T_i +
\MP_\alpha^{\phantom{\mu} a A} T_{a A}, 
\ee
with
\be
\label{e-5-1-17b}
\MQ_\mu^{\phantom{\mu} I} = \partial_\mu \varphi^\alpha \MA_\alpha^{\phantom{\alpha} I} ,\qquad \MQ_\mu^{\phantom{\mu} i} = \partial_\mu \varphi^\alpha \MA_\alpha^{\phantom{\alpha} i} ,\qquad \MP_\mu^{\phantom{\mu} a A} = \partial_\mu \varphi^\alpha \MV_\alpha^{\phantom{\alpha} a A}.
\ee
As explained in Section \ref{sec-2-3}, the covariant derivatives of fields carrying $H$ and $\mathrm{USp}(2)$ indices must be modified by couplings to the composite connections $\MQ_\mu$. Explicitly, for the supersymmetry spinor $\epsilon$ (one $\mathrm{USp}(2)$ index), the hyperino $\psi$ (one $H$ index) and the gauginos $\lambda^{\hat{I}}$ (one $H \times \mathrm{USp}(2)$ adjoint index and one $\mathrm{USp}(2)$ fundamental index) we have
\bea
\label{e-5-1-18}
\MD_\mu \epsilon &=& D_\mu \epsilon + \partial_\mu \varphi^\alpha \MA_\alpha^{\phantom{\alpha} i} T_i \epsilon, \nn\\
\MD_\mu \psi &=& D_\mu \psi + \partial_\mu \varphi^\alpha \MA_\alpha^{\phantom{\alpha} I} T_I \psi,\nn\\
\MD_\mu \lambda^{\hat{I}} &=& D_\mu \lambda^{\hat{I}} + \partial_\mu \varphi^\alpha \MA_\alpha^{\phantom{\alpha} i} T_i \lambda^{\hat{I}}.
\eea
The reason for the absence of the composite $H$--connection in the covariant derivative of the gauginos $\lambda^{\hat{I}}$ is a technical one and originates \cite{Nishino:1986dc} from the requirement of supersymmetry of the action. Given these covariant derivatives, we can define the $H$ and $\mathrm{USp}(2)$ curvatures, $\MF_{\alpha\beta}^{\phantom{\alpha\beta} I}$ and $\MF_{\alpha\beta}^{\phantom{\alpha\beta} i}$ through the commutators
\bea
\label{e-5-1-19}
[ \MD_\mu , \MD_\nu ] \epsilon &=& \frac{1}{4} R_{\mu\nu\rho\sigma} \Gamma^{\rho\sigma} \epsilon + \partial_\mu \varphi^\alpha \partial_\nu \varphi^\beta \MF_{\alpha\beta}^{\phantom{\alpha\beta} i} T_i \epsilon, \nn\\
{}[ \MD_\mu , \MD_\nu ] \psi &=& \frac{1}{4} R_{\mu\nu\rho\sigma} \Gamma^{\rho\sigma} \psi + \partial_\mu \varphi^\alpha \partial_\nu \varphi^\beta \MF_{\alpha\beta}^{\phantom{\alpha\beta} I} T_I \psi .
\eea
Another useful geometrical quantity is the triplet of complex structures
\be
\label{e-5-1-20}
J_{\alpha\beta}^{\phantom{\alpha\beta} i} = ( \MV_{\alpha a}^{\phantom{\alpha a} A} \MV_\beta^{\phantom{\beta} a B} - \MV_{\alpha a}^{\phantom{\alpha a} B} \MV_\beta^{\phantom{\beta} a A} ) (T^i)_{AB},
\ee
which satisfy the $\mathrm{USp}(2)$ algebra. This will be shown to be proportional to the curvature $\MF_{\alpha\beta}^{\phantom{\alpha\beta} i}$, as a result of supersymmetry.

The \emph{rigidly} supersymmetric theory of the hypermultiplets is described by the sigma-model Lagrangian 
\be
\label{e-5-1-21}
e^{-1} \mathcal{L}^{(0)}_{H} = - \frac{1}{2} g_{\alpha\beta}(\varphi) \partial_\mu \varphi^\alpha \partial^\mu \varphi^\beta - \frac{1}{2} \bar{\psi}^a \Gamma^\mu \MD_\mu \psi_a,
\ee
which is invariant under the transformations
\be
\label{e-5-1-22}
\delta \varphi^\alpha = \MV^\alpha_{\phantom{\alpha} a A} \bar{\psi}^a \epsilon^A ,\qquad\qquad \delta \psi^a =
 \Gamma^\mu \MP_\mu^{\phantom{\mu} a A} \epsilon_A.
\ee
However, as usual in sigma models, this invariance does not hold for any choice of the scalar manifold $\MM$ \cite{Bagger:1982fn}. First, we note that $\varphi^\alpha$ takes its value in the tangent space $T ( \MM )$ of $\MM$ while $\psi^a$ and $\epsilon^A$ take their values in $H$ and $\mathrm{USp}(2)$ respectively. Consistency of the first transformation law in (\ref{e-5-1-22}) requires then that $T ( \MM ) = H \times \mathrm{USp}(2)$. Since $T ( \MM )$ is real, $H$ must be pseudoreal, the generic choice being $H \subset \mathrm{USp}(2 n_H)$. Second, cancellation of the $\bar{\psi}_a \epsilon$ variations requires that the coset vielbein $\MV^\alpha_{\phantom{\alpha} a A}$ be covariantly constant,
\be
\label{e-5-1-23}
\MD_\beta \MV^\alpha_{\phantom{\alpha} a A} = 0.
\ee
This property implies the following identities
\bea
\label{e-5-1-24}
\MV^\alpha_{\phantom{\alpha} a A} \MV^{\beta a B} + \MV^\beta_{\phantom{\beta} a A} \MV^{\alpha a B} &=& g^{\alpha\beta} \delta_A^B ,\nn\\
\MV^\alpha_{\phantom{\alpha} a A} \MV^{\beta b A} + \MV^\beta_{\phantom{\beta} a A} \MV^{\alpha b A} &=& \frac{1}{n_H} g^{\alpha\beta} \delta_a^b, \nn\\
g_{\alpha\beta} \MV^\alpha_{\phantom{\alpha} a A} \MV^\beta_{\phantom{\beta} b B} &=& \epsilon_{AB} \Omega_{ab}.
\eea
where $\Omega_{ab}$ is the antisymmetric tensor of $\mathrm{USp}(2 n_H)$. Furthermore, it results in the integrability condition
\be
\label{e-5-1-25}
\MR_{\alpha\beta\gamma\delta} \MV^\delta_{\phantom{\delta} a A} \MV^\gamma_{\phantom{\gamma} b B} =  \epsilon_{AB} \MF_{\alpha\beta}^{\phantom{\alpha\beta} I} (T_I)_{ab} + \Omega_{ab} \MF_{\alpha\beta}^{\phantom{\alpha\beta} i} (T_i)_{AB}. 
\ee
relating the curvature $\MR_{\alpha\beta\gamma\delta}$ of $\MM$ to the $\mathrm{USp}(2 n_H)$ and $\mathrm{USp}(2)$ curvatures. In rigid supersymmetry, the spinor $\epsilon$ is constant and the second of (\ref{e-5-1-19}) implies that $\MF_{\alpha\beta}^{\phantom{\alpha\beta} i} = 0$, i.e. that the $\mathrm{USp}(2)$ connection is flat. Plugging this into (\ref{e-5-1-25}), we see that the holonomy of $\MM$ is contained in $\mathrm{USp}(2 n_H)$, that is, rigid supersymmetry requires $\MM$ to be a hyperK\"ahler manifold.

Let us now construct the \emph{locally} supersymmetric theory. In the standard way, we introduce the appropriate interaction of the gravitino with the hypermultiplet supercurrent,
\be
\label{e-5-1-26}
\mathcal{L}^{(1)}_{H} = e \bar{\psi}^A_\mu \Gamma^\nu \Gamma^\mu \psi^a \MP_{\nu a A}. 
\ee
whose variation cancels the $\bar{\psi}_a \Gamma \MV \partial \phi \MD \epsilon$ variation of $\mathcal{L}^{(0)}_{H}$ in the usual manner. The variation of $\mathcal{L}^{(1)}_{H}$ under $\delta \psi_\mu \sim \Gamma G \epsilon$ yields the uncancelled term
\be
\label{e-5-1-27}
\Delta^{(1)}_{H} = - \frac{1}{12} e e^\phi \bar{\epsilon}^A \Gamma^\mu \Gamma^{\nu\rho\sigma} \psi^a
\MP_{\mu a A} G_{\nu\rho\sigma},
\ee
which cancels if we introduce the interaction
\bea
\label{e-5-1-28}
\mathcal{L}^{(2)}_{H} = - \frac{1}{24} e e^\phi \bar{\psi}^a \Gamma^{\mu\nu\rho} \psi_a G_{\mu\nu\rho}.
\eea
Another uncancelled term arises from the $\delta \psi_a$ variation of $\mathcal{L}^{(1)}_{H}$. It is given by
\be
\label{e-5-1-29}
\Delta^{(2)}_{H} = - e \bar{\psi}_\mu \Gamma^{\mu\nu\rho} \partial_\nu \varphi^\alpha \partial_\rho \varphi^\beta J_{\alpha\beta}^{\phantom{\alpha\beta} i} T_i \epsilon.
\ee
To cancel this term, we note that the modified commutator (\ref{e-5-1-19}) implies that the gravitino kinetic term in $\ML_{GT}$ acquires the extra term
\be
\label{e-5-1-30}
\Delta^{(3)}_{H} = \frac{1}{2} e \bar{\psi}_\mu \Gamma^{\mu\nu\rho} \partial_\nu \varphi^\alpha \partial_\rho \varphi^\beta \MF_{\alpha\beta}^{\phantom{\alpha\beta} i} T_i \epsilon,
\ee
This is exactly equal and opposite to $\Delta^{(2)}_{H}$, provided that the $\mathrm{USp}(2)$ curvature $\MF_{\alpha\beta}^{\phantom{\alpha\beta} i}$ is related to the complex structure $J_{\alpha\beta}^{\phantom{\alpha\beta} i}$ according to 
\be
\label{e-5-1-31}
\MF_{\alpha\beta}^{\phantom{\alpha\beta} i} = 2 J_{\alpha\beta}^{\phantom{\alpha\beta} i}.
\ee
This tells us that now the holonomy of $\MM$ must be contained in $\mathrm{USp}(2 n_H) \times \mathrm{USp}(2)$ with nonzero $\mathrm{USp}(2)$ curvature, that is, local supersymmetry requires $\MM$ to be a quaternionic manifold. The complete locally supersymmetric Lagrangian for neutral hypermultiplets is given by
\bea
\label{e-5-1-32}
e^{-1} \mathcal{L}_{H} &=& - \frac{1}{2} g_{\alpha\beta}(\varphi) \partial_\mu \varphi^\alpha \partial^\mu \varphi^\beta - \frac{1}{2} \bar{\psi}^a \Gamma^\mu \MD_\mu \psi_a + \bar{\psi}^A_\mu \Gamma^\nu \Gamma^\mu \MP_{\nu a A} \psi^a \nn\\ && - \frac{1}{24} e^\phi \bar{\psi}^a \Gamma^{\mu\nu\rho} \psi_a G_{\mu\nu\rho}. 
\eea

\subsubsection{The action for charged hypermultiplets}

To complete the construction of the theory, we must generalize the above for the case where the hypermultiplets are charged under the Yang-Mills group. Without loss of generality, we will here restrict to the case where $\MG_h = \mathrm{USp}(2 n_H)$ and $\MG_r = \mathrm{USp}(2)$ and we will construct the complete locally supersymmetric theory. 

To gauge the theory, we consider an $\mathrm{USp}(2 n_H) \times \mathrm{USp}(2)$ isometry transformation on the scalar manifold. Under this transformation, the hyperscalars transform as
\be
\label{e-5-1-33}
\delta \varphi^\alpha = \Lambda^{\hat{I}} \xi^\alpha_{\hat{I}} ,
\ee
where $\xi^\alpha_{\hat{I}}$ are $\mathrm{USp}(2 n_H) \times \mathrm{USp}(2)$ Killing vectors. A convenient basis for these vectors is
\be
\label{e-5-1-34}
\xi^\alpha_{\hat{I}} = (T_{\hat{I}} \varphi)^\alpha.
\ee
To promote this isometry to a local symmetry, we introduce the $\mathrm{USp}(2 n_H) \times \mathrm{USp}(2)$ gauge fields $A_\mu^{\hat{I}} = (A_\mu^I,A_\mu^i)$, transforming as
\be
\label{e-5-1-35}
\delta A_\mu^{\hat{I}} = D_\mu \Lambda^{\hat{I}},
\ee
whose field strengths, defined in the usual way, are denoted by $F_{\mu\nu}^{\hat{I}} = (F_{\mu\nu}^I,F_{\mu\nu}^i)$. We then replace the ordinary derivative acting on the hyperscalars by the covariant derivative
\be
\label{e-5-1-36}
\MD_\mu \varphi^\alpha = \partial_\mu \varphi^\alpha - A_\mu^{\hat{I}} \tilde{\xi}^\alpha_{\hat{I}} ,
\ee
where $\tilde{\xi}^\alpha_{\hat{I}} = ( g \xi^\alpha_I, g' \xi^\alpha_i )$ with $g$ and $g'$ being the $\mathrm{USp}(2 n_H)$ and $\mathrm{USp}(2)$ gauge couplings. Appropriately, the Maurer-Cartan form (\ref{e-5-1-17a}) is replaced by the gauged version
\be
\label{e-5-1-36a}
L^{-1} \MD_\mu L = \MQ_\mu^{\phantom{\alpha} I} T_I + \MQ_\mu^{\phantom{\alpha} i} T_i + \MP_\mu^{\phantom{\mu} a A} T_{a A}, 
\ee
with
\be
\label{e-5-1-36b}
\MQ_\mu^{\phantom{\mu} I} = \MD_\mu \varphi^\alpha \MA_\alpha^{\phantom{\alpha} I} ,\qquad \MQ_\mu^{\phantom{\mu} i} = \MD_\mu \varphi^\alpha \MA_\alpha^{\phantom{\alpha} i} ,\qquad \MP_\mu^{\phantom{\mu} a A} = \MD_\mu \varphi^\alpha \MV_\alpha^{\phantom{\alpha} a A},
\ee
and the covariant derivatives (\ref{e-5-1-18}) are appropriately modified by using the composite connections of the gauged theory and adding gauge couplings. Explicitly, we have
\bea
\label{e-5-1-37}
\MD_\mu \epsilon &=& D_\mu \epsilon + \MD_\mu \varphi^\alpha \MA_\alpha^{\phantom{\alpha} i} T_i \epsilon + g' A_\mu^i T_i \epsilon, \nn\\
\MD_\mu \psi &=& D_\mu \psi + \MD_\mu \varphi^\alpha \MA_\alpha^{\phantom{\alpha} I} T_I \psi + g A_\mu^I T_I \psi, \nn\\
\MD_\mu \lambda^I &=& D_\mu \lambda^I + \MD_\mu \varphi^\alpha \MA_\alpha^{\phantom{\alpha} i} T_i \lambda^I + g' A_\mu^i T_i \lambda^I + g f^I_{\phantom{I} JK} A_\mu^J \lambda^K, \nn\\
\MD_\mu \lambda^i &=& D_\mu \lambda^i + \MD_\mu \varphi^\alpha \MA_\alpha^{\phantom{\alpha} j} T_j \lambda^i + g' A_\mu^j T_j \lambda^i + g' \epsilon^i_{\phantom{i} jk} A_\mu^j \lambda^k.
\eea
where $f^I_{\phantom{I} JK}$ are the $\mathrm{USp}(2 n_H)$ structure constants. Another important building block of the gauge theory is the ``prepotential'' $C^{i \hat{I}}$, given by
\be
\label{e-5-1-38}
C^{i I} = g \MA_\alpha^{\phantom{\alpha} i} \xi^{\alpha I} ,\qquad C^{ij} = g' ( \MA_\alpha^{\phantom{\alpha} i} \xi^{\alpha j} - \delta^{ij} ),
\ee
and satisfying the identity \cite{Nishino:1986dc}
\be
\label{e-5-1-39}
\MD_\mu C^{i \hat{I}} = \MD_\mu \varphi^\alpha \MD_\alpha C^{i \hat{I}} = 2 ( \MD_\mu \varphi^\alpha ) J_{\alpha\beta}^{\phantom{\alpha\beta} i} \tilde{\xi}^{\beta \hat{I}}.
\ee
This quantity appears in the generalization of the second commutator in (\ref{e-5-1-19}) for the gauged theory, namely
\be
\label{e-5-1-40}
[ \MD_\mu , \MD_\nu ] \epsilon = \frac{1}{4} R_{\mu\nu\rho\sigma} \Gamma^{\rho\sigma} \epsilon + \MD_\mu \varphi^\alpha \MD_\nu \varphi^\beta \MF_{\alpha\beta}^{\phantom{\alpha\beta} i} T_i \epsilon - F_{\hat{I} \mu\nu} C^{i\hat{I}} T_i \epsilon.
\ee

Having introduced the necessary formalism, we are ready to derive the Lagrangian for the gauged theory. Our first step is to replace all derivatives in the Lagrangians (\ref{e-5-1-3}), (\ref{e-5-1-16}) and (\ref{e-5-1-21}) and transformation rules (\ref{e-5-1-4}) and (\ref{e-5-1-22}) by gauge-covariant ones. So, regarding the hypermultiplets, we obtain the Lagrangian 
\bea
\label{e-5-1-41}
e^{-1} \mathcal{L}_{H} &=& - \frac{1}{2} g_{\alpha\beta}(\varphi) \MD_\mu \varphi^\alpha \MD^\mu \varphi^\beta - \frac{1}{2}  \bar{\psi}^a \Gamma^\mu \MD_\mu \psi_a + \bar{\psi}^A_\mu \Gamma^\nu \Gamma^\mu \MP_{\nu a A} \psi^a \nn\\
&& - \frac{1}{24} e^\phi \bar{\psi}^a \Gamma^{\mu\nu\rho} \psi_a G_{\mu\nu\rho},
\eea
and the supersymmetry transformations
\be
\label{e-5-1-42}
\delta \varphi^\alpha = \MV^\alpha_{\phantom{\alpha} a A} \bar{\psi}^a \epsilon^A,\qquad\qquad \delta \psi^a =
 \Gamma^\mu \MP_\mu^{\phantom{\mu} a A} \epsilon_A .
\ee
The combination $\mathcal{L}_{GT} + \mathcal{L}_V + \mathcal{L}_H$ obtained this way is not locally supersymmetric though. The reason is that $\MD_\mu \epsilon$ and $\MD_\mu \varphi^\alpha$ now include extra contributions involving the $\mathrm{USp}(2)$ gauge fields; these do not affect the supersymmetry variations of the various interaction terms, but they do modify the variations of the gravitino and hyperino kinetic terms. For the gravitino kinetic term, the modified commutator (\ref{e-5-1-40}) implies that its supersymmetry variation acquires the extra terms
\be
\label{e-5-1-43}
\Delta^{(1)}_G = \frac{1}{2} e \bar{\psi}_{\mu} \Gamma^{\mu\nu\rho} \MD_\nu \varphi^\alpha
\MD_\rho \varphi^\beta \MF_{\alpha\beta}^{\phantom{\alpha\beta} i} T_i \epsilon,
\ee
and
\be
\label{e-5-1-44}
\Delta^{(2)}_G = \frac{1}{2} e C^{i\hat{I}} \bar{\psi}_\mu \Gamma^{\mu\nu\rho} F_{\hat{I} \nu\rho} T_i \epsilon. 
\ee
By exactly the same considerations, it is easy to see that the hyperino kinetic term also gives rise to the extra variation
\be
\label{e-5-1-45}
\Delta^{(3)}_G = - \frac{1}{2} e \bar{\psi}_a \Gamma^{\mu\nu} F^{\hat{I}}_{\mu\nu} \MV_\alpha^{\phantom{\alpha} a A} \tilde{\xi}^\alpha_{\hat{I}} \epsilon.
\ee
We immediately see that $\Delta^{(1)}_G$ exactly cancels the covariant version of the variation $\Delta^{(2)}_{H}$. On the other hand, the variations $\Delta^{(2)}_G$ and $\Delta^{(3)}_G$ require additional terms for their cancellation. Starting from $\Delta^{(2)}_G$, the possible sources for its cancellation may be the modification of the gaugino supersymmetry transformation law by the extra term
\be
\label{e-5-1-46}
\delta' \lambda^{\hat{I}} = a e^{\phi/2} v_X^{-1} ( C^{i \hat{I}} )_X T_i \epsilon,
\ee
or an additional interaction of the form
\be
\label{e-5-1-47}
\mathcal{L}^{(1)}_G = b e e^{\phi/2} \bar{\psi}_\mu \Gamma^\mu T_i \lambda_{\hat{I}} C^{i \hat{I}}. 
\ee
Here, $a$ and $b$ are two coefficients, which can be determined by considering the $\bar{\psi} \Gamma F \epsilon$ terms. The variations of this type arising from the $\delta' \lambda$ variation of $\mathcal{L}_V$ is
\be
\label{e-5-1-48}
\Delta^{(4)}_G = \frac{a}{2 \sqrt{2}} e \left( \bar{\psi}_\mu \Gamma^{\mu\nu\rho} T_i \epsilon F_{\hat{I} \nu\rho} C^{i \hat{I}} - 2 \bar{\psi}^\mu \Gamma^\nu T_i \epsilon F_{\hat{I} \mu\nu} C^{i \hat{I}} \right),
\ee
while the $\delta \lambda$ variation of $\mathcal{L}^{(1)}_G$ gives
\be
\label{e-5-1-49}
\Delta^{(5)}_G = - \frac{b}{2 \sqrt{2}} e \left( \bar{\psi}_\mu \Gamma^{\mu\nu\rho} T_i \epsilon F_{\hat{I} \nu\rho} C^{i \hat{I}} + 2 \bar{\psi}^\mu \Gamma^\nu T_i \epsilon F_{\hat{I} \mu\nu} C^{i \hat{I}} \right).
\ee
We observe that the requirement for cancellation of the terms proportional to $\bar{\psi}_\mu \Gamma^{\mu\nu\rho} \epsilon$ and $\bar{\psi}^\mu \Gamma^\nu \epsilon$ fixes the coefficients $a$ and $b$ to
\be
\label{e-5-1-50}
a = - \frac{1}{\sqrt{2}} ,\qquad\qquad b = \frac{1}{\sqrt{2}}.
\ee
Next, let us consider the $\delta' \lambda$ variation of the gaugino kinetic term in $\mathcal{L}_V$. Doing an integration by parts and using (\ref{e-5-1-39}), we obtain
\be
\label{e-5-1-51}
\Delta^{(6)}_G = \frac{1}{2\sqrt{2}} e e^{\phi/2} \left( 2 \bar{\lambda}_{\hat{I}} \Gamma^\mu T_i \MD_\mu \epsilon C^{i \hat{I}} + 2 \bar{\lambda}^{\hat{I}} \Gamma^\mu \MF_{\alpha\beta}^{\phantom{\alpha\beta} i} \MD_\mu \varphi^\alpha \xi^\beta_{\hat{I}} T_i \epsilon 
+ \bar{\lambda}_{\hat{I}} \Gamma^\mu T_i \epsilon \partial_\mu \phi C^{i \hat{I}} \right). 
\ee
On the other hand, using $\Gamma^\mu \Gamma^{\nu\rho\sigma} \Gamma_\mu = 0$, we find that the $\delta \psi_\mu$ variation of $\mathcal{L}^{(1)}_G$ is given by
\be
\label{e-5-1-52}
\Delta^{(7)}_G = - \frac{1}{\sqrt{2}} e e^{\phi/2} \bar{\lambda}_{\hat{I}} \Gamma^\mu T_i \MD_\mu \epsilon C^{i \hat{I}},
\ee
and it cancels the first term of $\Delta^{(6)}_G$. The second term can cancel by the $\delta \psi^a$ variation of the new term
\be
\label{e-5-1-53}
\mathcal{L}^{(2)}_G = - \sqrt{2} e e^{\phi/2} \bar{\lambda}^{\hat{I}}_A \psi_a \MV_\alpha^{\phantom{\alpha} a A} \tilde{\xi}^\alpha_{\hat{I}} ,
\ee
whose $\delta \lambda$ variation cancels $\Delta^{(3)}_G$. Two other uncancelled terms are the $\delta' \lambda$ variations of the $\bar{\lambda} \Gamma \chi F$ and $\bar{\lambda} \Gamma \lambda G$ terms of $\mathcal{L}_V$, given by
\be
\label{e-5-1-54}
\Delta^{(8)}_G = \frac{1}{4} e \bar{\chi} \Gamma^{\mu\nu} T_i \epsilon F_{\hat{I} \mu\nu} C^{i \hat{I}},
\ee
and
\be
\label{e-5-1-55}
\Delta^{(9)}_G = - \frac{1}{12 \sqrt{2}} e e^{3\phi / 2} \bar{\lambda}_{\hat{I}} \Gamma^{\mu\nu\rho} T_i \epsilon  G_{\mu\nu\rho} C^{i \hat{I}},
\ee
respectively. $\Delta^{(8)}_G$ is cancelled by the $\delta \lambda$ variation of yet another new term
\be
\label{e-5-1-56}
\mathcal{L}^{(3)}_G = \frac{1}{\sqrt{2}} e e^{\phi/2} \bar{\chi} T_i \lambda_{\hat{I}} C^{i \hat{I}},
\ee
whose $\delta \chi$ variation is given by
\be
\label{e-5-1-57}
\Delta^{(10)}_G = - \frac{1}{2 \sqrt{2}} e e^{\phi/2} \bar{\lambda}_{\hat{I}} \Gamma^\mu T_i \epsilon \partial_\mu \phi C^{i \hat{I}} + \frac{1}{12 \sqrt{2}} e e^{3\phi/2} \bar{\lambda}_{\hat{I}} \Gamma^{\mu\nu\rho} T_i \epsilon G_{\mu\nu\rho} C^{i \hat{I}},
\ee
so that its first part cancels the third term of $\Delta^{(6)}_G$ and its second part cancels $\Delta^{(9)}_G$. What remains to be cancelled are the $\delta' \lambda$ variations of $\mathcal{L}^{(1)}_G$ and $\mathcal{L}^{(3)}_G$, given by
\be
\label{e-5-1-58}
\Delta^{(11)}_G = \frac{1}{8} e e^\phi v_X^{-1} \bar{\psi}_\mu \Gamma^\mu \epsilon ( C^{i \hat{I}} C_{i \hat{I}} )_X,
\ee
and
\be
\label{e-5-1-59}
\Delta^{(12)}_G = \frac{1}{8} e e^\phi v_X^{-1} \bar{\chi} \epsilon ( C^{i \hat{I}} C_{i \hat{I}} )_X,
\ee
respectively. To cancel them, we introduce the term
\be
\label{e-5-1-60}
\mathcal{L}^{(4)}_G = - \frac{1}{8} e e^\phi v_X^{-1} ( C^{i \hat{I}} C_{i \hat{I}} )_X,
\ee
whose $\delta e$ and $\delta \phi$ variation is given by
\be
\label{e-5-1-61}
\Delta^{(13)}_G = - \frac{1}{8} e e^\phi v_X^{-1} \bar{\psi}_\mu \Gamma^\mu \epsilon ( C^{i \hat{I}} C_{i \hat{I}} )_X - \frac{1}{8} e e^\phi v_X^{-1} \bar{\chi} \epsilon ( C^{i \hat{I}} C_{i \hat{I}} )_X,
\ee
and indeed yields the desired cancellation.

To summarize, the terms that should be added to $\mathcal{L}_{GT} + \mathcal{L}_V + \mathcal{L}_{GS} + \mathcal{L}_{H}$ in order to restore local supersymmetry in case $\mathrm{USp}(2)$ is gauged are the following
\be
\label{e-5-1-62}
e^{-1} \mathcal{L}_G = \frac{1}{\sqrt{2}} e^{\phi/2} \left( \bar{\psi}_\mu \Gamma^\mu T_i \lambda_{\hat{I}} C^{i \hat{I}} + \bar{\chi} T_i \lambda_{\hat{I}} C^{i \hat{I}} \right) - \sqrt{2} e^{\phi/2} \bar{\lambda}^{\hat{I}}_A \psi_a V_\alpha^{\phantom{\alpha} a A} \tilde{\xi}^\alpha_{\hat{I}} - \frac{1}{8} e^\phi v_X^{-1} ( C^{i \hat{I}} C_{i \hat{I}} )_X.
\ee
The important fact emerging from the above rather technical discussion is that the theory includes a scalar potential given by the last term of (\ref{e-5-1-62}). The appearance of this potential has interesting consequences for the allowed compactifications of the theory.

\subsection{The complete Lagrangian}
\label{sec-5-1-4} 

We are now in a position to summarize the above results. By combining (\ref{e-5-1-3}), (\ref{e-5-1-16}), (\ref{e-5-1-41}) and (\ref{e-5-1-62}), we obtain the full Lagrangian
\bea
\label{e-5-1-63}
e^{-1} \ML &=& \frac{1}{4} R - \frac{1}{12} e^{2 \phi} G_{\mu\nu\rho} G^{\mu\nu\rho} - \frac{1}{4} \partial_\mu \phi \partial^\mu \phi - \frac{1}{2} \bar{\psi}_\mu \Gamma^{\mu\nu\rho} \MD_\nu \psi_\rho - \frac{1}{2} \bar{\chi} \Gamma^\mu \MD_\mu \chi
\nn\\
&&- \frac{1}{4} e^{-\phi} v_X ( F^{\hat{I}}_{\mu\nu} F_{\hat{I}}^{\mu\nu} )_X - \frac{1}{2} v_X ( \bar{\lambda}^{\hat{I}} \Gamma^\mu \MD_\mu \lambda_{\hat{I}} )_X - \frac{1}{2} g_{\alpha\beta}(\varphi) \MD_\mu \varphi^\alpha \MD^\mu \varphi^\beta - \frac{1}{2}  \bar{\psi}^a \Gamma^\mu \MD_\mu \psi_a \nn\\
&&- \frac{1}{24} e^\phi \left[ \bar{\psi}^\lambda \Gamma_{[\lambda} \Gamma^{\mu\nu\rho} \Gamma_{\sigma]} \bar{\psi}^\sigma - 2 \bar{\chi} \Gamma^\lambda \Gamma^{\mu\nu\rho} \psi_\lambda - \bar{\chi} \Gamma^{\mu\nu\rho} \chi - v_X ( \bar{\lambda}^{\hat{I}} \Gamma^{\mu\nu\rho} \lambda_{\hat{I}} )_X + \bar{\psi}^a \Gamma^{\mu\nu\rho} \psi_a \right] G_{\mu\nu\rho} 
\nn\\ 
&&- \frac{1}{2 \sqrt{2}} e e^{-\phi/2} v_X ( \bar{\psi}_\mu \Gamma^{\nu\rho} \Gamma^\mu \lambda^{\hat{I}} F_{\hat{I} \nu\rho} )_X - \frac{1}{2 \sqrt{2}} e e^{-\phi/2} v_X ( \bar{\lambda}^{\hat{I}} \Gamma^{\mu\nu} \chi F_{\hat{I} \mu\nu} )_X  \nn\\
&&- \frac{1}{2} \bar{\chi} \Gamma^\nu \Gamma^\mu \psi_\nu \partial_\mu \phi + \bar{\psi}^A_\mu \Gamma^\nu \Gamma^\mu \MP_{\nu a A} \psi^a \nn\\
&&+ \frac{1}{8} e^{-1} \epsilon^{\mu\nu\rho\sigma\tau\upsilon} B_{\mu\nu} v_X ( F^{\hat{I}}_{\rho\sigma} F_{ \hat{I} \tau\upsilon} )_X \nn\\
&&+ \frac{1}{\sqrt{2}} e^{\phi/2} \bar{\psi}_\mu \Gamma^\mu T_i \lambda_{\hat{I}} C^{i \hat{I}} + \frac{1}{\sqrt{2}} e^{\phi/2} \bar{\chi} T_i \lambda_{\hat{I}} C^{i \hat{I}} - \sqrt{2} e^{\phi/2} \bar{\lambda}^{\hat{I}}_A \psi_a V_\alpha^{\phantom{\alpha} a A} \tilde{\xi}^\alpha_{\hat{I}} \nn\\
&& - \frac{1}{8} e^\phi v_X^{-1} ( C^{i \hat{I}} C_{i \hat{I}} )_X  + (\mathrm{Fermi})^4. 
\eea 
which is invariant under the local supersymmetry transformations
\bea
\label{e-5-1-64}
\delta e^a_\mu &=& \bar{\epsilon} \Gamma^a \psi_\mu, 
\nn\\
\delta B_{\mu\nu} &=& e^{-\phi} \left( - \bar{\epsilon} \Gamma_{[\mu} \psi_{\nu]} + \frac{1}{2} \bar{\epsilon} \Gamma_{\mu\nu} \chi \right),
\nn\\
\delta \phi &=& \bar{\epsilon} \chi, 
\nn\\
\delta \psi_\mu &=& \MD_\mu \epsilon + \frac{1}{24} e^\phi \Gamma^{\nu\rho\sigma} \Gamma_\mu G_{\nu\rho\sigma} \epsilon, \nn\\
\delta \chi &=& \frac{1}{2} \Gamma^\mu \partial_\mu \phi \epsilon - \frac{1}{12} e^\phi \Gamma^{\mu\nu\rho} G_{\mu\nu\rho} \epsilon, \nn\\
\delta A_\mu^{\hat{I}} &=& \frac{1}{\sqrt{2}} e^{\phi/2} \bar{\epsilon} \Gamma_\mu \lambda^{\hat{I}},\nn\\
\delta \lambda^{\hat{I}} &=& - \frac{1}{2 \sqrt{2}} e^{-\phi/2} \Gamma^{\mu\nu} \epsilon F^{\hat{I}}_{\mu\nu} - \frac{1}{\sqrt{2}} e^{\phi/2} v_X^{-1} ( C^{i \hat{I}} )_X T_i \epsilon, \nn\\
\delta \varphi^\alpha &=& \MV^\alpha_{\phantom{\alpha} a A} \bar{\psi}^a \epsilon^A,\nn\\
\delta \psi^a &=& \Gamma^\mu \MP_\mu^{\phantom{\mu} a A} \epsilon_A.
\eea

The Lagrangian (\ref{e-5-1-63}) correspond to one of the two possible formulations of $D=6$, $N=2$ supergravity, as explained in \cite{Nishino:1997ff}; the Lagrangian derived in the original papers \cite{Nishino:1984gk,Nishino:1986dc} is the dual to the one considered here. The main differences are (i) the reversed dilaton signs in the kinetic terms of the vector fields, (ii) the absence of the Green-Schwarz term and (iii) a modification of the $B_{\mu\nu}$ supersymmetry transformation law and to its field strength. Regarding the ten-dimensional origin of these theories (only for the case when $\mathrm{USp}(2)$ R-symmetry is not gauged), the Lagrangian derived here originates from Chamseddine's 6--form version of $D=10$, $N=1$ supergravity, while its dual version originates from Chapline and Manton's 2--form version.

\section{Anomaly Cancellation in $D=6$, $N=2$ Supergravity}
\label{sec-5-2} 

After the above general discussion on $D=6$, $N=2$ supergravity theories, we are ready to examine the issue of anomaly cancellation in these models \cite{Salam:1985mi, Bergshoeff:1986hv,Ketov:1989an,Ketov:1990jr,Erler:1993zy,Schwarz:1995zw}. In particular, we will discuss the conditions and the available mechanisms for the cancellation of local anomalies as well as the conditions for the absence of global anomalies in the context of ungauged and gauged supergravities. The discussion will be fairly detailed and self-contained, with the aim of providing a reference for future work on this topic.

\subsection{Local anomalies}
\label{sec-5-2-1} 

Let us first examine the local anomalies. Starting from gravitational anomalies, we represent the total gravitational anomaly of the theory by the anomaly 8--form
\be
\label{e-5-2-1}
I_{8} (R) = \frac{n_H - n_V + 29 n_T - 273}{360} \tr R^4 + \frac{n_H - n_V - 7 n_T + 51}{288} (\tr R^2)^2 .
\ee

Passing to the gauge and mixed anomalies, we have to introduce some notation. For our analysis, it is convenient to rewrite the Yang-Mills gauge group of the theory as $\MG = \MG_s \times \MG_r \times \MG_a$, where (i) $\MG_s$ is a semisimple group containing factors from $H$ and $S$ given by the product $\prod_x \MG_x$ where the $\MG_x$'s are simple, (ii) $\MG_r$ is the R-symmetry factor arising in gauged theories and can be either $\mathrm{USp}(2)$ or a $\mathrm{U}(1)$ subgroup thereof and (iii) $\MG_a$ is an abelian subgroup of $S$ (abelian factors arising from $H$ would only make sense if resulting from a fundamental model so that the charges would be fixed). Given these definitions, we let $F_x$ and $F_r$ be the field strengths associated with $\MG_x$ and $\MG_r$. We also let $n_{x,i}$ and $n_{x\beta,ij}$ denote the numbers of hypermultiplets transforming in the representation $\MR_i$ of $\MG_x$ and in $(\MR_i,\MR_j)$ of $\MG_x \times \MG_y$. Then, using the formulas in Appendix A, we write the gauge anomaly polynomial as
\bea
\label{e-5-2-2}
I_8 (F) &=& - \frac{2}{3} \sum_x \bigl( \Tr F_x^4 - \sum_i n_{x,i} \tr_i F_x^4 \bigr) + 4 \sum_{x < y} \sum_{i,j} n_{xy,ij} \tr_i F_x^2 \tr_j F_y^2 \nn\\ && - \frac{2}{3} \bigl[ \tr' F_r^4 + ( \dim \MG_s + \dim \MG_a + 5 - n_T ) \tr F_r^4 \bigr] \nn\\ &&- 4 \Tr F_x^2 \tr' F_r^2 ,
\eea
where ``$\Tr$'' and ``$\tr_i$'' stand for the traces in the adjoint and $\MR_i$ of $\MG_x$ while ``$\tr$'' and ``$\tr'$'' stand for the traces in the fundamental of $\MG_r$ and in the representation of the gauginos. The four terms in (\ref{e-5-2-2}) are recognized as (i) the contribution of the gauginos and hyperinos to the anomaly under the pure $\MG_x$ factors, (ii) the contribution of the hyperinos to the anomaly under the products $\MG_x \times \MG_y$, (iii) the contribution of the $\MG_r$, $\MG_s$ and $\MG_a$ gauginos, gravitino and tensorinos to the anomaly under $\MG_r$ and (iv) the contribution of the gauginos to the anomaly under $\MG_x \times \MG_r$. In a similar manner, we find the mixed anomaly
\bea
\label{e-5-2-3}
I_8 (F,R) &=& \frac{1}{6} \tr R^2 \sum_x \bigl( \Tr F_x^2 - \sum_i n_{x,i} \tr_i F_x^2 \bigr) \nn\\ &&+ \frac{1}{6} \tr R^2 \bigl[ \tr' F_r^2 + ( \dim \MG_s + \dim \MG_a - 19 - n_T ) \tr F_r^2 \bigr].
\eea
Eqs. (\ref{e-5-2-2}) and (\ref{e-5-2-3}) can be brought into a more convenient form by expressing all traces in terms of a single representation, which we take to be the fundamental. For each $\MG_x$ we will have expressions of the form
\be
\label{e-5-2-4}
\tr_i F_x^4 = a_{x,i} \tr F_x^4 + b_{x,i} ( \tr F_x^2 )^2 ,\qquad \tr_i F_x^2 = c_{x,i} \tr F_x^2,
\ee
where the various group- and representation-dependent coefficients $a_{x,i}$, $b_{x,i}$ and $c_{x,i}$ are given in Appendix \ref{app-d}. Similarly, for $\MG_r$ we will have
\bea
\label{e-5-2-5}
&\tr' F_r^4 = b'_r ( \tr F_r^2 )^2 ,\qquad \tr F_r^4 = b_r ( \tr F_r^2 )^2, \nn\\ &  \tr' F_r^2 = c'_r \tr F_r^2,
\eea
where
\bea
\label{e-5-2-6}
b'_r = \frac{83}{2} ,\qquad b_r = \frac{1}{2} ,\qquad c'_r = 11;&& \qquad \text{if }\MG_r = \mathrm{USp}(2),\nn\\
b'_r = b_r = c'_r = 1;&& \qquad \text{if }\MG_r =
\mathrm{U}(1).
\eea
Using (\ref{e-5-2-5}) and (\ref{e-5-2-6}), introducing the quantities
\bea
\label{e-5-2-7}
A_x \equiv a_{x,\MA} - \sum_i n_{x,i} a_{x,i}, &&
\nn\\
B_x \equiv b_{x,\MA} - \sum_i n_{x,i} b_{x,i}, &&\qquad B_r \equiv b'_r + ( \dim \MG_s + \dim \MG_a + 5 - n_T ) b_r , \nn\\
C_x \equiv c_{x,\MA} - \sum_i n_{x,i} c_{x,i}, &&\qquad C_r \equiv c'_r + \dim \MG_s + \dim \MG_a - 19 - n_T, \nn\\ 
C_{xy} \equiv \sum_{i,j} n_{xy,ij} c_{x,i} c_{y,j}, &&\qquad C_{x,r} \equiv - c_{x,\MA},
\eea
and employing the extended index $X$, we write the gauge and mixed anomaly polynomials of the theory in the compact forms
\be
\label{e-5-2-8}
I_8(F) = - \frac{2}{3} \sum_x A_x \tr F_x^4 - \frac{2}{3} \sum_A B_X (\tr F_X^2)^2 + 4 \sum_{X < Y} C_{XY} \tr F_X^2 \tr F_Y^2,
\ee
and
\be
\label{e-5-2-9}
I_8(F,R) = \frac{1}{6} \tr R^2 \sum_X C_X \tr F_X^2.
\ee
Combining (\ref{e-5-2-1}), (\ref{e-5-2-8}) and (\ref{e-5-2-9}), we finally find the total anomaly polynomial
\bea
\label{e-5-2-10}
I_8 &=& \frac{n_H - n_V + 29 n_T - 273}{360} \tr R^4 + \frac{n_H - n_V - 7 n_T + 51}{288} (\tr R^2)^2 \nn\\ &&+ \frac{1}{6} \tr R^2 \sum_X C_X \tr F_X^2 \nn\\ &&- \frac{2}{3} \sum_x A_x \tr F_x^4 - \frac{2}{3} \sum_X B_X (\tr F_X^2)^2 + 4 \sum_{X < Y} C_{XY} \tr F_X^2 \tr F_Y^2.
\eea
If the total anomaly is to cancel through a Green-Schwarz--type mechanism, the above polynomial must factorize. A necessary condition for this is that all irreducible $\tr R^4$ and $\tr F_x^4$ terms in (\ref{e-5-2-10}) must vanish. Regarding the $\tr R^4$ term, the fact that $\mathrm{SO}(5,1)$ possesses a fourth-order Casimir implies that the coefficient of this term must vanish. This way, we are led to our first constraint
\be
\label{e-5-2-11}
n_H - n_V = 273 - 29 n_T,
\ee
which clearly shows that the presence of hypermultiplets is necessary for anomaly cancellation at least for $n_T \leqslant 9$. Passing to the $\tr F_x^4$ terms, their vanishing requires that
\be
\label{e-5-2-12}
A_x = 0 ;\qquad \textrm{for all } x.
\ee
This can be achieved either (i) if the relevant representations of $\MG_x$ have no fourth-order invariants ($a_{x,i}=0$ for all $i$) or (ii) if the $n_{x,i}$'s are chosen appropriately. Provided that (\ref{e-5-2-11}) and (\ref{e-5-2-12}) hold, the anomaly polynomial reads
\be
\label{e-5-2-13}
I_8 = K (\tr R^2)^2 + \frac{1}{6} \tr R^2 \sum_X C_X \tr F_X^2 - \frac{2}{3} \sum_X B_X (\tr F_X^2)^2 + 4 \sum_{X < Y} C_{XY} \tr F_X^2 \tr F_Y^2.
\ee
where we introduced the quantity
\be
\label{e-5-2-14}
K = \frac{9 - n_T}{8}.
\ee

To make a general analysis of the factorization properties of this polynomial, it helps to treat the Lorentz group in an equal footing with the other gauge groups by defining $F_0 = R$ as in \cite{Randjbar-Daemi:1985wc}. Introducing a summation index $I=0,1,\ldots,N$, we can then represent the anomaly polynomial in the concise form
\be
\label{e-5-2-15}
I_8 = G^{IJ} \tr F_I^2 \tr F_J^2,
\ee
where $\mathbf{G}$ is a real symmetric $(N+1) \times (N+1)$ matrix with entries
\be
\label{e-5-2-16}
G^{00} = K ,\qquad G^{0 X} = \frac{C_X}{12} ,\qquad G^{XX} = - \frac{2 B_X}{3} ,\qquad G^{XY} = 2 C_{XY} \text{ ($X \ne Y$)}.
\ee
The anomaly cancellation conditions depend on the properties of the matrix $\MBFG$ as well as on the number $n_T$ of available tensor multiplets. The two possible mechanisms are as follows.
\begin{enumerate}
\item \emph{Green-Schwarz mechanism}. For an arbitrary number of tensor multiplets\footnote{In discussing the $n_T \ne 1$ case, we ignore subtleties related to the construction of actions for (anti-)self-dual 2--forms.}, anomalies may be cancelled by the standard Green-Schwarz mechanism. In order for that mechanism to be applicable, the matrix $\mathbf{G}$ must be a matrix of rank $r \leqslant 2$; if $r=2$, the (real) nonzero eigenvalues $\lambda_m$, $m=0,1$, must satisfy $\lambda_0 \lambda_1 < 0$. For $r=2$, we may define $c^{mI}$ as the eigenvectors corresponding to $\lambda_m$ multiplied by $|\lambda_m|^{1/2}$ and write the similarity transformation of $\mathbf{G}$ in the form
\be
\label{e-5-2-17}
G^{IJ} = \epsilon \eta_{mn} c^{mI} c^{nJ} = \frac{1}{2} ( u^I \tilde{u}^J + \tilde{u}^I u^J ),
\ee
where $\epsilon$ is the sign of $\lambda_0$, $\eta_{mn}=\diag(1,-1)$ is the $\mathrm{SO}(1,1)$--invariant tensor and
\be
\label{e-5-2-18}
u^I \equiv \epsilon( c^{0I} - c^{1I} ) ,\qquad \tilde{u}^I \equiv c^{0I} + c^{1I}.
\ee
Using (\ref{e-5-2-15}), we can write the anomaly polynomial in the factorized form
\be
\label{e-5-2-19}
I_8 = \epsilon \eta_{mn} c^{mI} c^{nJ} \tr F_I^2 \tr F_J^2 = u^I \tr F_I^2 \tilde{u}^J \tr F_J^2.
\ee
This anomaly can cancel by the standard Green-Schwarz mechanism. Letting $B_2^{(0)} = B_2^+$ be the self-dual 2--form of the supergravity multiplet and $B_2^{(1)}$ be any one of the anti-self-dual 2--forms in the tensor multiplets and setting $B_2 = B_2^{(0)} + B_2^{(1)}$, we construct the Green-Schwarz term
\be
\label{e-5-2-20}
S_{GS} \sim \int u^I B_2 \tr F_I^2,
\ee
and we modify the gauge/Lorentz transformation law of the $B_2$'s to
\be
\label{e-5-2-21}
\delta B_2 \sim \tilde{u}^I \omega^1_{2,I},
\ee
where $\omega^1_{2,I}$ is related to $\tr F_I^2$ by descent. The variation of (\ref{e-5-2-19}) under (\ref{e-5-2-20}) exactly cancels the anomaly of the theory. For $r=1$, one may repeat the discussion with the
appropriate $c^{mI}$ set to zero.

\item \emph{Generalized Green-Schwarz mechanism}. In the case $n_T > 1$, there exists a generalization of the Green-Schwarz mechanism, found by Sagnotti \cite{Sagnotti:1992qw}, which allows for anomaly cancellation under weaker constraints. For that mechanism to apply, the matrix $\mathbf{G}$ must be a matrix of rank $r \leqslant
n_T+1$ whose nonzero eigenvalues $\lambda_m$, $m=0,\ldots,r-1$, include an eigenvalue $\lambda_0$ such that $\lambda_0 \lambda_m < 0$ for $m > 0$. For $r=n_T+1$, we may define $c^{mI}$ as before and we write the similarity transformation of $\mathbf{G}$ as
\be
\label{e-5-2-22}
G^{IJ} = \epsilon \eta_{mn} c^{mI} c^{nJ} = \frac{1}{2} \sum_{i=1}^{n_T} ( u^{iI} \tilde{u}^{iJ} + \tilde{u}^{iI} u^{iJ} ),
\ee
where now $\eta_{mn}=\diag(1,-1,\ldots,-1)$ is the $\mathrm{SO}(1,n_T)$--invariant metric and
\bea
\label{e-5-2-23}
u^{iI} \equiv \epsilon\left( \frac{c^{0I}}{\sqrt{n_T}} - c^{iI} \right) ,\qquad \tilde{u}^{iI} \equiv \frac{c^{0I}}{\sqrt{n_T}} + c^{iI}.
\eea
This way, the anomaly polynomial is written as a sum of factorized terms,
\be
\label{e-5-2-24}
I_8 = \epsilon \eta_{mn} c^{mI} c^{nJ} \tr F_I^2 \tr F_J^2 = \sum_{i=1}^{n_T} u^{iI} \tilde{u}^{iJ} \tr F_I^2 \tr F_J^2.
\ee
This anomaly can cancel by a generalization of the Green-Schwarz mechanism. Letting $B_2^{(0)} = B_2^+$ and $B_2^{(i)}$ be the anti-self-dual 2--forms in the tensor multiplets, we construct the $\mathrm{SO}(1,n_T)$--invariant generalized Green-Schwarz term \cite{Sagnotti:1992qw,Riccioni:1998th}
\be
\label{e-5-2-25}
S_{GS} \sim \int \epsilon \eta_{mn} c^{mI} B_2^{(n)} \tr F_I^2,
\ee
and we modify the gauge/Lorentz transformation law of the $B_2$'s to
\be
\label{e-5-2-26}
\delta B_2^{(m)} \sim c^{mI} \omega^1_{2,I}.
\ee
Again, for $r<n_T+1$, one may repeat the discussion with the appropriate $c^{mI}$'s set to zero.
\end{enumerate}

Here, we will only consider theories whose anomalies cancel by the standard Green-Schwarz mechanism. To examine the conditions for anomaly cancellation, it is very useful to state them in a more explicit form. To do so, we compare the general form (\ref{e-5-2-16}) of $\mathbf{G}$ with the expression (\ref{e-5-2-17}). Comparison of the $G^{00}$, $G^{0 X}$, $G^{X X}$ and $G^{X Y}$ terms leads respectively to the conditions
\bea
\label{e-5-2-27}
&u^0 \tilde{u}^0 = K, \\
\label{e-5-2-28}
&u^0 \tilde{u}^X + \tilde{u}^0 u^X = \frac{C_X}{6} ,\qquad u^X \tilde{u}^X = - \frac{2 B_A}{3}; \qquad &\textrm{for all } A,\\
\label{e-5-2-29}
&u^X \tilde{u}^Y + \tilde{u}^X u^Y = 4 C_{XY} ;\qquad &\textrm{for all } X < Y.
\eea
To begin, we note that we can set $u^0 = K$ and $v^0 = 1$ without loss of generality. To proceed, we have to distinguish between the cases $n_T\ne9$ ($K \ne 0$) and $n_T=9$ ($K=0$).
\begin{itemize}
\item $n_T \ne 9$. In this case, Eqs. (\ref{e-5-2-28}) imply that $u^X$ and $K \tilde{u}^X$ must be roots of the equation
\be
\label{e-5-2-30}
x^2 - \frac{C_X}{6} x - \frac{2 K B_X}{3} = 0,
\ee
and, in order for them to be real, we must have
\be
\label{e-5-2-31}
C_X^2 + 96 K B_X \geqslant 0;\qquad \textrm{for all } X.
\ee
Finally, Eq. (\ref{e-5-2-29}) leads to the condition
\be
\label{e-5-2-32}
C_X C_Y \pm \sqrt{ ( C_X^2 + 96 K B_X )( C_Y^2 + 96 K B_Y )} = 288 K C_{XY},
\ee
for at least one choice for the $u^X$'s and $K \tilde{u}^X$'s as roots of (\ref{e-5-2-30}), the plus or minus sign depending on the particular choice. E.g. in the case of three groups, (\ref{e-5-2-29}) is satisfied when (\ref{e-5-2-32}) holds for each pair $XY =(12,13,23)$ with either one of the sign combinations $(-,-,-)$, $(-,+,+)$,
$(+,-,+)$ and $(+,+,-)$.

\item $n_T = 9$. In that case, the first of Eqs. (\ref{e-5-2-28}) determines $u^X = C_X/6$, the second of Eqs. (\ref{e-5-2-28}) gives
\be
\label{e-5-2-33}
C_X \tilde{u}^X = - 4 B_X;\qquad \textrm{for all } X,
\ee
and Eq. (\ref{e-5-2-29}) gives
\be
\label{e-5-2-34}
C_X \tilde{u}^Y + C_Y \tilde{u}^X = 24 C_{XY};\qquad \textrm{for all } X < Y.
\ee
Eqs. (\ref{e-5-2-33}) and (\ref{e-5-2-34}) together form an overdetermined linear system of $N(N+1)/2$ equations for $N$ unknowns. In the general case, the system has the form $\mathbf{A} \mathbf{\tilde{u}} = \mathbf{b}$ with
\be
\label{e-5-2-35}
\mathbf{A} = \left(
\begin{array}{ccccc}
C_1 & 0 & \cdots & 0 & 0 \\
0 & C_2 & \cdots & 0 & 0 \\
\vdots & \vdots & \ddots & \vdots & \vdots \\
0 & 0 & \cdots & C_{N-1} & 0 \\
0 & 0 & \cdots & 0 & C_N \\
C_2 & C_1 & \ldots & 0 & 0 \\
\vdots & \vdots & \ddots & \vdots & \vdots \\
C_{N-1} & 0 & \ldots & C_1 & 0 \\
C_N & 0 & \ldots & 0 & C_1 \\
\vdots & \ddots &  & \vdots & \vdots \\
\vdots & \vdots & \ddots & \vdots & \vdots \\
0 & 0 & \ldots & C_N & C_{N-1}
\end{array} \right),\quad
\mathbf{\tilde{u}} = \left(\begin{array}{c}
\tilde{u}^1 \\
\tilde{u}^2 \\
\vdots\\
\tilde{u}^{N-1} \\
\tilde{u}^N
\end{array} \right) ,\quad
\mathbf{b} = \left(
\begin{array}{c}
-4 B_1 \\
-4 B_2 \\
\vdots\\
-4 B_{N-1} \\
-4 B_N \\
24 C_{12} \\
\vdots\\
24 C_{1,N-1} \\
24 C_{1N} \\
\vdots\\
\vdots\\
24 C_{N-1,N}
\end{array} \right),
\ee
and the constraints determining whether it has solutions are given by
\be
\label{e-5-2-36}
\det \MBFC = 0; \qquad \text{ for every $(N+1)\times(N+1)$ submatrix $\mathbf{C}$ of $(\mathbf{A},\mathbf{b})$}.
\ee
From (\ref{e-5-2-35}), we can see that when we have $B_X=C_X=0$ for all $X \ne \bar{X}$ and $C_{XY}=C_{YX}=0$ for $X,Y \ne \bar{X}$ where $\bar{X}$ is a given value of the index $X$, the system reduces to $N$ independent equations and has always a solution. For ungauged supergravities, this corresponds to the case where the hypermultiplets transform in the adjoint representation of $N-1$ group factors in $\MG_s$ and in an arbitrary representation of the remaining factor. For gauged supergravities, this corresponds to the case where the hypermultiplets transform in the adjoint of $\MG_s$; these were the solutions found in \cite{Salam:1985mi}.
\end{itemize}
To summarize, the requirement of Green-Schwarz cancellation of local anomalies in $D=6$, $N=2$ supergravity boils down, for $n_T \ne 9$, to the four conditions (\ref{e-5-2-11}), (\ref{e-5-2-12}), (\ref{e-5-2-31}) and (\ref{e-5-2-32}). The first condition fixes the number of hypermultiplets in terms of the gauge group. The second condition either holds identically (in the absence of fourth-order Casimirs) or constrains the numbers of representations (in the presence of fourth-order Casimirs). The third condition is an inequality whose main effect is to forbid higher representations (for which $B_x$ can attain large negative values). Finally, the fourth condition imposes a very stringent constraint on the numbers of representations; it is this latter condition that seriously reduces the number of possible models in the case when product groups are considered. In the special case $n_T=9$, the last two conditions are replaced by (\ref{e-5-2-36}).

One readily sees that the above anomaly cancellation conditions are quite weaker than those in the 10D case. Indeed, comparing the two cases we see that (i) the condition for the cancellation of irreducible gravitational anomalies does not uniquely fix the dimension of the gauge group but it just sets an upper bound on the number of non-singlet hypermultiplets and (ii) the factorization condition does not determine how the highest-order traces in the gauge anomaly must factorize but instead leads to two weaker constraints. As a result, these conditions admit a large number of solutions for the gauge group and the possible hypermultiplet representations and, in fact, a complete classification is an intractable task. The search for more solutions than those already found in the literature was the main motivation for the work presented in this chapter.

\subsection{Global anomalies}
\label{sec-5-2-2} 

Besides the perturbative anomalies described above, there is also the possibility that the theory may suffer from global anomalies, as discussed in \ref{sec-3-6}. We recall that, in the 6D case under consideration, such anomalies arise when there is a gauge group factor $\MG$ with nontrivial $\pi_6 (\MG)$, the possible cases being $G_2$, $\mathrm{SU}(3)$ and $\mathrm{SU}(2)$. For the case of $D=6$, $N=2$ supergravity, the conditions for the absence of global anomalies in the presence of a factor $\MG_x=G_2,\mathrm{SU}(3),\mathrm{SU}(2)$ in the $\MG_s$ part of the gauge group are given in \cite{Bershadsky:1997sb} and they amount to the following integrality constraints
\begin{align}
\label{e-5-2-37}
&\MG_x=G_2 :            &&1 - 4 \sum_i n_{x,i} b_{x,i} = 0 \mod 3, \nn\\
&\MG_x=\mathrm{SU}(3) : &&- 2 \sum_i n_{x,i} b_{x,i} = 0 \mod 6, \nn\\
&\MG_x=\mathrm{SU}(2) : &&4 - 2 \sum_i n_{x,i} b_{x,i} = 0 \mod 6.
\end{align}
where $n_{x,i}$ and $b_{x,i}$ are defined in \S\ref{sec-5-2-1}. Moreover, when the whole $\mathrm{USp}(2)_R \cong \mathrm{SU}(2)_R$ is gauged, there are also global R-symmetry anomalies. The condition for their absence is given by
\begin{align}
\label{e-5-2-38}
&\MG_r=\mathrm{USp}(2) : && 4 + \dim \MG_s + \dim \MG_a - n_T = 0 \mod 6.
\end{align}
Eqs. (\ref{e-5-2-37}) and (\ref{e-5-2-38}) must be solved together with the local anomaly cancellation conditions of the previous subsection in order to determine the possible anomaly-free models.

\section{Known Anomaly-Free Models}
\label{sec-5-3} 

In this section, we shall review some of the simplest anomaly-free $D=6$, $N=2$ supergravity models realized in the context of superstring compactifications, with the aim of describing concrete realizations of the anomaly-free theories discussed in the previous section. As our examples, we will consider the well-known compactifications of the $\mathrm{SO}(32)$ and $E_8 \times E_8$ heterotic string theories on $K3$. In the $\mathrm{SO}(32)$ case, we will also consider the construction of Witten where the symmetry is non-perturbatively enhanced by small instantons.

\subsection{Ungauged theories: Heterotic string theory on $K3$}
\label{sec-5-3-1} 

The simplest anomaly-free ungauged supergravities in 6D are those obtained by the reduction of the two anomaly-free 10D heterotic string theories on the $K3$ manifold. In what follows, we will outline some of the simplest such constructions, namely the straightforward $K3$ reduction of the $\mathrm{SO}(32)$ and $E_8 \times E_8$ heterotic theories as well as the mechanism of symmetry enhancement by small instantons.

\subsubsection{Generalities}

The $K3$ surface is a four dimensional complex K\"ahler manifold of $\mathrm{SU}(2)$ holonomy, that is, a Calabi-Yau 2--fold. Due to this latter property, compactifications of $D=10$, $N=1$ supergravity on this manifold give rise to minimal $D=6$, $N=2$ theories in the same way as the compactifications on Calabi-Yau 3--folds give rise to minimal $D=4$, $N=1$ theories. Here, we will review certain essential facts about the $K3$ surface and heterotic string compactifications on this surface. 

The main topological quantities of interest in the case of $K3$ are the second terms in the Chern class and Chern character and the first terms in the Pontrjagin class and $A$--roof genus, whose explicit forms (see Appendix C) are
\be
\label{e-5-3-1}
c_2 = \frac{1}{16 \pi^2} \tr R^2 ,\quad \cch_2 = - \frac{1}{16 \pi^2} \tr R^2 ,\quad p_1 = - \frac{1}{8 \pi^2} \tr R^2 ,\quad \widehat{A}_1 = \frac{1}{192 \pi^2} \tr R^2.
\ee
To obtain topological integrals, one uses the formula (see e.g. \cite{Eguchi:1980jx}) 
\be
\label{e-5-3-2}
\frac{1}{96 \pi^2} \int_{K3} \tr R^2 = 4,
\ee
which yields the Euler characteristic and the Hirzebruch signature
\be
\label{e-5-3-3}
\chi = \int_{K3} c_2 =  24 ,\qquad \tau = \frac{1}{3} \int_{K3} p_1 = -16.
\ee

From this information, one can determine the zero-mode spectrum of wave operators on $K3$. Starting from the Laplacian, the number of harmonic $p$--forms is given by the Betti numbers
\be
\label{e-5-3-4}
b_0 = b_4 = 1 ,\qquad b_1 = b_3 =0 ,\qquad b_2^+ = 3,b_2^- = 19,
\ee
where $b_2^+$ ($b_2^-$) stands for the number of (anti-)self-dual harmonic 2--forms. For the Lichnerowicz operator, the number of zero modes is  
\be
\label{e-5-3-5}
n^L = 1 + b_2^+ b_2^- = 58.
\ee
For the Dirac and Rarita-Schwinger operator, the net number of positive-chirality zero modes is found from the Atiyah-Singer index theorem. For the Dirac operator in the absence of gauge fields, one has
\be
\label{e-5-3-6}
n^{1/2} = \int_{K3} \widehat{A}_1 = 2.
\ee
For the Dirac operator in the presence of gauge fields, this is modified to
\be
\label{e-5-3-7}
n^{1/2} = \int_{K3} \left[ \left( 1 + \widehat{A}_1 (R) \right) \left( \dim \MS + \cch_2(F) \right) \right]_{\mathrm{vol}} = 2 \left( \dim \MS - 12 \frac{\int_{K3} \tr_{\MS} F^2}{\int_{K3} \tr R^2} \right).
\ee
where $\MS$ is the representation of the fermions under the gauge group. Finally, for the Rarita-Schwinger operator, we have
\be
\label{e-5-3-8}
n^{3/2} = \int_{K3} \left[ \left( 1 + \widehat{A}_1 \right) \left( 4 + p_1 \right) \right]_{\mathrm{vol}} = \int_{K3} \left( 4 \widehat{A}_1 + p_1 \right) = -40.
\ee

After these above preliminaries, let us now examine reduction of heterotic string theory on $K3$. We have seen in \S\ref{sec-4-1-2} that anomaly cancellation requires replacing the usual 3--form field strength by the modified field strength $\tilde{G}_3$ given in (\ref{e-4-1-26}). This field strength satisfies the modified Bianchi identity
\be
\label{e-5-3-9}
\dd \tilde{G}_3 = \tr R^2 - \tr F^2.
\ee
The requirement that $\tilde{G}_3$ be globally well-defined \cite{Witten:1984dg} implies that $\dd \tilde{G}_3$ should integrate to zero over any 4--cycle in 10D spacetime. So, assuming that we turn on the background fields $R_0$ and $F_0$ only inside $K3$, we must have
\be
\label{e-5-3-10}
\int_{K3} \left( \tr R_0^2 - \tr F_0^2 \right) = 0.
\ee
This is obviously satisfied if we embed the $\mathrm{SU}(2)$ holonomy group of $K3$ in the gauge group; the instanton number of the resulting configuration is
\be
\label{e-5-3-11}
n^{inst} = \frac{1}{16 \pi^2} \int_{K3} \tr F_0^2 = 24,
\ee
as it should. After this embedding, the 10D gauge group $\MG_{10}$ decomposes as
\be
\label{e-5-3-12}
\MG_{10} \supset \MG_6 \times \mathrm{SU}(2),
\ee
where $\MG_6$ is to be identified with the 6D gauge group. Accordingly, the adjoint representation of $\MG_{10}$ branches into a sum of the form
\be
\label{e-5-3-13}
\MA_{10} \to \sum_i (\MR_i,\MS_i),
\ee
where $\MR_i$ and $\MS_i$ stand for representations $\MG_6$ and $\mathrm{SU}(2)$ respectively.

Let us now examine how 10D fields of various type reduce on $K3$. Starting from the fermions of the supergravity multiplet, we know from Eqs. (\ref{e-5-3-6}) and (\ref{e-5-3-8}) that the net numbers of negative-chirality zero modes of the Dirac and Rarita-Schwinger operators on $K3$ are given by
\be
\label{e-5-3-14}
n^{1/2} = 2 ,\qquad n^{3/2} = - 40,
\ee
respectively. We also note that the 10D chirality projections correlate the 6D and 4D chiralities according to
\be
\label{e-5-3-15}
\Gamma_7 \Gamma_5 = \Gamma_{11},
\ee
and that the 10D fermions are Majorana-Weyl while the 6D ones are Weyl so the above numbers must be divided by two. Thus, we see that the 10D positive-chirality gravitino gives rise to 1 6D negative-chirality Weyl gravitino plus 20 6D positive-chirality hyperinos while the 10D negative-chirality dilatino gives 1 6D positive-chirality tensorino. Although supersymmetry guarantees that the number of 6D bosonic zero modes will be the same, let us verify this explicitly. The 10D graviton gives rise to one 6D graviton, no vectors (since $K3$ has no isometries) and $n^L=58$ scalars. Similarly, $B_{MN}$ reduces to $b_0=1$ 2--form $B_{\mu\nu}$, no vectors (since $b_1=0$) and $b_2=22$ scalars while $\phi$ reduces trivially. Putting everything together and consulting (\ref{e-5-1-1}), we see that the above fields arrange themselves into supersymmetry multiplets, namely one supergravity multiplet, one tensor multiplet and 20 hypermultiplets. Regarding the hypermultiplets, we note that they are inert under the gauge group and that, by standard Kaluza-Klein arguments, the associated hyperscalars parameterize the coset $\mathrm{SO}(20,4) / \mathrm{SO}(20) \times \mathrm{SO}(3) \times \mathrm{USp}(2)$ included in the discussion following Eq. (\ref{e-5-1-2}).

The analysis of the reduction of the 10D vector multiplet is slightly more involved. For the fermions, one now needs to use Eq. (\ref{e-5-3-7}), which shows that the net number of negative-chirality zero mode fermions transforming in $\MS_i$ of $\mathrm{SU}(2)$ is given by
\be
\label{e-5-3-16}
n_i^{1/2} = 2 \left( \dim \MS_i - 12 c_{\mathrm{SU}(2),i} \right),
\ee
where in the last line we have made use of (\ref{e-5-3-10}) and where the coefficient $c_{\mathrm{SU}(2),i}$ is defined according to (\ref{e-5-2-4}). So, for each $i$ the 10D positive-chirality gaugino gives $| n_i^{1/2} |$ 6D spinors transforming in $\MR_i$ of $\MG_6$. These are classified either as negative-chirality gauginos (if $n_i^{1/2} > 0$) or as positive-chirality hyperinos (if $n_i^{1/2} < 0$). By a similar discussion, one may verify that the bosonic degrees of freedom work out as required. Specific examples will be discussed below.

\subsubsection{The $\mathrm{SO}(32)$ Theory}

As a first definite example of the above, let us consider the reduction of the $\mathrm{SO}(32)$ theory on $K3$. Embedding the $\mathrm{SU}(2)$ spin connection in the gauge group as described above, we decompose $\mathrm{SO}(32)$ according to the maximal subgroup,
\be
\label{e-5-3-17}
\mathrm{SO}(32) \supset \left[ \mathrm{SO}(28) \times \mathrm{SU}(2) \right] \times \mathrm{SU}(2),
\ee
the factor in brackets corresponding to $\MG_6$. Under this decomposition, the adjoint $\mathbf{496}$ decomposes as
\be
\label{e-5-3-18}
\mathbf{496} \to (\mathbf{378},\mathbf{1};\mathbf{1}) + (\mathbf{1},\mathbf{3};\mathbf{1}) + (\mathbf{28},\mathbf{2};\mathbf{2}) + (\mathbf{1},\mathbf{1};\mathbf{3}).
\ee
Straightforward application of (\ref{e-5-3-16}) for the $\mathrm{SU}(2)$ representations in (\ref{e-5-3-18}) gives then $n_{\mathbf{1}}^{1/2}=1$, $n_{\mathbf{2}}^{1/2}=-10$ and $n_{\mathbf{3}}^{1/2}=-45$. So, we have one negative-chirality gaugino in the adjoint $(\mathbf{378},\mathbf{1}) + (\mathbf{1},\mathbf{3})$ of $\mathrm{SO}(28) \times \mathrm{SU}(2)$, 10 hyperinos in the $(\mathbf{28},\mathbf{2})$, 45 singlet hyperinos and, of course, their bosonic superpartners. Combining these with the multiplets from the 10D supergravity multiplet we have, in total, one supergravity multiplet, one tensor multiplet, one vector multiplet in the $(\mathbf{378},\mathbf{1}) + (\mathbf{1},\mathbf{3})$, 10 hypermultiplets in the $(\mathbf{28},\mathbf{2})$ and 65 singlet hypermultiplets.

Since the above 6D theory descends from an anomaly-free 10D theory, it is anomaly-free by construction. However, let us check this explicitly. The total numbers of vectors and hypers are $n_V = 378+3 = 381$ and $n_H = 10 \cdot 56 + 65 = 625$ and so we have $n_H = n_V + 244$. For the irreducible $\mathrm{SO}(28)$ factor, we also have $A_{\mathrm{SO}(28)} = 20 - 20 = 0$. Similarly, we find $B_{\mathrm{SO}(28)} = 3 - 20 \cdot 0 = 3$, $C_{\mathrm{SO}(28)} = 26 - 20 \cdot 1 = 6$, $B_{\mathrm{SU}(2)} = 8 - 280 \cdot \frac{1}{2} = -132$, $C_{\mathrm{SU}(2)} = 4 - 280 \cdot 1 = - 276$ and $C_{\mathrm{SO}(28),\mathrm{SU}(2)} = 10$ and the inequalities (\ref{e-5-2-31}) are satisfied, as is the equality (\ref{e-5-2-32}) (with plus sign). So, all anomaly cancellation conditions are indeed satisfied.

\subsubsection{The $E_8 \times E_8$ Theory}

The $E_8 \times E_8$ theory is treated in an entirely analogous manner. In contrast to the previous case however, it is possible to consider embeddings in which a part of the $K3$ instanton charge is embedded into the first $E_8$ factor and the remaining part is embedded into the second.

In the simplest case where all $K3$ instanton charge is embedded into, say, the second $E_8$, the maximal subgroup is given by
\be
\label{e-5-3-19}
E_8 \times E_8 \supset \left[ E_8 \times E_7 \right] \times \mathrm{SU}(2)
\ee
and the adjoint decomposes as
\be
\label{e-5-3-20}
(\mathbf{248},\mathbf{1}) + (\mathbf{1},\mathbf{248})\to (\mathbf{1},\mathbf{133};\mathbf{1}) + (\mathbf{248},\mathbf{1};\mathbf{1}) + (\mathbf{1},\mathbf{56};\mathbf{2}) + (\mathbf{1},\mathbf{1};\mathbf{3}).
\ee
The multiplicities $n_{i}^{1/2}$ are as before and so we have, in total, one supergravity multiplet, one tensor multiplet, one vector multiplet in the $(\mathbf{133},\mathbf{1}) + (\mathbf{1},\mathbf{248})$, 10 hypermultiplets in the $(\mathbf{56},\mathbf{1})$ and 65 singlet hypermultiplets. Since the hypermultiplets are all neutral under $E_8$, the latter is a ``hidden sector'' group. The theory is anomaly-free as can be easily verified.

In the more general case, one may embed $n_1$ units of instanton charge in the first $E_8$ and $n_2$ units in the second $E_8$ under the condition that $n_1+n_2=24$. Now, the maximal-subgroup decomposition is
\be
\label{e-5-3-21}
E_8 \times E_8 \supset \left[ E_7 \times E_7 \right] \times \mathrm{SU}(2)_1 \times \mathrm{SU}(2)_2
\ee
and the adjoint decomposes as
\bea
\label{e-5-3-22}
(\mathbf{248},\mathbf{1}) + (\mathbf{1},\mathbf{248}) &\to& (\mathbf{133},\mathbf{1};\mathbf{1},\mathbf{1}) + (\mathbf{1},\mathbf{133};\mathbf{1},\mathbf{1}) + (\mathbf{56},\mathbf{1};\mathbf{2},\mathbf{1}) + (\mathbf{1},\mathbf{56};\mathbf{1},\mathbf{2}) \nn\\ 
&&+ (\mathbf{1},\mathbf{1};\mathbf{3},\mathbf{1}) + (\mathbf{1},\mathbf{1};\mathbf{1},\mathbf{3}).
\eea
By a straightforward modification of the index-theorem formula (\ref{e-5-3-16}), one easily finds $n_{(\mathbf{2},\mathbf{1})}^{1/2} = \frac{4 - n_1}{2}$ and $n_{(\mathbf{1},\mathbf{2})}^{1/2} = \frac{4 - n_2}{2}$. By similar considerations as before, we then have one supergravity multiplet, one tensor multiplet, one vector multiplet in the $(\mathbf{133},\mathbf{1}) + (\mathbf{1},\mathbf{133})$, $\frac{n_1-4}{2}$ hypermultiplets in the $(\mathbf{56},\mathbf{1})$, $\frac{n_2-4}{2}$ hypermultiplets in the $(\mathbf{1},\mathbf{56})$ and 62 singlet hypermultiplets. Again the theory is anomaly-free.

\subsubsection{Symmetry enhancement by small instantons}

It has been shown by Witten \cite{Witten:1995gx} that, in the case of the $\mathrm{SO}(32)$ theory, the gauge symmetry of the theory can be further enhanced by a non-perturbative mechanism involving instantons collapsing to zero size. From the point of view of $\mathrm{SO}(32)$ heterotic string theory, these instantons have an interpretation as solitonic 5--branes and, assuming that they are localized at different points, each one contributes an extra $\mathrm{USp}(2)$ gauge symmetry factor. One can also use heterotic/Type I duality to pass to a dual description in terms of $\mathrm{SO}(32)$ Type I theory where the instantons correspond to D5--branes with symplectic symmetry (due to the fact that Type I strings are unoriented). In the Type I description, it is clear that a configuration that consists of $N$ coincident instantons shrinking to zero size gives rise to an extra $\mathrm{USp}(2N)$ gauge symmetry. Since the total $K3$ instanton number is 24, the largest extra symmetry one can get is $\mathrm{USp}(48)$. For $N < 24$, one has to embed $24-N$ units of instanton charge in $\mathrm{SO}(32)$ in order that (\ref{e-5-3-11}) be satisfied; the full symmetry group is thus $\mathrm{SO}(N+8) \times \mathrm{USp}(2N)$.

Let us now analyze the spectrum of the resulting theory. For the gravitational sector the analysis goes on as before and we have one graviton, one 2--form, one dilaton and 20 hypermultiplets, all singlets under the gauge group. For the Yang-Mills sector, it is here more advantageous to work out the bosonic excitations in the Type I string description. Open strings in the field of the D5-brane are naturally divided into strings not attached to the brane (NN), strings with one end attached to the brane (DN) and strings with both ends attached to the brane (DD), with N and D referring to Neumann and Dirichlet boundary conditions respectively. The NN strings originally carry two ordinary $\mathrm{SO}(32)$ Chan-Paton factors but, since $\mathrm{SO}(32)$ is broken to $\mathrm{SO}(N+8)$ by the embedding, their spectrum consists of vector excitations that transform in $(\mathbf{\frac{(N+8)(N+7)}{2}},\mathbf{1})$ plus $\frac{(24-N)(21-N)}{2}$ scalar excitations that are singlets. The DN strings originally carry an ordinary $\mathrm{SO}(32)$ Chan-Paton factor at their N end and a symplectic $\mathrm{USp}(2N)$ factor at their D end and, for the same reason as before, the strings carrying $\mathrm{SO}(N+8)$ Chan-Paton factors transform in $\frac{1}{2}(\mathbf{N+8},\mathbf{2N})$ while there are also $24-N$ strings in $\frac{1}{2}(\mathbf{1},\mathbf{2N})$ (here, the $\frac{1}{2}$ factors are due to the fact that $\mathbf{2N}$ is pseudoreal). Finally, for the DD strings, the $\Omega$ projection to unoriented states implies that the vector excitations $A_\mu$ and the scalar excitations $\varphi_m$ are described by vertex operators of the type $\MV_A \sim A_\mu \partial_\tau X^\mu$ ($\Omega$--odd) and $\MV_\varphi \sim \varphi_m \partial_\sigma X^m$ ($\Omega$--even) respectively. The vectors transform of course in the adjoint $(\mathbf{1},\mathbf{N(2N+1)})$ which corresponds to a second-rank symmetric tensor of $\mathrm{USp}(2N)$. For the scalars, the fact that $\MV_\varphi$ contains a normal derivative instead of a tangential one implies that they transform in the second-rank antisymmetric representation; since the latter is reducible into a ``traceless'' part plus a singlet, the DD scalars transform in $(\mathbf{1},\mathbf{N(2N-1)-1}) + (\mathbf{1},\mathbf{1})$. Combining the above results, we have one supergravity multiplet, one tensor multiplet, one vector multiplet in the $(\mathbf{\frac{(N+8)(N+7)}{2}},\mathbf{1}) + (\mathbf{1},\mathbf{N(2N+1)})$, half a hypermultiplet in the $(\mathbf{N+8},\mathbf{2N})$, $\frac{24-N}{2}$ hypermultiplets in the $(\mathbf{1},\mathbf{2N})$, one hypermultiplet in the $(\mathbf{1},\mathbf{N(2N-1)-1})$ plus $21 + \frac{(24-N)(21-N)}{2} $ singlet hypermultiplets.

By plugging the above numbers in the anomaly cancellation conditions, it is straightforward to verify that they are all satisfied. This is a very important check of the consistency of Witten's construction which is based in string-theoretical arguments.

\subsection{Gauged theories: The $E_7 \times E_6 \times \mathrm{U}(1)$ model}
\label{sec-5-3-2} 

Besides the string-theoretical models just considered, there also exist anomaly-free models constructed purely from the 6D supergravity viewpoint, engineered so that the anomaly cancellation conditions are met. The best-known of these models and was the one constructed by Randjbar-Daemi, Salam, Sezgin and Strathdee \cite{Randjbar-Daemi:1985wc} in 1985. The gauge group of the theory is taken to be $E_7 \times E_6 \times \mathrm{U}(1)_R$, and the representations of the fermions are chosen appropriately so that the anomaly cancellation conditions are satisfied.

To sketch the construction of the model, we first note note that the dimension of the gauge group is $n_V = 133+78+1 = 212$. The anomaly cancellation conditions demand then that the number of hypermultiplets be equal to $n_H = 212+244 = 456$. A natural choice is then to identify them with half a hypermultiplet in the pseudoreal representation $\mathbf{912}$ of $E_7$ and to take them to be singlets under $E_6$. Recalling that, under $\mathrm{U}(1)$, the gravitino, tensorino and gauginos have unit charge while the hyperinos are inert, we see that the fermionic spectrum of the theory contains one gravitino in the $(\mathbf{1},\mathbf{1})_1$, one tensorino in the $(\mathbf{1},\mathbf{1})_1$, gauginos in the $(\mathbf{133},\mathbf{1})_1 + (\mathbf{1},\mathbf{78})_1 + (\mathbf{1},\mathbf{1})_1$ and hyperinos in the $(\mathbf{912},\mathbf{1})_0$, the subscripts indicating $\mathrm{U}(1)_R$ charges. Further aspects of this model will be described in \S\ref{sec-7-3-2}.

Repeating the familiar analysis, this time adapted for the presence of the $\mathrm{U}(1)$ R-symmetry factor, one may verify that all anomaly cancellation conditions are satisfied. However, in this particular case, the gauge group is a product of three factors and the anomaly cancellation conditions are far more stringent than in the cases discussed above. The fact that there even exists a non-trivial model satisfying all of these conditions is truly remarkable.

Actually, the model outlined above is the first anomaly-free 6D model found from pure 6D supergravity considerations and, until recently, was the only known anomaly-free gauged model with $n_T = 1$ and without ``unnatural'' field content. However, as we shall see below, there actually exist a few other non-trivial consistent models of this type.

\section{Searching for Anomaly-Free Theories}
\label{sec-5-4} 

As stated earlier on, the anomaly cancellation conditions of $D=6$, $N=2$ supergravity are weaker than those in the 10D case, mainly due to the existence of the massless hypermultiplets that may transform in arbitrary representations of the gauge group. As a result, the classification of the theories satisfying these conditions is a far more complicated task than the analogous problem in 10D. In this section, we turn to the problem of obtaining solutions to these conditions in a systematic manner.

The search for consistent six-dimensional $N=2$ supergravities is greatly motivated by a number of reasons, namely (i) their shared properties with ten-dimensional $N=2$ supergravities, (ii) their use as toy models for the study of complicated phenomena such as flux compactifications, (iii) their connection, in the gravity-decoupling limit, to the much-studied $N=2$ supersymmetric gauge theories in four dimensions, (iv) the possibility of vectorlike \cite{Salam:1984cj} or chiral \cite{Randjbar-Daemi:1985wc} compactifications of the gauged theories down to flat four-dimensional space using a gauge field residing in an internal $\mathbf{S}^2$ and (v) the partial solution they provide to the cosmological constant problem in both ungauged \cite{Kehagias:2004fb,Nair:2004yu,Randjbar-Daemi:2004ni} and gauged \cite{Aghababaie:2003wz,Aghababaie:2003ar,Burgess:2004dh} cases.

Regarding the case of ungauged supergravities, most models found so far correspond to heterotic string compactifications on $K3$ \cite{Green:1985bx}, possibly involving symmetry enhancement either from the Gepner points of orbifold realizations of $K3$ \cite{Erler:1993zy} or by small instantons \cite{Schwarz:1995zw,Witten:1995gx}, as well as chains of models obtained from the above ones by Higgsing; some simple cases have been reviewed in \S\ref{sec-5-3-1}. In \cite{Schwarz:1995zw}, a few more models were found by directly solving the anomaly-cancellation conditions. Finally, many series of models were constructed \cite{Bershadsky:1996nh,Bershadsky:1997sb} using geometric engineering via F-theory. Another class of models corresponds to flat-space 6D gauge theories, where anomaly cancellation is related to the existence of non-trivial RG fixed points \cite{Seiberg:1996qx,Danielsson:1997kt}. Although, the number of known anomaly-free 6D ungauged supergravities is quite large, it is certainly useful to tabulate the simplest of them and it is interesting to search whether there are more anomaly-free models or chains of models than those already found.

Turning to the gauged case, the only known anomaly-free model until the work related to this thesis were the $E_7 \times E_6 \times \textrm{U}(1)_R$ model of \cite{Randjbar-Daemi:1985wc}, briefly described in \S\ref{sec-5-3-2}. There are also a few models \cite{Salam:1985mi} involving extra ``drone'' $\textrm{U}(1)$'s. These models have been found from purely supergravity considerations, guided by the requirement of anomaly cancellation. The uniqueness of these models and their interesting physical properties provide a great motivation for investigating whether more models of this type exist.

In the following sections, we partially address the two problems mentioned in the preceding two paragraphs. In particular, we conduct a systematic search for 6D supergravity models satisfying the anomaly cancellation conditions stated above. Since a complete classification seems to be very difficult, we will make several assumptions, expected to hold for many models of potential physical interest. The restrictions to be imposed are the following.
\begin{enumerate}
\item The theory contains only one tensor multiplet, $n_T=1$.
\item The semisimple gauge group factor $\MG_s$ is a product of up to two simple groups.
\item The hypermultiplets may transform in a set of low-dimensional representations of the simple factors in $\MG_s$. The representations to be considered are shown on Table \ref{t-5-1}.
\item For ungauged theories, the allowed exceptional groups are $E_8$, $E_7$, $E_6$ and $F_4$ and the allowed classical groups are $\mathrm{SU}(5 \leqslant N \leqslant 32)$, $\mathrm{SO}(10 \leqslant N \leqslant 64)$ and $\mathrm{Sp}(4 \leqslant N \leqslant 32)$. At most one simple factor in $\MG_s$ may be a classical group. The abelian factor $\MG_a$ is empty.
\item For gauged theories, all exceptional groups are allowed while the allowed classical groups are as before. At most one simple factor in $\MG_s$ may be a classical group. The abelian factor $\MG_a$ can be non-trivial.
\end{enumerate}
\begin{table}[!t]
\label{t-5-1}
\begin{center}
\begin{tabular}{|c||l|l|}
\hline
Group & Low-dimensional Irreps & Comments\\
\hline
\hline
$E_8$ & $\mathbf{248}$ &\\
\hline
$E_7$ & $\mathbf{56}^*,\mathbf{133},\mathbf{912}^*$ & ${}^*$pseudoreal \\
\hline
$E_6$ & $\mathbf{27},\mathbf{78},\mathbf{351},\mathbf{351'},\mathbf{650}$ &\\
\hline
$F_4$ & $\mathbf{26},\mathbf{52},\mathbf{273},\mathbf{324}$ &\\
\hline
$G_2$ & $\mathbf{7},\mathbf{14},\mathbf{27},\mathbf{64}$ &\\
\hline
\hline
$\mathrm{SU}(N)$ & $\mathbf{N},\mathbf{N^2-1},\mathbf{\frac{N(N-1)}{2}},\textstyle{\mathbf{\frac{N(N+1)}{2}}}$ &\\
\hline
$\mathrm{SO}(N)$ & $\mathbf{N}, \mathbf{\frac{N(N-1)}{2}}, \mathbf{2^{\lfloor \frac{N+1}{2} \rfloor - 1}}^*$ & ${}^*$pseudoreal if $N = 3,4,5 \mod 8$ \\
\hline
$\mathrm{USp}(2N)$ & $\mathbf{2N}^*,\mathbf{N(2N+1)},\mathbf{N(2N-1)-1}$ & ${}^*$pseudoreal \\
\hline
\end{tabular}
\end{center}
\caption{The possible simple gauge groups and their low-dimensional representations.}
\end{table}
All of these assumptions have been made on a purely practical basis. In particular, the lower bounds on the group rank as well as the restriction to at most one classical group factor were imposed because the proliferation of possible models in the case these assumptions were relaxed would make the exhaustive search for anomaly-free models and their classification an intractable task.

In the next two sections, we will present the complete lists of anomaly-free models under these conditions, starting from the case of ungauged supergravities and proceeding to the case of gauged supergravities. In the course, we will identify as many of the known models as possible and we will comment on their construction, their origin and their properties. The results presented should be read according to the convention that each representation, designated by its dimension, corresponds to all representations with the same dimension and second and fourth indices, i.e. to all representations related by symmetries such as complex conjugation and triality. Accordingly, the corresponding numbers of multiplets for a representation are understood as the total numbers of multiplets in these representations. For example, in the case of $E_6$, the notation $\mathbf{27}$ refers to the two conjugate representations $\mathbf{27}$ and $\overline{\mathbf{27}}$ and the field content $n \cdot \mathbf{27}$ is understood as all combinations of the form $n_1 \cdot \mathbf{27} + n_2 \cdot \overline{\mathbf{27}}$ with $n_1 + n_2 = n$. Also, the numbers of singlet hypermultiplets for each model will not be displayed explicitly.

Finally, there are two issues referring to the reality properties of the representations under consideration. First, when there appear pseudoreal representations, one may allow the corresponding numbers of hypermultiplets to take half-integer values as well. For example, in the case of $E_7$, the notation $\frac{1}{2} \cdot \mathbf{56}$ refers to ``half'' a hypermultiplet in the pseudoreal representation $\mathbf{56}$, also understood as one hypermultiplet in the minimal representation $\mathbf{28}$. Second, in the case where there appear complex representations, CPT invariance requires that these representations occur in complex-conjugate pairs; it is only these representations that will be considered here.

\section{Anomaly-free Ungauged Supergravities}
\label{sec-5-5} 

In this section, we begin our search by considering the case of ungauged supergravities, i.e. the case where the gauge group does not involve an R-symmetry subgroup. As mentioned in the introduction, the number of these models is expected to be quite large; it turns out that this is indeed the case. In the course of the search, we recover various models already found in the literature, and we find some models not previously identified.

\subsection{Simple groups}
\label{sec-5-5-1}

Let us start from the case of one simple gauge group. In this case, the conditions to be solved are Eq. (\ref{e-5-2-11}) for the cancellation of the irreducible gravitational anomaly, Eq. (\ref{e-5-2-12}) for the cancellation of the irreducible gauge anomaly (when applicable) and the factorization condition (\ref{e-5-2-31}). Below, we present all possible models satisfying these conditions under the assumptions introduced at the end of Section 2. To make the discussion more pedagogical, we illustrate the procedure in detail.

For the exceptional groups, the only conditions to be solved are Eqs. (\ref{e-5-2-11}) and (\ref{e-5-2-31}). Noting that the number of singlets must be a nonnegative integer, we see that the first condition constrains the number of charged hypermultiplets according to
\be
\label{e-5-5-1}
\sum_i n_{i} \dim \MR_i \leqslant \dim \MG + 244.
\ee
Also, the second condition takes the explicit form
\be
\label{e-5-5-2}
(c_{\MA} - \sum_i n_i c_i )^2 + 96 (b_{\MA} - \sum_i n_i b_i) \geqslant 0,
\ee
where the subscript ``$\MA$'' refers to the adjoint. Finally, since $G_2$, $\mathrm{SU}(3)$ and $\mathrm{SU}(2)$ are excluded from the search, we need not examine global anomalies. One can immediately see that Eqs. (\ref{e-5-3-1}) and (\ref{e-5-3-2}) are automatically satisfied when there is a hypermultiplet in the adjoint plus $244$ singlets or when all hypermultiplets are singlets; such solutions will be considered as trivial and will not be displayed. Our results are shown below.

\begin{enumerate}
\item $E_8$. For the $E_8$ gauge group we must have $n_H = 248 + 244 = 492$ and the only available representation is the adjoint. Since the hypermultiplets can fit in at most one adjoint, the only solutions are the trivial ones.
\item $E_7$. Since this is the first non-trivial case to be considered, we will present it in some detail. For the $E_7$ gauge group we must have $n_H = 133 + 244 = 377$ and the available representations are the adjoint $\mathbf{133}$ and the pseudoreal fundamental $\mathbf{56}$. So, the condition (\ref{e-5-3-1}) translates to
\be
133 n_{\mathbf{133}} + 56 n_{\mathbf{56}} \leqslant 377,
\ee
and is satisfied by the following matter content
\begin{align*}
&\textstyle{\frac{n}{2}} \cdot \mathbf{56} ; && n=0,\ldots,13, \\
&\mathbf{133} + \textstyle{\frac{n}{2}} \cdot \mathbf{56} ; && n=0,\ldots,8 \\
&2 \cdot \mathbf{133} + \textstyle{\frac{n}{2}} \cdot \mathbf{56}; &&
n=0,\ldots,3,
\end{align*}
plus the appropriate numbers of singlets. However, the second condition (\ref{e-5-3-2}), namely
\be 
( 3 - 3 n_{\mathbf{133}} - n_{\mathbf{56}} )^2 + 4 ( 4 - 4 n_{\mathbf{133}} - n_{\mathbf{56}} ) \geqslant 0,
\ee
further restricts the possible solutions to
\begin{align}
&\mathrm{(a) }\textstyle{\frac{n}{2}} \cdot \mathbf{56} ; && n=0,\ldots,13, \nn\\
&\mathrm{(b) }\mathbf{133} + 4 \cdot \mathbf{56}.
\end{align}
Regarding the models (a), one may make a shift of $n$ to $n_1 = n+4$ and rewrite them as $\textstyle{\frac{n_1-4}{2}} \cdot \mathbf{56}$. These models are then recognized as those resulting from the $E_8 \times E_8$ heterotic string on $K3$ by embeddding $n_1$ units of instanton charge in an $\mathrm{SU}(2)$ subgroup of the first $E_8$ (and ignoring the other $E_8$). These theories are the first ones in a chain of theories related to each other by successive Higgsing; in terms of theories to be discussed here, the relevant parts of the chain are $E_7\mathrm{(a)} \to E_6\mathrm{(a)} \to F_4\mathrm{(a)}\to \ldots$ and $E_7\mathrm{(a)} \to \mathrm{SO}(11) \mathrm{(b)} \to \mathrm{SO}(10) \mathrm{(b)} \to \ldots$.
\item $E_6$. Now, we have $n_H=322$ and the available representations are $\mathbf{27}$ and $\mathbf{78}$. Proceeding as before, we find the solutions
\begin{align}
&\mathrm{(a) }2 n \cdot \mathbf{27} ;&& n=1,\ldots,5, \nn\\
&\mathrm{(b) }4 \cdot \mathbf{78}, \nn\\
&\mathrm{(c) }\mathbf{78} + 8 \cdot \mathbf{27}, \nn\\
&\mathrm{(d) }2 \cdot \mathbf{78} + 6 \cdot \mathbf{27},
\end{align}
where, in addition, we imposed the requirement of CPT invariance which demands an even number of $\mathbf{27}$'s, understood as $2 n \cdot \mathbf{27} \to n \cdot \mathbf{27} + n \cdot \overline{\mathbf{27}}$.
\item $F_4$. Now, we have $n_H=296$ and the available representations are $\mathbf{26}$, $\mathbf{52}$ and
$\mathbf{273}$. The possible solutions are
\begin{align}
&\mathrm{(a) }n \cdot \mathbf{26}; && n=0,\ldots,11, \nn\\
&\mathrm{(b) }\mathbf{52} + 8 \cdot \mathbf{26}, \nn\\
&\mathrm{(c) }n \cdot \mathbf{52} + (11-2n) \cdot \mathbf{26}; &&
n=1,\ldots,5.
\end{align}
\end{enumerate}

For the classical groups, there is the extra condition (\ref{e-5-2-12}) which we write explicitly as
\be
\label{e-5-5-3}
\sum_i n_i a_i = a_{\MA}
\ee
Again, there exist trivial solutions, corresponding to a hypermultiplet in the adjoint plus $244$ singlets, that will not be displayed. The search for anomaly-free models can be conducted as before and the results are summarized as follows.

\begin{enumerate}

\item $\mathrm{SU}(N)$:
\begin{align}
&5 \leqslant N \leqslant 18: &&\mathrm{(a) }\left[ 2N - 2n (N-8) \right] \cdot \mathbf{N} + 2n \cdot \textstyle{\mathbf{\frac{N(N-1)}{2}}}. \\
&N=8: &&\mathrm{(b) }\mathbf{63} + 8 \cdot \mathbf{28}.\\
&N=7: &&\mathrm{(b) }\mathbf{48} + 8 \cdot \mathbf{7} + 8 \cdot \mathbf{21}.\\
&N=6: && \mathrm{(b) } \mathbf{35} + 16 \cdot \mathbf{6} + 8 \cdot \mathbf{15}, \nn\\
      &&&\mathrm{(c) } 2 \cdot \mathbf{35} + 8 \cdot \mathbf{6} + 10 \cdot \mathbf{15},\nn\\
      &&&\mathrm{(d) }8 \cdot \mathbf{6} + 2 \cdot \mathbf{21} + 12 \cdot \mathbf{15}.\\
&N=5: &&  \mathrm{(b) }\mathbf{24} + 24 \cdot \mathbf{5} + 8 \cdot \mathbf{10},\nn\\
      &&& \mathrm{(c) }2 \cdot \mathbf{24} + 20 \cdot \mathbf{5} + 10 \cdot \mathbf{10},\nn\\
      &&& \mathrm{(d) }4 \cdot \mathbf{24} + 6 \cdot \mathbf{5} + 12 \cdot \mathbf{10},\nn\\
      &&& \mathrm{(e) }2 \cdot \mathbf{24} + 6 \cdot \mathbf{5} + 2 \cdot \mathbf{15} + 14 \cdot \mathbf{10},\nn\\
      &&& \mathrm{(f) }6 \cdot \mathbf{5} + 4 \cdot \mathbf{15} + 14 \cdot \mathbf{10},\nn\\
      &&& \mathrm{(g) }20 \cdot \mathbf{5} + 2 \cdot \mathbf{15} + 12 \cdot \mathbf{10}.
\end{align}
In the first series of solutions, $n$ is restricted to all integer values such that all multiplicities, including the $243 - N^2 + n \textstyle{\frac{N^2 - 15N}{2}}$ singlets, are nonnegative.
\item $\mathrm{SO}(N)$:
\begin{align}
&\!\!\!\!\!\!\!\! 10 \leqslant N \leqslant 30: &&\mathrm{(a) } (N-8) \cdot \mathbf{N}. \\
&\!\!\!\!\!\!\!\! 10 \leqslant N \leqslant 14: &&\mathrm{(b) } \textstyle{\mathbf{\frac{N(N-1)}{2}}} + 8 \cdot \mathbf{N} +
2^{8 - \lfloor \frac{N+1}{2} \rfloor} \cdot \mathbf{2^{\lfloor \frac{N+1}{2} \rfloor - 1}}. \\
&\!\!\!\!\!\!\!\! N=14: && \mathrm{(c) } (4n+6) \cdot \mathbf{14} + n \cdot \mathbf{64}; && n=1,2. \\
&\!\!\!\!\!\!\!\! N=13: && \mathrm{(c) } (2n+5) \cdot \mathbf{13} + \textstyle{\frac{n}{2}} \cdot \mathbf{64}; && n=1,4. \\
&\!\!\!\!\!\!\!\! N=12: && \mathrm{(c) } (n+4) \cdot \mathbf{12} + \textstyle{\frac{n}{2}} \cdot \mathbf{32}; && n=1,\ldots,9, \nn\\
       &&&\mathrm{(d) } 2 \cdot \mathbf{66} + 4 \cdot \mathbf{12} + 4 \cdot \mathbf{32}. \\
&\!\!\!\!\!\!\!\! N=11: && \mathrm{(c) } (n+3) \cdot \mathbf{11} + \textstyle{\frac{n}{2}} \cdot \mathbf{32}; && n=1,\ldots,9, \nn\\
       &&&\mathrm{(d) } n \cdot \mathbf{55} + (13-4n) \cdot \mathbf{11} + \textstyle{\frac{10-n}{2}} \cdot                        \mathbf{32}; && n=1,\ldots,3. \\
&\!\!\!\!\!\!\!\! N=10: && \mathrm{(c) } (n+2) \cdot \mathbf{10} + n \cdot \mathbf{16}; && n=1,\ldots,10, \nn\\
       &&&\mathrm{(d) } n \cdot \mathbf{45} + (12-3n) \cdot \mathbf{10} + (10-n) \cdot \mathbf{16}; && n=1,\ldots,4.
\end{align}
Let us try to identify some known models.

\begin{itemize}
\item In the first series of models, the $\mathrm{SO}(28)\mathrm{(a)}$ model is identified with the theory obtained from the $\mathrm{SO}(32)$ heterotic string on $K3$ by embedding all 24 units of $K3$ instanton charge into one of the $\mathrm{SU}(2)$ factors in the decomposition $\mathrm{SO}(32) \supset \mathrm{SO}(28) \times \mathrm{SU}(2) \times \mathrm{SU}(2)$ and breaking the other $\mathrm{SU}(2)$ factor by Higgsing. By further Higgsing of this theory, one obtains all the $N < 28$ theories. Note that our list also contains models for $N=29,30$ which cannot be realized in a compactification context.

\item The $\mathrm{SO}(12)\mathrm{(c)}$ models are identified with the theories resulting from the $E_8 \times E_8$ heterotic string on $K3$, this time by embedding $n+4$ units of instanton charge in an $\mathrm{SU}(2) \times \mathrm{SU}(2)$ subgroup of the first $E_8$. These theories are also the first ones in a Higgs chain; in terms of the theories to be discussed here, the relevant parts of the chain are $\mathrm{SO}(12)\mathrm{(c)} \to \mathrm{SO}(11) \mathrm{(c)} \to \mathrm{SO}(10) \mathrm{(c)} \to \ldots$ and $\mathrm{SO}(12)\mathrm{(c)} \to \mathrm{SU}(6) \mathrm{(a)} \to \mathrm{SU}(5) \mathrm{(a)} \to \ldots$. The $E_7\mathrm{(a)}$ and $\mathrm{SO}(12) \mathrm{(c)}$ models together form the top of the ``Higgs tree'' that contains all possible chains of theories that can be obtained from them by Higgsing. All these chains were constructed in \cite{Bershadsky:1996nh} by geometric engineering via F-theory.

\item The $\mathrm{SO}(13)\mathrm{(c)}$ models can also be realized \cite{Bershadsky:1997sb} from the $E_8 \times E_8$ heterotic string on $K3$ by considering the decomposition $E_8 \supset \mathrm{SO}(16) \supset \mathrm{SO}(13) \times \mathrm{SU}(2)$ and embedding $n+4$ units of instanton charge in $\mathrm{SU}(2)$.
\end{itemize}

\item $\mathrm{USp}(2N)$:
\begin{align}
&\!\!\!\!\!\!\!\!\!\!\!\!\!\!\!\! 4 \leqslant N \leqslant 9: && \mathrm{(a) } \left[ (2N+8) - n(2N-8) \right] \cdot \mathbf{2 N} + n \cdot \mathbf{N(2N-1)-1},\\
&\!\!\!\!\!\!\!\!\!\!\!\!\!\!\!\! N=4: &&  \mathrm{(b) }\mathbf{36} + (n+8) \cdot \mathbf{27}; && n=0,1.
\end{align}
In the first series of solutions, $n$ is restricted to all integer values such that all multiplicities, including the $244 - 4 N^2 - 16N + n(6 N^2 - 17 N - 1)$ singlets, are nonnegative.

The first series of models has been identified in the literature \cite{Bershadsky:1996nh} as models with perturbatively enhanced symmetry resulting from F-theory compactifications on elliptic Calabi-Yau 3-folds based on the Hirzebruch surface; in this description, they originate from an $A_{2N-1}$ singularity on the coordinate of the $\mathbb{CP}^1$ fiber in that surface. The cases $n=0,1$ in this series were also given a gauge-theory interpretation \cite{Intriligator:1997kq} in terms of Type I D5--branes ($\mathrm{SO}(32)$ small instantons) placed at a $\mathbb{Z}_2$ orbifold singularity. For $n=1$, where the field content is given by
\begin{align}
&16 \cdot \mathbf{2N} + \mathbf{N(2N-1)-1},
\end{align}
the theory is on the Higgs branch. For $n=0$, where the field content is
\begin{align}
&(2N+8) \cdot \mathbf{2N},
\end{align}
the positions of all instantons are fixed, the blowing-up mode is zero and the theory rests on a non-trivial RG fixed point at the origin of the Coulomb branch.

\end{enumerate}

\subsection{Products of two simple groups}
\label{sec-5-5-2}

We now pass to the more complicated task of identifying anomaly-free models where the gauge group contains two simple group factors, $\MG = \MG_1 \times \MG_2$. This time, Eq. (\ref{e-5-2-12}) (when applicable) and Eq. (\ref{e-5-2-31}) must hold for each one of $\MG_1$ and $\MG_2$, while we also have the strict equality (\ref{e-5-2-32}) involving both group factors. Before we begin, we note that each of the simple-group solutions for, say, $\MG_1$ can be extended to a solution for $\MG_1 \times \MG_2$ by simply adding one adjoint of $\MG_2$. Such ``reducible'' solutions will not be written out explicitly.

We start our search from the case where both $\MG_1$ and $\MG_2$ are exceptional groups, in which case there are no fourth-order Casimirs. The largest-rank group of this type is $E_8 \times E_8$, which is one of the possible gauge groups of heterotic string theory; it is easily seen that this group admits only the trivial solutions. The group $E_8 \times E_7$ ($E_7 \times E_7$) is that obtained from the reduction of the $E_8 \times E_8$ heterotic string on $K3$ using the standard (non-standard) embedding(s) of the $K3$ instanton charge. So, in this search, we expect to obtain all solutions corresponding to these embeddings as well as the chains produced from these solutions by Higgsing. The solutions found are shown below.

\begin{enumerate}
\item $E_8 \times E_7$:
\begin{align}
&\mathrm{(a) }10 (\mathbf{1},\mathbf{56}), \nn\\
&\mathrm{(b) }\textstyle{\frac{3}{2}} (\mathbf{1},\mathbf{56}) + 4 (\mathbf{1},\mathbf{133}).
\end{align}
The first model on the list is the well-known model obtained from the reduction of the $E_8 \times E_8$ heterotic string on $K3$ using the standard embedding (24 units of instanton charge in one $E_8$).

\item $E_8 \times E_6$:
\begin{align}
& 18 (\mathbf{1},\mathbf{27}).
\end{align}
This solution, written in full as $9 (\mathbf{1},\mathbf{27}) + 9 (\mathbf{1},\overline{\mathbf{27}})$, may be obtained from the $E_8 \times E_7\mathrm{(a)}$ models by Higgsing. The chain of Higgsing continues to further subgroups.

\item $E_8 \times F_4$:
\begin{align}
&\mathrm{(a) } 17 (\mathbf{1},\mathbf{26}), \nn\\
&\mathrm{(b) } 4 (\mathbf{1},\mathbf{52}) + 12 (\mathbf{1},\mathbf{26}).
\end{align}
\item $E_7 \times E_7$:
\begin{align}
&\mathrm{(a) } \textstyle{\frac{n}{2}} (\mathbf{56},\mathbf{1}) + \textstyle{\frac{16-n}{2}} (\mathbf{1},\mathbf{56}); && n=0,\ldots,8, \nn\\
&\mathrm{(b) } \textstyle{\frac{9}{2}} (\mathbf{56},\mathbf{1}) + n (\mathbf{133},\mathbf{1}) + 2 (\mathbf{1},\mathbf{56}); && n=0,1.
\end{align}
The first class of models on the list may be written in the more suggestive form $\textstyle{\frac{n_1-4}{2}} (\mathbf{56},\mathbf{1}) + \textstyle{\frac{n_2-4}{2}} (\mathbf{1},\mathbf{56})$, $n_1+n_2=24$ and they are recognized as the models constructed by reduction of the $E_8 \times E_8$ theory on $K3$ with $n_1$ and $n_2$ units of instanton charge embedded in the first and second $E_8$ respectively.

\item $E_7 \times E_6$:
\begin{align}
&\mathrm{(a) } n (\mathbf{56},\mathbf{1}) + (14-2n) (\mathbf{1},\mathbf{27}); && n=0,\ldots,7, \nn\\
&\mathrm{(b) } \textstyle{\frac{9}{2}} (\mathbf{56},\mathbf{1}) + 2 (\mathbf{1},\mathbf{27}), \nn\\
&\mathrm{(c) } \textstyle{\frac{9}{2}} (\mathbf{56},\mathbf{1}) + (\mathbf{133},\mathbf{1}) + 2 (\mathbf{1},\mathbf{27}), \nn\\
&\mathrm{(d) } 3(\mathbf{133},\mathbf{1}) + 2 (\mathbf{1},\mathbf{27}), \nn\\
&\mathrm{(e) } \textstyle{\frac{n+2}{2}} (\mathbf{56},\mathbf{1}) + (5-n) (\mathbf{1},\mathbf{78}) + 2n (\mathbf{1},\mathbf{27}); && n=0,\ldots,2.
\end{align}
The first class of models are obtained from the $E_7 \times E_7 \mathrm{(a)}$ models by Higgsing. The chain of Higgsing continues further on.

\item $E_7 \times F_4$:
\begin{align}
&\mathrm{(a) } \textstyle{\frac{n}{2}} (\mathbf{56},\mathbf{1}) + (13-n) (\mathbf{1},\mathbf{26}); && n=0,\ldots,13, \nn\\
&\mathrm{(b) } 2 (\mathbf{56},\mathbf{1}) + 6 (\mathbf{1},\mathbf{26}), \nn\\
&\mathrm{(c) } \textstyle{\frac{9}{2}} (\mathbf{56},\mathbf{1}) + (\mathbf{1},\mathbf{26}), \nn\\
&\mathrm{(d) } 2 (\mathbf{56},\mathbf{1}) + (\mathbf{1},\mathbf{52}) + 9 (\mathbf{1},\mathbf{26}), \nn\\
&\mathrm{(e) } 2 (\mathbf{56},\mathbf{1}) + 3 (\mathbf{1},\mathbf{52}) + 6 (\mathbf{1},\mathbf{26}), \nn\\
&\mathrm{(f) } 2 (\mathbf{56},\mathbf{1}) + 6 (\mathbf{1},\mathbf{52}), \nn\\
&\mathrm{(g) } (\mathbf{133},\mathbf{1}) + \textstyle{\frac{9}{2}} (\mathbf{56},\mathbf{1}) + (\mathbf{1},\mathbf{26}), \nn\\
&\mathrm{(h) } 3 (\mathbf{133},\mathbf{1}) + (\mathbf{1},\mathbf{26}), \nn\\
&\mathrm{(i) } n (\mathbf{1},\mathbf{52}) + (9-n) (\mathbf{1},\mathbf{26}); && n=1,\ldots,6.
\end{align}

\item $E_6 \times E_6$:
\begin{align}
&\mathrm{(a) } 2n (\mathbf{27},\mathbf{1}) + (12-2n) (\mathbf{1},\mathbf{27}); &&n=0,\ldots,6, \nn \\
&\mathrm{(b) } 5 (\mathbf{78},\mathbf{1}); \nn\\
&\mathrm{(c) } 2 (\mathbf{27},\mathbf{1}) + 4 (\mathbf{1},\mathbf{27}) + 3 (\mathbf{1},\mathbf{78}).
\end{align}

\item $F_4 \times F_4$:
\begin{align}
&\mathrm{(a) } n (\mathbf{26},\mathbf{1}) + (10-n) (\mathbf{1},\mathbf{26}); && n=0,\ldots,5 \nn\\
&\mathrm{(b) } n (\mathbf{26},\mathbf{1}) + (4-n) (\mathbf{1},\mathbf{52}) + (n+5) (\mathbf{1},\mathbf{26}); &&n=0,\ldots,4. \nn\\
&\mathrm{(c) } (\mathbf{26},\mathbf{1}) + 6 (\mathbf{1},\mathbf{52}), \nn\\
&\mathrm{(d) } (\mathbf{26},\mathbf{1}) + (\mathbf{1},\mathbf{52}) + 9 (\mathbf{1},\mathbf{26}).
\end{align}

\end{enumerate}

We finally proceed to the case where $\MG_1$ is an exceptional group while $\MG_2$ is classical. In this case,  $\MG_2$ does have fourth-order Casimirs and so we also have the extra condition (\ref{e-5-2-12}) for this factor. The models found are the following.

\begin{enumerate}
\item $E_8 \times \mathrm{SU}(N)$:
\begin{align}
& 5 \leqslant N \leqslant 8: &&(112-12N) (\mathbf{1},\mathbf{N} ) + 14 \left(
\mathbf{1},\textstyle{\mathbf{\frac{N(N-1)}{2}}} \right).
\end{align}

\item $E_8 \times \mathrm{SO}(N)$:
\begin{align}
& N=14: && 22 (\mathbf{1},\mathbf{14}) + 4 (\mathbf{1},\mathbf{64}).\\
& N=13: && 21 (\mathbf{1},\mathbf{13}) + 4 (\mathbf{1},\mathbf{64}).\\
& N=12: && \mathrm{(a) } 20 (\mathbf{1},\mathbf{12}) + 8 (\mathbf{1},\mathbf{32}),\nn\\
                             &&&\mathrm{(b) } 4 (\mathbf{1},\mathbf{66}) + 3 (\mathbf{1},\mathbf{12}) + \textstyle{\frac{15}{2}} (\mathbf{1},\mathbf{32}).\\
& N=11: && \mathrm{(a) } 19 (\mathbf{1},\mathbf{11}) + 8 (\mathbf{1},\mathbf{32}),\nn\\
                             &&&\mathrm{(b) } 4 (\mathbf{1},\mathbf{55}) + 6 (\mathbf{1},\mathbf{11}) +  \frac{15}{2} (\mathbf{1},\mathbf{32}).\\
& N=10: && \mathrm{(a) }  18 (\mathbf{1},\mathbf{10}) + 16 (\mathbf{1},\mathbf{16}),\nn\\
                             &&&\mathrm{(b) } 4 (\mathbf{1},\mathbf{45}) + 9 (\mathbf{1},\mathbf{10}) +  15 (\mathbf{1},\mathbf{16}).
\end{align}

\item $E_8 \times \mathrm{USp}(2N)$:
\begin{align}
& N=4: &&16 (\mathbf{1},\mathbf{8} ) + 13 \left( \mathbf{1},\mathbf{27} \right).
\end{align}

\item $E_7 \times \mathrm{SU}(N)$:
\begin{align}
&\!\!\!\!\!\!\!\!\!\!\!\!\!\!\!\! N=12: && 2 (\mathbf{56},\mathbf{1}) + 6
(\mathbf{1},\mathbf{66}).\\
&\!\!\!\!\!\!\!\!\!\!\!\!\!\!\!\! N=11: && 2 (\mathbf{56},\mathbf{1}) + 2 (\mathbf{1},\mathbf{11}) + 6 (\mathbf{1},\mathbf{55}).\\
&\!\!\!\!\!\!\!\!\!\!\!\!\!\!\!\! 5 \leqslant N \leqslant 10: && \mathrm{(a) }  n (\mathbf{56},\mathbf{1}) +
\left[ 80-8N + 2n (N-8) \right] (\mathbf{1},\mathbf{N}) \nn\\
&&& + (10-2n) \left( \mathbf{1},\textstyle{\mathbf{\frac{N(N-1)}{2}}} \right); &&n=0,\ldots,5. \\
&\!\!\!\!\!\!\!\!\!\!\!\!\!\!\!\! N=5: && \mathrm{(b) } \textstyle{\frac{n_1}{2}} (\mathbf{56},\mathbf{1}) + 4 n_1
(\mathbf{1},\mathbf{5}) + (7-n_1-2n_2) (\mathbf{1},\mathbf{24}) \nn\\
&&&+ 2n_2 (\mathbf{1},\mathbf{15}) +
(20-2n_1+2n_2) (\mathbf{1},\mathbf{10}); \nn\\
&&& n_1=0-7 , n_2 =0- \left\lfloor
\textstyle{\frac{7-n_1}{2}} \right\rfloor.
\end{align}

\item $E_7 \times \mathrm{SO}(N)$:
\begin{align}
&\!\!\!\!\!\!\!\!\!\!\!\!\!\!\!\! 10 \leqslant N \leqslant 25: && \mathrm{(a) } \textstyle{\frac{9}{2}} (\mathbf{56},\mathbf{1}) + (N-8) (\mathbf{1},\mathbf{N}).\\
&\!\!\!\!\!\!\!\!\!\!\!\!\!\!\!\! 10 \leqslant N \leqslant 19: && \mathrm{(b) } 6 (\mathbf{56},\mathbf{1}) + (N-8) (\mathbf{1},\mathbf{N}).\\
&\!\!\!\!\!\!\!\!\!\!\!\!\!\!\!\! N=16: &&  \mathrm{(c) } 2 (\mathbf{56},\mathbf{1}) + 16 (\mathbf{1},\mathbf{16}) + (\mathbf{1},\mathbf{128}).\\
&\!\!\!\!\!\!\!\!\!\!\!\!\!\!\!\! N=15: &&  \mathrm{(c) } 2 (\mathbf{56},\mathbf{1}) + 15 (\mathbf{1},\mathbf{15}) + (\mathbf{1},\mathbf{128}).\\
&\!\!\!\!\!\!\!\!\!\!\!\!\!\!\!\! N=14: && \mathrm{(c) }  (4-2n) (\mathbf{56},\mathbf{1}) + (4n+10) (\mathbf{1},\mathbf{14}) \nn\\
                              &&&+ (n+1)(\mathbf{1},\mathbf{64}); &&n=0,\ldots,2. \\
&\!\!\!\!\!\!\!\!\!\!\!\!\!\!\!\! 10 \leqslant N \leqslant 13: && \mathrm{(c) } \textstyle{\frac{9}{2}} (\mathbf{56},\mathbf{1}) + (\mathbf{133},\mathbf{1}) + (N-8) (\mathbf{1},\mathbf{N}).\\
&\!\!\!\!\!\!\!\!\!\!\!\!\!\!\!\! N=13: && \mathrm{(d) }  (6-n) (\mathbf{56},\mathbf{1}) + (2n+5) (\mathbf{1},\mathbf{13}) + \textstyle{\frac{n}{2}} (\mathbf{1},\mathbf{64}); &&n=0,\ldots,6. \\
&\!\!\!\!\!\!\!\!\!\!\!\!\!\!\!\! N=12: &&  \mathrm{(d) } \textstyle{\frac{12-n}{2}} (\mathbf{56},\mathbf{1}) + (n+4) (\mathbf{1},\mathbf{12}) + \textstyle{\frac{n}{2}} (\mathbf{1},\mathbf{32}); &&n=0,\ldots,6. \\
&\!\!\!\!\!\!\!\!\!\!\!\!\!\!\!\! N=11: &&  \mathrm{(d) } \textstyle{\frac{12-n}{2}} (\mathbf{56},\mathbf{1}) + (n+3) (\mathbf{1},\mathbf{11}) + \textstyle{\frac{n}{2}} (\mathbf{1},\mathbf{32}); &&n=0,\ldots,6, \nn\\
                              &&&\mathrm{(e) } 3 (\mathbf{133},\mathbf{1}) + 3 (\mathbf{1},\mathbf{11}).\\
&\!\!\!\!\!\!\!\!\!\!\!\!\!\!\!\! N=10: && \mathrm{(d) } \textstyle{\frac{12-n}{2}} (\mathbf{56},\mathbf{1}) + (n+2) (\mathbf{1},\mathbf{10}) + n (\mathbf{1},\mathbf{16}); &&n=0,\ldots,12, \nn\\
                              &&&\mathrm{(e) } 3 (\mathbf{133},\mathbf{1}) + 2 (\mathbf{1},\mathbf{10}), \nn\\
                              &&&\mathrm{(f) } 2 (\mathbf{56},\mathbf{1}) + (\mathbf{1},\mathbf{45}) + 9 (\mathbf{1},\mathbf{10}) + 9 (\mathbf{1},\mathbf{16}), \nn\\
                              &&&\mathrm{(g) } \textstyle{\frac{n+2}{2}} (\mathbf{56},\mathbf{1}) + (5-n) (\mathbf{1},\mathbf{45}) \nn\\
                              &&&+ 2n (\mathbf{1},\mathbf{10}) + 8 (\mathbf{1},\mathbf{16}); &&n=0,\ldots,4.
\end{align}

\item $E_7 \times \mathrm{USp}(2N)$:
\begin{align}
&\!\!\!\!\!\!\!\!\!\!\!\!\!\!\!\! 4 \leqslant N \leqslant 12: && \mathrm{(a) }  2 (\mathbf{56},\mathbf{1}) + (24-2N) (\mathbf{1},\mathbf{2N}) \nn\\
                  &&&+ 2 (\mathbf{1},\mathbf{N(2N-1)-1}). \\
&\!\!\!\!\!\!\!\!\!\!\!\!\!\!\!\! N=6: && \mathrm{(b) }  \textstyle{\frac{5}{2}} (\mathbf{56},\mathbf{1}) + 4 (\mathbf{1},\mathbf{12}) + 4 (\mathbf{1},\mathbf{65}) .\\
&\!\!\!\!\!\!\!\!\!\!\!\!\!\!\!\! N=5: && \mathrm{(b) } \textstyle{\frac{9-n}{2}} (\mathbf{56},\mathbf{1}) + (18-2n) (\mathbf{1},\mathbf{10}) + n (\mathbf{1},\mathbf{44}); && n=0,\ldots,9.\\
&\!\!\!\!\!\!\!\!\!\!\!\!\!\!\!\! N=4: &&\mathrm{(b) } \textstyle{\frac{9-n}{2}} (\mathbf{56},\mathbf{1}) + 16 (\mathbf{1},\mathbf{8}) + n (\mathbf{1},\mathbf{27}); && n=0,\ldots,9, \nn\\
                                      &&&\mathrm{(c) } 2 (\mathbf{56},\mathbf{1}) + (\mathbf{1},\mathbf{36}) + 9 (\mathbf{1},\mathbf{27}) .
\end{align}

\item $E_6 \times \mathrm{SU}(N)$:
\begin{align}
&\!\!\!\!\!\!\!\!\!\!\!\!\!\!\!\! N = 12: && 2 (\mathbf{27},\mathbf{1} ) + 6 (
\mathbf{1},\mathbf{66} ). \\
&\!\!\!\!\!\!\!\!\!\!\!\!\!\!\!\! N = 11: && \mathrm{(a) } 2 (\mathbf{27},\mathbf{1} ) + 4 ( \mathbf{1},\mathbf{11} ) + 6 ( \mathbf{1},\mathbf{55} ), \nn\\
                          &&&\mathrm{(b) } 4 (\mathbf{27},\mathbf{1} ) + 10 ( \mathbf{1},\mathbf{11} ) + 4 ( \mathbf{1},\mathbf{55} ). \\
&\!\!\!\!\!\!\!\!\!\!\!\!\!\!\!\! 5 \leqslant N \leqslant 10: && \mathrm{(a) }  2n (\mathbf{27},\mathbf{1}) + \left[ 64-6N + 2n (N-8) \right] (\mathbf{1},\mathbf{N}) \nn\\
                                  &&& + (8-2n) \left( \mathbf{1},\textstyle{\mathbf{\frac{N(N-1)}{2}}} \right); &&n=0,\ldots,4. \\
&\!\!\!\!\!\!\!\!\!\!\!\!\!\!\!\! N = 5: && \mathrm{(b) } (2n+1) (\mathbf{1},\mathbf{24} ) + 8 ( \mathbf{1},\mathbf{5} ) \nn\\
          &&& + (4-2n) ( \mathbf{1},\mathbf{15} ) + (20-2n) ( \mathbf{1},\mathbf{10} ); &&n=0,\ldots,2, \nn\\
                                         &&&\mathrm{(c) } 2 (\mathbf{27},\mathbf{1} ) + (\mathbf{1},\mathbf{24} ) + 16 ( \mathbf{1},\mathbf{5} ) \nn\\
                                         &&&+ 2 ( \mathbf{1},\mathbf{15} ) + 14 ( \mathbf{1},\mathbf{10} ), \nn\\
                                         &&&\mathrm{(d) } 2 (\mathbf{27},\mathbf{1} ) + 3(\mathbf{1},\mathbf{24} ) + 16 ( \mathbf{1},\mathbf{5} ) + 12 ( \mathbf{1},\mathbf{10} ), \nn\\
                                         &&&\mathrm{(e) } 4 (\mathbf{27},\mathbf{1} ) + (\mathbf{1},\mathbf{24} ) + 24 ( \mathbf{1},\mathbf{5} ) + 8 ( \mathbf{1},\mathbf{10} ), \nn\\
                                         &&&\mathrm{(f) } (\mathbf{78},\mathbf{1} ) + 8 (\mathbf{27},\mathbf{1} ) + 10 ( \mathbf{1},\mathbf{5} ).
\end{align}

\item $E_6 \times \mathrm{SO}(N)$:
\begin{align}
&\!\!\!\!\!\!\!\!\!\!\!\!\!\!\!\! 10 \leqslant N \leqslant 20: && \mathrm{(a) } 10 (\mathbf{27},\mathbf{1}) + (N-8) (\mathbf{1},\mathbf{N}).\\
&\!\!\!\!\!\!\!\!\!\!\!\!\!\!\!\! N=16: &&  \mathrm{(b) } 2 (\mathbf{27},\mathbf{1}) + 16 (\mathbf{1},\mathbf{16}) + (\mathbf{1},\mathbf{128}).\\
&\!\!\!\!\!\!\!\!\!\!\!\!\!\!\!\! N=15: &&  \mathrm{(b) } 2 (\mathbf{27},\mathbf{1}) + 15 (\mathbf{1},\mathbf{15}) + (\mathbf{1},\mathbf{128}).\\
&\!\!\!\!\!\!\!\!\!\!\!\!\!\!\!\! N=14: && \mathrm{(b) }  (6-4n) (\mathbf{27},\mathbf{1}) + (4n+10) (\mathbf{1},\mathbf{14}) \nn\\
                                        &&&+ (n+1) (\mathbf{1},\mathbf{64}); &&n=0,1. \\
&\!\!\!\!\!\!\!\!\!\!\!\!\!\!\!\! N=13: && \mathrm{(b) }  (10-2n) (\mathbf{27},\mathbf{1}) + (2n+5) (\mathbf{1},\mathbf{13}) + \textstyle{\frac{n}{2}} (\mathbf{1},\mathbf{64}); &&n=0,\ldots,5, \nn\\
&\!\!\!\!\!\!\!\!\!\!\!\!\!\!\!\! N=12: &&  \mathrm{(b) } (10-n) (\mathbf{27},\mathbf{1}) + (2n+4) (\mathbf{1},\mathbf{12}) + n (\mathbf{1},\mathbf{32}); &&n=0,\ldots,5, \nn\\
                                        &&& \mathrm{(c) } 2 (\mathbf{27},\mathbf{1}) + 9 (\mathbf{1},\mathbf{12}) + \textstyle{\frac{5}{2}} (\mathbf{1},\mathbf{32}), \nn\\
                                        &&& \mathrm{(d) } 2 (\mathbf{27},\mathbf{1}) + (\mathbf{1},\mathbf{66}) + 9 (\mathbf{1},\mathbf{12}) + \textstyle{\frac{9}{2}} (\mathbf{1},\mathbf{32}) , \nn\\
                                        &&& \mathrm{(e) } 2 (\mathbf{27},\mathbf{1}) + 3 (\mathbf{1},\mathbf{66}) + 4 (\mathbf{1},\mathbf{32}). \\
&\!\!\!\!\!\!\!\!\!\!\!\!\!\!\!\! N=11: &&  \mathrm{(b) } (10-2n) (\mathbf{27},\mathbf{1}) + (2n+3) (\mathbf{1},\mathbf{11}) + n (\mathbf{1},\mathbf{32}); &&n=0,\ldots,5, \nn\\
                                        &&& \mathrm{(c) } 2 (\mathbf{27},\mathbf{1}) + 8 (\mathbf{1},\mathbf{11}) + \textstyle{\frac{5}{2}} (\mathbf{1},\mathbf{32}), \nn\\
                                        &&& \mathrm{(d) } 2 (\mathbf{27},\mathbf{1}) + (\mathbf{1},\mathbf{55}) + 9 (\mathbf{1},\mathbf{11}) + \textstyle{\frac{9}{2}} (\mathbf{1},\mathbf{32}) , \nn\\
                                        &&& \mathrm{(e) } 2 (\mathbf{27},\mathbf{1}) + 3 (\mathbf{1},\mathbf{55}) + 2 (\mathbf{1},\mathbf{11}) + 4 (\mathbf{1},\mathbf{32}), \nn\\
                                        &&&\mathrm{(f) } 3 (\mathbf{78},\mathbf{1}) + 4 (\mathbf{27},\mathbf{1})+ 3 (\mathbf{1},\mathbf{11}).\\
&\!\!\!\!\!\!\!\!\!\!\!\!\!\!\!\! N=10: && \mathrm{(b) } (10-2n) (\mathbf{27},\mathbf{1}) + (n+2) (\mathbf{1},\mathbf{10}) + 2n (\mathbf{1},\mathbf{16}); &&n=0,\ldots,5, \nn\\
                                        &&&\mathrm{(c) } 5 (\mathbf{1},\mathbf{45}) + 8 (\mathbf{1},\mathbf{16}), \nn\\
                                        &&&\mathrm{(d) } 2 (\mathbf{27},\mathbf{1}) + (\mathbf{1},\mathbf{45}) + 9 (\mathbf{1},\mathbf{10}) + 9 (\mathbf{1},\mathbf{16}) , \nn\\
                                        &&&\mathrm{(e) } 2 (\mathbf{27},\mathbf{1}) + 3 (\mathbf{1},\mathbf{45}) + 4 (\mathbf{1},\mathbf{10}) + 8 (\mathbf{1},\mathbf{16}), \nn\\
                                        &&&\mathrm{(f) } (3-n) (\mathbf{78},\mathbf{1}) + (2n+4) (\mathbf{27},\mathbf{1}) \nn\\
                                        &&&+ (n+2) (\mathbf{1},\mathbf{10}) + n (\mathbf{1},\mathbf{16}); &&n=0,\ldots,2.
\end{align}

\item $E_6 \times \mathrm{USp}(2N)$:
\begin{align}
&\!\!\!\!\!\!\!\!\!\!\!\!\!\!\!\!  4 \leqslant N \leqslant 12: && \mathrm{(a) }  2 (\mathbf{27},\mathbf{1}) + (24-2N) (\mathbf{1},\mathbf{2N}) \nn\\
                  &&&+ 2 (\mathbf{1},\mathbf{N(2N-1)-1}). \\
&\!\!\!\!\!\!\!\!\!\!\!\!\!\!\!\!  N=6: && \mathrm{(b) }  2 (\mathbf{27},\mathbf{1}) + 5 (\mathbf{1},\mathbf{65}) , \nn\\
       &&&\mathrm{(c) }  4 (\mathbf{27},\mathbf{1}) + 8 (\mathbf{1},\mathbf{12}) + 3 (\mathbf{1},\mathbf{65}) . \\
&\!\!\!\!\!\!\!\!\!\!\!\!\!\!\!\!  N=5: &&  \mathrm{(b) } (6-2n) (\mathbf{27},\mathbf{1}) + (16-4n) (\mathbf{1},\mathbf{10}) \nn\\
&&&+ (2n+1) (\mathbf{1},\mathbf{44}); && n=0,\ldots,3.\\
&\!\!\!\!\!\!\!\!\!\!\!\!\!\!\!\!  N=4: && \mathrm{(b) } (6-2n) (\mathbf{27},\mathbf{1}) + 16 (\mathbf{1},\mathbf{8}) \nn\\
       &&&+ (2n+1) (\mathbf{1},\mathbf{27}); && n=0,\ldots,3, \nn\\
       &&&\mathrm{(c) } 2 (\mathbf{27},\mathbf{1}) + (\mathbf{1},\mathbf{36}) + 9 (\mathbf{1},\mathbf{27}) .
\end{align}

\item $F_4 \times \mathrm{SU}(N)$:
\begin{align}
&\!\!\!\!\!\!\!\!\!\!\!\!\!\!\!\! N = 12: && \mathrm{(a) } (\mathbf{26},\mathbf{1} ) + 6 ( \mathbf{1},\mathbf{66} ), \nn\\
                          &&&\mathrm{(b) } 3 (\mathbf{26},\mathbf{1} ) + 8 ( \mathbf{1},\mathbf{12} ) + 4 ( \mathbf{1},\mathbf{66} ). \\
&\!\!\!\!\!\!\!\!\!\!\!\!\!\!\!\! N = 11: && \mathrm{(a) } (\mathbf{26},\mathbf{1} ) + 4 ( \mathbf{1},\mathbf{11} ) + 6 ( \mathbf{1},\mathbf{55} ), \nn\\
                          &&&\mathrm{(b) } 3 (\mathbf{26},\mathbf{1} ) + 10 ( \mathbf{1},\mathbf{11} ) + 4 ( \mathbf{1},\mathbf{55} ), \nn\\
                          &&&\mathrm{(c) } 5 (\mathbf{26},\mathbf{1} ) + 16 ( \mathbf{1},\mathbf{11} ) + 2 ( \mathbf{1},\mathbf{55} ). \\
&\!\!\!\!\!\!\!\!\!\!\!\!\!\!\!\! 5 \leqslant N \leqslant 10: && \mathrm{(a) } (2n+1) (\mathbf{27},\mathbf{1}) + \left[ 48-4N + 2n (N-8) \right] (\mathbf{1},\mathbf{N}) \nn\\
                                  &&& + (6-2n) \left( \mathbf{1},\textstyle{\mathbf{\frac{N(N-1)}{2}}} \right); &&n=0,\ldots,3. \\
&\!\!\!\!\!\!\!\!\!\!\!\!\!\!\!\! N = 6: && \mathrm{(b) } 3 (\mathbf{26},\mathbf{1} ) +  ( \mathbf{1},\mathbf{35} ) +  16 ( \mathbf{1},\mathbf{6} ) + 8 ( \mathbf{1},\mathbf{15} ). \\
&\!\!\!\!\!\!\!\!\!\!\!\!\!\!\!\! N = 5: && \mathrm{(b) } 2n (\mathbf{1},\mathbf{24} ) + 12 ( \mathbf{1},\mathbf{5} ) \nn\\
          &&& + (4-2n) ( \mathbf{1},\mathbf{15} ) + (18-2n) ( \mathbf{1},\mathbf{10} ); &&n=0,\ldots,2, \nn\\
          &&&\mathrm{(c) } (\mathbf{26},\mathbf{1} ) + (\mathbf{1},\mathbf{24} ) + 16 ( \mathbf{1},\mathbf{5} )\nn\\
          &&& + 2 ( \mathbf{1},\mathbf{15} ) + 14 ( \mathbf{1},\mathbf{10} ), \nn\\
          &&&\mathrm{(d) } (4-n) (\mathbf{26},\mathbf{1} ) + n (\mathbf{1},\mathbf{24} ) \nn\\
          &&& + (28-4n) ( \mathbf{1},\mathbf{5} ) + (2n+6) ( \mathbf{1},\mathbf{10} ); &&n=0,\ldots,3, \nn\\
          &&&\mathrm{(e) } 2 (\mathbf{26},\mathbf{1} ) + 20 ( \mathbf{1},\mathbf{5} ) + 2 ( \mathbf{1},\mathbf{15} ) + 12 ( \mathbf{1},\mathbf{10} ), \nn\\
          &&&\mathrm{(f) } (\mathbf{52},\mathbf{1} ) + 8 (\mathbf{27},\mathbf{1} ) + 10 ( \mathbf{1},\mathbf{5} ).
\end{align}

\item $F_4 \times \mathrm{SO}(N)$:
\begin{align}
&\!\!\!\!\!\!\!\!\!\!\!\!\!\!\!\! 10 \leqslant N \leqslant 20: && \mathrm{(a) } 9 (\mathbf{26},\mathbf{1}) + (N-8) (\mathbf{1},\mathbf{N}).\\
&\!\!\!\!\!\!\!\!\!\!\!\!\!\!\!\! 10 \leqslant N \leqslant 25: && \mathrm{(b) } 6 (\mathbf{26},\mathbf{1}) + (N-8) (\mathbf{1},\mathbf{N}).\\
&\!\!\!\!\!\!\!\!\!\!\!\!\!\!\!\! N=16:  && \mathrm{(c) }  (\mathbf{26},\mathbf{1}) + 16 (\mathbf{1},\mathbf{16}) + (\mathbf{1},\mathbf{128}), \nn\\
                              &&&\mathrm{(d) } (\mathbf{52},\mathbf{1}) + 9 (\mathbf{26},\mathbf{1}) + 8 (\mathbf{1},\mathbf{16}) . \\
&\!\!\!\!\!\!\!\!\!\!\!\!\!\!\!\! N=15: && \mathrm{(c) } (\mathbf{26},\mathbf{1}) + 15 (\mathbf{1},\mathbf{15}) + (\mathbf{1},\mathbf{128}), \nn\\
                              &&&\mathrm{(d) } (\mathbf{52},\mathbf{1}) + 9 (\mathbf{26},\mathbf{1}) + 7 (\mathbf{1},\mathbf{15}) . \\
&\!\!\!\!\!\!\!\!\!\!\!\!\!\!\!\! N=14: && \mathrm{(c) }  (\mathbf{26},\mathbf{1}) + 14 (\mathbf{1},\mathbf{14}) + 2 (\mathbf{1},\mathbf{64}), \nn\\
                              &&&\mathrm{(d) } 5 (\mathbf{26},\mathbf{1}) + 10 (\mathbf{1},\mathbf{14}) + (\mathbf{1},\mathbf{64}), \nn\\
                              &&&\mathrm{(e) } (\mathbf{52},\mathbf{1}) + 9 (\mathbf{26},\mathbf{1}) + 6 (\mathbf{1},\mathbf{14}). \\
&\!\!\!\!\!\!\!\!\!\!\!\!\!\!\!\! 10 \leqslant N \leqslant 13: && \mathrm{(c) } 9 (\mathbf{26},\mathbf{1}) + (\mathbf{52},\mathbf{1}) + (N-8) (\mathbf{1},\mathbf{N}).\\
&\!\!\!\!\!\!\!\!\!\!\!\!\!\!\!\! N=13: && \mathrm{(d) } (7-2n) (\mathbf{26},\mathbf{1}) + (2n+7) (\mathbf{1},\mathbf{13}) \nn\\
                                        &&&+ \textstyle{\frac{n}{2}} (\mathbf{1},\mathbf{64}); &&n=0,\ldots,3. \\
&\!\!\!\!\!\!\!\!\!\!\!\!\!\!\!\! N=12: && \mathrm{(d) } (8-n) (\mathbf{26},\mathbf{1}) + (n+5) (\mathbf{1},\mathbf{12}) \nn\\
                                        &&&+ \textstyle{\frac{n}{2}} (\mathbf{1},\mathbf{32}); &&n=0,\ldots,8, \nn\\
                              &&&\mathrm{(e) } 13 (\mathbf{1},\mathbf{12}) + \textstyle{\frac{9}{2}} (\mathbf{1},\mathbf{32}), \nn\\
                              &&&\mathrm{(f) } (\mathbf{26},\mathbf{1}) + 3 (\mathbf{1},\mathbf{66}) + 4 (\mathbf{1},\mathbf{32}),\nn\\
                              &&&\mathrm{(g) } (\mathbf{26},\mathbf{1}) + 9 (\mathbf{1},\mathbf{12}) + \textstyle{\frac{5}{2}} (\mathbf{1},\mathbf{32}) ,\nn\\
                              &&&\mathrm{(h) } (\mathbf{26},\mathbf{1}) + (\mathbf{1},\mathbf{66}) + 9 (\mathbf{1},\mathbf{12}) + \textstyle{\frac{9}{2}} (\mathbf{1},\mathbf{32}), \nn\\
                              &&&\mathrm{(i) } 2 (\mathbf{26},\mathbf{1}) + 2 (\mathbf{1},\mathbf{66}) + 4 (\mathbf{1},\mathbf{12}) + 4 (\mathbf{1},\mathbf{32}),\nn\\
                              &&&\mathrm{(j) } 2 (\mathbf{52},\mathbf{1}) + 7 (\mathbf{26},\mathbf{1}) + 5 (\mathbf{1},\mathbf{12}) + \textstyle{\frac{1}{2}} (\mathbf{1},\mathbf{32}),\nn\\
                              &&&\mathrm{(k) } 3 (\mathbf{52},\mathbf{1}) + 6 (\mathbf{26},\mathbf{1}) + 4 (\mathbf{1},\mathbf{12}),\nn\\
                              &&&\mathrm{(l) } 6 (\mathbf{52},\mathbf{1}) + 4 (\mathbf{1},\mathbf{12}).\\
&\!\!\!\!\!\!\!\!\!\!\!\!\!\!\!\! N=11: && \mathrm{(d) } (8-n) (\mathbf{26},\mathbf{1}) + (n+4) (\mathbf{1},\mathbf{11}) + \textstyle{\frac{n}{2}} (\mathbf{1},\mathbf{32}); &&n=0,\ldots,8, \nn\\
                              &&&\mathrm{(e) } 12 (\mathbf{1},\mathbf{11}) + \textstyle{\frac{9}{2}} (\mathbf{1},\mathbf{32}), \nn\\
                              &&&\mathrm{(f) } (\mathbf{26},\mathbf{1}) + 8 (\mathbf{1},\mathbf{11}) + \textstyle{\frac{5}{2}} (\mathbf{1},\mathbf{32}),\nn\\
                              &&&\mathrm{(g) } (\mathbf{26},\mathbf{1}) + (\mathbf{1},\mathbf{55}) + 9 (\mathbf{1},\mathbf{11}) + \textstyle{\frac{9}{2}} (\mathbf{1},\mathbf{32}) ,\nn\\
                              &&&\mathrm{(h) } (n+1) (\mathbf{26},\mathbf{1}) + (3-n) (\mathbf{1},\mathbf{55})\nn\\
                              &&& + (3n+2) (\mathbf{1},\mathbf{11}) + 4 (\mathbf{1},\mathbf{32}); &&n=0,\ldots,2, \nn\\
                              &&&\mathrm{(i) } 9 (\mathbf{26},\mathbf{1}) + 6 (\mathbf{1},\mathbf{11}) + \textstyle{\frac{3}{2}} (\mathbf{1},\mathbf{32}),\nn\\
                              &&&\mathrm{(j) } (3-n) (\mathbf{52},\mathbf{1}) + (n+6) (\mathbf{26},\mathbf{1})\nn\\
                              &&& + (n+3) (\mathbf{1},\mathbf{11}) + \textstyle{\frac{n}{2}} (\mathbf{1},\mathbf{32}); &&n=0,\ldots,2, \nn\\
                              &&&\mathrm{(k) } 6 (\mathbf{52},\mathbf{1}) + 3 (\mathbf{1},\mathbf{11}).\\
&\!\!\!\!\!\!\!\!\!\!\!\!\!\!\!\! N=10: && \mathrm{(d) } (8-n) (\mathbf{26},\mathbf{1}) + (n+3) (\mathbf{1},\mathbf{11}) \nn\\
                                        &&&+ (n+1) (\mathbf{1},\mathbf{16}); &&n=0,\ldots,8, \nn\\
                              &&&\mathrm{(e) } 4 (\mathbf{1},\mathbf{45}) + 2 (\mathbf{1},\mathbf{10}) + 8 (\mathbf{1},\mathbf{16}), \nn\\
                              &&&\mathrm{(f) } (\mathbf{26},\mathbf{1}) + 7 (\mathbf{1},\mathbf{10}) + 5 (\mathbf{1},\mathbf{16}),\nn\\
                              &&&\mathrm{(g) } (\mathbf{26},\mathbf{1}) + (\mathbf{1},\mathbf{45}) + 9 (\mathbf{1},\mathbf{10}) + 9 (\mathbf{1},\mathbf{16}) ,\nn\\
                              &&&\mathrm{(h) } (n+1) (\mathbf{26},\mathbf{1}) + (3-n) (\mathbf{1},\mathbf{45}) \nn\\
                              &&& + (2n+4) (\mathbf{1},\mathbf{10}) + 8 (\mathbf{1},\mathbf{16}), \nn\\
                              &&&\mathrm{(i) } 4 (\mathbf{26},\mathbf{1}) + 10 (\mathbf{1},\mathbf{10}) + 8 (\mathbf{1},\mathbf{16}),\nn\\
                              &&&\mathrm{(j) } 9 (\mathbf{26},\mathbf{1}) + 5 (\mathbf{1},\mathbf{10}) + 3 (\mathbf{1},\mathbf{16}),\nn\\
                              &&&\mathrm{(k) } (3-n) (\mathbf{52},\mathbf{1}) + (n+6) (\mathbf{26},\mathbf{1})\nn\\
                              &&& + (n+2) (\mathbf{1},\mathbf{10}) + n (\mathbf{1},\mathbf{16}); &&n=0,\ldots,2, \nn\\
                              &&&\mathrm{(l) } 6 (\mathbf{52},\mathbf{1}) + 2 (\mathbf{1},\mathbf{10}).
\end{align}

\item $F_4 \times \mathrm{USp}(2N)$:
\begin{align}
&\!\!\!\!\!\!\!\!\!\!\!\!\!\!\!\! 4 \leqslant N \leqslant 12: && \mathrm{(a) } (\mathbf{26},\mathbf{1}) + (24-2N) (\mathbf{1},\mathbf{2N}) \nn\\
                                                              &&&+ 2 (\mathbf{1},\mathbf{N(2N-1)-1}). \\
&\!\!\!\!\!\!\!\!\!\!\!\!\!\!\!\! N=6: && \mathrm{(b) }  2 (\mathbf{26},\mathbf{1}) + 4 (\mathbf{1},\mathbf{12}) + 4 (\mathbf{1},\mathbf{65}) , \nn\\
                                       &&&\mathrm{(c) }  3 (\mathbf{26},\mathbf{1}) + 8 (\mathbf{1},\mathbf{12}) + 3 (\mathbf{1},\mathbf{65}) . \\
&\!\!\!\!\!\!\!\!\!\!\!\!\!\!\!\! N=5: && \mathrm{(b) }  (6-n) (\mathbf{26},\mathbf{1}) + (18-2n) (\mathbf{1},\mathbf{10})+ n (\mathbf{1},\mathbf{44}); && n=0,\ldots,6.\\
&\!\!\!\!\!\!\!\!\!\!\!\!\!\!\!\! N=4: && \mathrm{(b) } (6-n) (\mathbf{26},\mathbf{1}) + 16 (\mathbf{1},\mathbf{8}) + n (\mathbf{1},\mathbf{27}); && n=0,\ldots,3, \nn\\
                                       &&&\mathrm{(c) } 3 (\mathbf{26},\mathbf{1}) + (\mathbf{1},\mathbf{36}) + 8 (\mathbf{1},\mathbf{27}) .
\end{align}
\end{enumerate}

Before concluding this section we note that, although we have not been able to make a thorough search for anomaly-free models when the gauge group is a product of two classical groups, a non-systematic search did not reveal any interesting models apart from those reported by Schwarz in \cite{Schwarz:1995zw} and some models related to them by Higgsing. For the sake of completeness, we list the basic models below.
\begin{enumerate}
\item $\mathrm{SU}(N) \times \mathrm{SU}(N)$. There exists the infinite class of models
\begin{align}
2 (\mathbf{N},\mathbf{N}).
\end{align}
\item $\mathrm{SO}(N+8) \times \mathrm{USp}(2N)$ with $0 \leqslant N \leqslant 24$. There exist the well-known small-instanton models
\begin{align}
\textstyle{\frac{1}{2}} (\mathbf{N+8},\mathbf{2N}) + \textstyle{\frac{24-N}{2}} (\mathbf{1},\mathbf{2N}) + (\mathbf{1},\mathbf{N(2N-1)-1}).
\end{align}
\item $\mathrm{SO}(2N+8) \times \mathrm{USp}(2N)$. There exists the infinite class of models
\begin{align}
(\mathbf{2N+8},\mathbf{2N}).
\end{align}
\end{enumerate}
The reader is referred to \cite{Schwarz:1995zw,Witten:1995gx} for more details on these models.

\section{Anomaly-free Gauged Supergravities}
\label{sec-5-6} 

In this section, we continue our search, turning to the case of gauged supergravities where the R-symmetry group or a $\mathrm{U}(1)_R$ subgroup thereof is gauged. The search for such models is of considerable interest due to the fact that these theories can spontaneously compactify on $\MBBR^4 \times \mathbf{S}^2$ through a magnetic monopole background, leading to four-dimensional theories. In the case where the magnetic monopole is embedded in the R-symmetry group, stability \cite{Randjbar-Daemi:1983bw} of the compactification is ensured and the 4D theory is vectorlike. However, under certain conditions, it is also possible to embed the monopole in one of the other gauge group factors and obtain a chiral 4D spectrum. The aforementioned facts, as well as other interesting properties of the gauged models, provide enough motivation for looking for more consistent theories of this type. In fact, it is the search for anomaly-free gauged supergravities that motivated the work presented in this Chapter: given the fact that there is no known construction of such theories following from standard string/M-theory compactifications, the only way to identify consistent theories of this type is to directly solve the anomaly cancellation conditions.

The search for the gauged theories can be carried out in the same manner as before, this time including an extra $\mathrm{USp}(2)_R$ or $\mathrm{U}(1)_R$ factor in the gauge group. So, the gauge group is now $\MG_s \times \MG_r$ and the new conditions that have to be satisfied are Eq. (\ref{e-5-2-31}) for the $\MG_r$ factor plus Eq. (\ref{e-5-2-32}) for the gauginos that transform nontrivially under both $\MG_s$ and $\MG_r$; using (\ref{e-5-2-6}) and (\ref{e-5-2-7}), we easily see that the first of these conditions is identically satisfied, leaving the second condition as the only non-trivial one. However, this last condition amounts to a set of strict equalities and, moreover, regarding the equalities involving $\MG_r$, the fact that the representations (or charges) of the fermions under this factor are fixed leaves little freedom for satisfying these constraints. So, one is led to expect that the gauged anomaly-free models will be very few.

The results of our search show that this is indeed the case. In the case of a gauge group of the type $\MG_1 \times \MG_r$, there is one equality constraint of the type (\ref{e-5-2-32}). In our search, we have not found any model solving the anomaly cancellation conditions. Passing to the case of a gauge group of the type $\MG_1 \times \MG_2 \times \MG_r$, there are three equality constraints of the type (\ref{e-5-2-32}) which are expected to seriously restrict the number of possible solutions. For the case where the whole $\mathrm{USp}(2)_R$ is gauged, we have found no solution. For the case where a $\mathrm{U}(1)_R$ subgroup is gauged, we have found the following models.
\begin{enumerate}
\item $E_7 \times E_6 \times \mathrm{U}(1)_R$ with the hypermultiplets transforming in
\begin{align}
\label{e-5-6-1}
\textstyle{\frac{1}{2}} (\mathbf{912},\mathbf{1}),
\end{align}
without singlet hypermultiplets. This is a well-known model, first found by Randjbar-Daemi, Salam, Sezgin and Strathdee in 1985. The important property of this model is that, besides the compactification with the monopole embedded in $\mathrm{U}(1)_R$, it also admits a compactification with the monopole embedded in the ``hidden'' $E_6$, leading to an $\mathrm{SO}(10) \times \mathrm{SU}(2)_{KK}$ four-dimensional theory with chiral fermions. 

\item $E_7 \times G_2 \times \mathrm{U}(1)_R$ with the hypermultiplets transforming in
\begin{align}
\label{e-5-6-2}
\textstyle{\frac{1}{2}} (\mathbf{56},\mathbf{14}),
\end{align}
again without singlets. This is a recently-found model, whose existence was reported in \cite{Avramis:2005qt} where the absence of both local and global anomalies was analytically demonstrated and various properties of the resulting supergravity theory were investigated.

\item $F_4 \times \mathrm{Sp}(9) \times \mathrm{U}(1)_R$ with the hypermultiplets transforming in
\begin{align}
\label{e-5-6-3}
\textstyle{\frac{1}{2}} (\mathbf{52},\mathbf{18}),
\end{align}
again without singlets. This is a new model, first reported in \cite{Avramis:2005hc}. 
\end{enumerate}

The structure of the models found is truly very interesting. In particular, they have the shared features that (i) the hypermultiplets transform in non-trivial representations (and, in the latter two cases, in product representations), (ii) there are no singlet hypermultiplets and (iii) the representations involve half-hypermultiplets. Moreover, as mentioned before, the cancellation of anomalies in these models is very delicate as can be verified by the explicit calculations of \cite{Randjbar-Daemi:1985wc} and \cite{Avramis:2005qt} for the former two. These facts might serve as indications that these gauged models are somehow related to critical string theory or M-theory by means of some mechanism. However, although some progress has been made \cite{Cvetic:2003xr} regarding the archetypal Salam-Sezgin model, the origin of the models considered here remains mysterious up to date.

An investigation of the 4D compactifications of the gauged theories, including discussions of the fermion spectra, supersymmetry breaking and classical stability, will be presented in Chapter \ref{chap-7}. \\

\begin{small}
As mentioned in Section \ref{sec-5-4}, in the gauged case we have also allowed for an abelian gauge group factor $\MG_a$ that does not act on hypermultiplets. In the presence of such a factor, the gauge group includes ``drone'' $\mathrm{U}(1)$'s under which all hypermultiplets and gauginos are singlets. Although this possibility leads to new anomaly-free models, the usual viewpoint is that turning on a large number of  $\mathrm{U}(1)$'s so that the gravitational and R-symmetry anomalies are tuned to give a factorizable polynomial is quite \emph{ad hoc} and so these models are considered to be less important than the previous ones. Nevertheless, for reasons of completeness, we will list these models, for the case where the factor $\MG_s$ is simple and the number of $\mathrm{U}(1)$'s is at most 50. The models found are the following.
\begin{enumerate}
\item $E_8 \times \mathrm{U}(1)^3 \times \mathrm{U}(1)_R$:
\begin{align}
\label{e-4-4}
2 \cdot \mathbf{248}.
\end{align}
\item $E_7 \times \mathrm{U}(1)^{14} \times \mathrm{USp}(2)_R$:
\begin{align}
\label{e-4-5}
2 \cdot \mathbf{133} + 2 \cdot \mathbf{56}.
\end{align}
\item $E_7 \times \mathrm{U}(1)^M \times \mathrm{U}(1)_R$:
\begin{align}
& M=14: && 7 \cdot \mathbf{56}. \\
& M=18: && \mathbf{133} + \textstyle{\frac{9}{2}} \cdot \mathbf{56}. \\
& M=22: && 2 \cdot \mathbf{133} + 2 \cdot \mathbf{56}.
\end{align}
The $E_7 \times \mathrm{U}(1)^{14} \times \mathrm{U}(1)_R$ model has no singlets and is related to the $E_7 \times G_2 \times \mathrm{U}(1)_R$ model of (\ref{e-5-6-2}) in the sense that the $G_2$ factor in the latter has been replaced by 14 $\mathrm{U}(1)$'s. The existence of the $E_7 \times \mathrm{U}(1)^{22} \times \mathrm{U}(1)_R$ model was first pointed out in the footnote of \cite{Randjbar-Daemi:1985wc}.
\item $E_6 \times \mathrm{U}(1)^{27} \times \mathrm{USp}(2)_R$:
\begin{align}
4 \cdot \mathbf{78}.
\end{align}
\item $E_6 \times \mathrm{U}(1)^M \times \mathrm{U}(1)_R$:
\begin{align}
& M=21: && 12 \cdot \mathbf{27}, \\
& M=29: && 2 \cdot \mathbf{78} + 6 \cdot \mathbf{27}, \\
& M=37: && 4 \cdot \mathbf{78}.
\end{align}
\item $\mathrm{SU}(N) \times \mathrm{U}(1)^M \times \mathrm{U}(1)_R$:
\begin{align}
& N=8, M=42: && \mathbf{63} + 8 \cdot \mathbf{28}. \\
& N=7, M=45: && \mathbf{48} + 8 \cdot \mathbf{7} + 8 \cdot \mathbf{21}. \\
& N=6, M=46: && \mathbf{35} + 16 \cdot \mathbf{6} + 8 \cdot \mathbf{15}.
\end{align}
\item $\mathrm{SU}(N) \times \mathrm{U}(1)^M \times \mathrm{U}(1)_R$:
\begin{align}
& 6 \leqslant N \leqslant 12, M = 8 + 12 N - N^2: && ( 48 - 4N ) \cdot \mathbf{N} + 6 \cdot \textstyle{\mathbf{\frac{N(N-1)}{2}}}. \\
&N=6,M=16: && 28 \cdot \mathbf{6} + 8 \cdot \mathbf{15}.
\end{align}
The first series of models have the same field content as the $\mathrm{SU}(N)$(a) ungauged theories found in \S\ref{sec-3-1} for $n=3$.
\item $\mathrm{SO}(N) \times \mathrm{U}(1)^M \times \mathrm{USp}(2)_R$:
\begin{align}
&N=10,M=12: && 12 \cdot \mathbf{11} + 10 \cdot \mathbf{16}.
\end{align}
\item $\mathrm{SO}(N) \times \mathrm{U}(1)^M \times \mathrm{U}(1)_R$:
\begin{align}
&\!\!\!\!\!\!\!\!\!\!\!\!\!\!\!\! N=16,M=3: && 2 \cdot \mathbf{120} + \mathbf{128}. \\
&\!\!\!\!\!\!\!\!\!\!\!\!\!\!\!\! N=15,M=10: && 2 \cdot \mathbf{105} + \mathbf{15} + \mathbf{128}. \\
&\!\!\!\!\!\!\!\!\!\!\!\!\!\!\!\! N=14,M=16: && 2 \cdot \mathbf{91} + 2 \cdot \mathbf{14} + 2 \cdot \mathbf{64}. \\
&\!\!\!\!\!\!\!\!\!\!\!\!\!\!\!\! N=13,M=21: && 2 \cdot \mathbf{78} + 3 \cdot \mathbf{13} + 2 \cdot \mathbf{64}. \\
&\!\!\!\!\!\!\!\!\!\!\!\!\!\!\!\! N=12,M=17+4n: && n \cdot \mathbf{66} + (14 - 5n) \cdot \mathbf{12} + \textstyle{\frac{10-n}{2}} \cdot \mathbf{32}; && n=0,\ldots,2. \\
&\!\!\!\!\!\!\!\!\!\!\!\!\!\!\!\! N=11,M=20+4n: && n \cdot \mathbf{55} + (13-4n) \cdot \mathbf{11} + \textstyle{\frac{10-n}{2}} \cdot                        \mathbf{32}; && n=0,\ldots,3, \nn\\
&\!\!\!\!\!\!\!\!\!\!\!\!\!\!\!\! N=11,M=36: && 12 \cdot \mathbf{11} + \textstyle{\frac{9}{2}} \cdot \mathbf{32}. \\
&\!\!\!\!\!\!\!\!\!\!\!\!\!\!\!\! N=10,M=22+4n: && n \cdot \mathbf{45} + (12-3n) \cdot \mathbf{10} + (10-n) \cdot \mathbf{16}; && n=0,\ldots,4.
\end{align}
\item $\mathrm{USp}(2N) \times \mathrm{U}(1)^M \times \mathrm{U}(1)_R$:
\begin{align}
&N=12,M=19: && 2 \cdot \mathbf{275}. \\
&N=11,M=42: && 2 \cdot \mathbf{22} + \mathbf{15} + \mathbf{230}. \\
&N=6,M=13: && 5 \cdot \mathbf{65}, \nn\\
&N=6,M=45: && 4 \cdot \mathbf{12} + 4 \cdot \mathbf{65}. \\
&N=5,M=24: && 8 \cdot \mathbf{10} + 5 \cdot \mathbf{44}.
\end{align}
\end{enumerate}
We see thus that allowing for the possibility of $\mathrm{U}(1)$'s acting trivially on the hypermultiplets, we obtain many anomaly-free gauged models, some of which are extensions of the ungauged models of \S\ref{sec-5-5}. Increasing the number of $\mathrm{U}(1)$'s leads to numerous other models. However, as stressed above, these models are considered of limited interest.
\end{small}

\section{Anomaly Cancellation in $D=7$, $N=2$ Supergravity on $\mathbf{S}^1 / \mathbb{Z}_2$}
\label{sec-6-1}

As the issue of Green-Schwarz anomaly cancellation in 10D supergravity can be extended to 6D supergravity, so can the Ho\v rava-Witten construction of anomaly-free theories living on the 10D fixed planes of $D=11$ supergravity on $\mathbf{S}^1 / \mathbb{Z}_2$ be extended to theories living on the 6D fixed planes of $D=7$, $N=2$ supergravity on $\mathbf{S}^1 / \mathbb{Z}_2$. The motivation of finding such theories is that the stringent anomaly cancellation conditions of six-dimensional theories may provide a natural way to specify the matter content on on the brane, in sharp contrast to the much-studied case of four-dimensional brane worlds in five dimensions. This scenario has been explored in \cite{Gherghetta:2002nq,Gherghetta:2002xf} for compactifications of the 3--form version of gauged $D=7$, $N=2$ supergravity without vector multiplets on $\mathbf{S}^1 / \mathbb{Z}_2$. In this section, we examine anomaly cancellation in another related context, namely for compactifications of the 2--form version of gauged $D=7$, $N=2$ supergravity with $N_V$ vector multiplets on $\mathbf{S}^1 / \mathbb{Z}_2$, and we state the anomaly cancellation conditions in this new context, based on \cite{Avramis:2004cn}. Since that paper has significant overlap with certain of the preceding sections, we will here restrict to a sketchy discussion, emphasizing only the issue of anomaly cancellation.

\subsection{Basics of $D=7$, $N=2$ supergravity}
\label{sec-6-1-1}

The minimal supersymmetry algebra in seven dimensions is the $D=7$, $N=2$ algebra, whose representations (see \S\ref{sec-2-1-2})are given by the following multiplets
\bea
\label{e-6-1-1}
\text{Gravity multiplet} \quad&:&\quad ( g_{MN}, A_{MNP}\textrm{ or }B_{MN} , \RA^{\phantom{M} A}_{M \phantom{A} B}, \phi, \psi^A_M, \chi^A ), \nn\\
\text{Vector multiplet} \quad&:&\quad ( \RA_M, \upvarphi^A_{\phantom{A} B}, \uplambda^A ), 
\eea
where all spinors are symplectic Majorana and the index $A=1,2$ takes values in the fundamental representation of $\mathrm{USp}(2)$. Here, we note that there are two alternative versions of the theory, one involving a 3--form potential $A_3$ and one involving a 2--form potential $B_2$. 

A general $D=7$, $N=2$ supergravity coupled to matter is constructed by combining the gravity multiplet with $N_V$ vector multiplets. The $3 N_V$ scalars in the vector multiplets parameterize the coset space
\be
\label{e-6-1-2}
\MM = \frac{\mathrm{SO}(N_V,3)}{\mathrm{SO}(N_V) \times \mathrm{USp}(2)}
\ee
while the $N$ gauginos transform in the fundamental of $\mathrm{SO}(N)$. The field content of the resulting theory is thus given by the reducible multiplet
\be
\label{e-6-1-3}
( g_{MN}, A_{MNP}\textrm{ or }B_{MN} ,  \RA^{\RI}_M, \upvarphi^\upalpha, \phi, \psi^A_M, \chi^A , \uplambda^{a A} )
\ee
where $a=1,\ldots,N_V$ takes values in the fundamental of $\mathrm{SO}(N_V)$ and labels the individual gauginos, $\RI=1,\ldots,N_V+3$ takes values in the fundamental of $\mathrm{SO}(N_V,3)$ and labels the full set of vector fields of the theory and $\upalpha=1,\ldots,3 N_V$ is an index for the scalar coset manifold. To construct gauged theories, one gauges a subgroup $G$ of the $\mathrm{SO}(N_V,3)$ isometry group of $\MM$, whose dimension must equal the number of vectors of the theory, $\dim G = N_V+3$, and whose structure constants must be subject to a certain restriction \cite{Bergshoeff:1985mr}. Such a theory has a complicated scalar potential which, nevertheless, is easily shown to admit a 7D Minkowski vacuum.

The 3--form version of the theory has been first constructed in \cite{Townsend:1983kk} (see also \cite{Giani:1984dw,Mezincescu:1984ta,Park:1988id}) while the 2--form version of the theory has been constructed in \cite{Bergshoeff:1985mr}. In our setup, our starting point is the 3--form version of the theory, defined on a 7D manifold with boundary. Starting from this theory, we apply a duality transformation of the standard type \cite{Cremmer:1979up,Nicolai:1981td} to pass to the 2--form version. Although normally this would yield just the theory of \cite{Bergshoeff:1985mr}, the presence of boundaries implies that we must, in addition, retain a certain boundary term that emerges during the duality transformation; this will eventually give rise to a Green-Schwarz anomaly cancellation term for the 6D supergravity living on the boundary of the 7D space. 

\subsection{Compactification on $\mathbf{S}^1 / \mathbb{Z}_2$ and anomalies}
\label{sec-6-2}

To obtain a chiral 6D theory from $D=7$, $N=2$ supergravity, we will consider compactification of the $x_7$ coordinate on the $\mathbf{S}^1 / \mathbb{Z}_2$ orbifold, in the spirit of Ho\v rava and Witten. The $\mathbb{Z}_2$ action is as usual $x_7\to -x_7$ and the two fixed points are at $x_7=0$ and $x_7=\pi R$. To find the spectrum of the resulting theory, we need to discuss the reduction of the various fields on $\mathbf{S}^1$, make appropriate parity assignments to the fields consistent with the 7D Lagrangian and supersymmetry transformation rules, and keep the fields that are even under $\mathbb{Z}_2$. As we will see, this will yield an anomalous spectrum, which necessitates the addition of 6D boundary multiplets in order for the anomalies of the theory to cancel.

The detailed procedure of the reduction of the fields on $\mathbf{S}^1$ was presented in \cite{Giani:1984dw}, whose results we quote here. The 7D fields decompose a la Kaluza-Klein according to
\bea
\label{e-6-2-4}
&\hat{g}_{MN} \to ( g_{\mu\nu} , A_\mu , \tilde{\xi} ) , \qquad \hat{B}_{MN} \to ( B_{\mu\nu} , B_\mu ) , \qquad \hat{A}_M^{\RI} \to ( A_\mu^{\RI} , A^{\RI} ) ,\qquad \hat{\upvarphi}^\upalpha,\hat{\phi} \to \upvarphi^\upalpha, \tilde{\phi} \nn\\ 
&\hat{\psi}_M \to \psi_\mu, \tilde{\uppsi} ,\qquad \hat{\chi} \to \tilde{\chi} ,\qquad \hat{\uplambda}^a \to \uplambda^a. 
\eea
where here we will use a hat to denote 7D fields. The Kaluza-Klein ansatz for the 7D metric reads
\be
\label{e-6-2-5}
d s_7^2 = e^{ - \tilde{\xi} / \sqrt{5} } d s_6^2 + e^{ 4\tilde{\xi} / \sqrt{5} } \left( d x_7 + A_\mu dx^\mu
\right)^2, 
\ee
the various 7D spinors reduce according to
\bea
\label{e-6-2-6}
&\hat{\psi}_{\mu} = e^{ - \tilde{\xi} / 4 \sqrt{5} } \left[ \psi_{\mu} - \frac{1}{2 \sqrt{5}} \left( \Gamma_\mu \Gamma_7 + 4 \sqrt{2} A_\mu \right) \tilde{\uppsi} \right], \nn\\
&\hat{\psi}_7 = \frac{2}{\sqrt{5}} e^{ \tilde{\xi} / 4\sqrt{5} } \tilde{\uppsi} ,\qquad \hat{\chi} = e^{ \tilde{\xi} / 4\sqrt{5}} \tilde{\chi} ,\qquad \hat{\uplambda}_a = e^{ \tilde{\xi} / 4\sqrt{5} } \uplambda_a, 
\eea
while antisymmetric tensors and scalars reduce in the usual way. The tilded 6D fields $( \tilde{\uppsi} , \tilde{\chi} )$ and $( \tilde{\xi} , \tilde{\phi} )$ are conveniently traded for the fields $( \uppsi_i , \chi_i )$ and $( \xi , \phi )$ defined by
\bea
\label{e-6-2-7}
&\tilde{\uppsi} = \frac{1}{\sqrt{5}} \left( 2 \uppsi - \Gamma_7 \chi \right) ,\qquad \tilde{\chi} = \frac{1}{\sqrt{5}} \left( 2 \chi + \Gamma_7 \uppsi \right), \nn\\
&\tilde{\xi} = \frac{1}{\sqrt{5}} \left( 2 \xi - \phi \right) ,\qquad \tilde{\phi} = \frac{1}{\sqrt{5}} \left( 2 \phi
+ \xi \right). 
\eea

We next turn to the $\mathbb{Z}_2$ parity assignments. As in the HW case, these are found by demanding that the 7D Lagrangian and supersymmetry transformation rules stay invariant under $x_7 \to -x_7$. By the same considerations as in \cite{Bagger:2002rw,Gherghetta:2002nq}, it is easy to see that, after projecting out the $\mathbb{Z}_2$-odd fields, the surviving bosonic fields are
\be
\label{e-6-2-8}
g_{\mu\nu} , B_{\mu\nu} , \RA^{\RI} , \upvarphi^\upalpha , \phi , \xi,
\ee
while the spinors are subject to the chirality constraints
\be
\label{e-6-2-9}
\Gamma_7 \left( \psi_\mu , \epsilon \right) = - \left( \psi_\mu , \epsilon \right) ,\qquad \Gamma_7 \left( \uppsi, \chi , \uplambda^a \right) = \left( \uppsi, \chi , \uplambda^a \right).
\ee
The surviving fields can be then arranged into multiplets of the $D=6$, $N=2$ supersymmetry algebra. To do the decomposition, we split $B_{\mu\nu}$ into a self-dual and an anti-self-dual part as $B_{\mu\nu} = B^+_{\mu\nu} + B^-_{\mu\nu}$, we group all scalars except from $\phi$ as $(A^I,\phi^\alpha,\xi)$ and we group all spin--$1/2$ fermions except from $\chi^+$ as $(\uppsi^+,\uplambda^{a+})$. Then, the fields surviving the $\mathbb{Z}_2$ projection can be arranged into $D=6$, $N=2$ multiplets as follows
\bea
\label{e-6-2-10}
\text{Gravity multiplet} \quad&:&\quad ( g_{\mu\nu} , B^+_{\mu\nu} , \psi^{A-}_\mu ),\nn\\
\text{Tensor multiplet} \quad&:&\quad ( B^-_{\mu\nu} , \phi , \chi^{A+} ), \nn\\
\text{$N_V+1$ hypermultiplets} \quad&:&\quad \left( (\RA^{\RI},\upvarphi^\upalpha,\xi), ((\uppsi^{A+},\uplambda^{aA+}) \right). 
\eea
where the $\pm$ superscripts on spinors indicate 6D chirality. 

The presence of chiral fermions on the fixed planes inevitably introduces anomalies which, since all vector fields are projected out, are purely gravitational. To arrive at a consistent theory, we have to follow the Ho\v rava-Witten recipe by adding boundary fields whose contribution to the anomalies will cancel those of the bulk fields by a Green-Schwarz--type mechanism. In contrast to the HW case where only one type of boundary multiplet was available, here the available types of boundary multiplets are vector multiplets, tensor multiplets and hypermultiplets, whose field content is given by
\bea
\label{e-6-2-12}
\text{Tensor multiplet} \quad&:&\quad ( B^{\prime -}_{\mu\nu} , \phi^\prime , \chi^{\prime A+} ), \nn\\
\text{Vector multiplet} \quad&:&\quad ( A_\mu , \lambda^{-} ), \nn\\
\text{Hypermultiplet}   \quad&:&\quad ( 4 \varphi , 2 \psi ).
\eea
Coupling these boundary multiplets to the bulk supergravity and demanding local supersymmetry introduces various bulk-boundary interaction terms and requires certain modifications of the boundary conditions of the bulk fields. In particular, local supersymmetry of the vector multiplet action requires a modified boundary condition on the bulk 3--form field strength $G_3$ which, inserted into the surface term remaining from the dualization, yields the Green-Schwarz term of the theory. Also, local supersymmetry of the hypermultiplet action requires a modified boundary condition on the bulk Yang-Mills field strengths which in turn induces the quaternionic structure on the manifold spanned by the 6D hyperscalars. The relevant analysis has been carried out in detail in \cite{Avramis:2004cn} and will not be reproduced here. We will only concern ourselves with the anomaly constraints of these bulk-boundary theories, to be described below.

\subsection{Anomaly cancellation}
\label{sec-6-4}

Given the spectrum of the theories living on the orbifold fixed planes, we are ready to discuss the issue of anomaly cancellation. Starting from local anomalies, the bulk theory reduced on $\mathbf{S}^1/\mathbb{Z}_2$ has gravitational anomalies as an obvious consequence of its chiral spectrum and, as in the HW case, these anomalies are equally distributed in the two fixed planes. Noting that the contributions coming from the bulk 2--form $B_2$ cancel each other, the bulk gravitational anomaly on a given fixed plane is half of that corresponding to a negative-chirality gravitino and $N_V+2$ positive-chirality spinors. Using the 6D anomaly polynomials (\ref{e-3-4-8}), we find
\be
\label{e-6-4-1}
I^{bulk}_8 (R) = \frac{1}{2} \left[ - I^{3/2}_{8} (R) + (N_V+2) I^{1/2}_{8} (R) \right] = \frac{N_V - 243}{720} \tr R^4 + \frac{N_V + 45}{576} (\tr R^2)^2 . 
\ee
The inclusion of extra boundary multiplets, namely $n_T$ tensor multiplets, $n_V$ vector multiplets and $n_H$ hypermultiplets, contributes extra terms to the gravitational anomaly and introduces gauge and mixed anomalies. The gravitational contribution is given by (\ref{e-5-2-1}) minus the contribution of the 6D gravity multiplet, that is
\be
\label{e-6-4-2}
I^{bdy}_8 (R) = \frac{n_H - n_V + 29 n_T}{360} \tr R^4 + \frac{n_H - n_V - 7 n_T}{288} (\tr R^2)^2 .
\ee
Similarly, the gauge and mixed contributions are found by modifying (\ref{e-5-2-2}) and (\ref{e-5-2-3}) to take account of the fact that the bulk gravitino does not couple to the boundary $\mathrm{USp}(2)$ R-symmetry as in the purely 6D case. Using exactly the same conventions as in Section \ref{sec-5-2}, we find the gauge anomaly
\bea
\label{e-6-4-3}
I^{bdy}_8 (F) &=& - \frac{2}{3} \sum_x \bigl( \Tr F_x^4 - \sum_i n_{x,i} \tr_i F_x^4 \bigr) + 4 \sum_{x < y} \sum_{i,j} n_{xy,ij} \tr_i F_x^2 \tr_j F_y^2 \nn\\ && - \frac{2}{3} \bigl[ \tr' F_r^4 + ( \dim \MG_s + \dim \MG_a - n_T ) \tr F_r^4 \bigr] \nn\\ &&- 4 \Tr F_x^2 \tr' F_r^2 ,
\eea
and the mixed anomaly
\bea
\label{e-6-4-4}
I^{bdy}_8 (F,R) &=& \frac{1}{6} \tr R^2 \sum_x \bigl( \Tr F_x^2 - \sum_i n_{x,i} \tr_i F_x^2 \bigr) \nn\\ &&+ \frac{1}{6} \tr R^2 \bigl[ \tr' F_r^2 + ( \dim \MG_s + \dim \MG_a - n_T ) \tr F_r^2 \bigr].
\eea
Introducing the anomaly coefficients as in (\ref{e-5-2-4})--(\ref{e-5-2-6}) and defining
\bea
\label{e-6-4-5}
A_x \equiv a_{x,\MA} - \sum_i n_{x,i} a_{x,i}, &&
\nn\\
B_x \equiv b_{x,\MA} - \sum_i n_{x,i} b_{x,i}, &&\qquad B_r \equiv b'_r + ( \dim \MG_s + \dim \MG_a - n_T ) b_r , \nn\\
C_x \equiv c_{x,\MA} - \sum_i n_{x,i} c_{x,i}, &&\qquad C_r \equiv c'_r + \dim \MG_s + \dim \MG_a - n_T, \nn\\ 
C_{xy} \equiv \sum_{i,j} n_{xy,ij} c_{x,i} c_{y,j}, &&\qquad C_{x,r} \equiv - c_{x,\MA},
\eea
we can write (\ref{e-6-4-3}) and (\ref{e-6-4-4}) in the compact form
\be
\label{e-6-4-6}
I^{bdy}_8(F) = - \frac{2}{3} \sum_x A_x \tr F_x^4 - \frac{2}{3} \sum_A B_X (\tr F_X^2)^2 + 4 \sum_{X < Y} C_{XY} \tr F_X^2 \tr F_Y^2,
\ee
and
\be
\label{e-6-4-7}
I^{bdy}_8(F,R) = \frac{1}{6} \tr R^2 \sum_X C_X \tr F_X^2.
\ee
Combining all contributions, we finally find the anomaly polynomial
\bea
\label{e-6-4-8}
I_8 &=& \frac{2 n_H - 2 n_V + 58 n_T + N_V - 243}{720} \tr R^4 + \frac{2 n_H - 2 n_V - 14 n_T + N_V + 45}{576} (\tr R^2)^2 \nn\\ &&+ \frac{1}{6} \tr R^2 \sum_X C_X \tr F_X^2 \nn\\ &&- \frac{2}{3} \sum_x A_x \tr F_x^4 - \frac{2}{3} \sum_A B_X (\tr F_X^2)^2 + 4 \sum_{X < Y} C_{XY} \tr F_X^2 \tr F_Y^2.
\eea
This time, the condition for the cancellation of the irreducible gravitational anomaly is given by
\be
\label{e-6-4-9}
2 n_H - 2 n_V + 58 n_T = 243 - N_V, 
\ee
and, since the LHS is an even integer, it implies that the presence of an odd number of bulk hypermultiplets is necessary for anomaly cancellation. The condition for the vanishing of irreducible gauge anomalies is again
\be
\label{e-6-4-10}
A_x = 0 ;\qquad \textrm{for all } x.
\ee
Provided that (\ref{e-6-4-9}) and (\ref{e-6-4-10}) hold, the anomaly polynomial reads
\be
\label{e-6-4-11}
I_8 = K (\tr R^2)^2 + \frac{1}{6} \tr R^2 \sum_X C_X \tr F_X^2 - \frac{2}{3} \sum_X B_X (\tr F_X^2)^2 + 4 \sum_{X < Y} C_{XY} \tr F_X^2 \tr F_Y^2,
\ee
this time with
\be
\label{e-6-4-12}
K = \frac{4 - n_T}{8}.
\ee
From now on, the analysis of the local anomaly cancellation conditions proceeds exactly as in \S\ref{sec-5-2-1}. To summarize, Green-Schwarz cancellation of local anomalies leads, for $n_T \ne 4$, to the conditions (\ref{e-6-4-9}), (\ref{e-6-4-10}), (\ref{e-5-2-31}) and (\ref{e-5-2-32}), while, in the special case $n_T=4$, the last two conditions are replaced by (\ref{e-5-2-36}).

Extend our considerations to take account of the presence of the two fixed planes and noting that the Green-Schwarz mechanism employed here involves a \emph{single} bulk $2$--form, we have to ensure (see e.g. \cite{Scrucca:2004jn}) that one of the two factors in the factorization equation is common to both planes, i.e. that we have $u^{(1) I} = u^{(2) I}$ or $\tilde{u}^{(1) I} = \tilde{u}^{(2) I}$. This condition obviously holds when the boundary matter and gauge groups are the same on both fixed planes, as in the HW model.

Finally, as in the purely 6D case, the theory may also include global gauge anomalies. For the case when the $\MG_s$ part of the gauge group includes a factor $\MG_x=G_2,\mathrm{SU}(3),\mathrm{SU}(2)$ in $\MG_s$, the conditions for the absence of global gauge anomalies are given by (\ref{e-5-2-37}). For the case where the whole $\mathrm{USp}(2)$ 6D R-symmetry is gauged, we also have the condition
\begin{align}
\label{e-6-4-13}
&\MG_r=\mathrm{USp}(2) : && 5 + \dim \MG_s + \dim \MG_a - n_T = 0 \mod 6.
\end{align}

\section{Discussion}
\label{sec-5-7} 

In this chapter, we have presented a detailed review of $D=6$, $N=2$ supergravity, we have stated the most general anomaly cancellation conditions in the absence of anomalous $\mathrm{U}(1)$'s and we have made a thorough search for anomaly-free models, within the limits set by certain restrictions on the possible gauge groups and their representations. The search was made for both the ungauged and gauged cases and all CPT-invariant hypermultiplet representations satisfying the anomaly cancellation conditions have been enumerated. We have also presented the anomaly cancellation conditions for a certain Ho\v rava-Witten--type compactification of $D=7$, $N=2$ supergravity on $\mathbf{S}^1 / \mathbb{Z}_2$, but we have not performed a search for anomaly-free models of this type.

Our results are summarized as follows. In the ungauged case, where there exist numerous solutions to the anomaly cancellation conditions, we have recovered most of the known models that have already been identified and constructed via various methods in the literature, plus a series of closely related models. We have also
found some models that have not been, to our knowledge, previously identified. Classifying these models and tracing their possible origin is outside the scope of the present work.

In the gauged case, where the anomaly cancellation conditions are far more restrictive than in the ungauged case, our search revealed the existence of just three models. The first is the well-known $E_7 \times E_6 \times \textrm{U}(1)_R$ model of \cite{Randjbar-Daemi:1985wc}, the second is an $E_7 \times G_2
\times \textrm{U}(1)_R$ model recently reported in \cite{Avramis:2005qt} and the third is an $F_4 \times
\textrm{Sp}(9) \times \textrm{U}(1)_R$ model reported in \cite{Avramis:2005hc}. All three models have an intriguing structure in the sense that the hypermultiplets transform in a single ``unusual'' representation of the gauge group with no singlets and, moreover, they satisfy the anomaly cancellation conditions in a ``miraculous'' manner. On the physical side, these models have very interesting properties, the most important one being the possibility of compactification to four dimensions through a monopole background with self-tuning of the cosmological constant. These compactifications however reveal (see Chapter 7) some phenomenological problems with these theories, for example the fact that the demand for stability of these compactifications leads to unrealistic spectra. Allowing for the presence of extra ``drone'' $\textrm{U}(1)$ factors, we have identified many more anomaly-free gauged models. However, the presence of the extra $\textrm{U}(1)$'s renders these models less elegant than those described earlier.

The search presented here can be extended towards several directions, the focus being on finding new consistent gauged theories. For instance, one may consider gauge groups that contain three or more simple factors. Also, one may consider theories with more than one tensor multiplet, where there exists the generalized Green-Schwarz mechanism that allows anomaly freedom under weaker constraints. One could finally consider adding extra $\textrm{U}(1)$ factors that act non-trivially on the hypermultiplets but, unless there is a physical principle that determines the $\textrm{U}(1)$ charges in some way, this is a very complicated task. We hope that the work presented here will motivate further work along these lines.
\chapter[Magnetic Compactifications]{Magnetic Compactifications and Supersymmetry Breaking}
\label{chap-7} 

The six-dimensional theories examined in Chapter \ref{chap-5} admit various types of compactifications to four dimensions. Considering the simplest compactifications possible, ungauged theories can be compactified on a torus leading to $D=4$, $N=2$ theories (they can also be "compactified" on a magnetized torus giving chiral theories with broken supersymmetry, but these solutions have gravitational and dilatonic tadpoles). Gauged theories on the other hand spontaneously compactify to 4D Minkowski space times a 2-sphere with a magnetic monopole background and this compactification is unique. The resulting theories may have vectorlike or chiral fermion spectra, tachyonic instabilites or absence thereof and partial or complete supersymmetry breaking. 

\section{Equations of Motion, Supersymmetry and the Scalar Potential}
\label{sec-7-1}

In order to seek specific solutions of the supergravity theories under consideration, we need to examine the Lagrangian and the equations of motion in more detail. To conform with standard conventions, we will use the dual Lagrangian to (\ref{e-5-1-63}) (where some dilaton factors have a reversed sign and the Green-Schwarz term is traded for a redefinition of the field-strength 3--form). After a further rescaling of the gauge fields, the bosonic part of the Lagrangian is written as
\bea
\label{e-7-1-1}
E^{-1} \ML &=& \frac{1}{4} R - \frac{1}{12} e^{2 \phi} G_{MNP} G^{MNP} - \frac{1}{4} \partial_M \phi \partial^M \phi - \frac{1}{4 g_X^2} e^{\phi} ( F^{\hat{I}}_{MN} F_{\hat{I}}^{MN} )_X \nn\\ 
&& - \frac{1}{2} g_{\alpha\beta}(\varphi) \MD_\mu \varphi^\alpha \MD^\mu \varphi^\beta - V(\varphi) .
\eea 
where $V(\varphi)$ is the hyperscalar potential which, for the $\mathrm{U}(1)_R$ gauging of interest, takes the particular form
\be
\label{e-7-1-2}
V(\varphi) = \frac{1}{8} e^{- \phi} [ g_x^2 ( C^{i I} C_{i I} )_x + g_1^2 C^i C_i ],
\ee
where the $\mathrm{USp}(2 n_H)$ and $\mathrm{U}(1)_R$ contributions are separated out, $x$ labels the various group factors in $\mathrm{USp}(2 n_H)$ and the ``prepotentials'' have the explicit form
\be
\label{e-7-1-3}
C^{i I} = \MA_\alpha^{\phantom{\alpha} i} (T^I \varphi)^\alpha ,\qquad C^{i} = \MA_\alpha^{\phantom{\alpha} i} (T^3 \varphi)^\alpha - \delta^{i3},
\ee
The equations of motion resulting from the Lagrangian (\ref{e-7-1-1}) can be written in the form
\bea
\label{e-7-1-4}
&&\!\!\!\!\!\!\!\!\!\!\!\! R_{MN} = e^{2 \phi} G_{MPQ} G_N^{\phantom{N}PQ} + \frac{2}{g_X^2} e^\phi ( F^{\hat{I}}_{MP} F_{\hat{I} N}^{\phantom{\hat{I} N} P} )_X + \partial_M \phi \partial_N \phi + 2 g_{\alpha\beta} \partial_M \varphi^\alpha \partial_N \varphi^\beta - \frac{1}{2} g_{MN} D^2 \phi, \\
\label{e-7-1-5}
&&\!\!\!\!\!\!\!\!\!\!\!\! \partial_M (e e^{2\phi} G^{MNP}) = 0, \\
\label{e-7-1-6}
&&\!\!\!\!\!\!\!\!\!\!\!\! D^2 \phi = \frac{1}{3} G_{MNP} G^{MNP} + \frac{1}{2 g_X^2} e^\phi ( F^{\hat{I}}_{MN} F_{\hat{I}}^{MN} )_X - 2 V(\varphi), \\
\label{e-7-1-7}
&&\!\!\!\!\!\!\!\!\!\!\!\! D_M (e e^\phi F^{\hat{I}MN}) = e^{2\phi} G^{NPQ} F^{\hat{I}}_{PQ} + \MD^N \varphi^\alpha \tilde{\xi}^{\hat{I}}_\alpha, \\
\label{e-7-1-8}
&&\!\!\!\!\!\!\!\!\!\!\!\! g_{\alpha\beta} \! \MD_M \! \MD^M \! \varphi^\beta = \frac{\partial V}{\partial \varphi^\alpha}.
\eea
Note that we have used the dilaton field equation to absorb all trace terms in the Einstein equation in its last term. The supersymmetry variations of the fermions are given by
\bea
\label{e-7-1-9}
\delta \psi_M &=& \MD_M \epsilon + \frac{1}{24} e^\phi \Gamma^{NPQ} \Gamma_M G_{NPQ} \epsilon, \nn\\
\delta \chi &=& \frac{1}{2} \Gamma^M \partial_M \phi \epsilon - \frac{1}{12} e^\phi \Gamma^{MNP} G_{MNP} \epsilon, \nn\\
\delta \lambda^{\hat{I}} &=& - \frac{1}{2 \sqrt{2} g_X^2} e^{\phi/2} \Gamma^{MN} (F^{\hat{I}}_{MN})_X \epsilon  - \frac{1}{\sqrt{2}} e^{- \phi/2} ( C^{i \hat{I}} )_X T_i \epsilon, \nn\\
\delta \psi^a &=& V_\alpha^{\phantom{\alpha} a A} \Gamma^M \MD_M \varphi^\alpha \epsilon_A.
\eea

Let us now find an explicit expression for the scalar potential, following \cite{Randjbar-Daemi:2004qr}. Our first task is to rewrite Eqs. (\ref{e-7-1-3}) for the prepotentials $C^{i \hat{I}}$ in terms of a coset representative $L$. In \cite{Percacci:1998ag}, it was shown that the appropriate expression is
\be
\label{e-7-1-10}
C^{iI} = 2 ( L^{-1} T^I L )_{x,AB} (T^i)^{AB} ,\qquad C^i = 2 ( L^{-1} T^3 L )_{AB} (T^i)^{AB}.
\ee 
To proceed, we must find a convenient form for the coset representative. Considering the generic $\mathrm{USp}(2 n_H,2) / \mathrm{USp}(2 n_H) \times \mathrm{USp}(2)$ case, we recall that the $\mathrm{USp}(2 n_H,2)$ isometry group is actually defined as the intersection $\mathrm{SU}(2n_H,2) \cap \mathrm{USp}(2 n_H+2)$. Its elements are therefore the set of $(2 n_H+2) \times (2 n_H+2)$ matrices $g$ that leave invariant the ordinary and symplectic metrics of signature $(+,\ldots,+,-)$,
\be 
\label{e-7-1-11}
\eta = \diag ( \mathbf{1}, \ldots, \mathbf{1},- \mathbf{1} ) ,\qquad J = \diag ( \sigma_2, \ldots, \sigma_2, - \sigma_2 ) ,
\ee 
i.e. the set of matrices that satisfy
\be 
\label{e-7-1-12}
g^\dag \eta g = \eta ,\qquad g^T J g = J.
\ee 
Each matrix $g$ can be written as $(n_H+1) \times (n_H+1)$ array of $2 \times 2$ matrices $g_{\bar{m}}^{\phantom{\bar{m}} \bar{n}}$, with $\bar{m}, \bar{n} = 0,1,\ldots,n_H$. Each of these $2 \times 2$ matrices can be shown to satisfy
\be 
\label{e-7-1-13}
\sigma_2 ( g_{\bar{m}}^{\phantom{\bar{m}} \bar{n}} )^* \sigma_2 = g_{\bar{m}}^{\phantom{\bar{m}} \bar{n}},
\ee 
and can thus be interpreted as a real quaternion. The elements of the $\mathrm{USp}(2 n_H) \times \mathrm{USp}(2)$ maximal compact subgroup are obtained by the restriction
\be 
\label{e-7-1-14}
g_{0}^{\phantom{0} \bar{m}}  = g_{\bar{m}}^{\phantom{\bar{m}} 0} = 0.
\ee 
Afther the above preliminaries, we can find an explicit parameterization for the scalar coset. We first introduce the $n_H \times 1$ array
\be 
\label{e-7-1-15}
\varphi = \left(
\begin{array}{c} \varphi_1 \\ \vdots \\ \varphi_{n_H} \end{array}
\right), 
\ee 
where each $\varphi_n$ is a real quaternion satisfying $\varphi_n=\sigma_2 \varphi_n^* \sigma_2$. As such, it can be written as $\varphi_n = a_n \mathbf{1} + \ii \vec{b}_n \cdot \vec{\sigma}$, where $a_n$ and $\vec{b}_n$ are real. It then follows that
\be 
\label{e-7-1-16}
\varphi^\dag \varphi = \sum_{n=1}^{n_H} ( a_n^2 + \vec{b}_n^2 ) \mathbf{1},
\ee 
i.e. that $\varphi^\dag \varphi$ is proportional to the identity. We can then consider the coset representative
\be 
\label{e-7-1-17}
L = \left(
\begin{array}{cc} 1 + \left( \frac{\sqrt{1 + \varphi^\dag \varphi} -1}{\varphi^\dag \varphi} \right) \varphi \varphi^\dag & \varphi \\ \varphi^\dag & \sqrt{1 + \varphi^\dag \varphi} \end{array}
\right), 
\ee 
where the factor inside parentheses is understood as a scalar. This representative is easily shown to satisfy  the defining relations (\ref{e-7-1-12}) and (\ref{e-7-1-12}). Using this representative, the prepotentials (\ref{e-7-1-10}) take the form 
\be
\label{e-7-1-18}
C^{iI} = 2 ( \varphi^\dag T^I \varphi )_{x,AB} (T^i)^{AB} ,\qquad C^i = 2 \left[ 1 + \tr (\varphi^\dag \varphi) \right] \delta^{i3}.
\ee 
and hence the scalar potential is given by the simple expression 
\be 
\label{e-7-1-19}
V(\varphi) = \frac{1}{4} e^{-\phi} \left\{ - g_x^2 \left[ ( \varphi^\dag T^I \varphi )_x \right]^2 + 2 g_1^2 \left[ 1 + \tr (\varphi^\dag \varphi) \right]^2 \right\}. 
\ee 

Considering first the case of an ungauged theory ($g_1 = 0$) and noting that $T^I$ are antihermitian, we see that the potential is positive-semidefinite with $V(\varphi)=0$ as its global minimum; this minimum is obviously attained at $\varphi=0$ and possibly in other configurations with $T^I \varphi = 0$ and $\varphi$ satisfying its equation of motion. Turning to the case of a gauged theory ($g_1 \ne 0$), we see that now the potential is strictly positive-definite; its global minimum is now unique, it is attained only when $\varphi = 0$, and it corresponds to an exponential potential for the dilaton
\be 
\label{e-7-1-20}
V_{\min} = \frac{1}{2} g_1^2 e^{-\phi}.
\ee 
For the case of a constant dilaton \emph{vev}, this corresponds to a positive cosmological constant. As we will see, it is this effective cosmological constant and the particular form of the dilaton coupling which picks up the $\mathbf{M}^4 \times \mathbf{S}^2$ vacuum as the unique maximally-symmetric solution of the gauged theory among other maximally-symmetric spaces \cite{Salam:1984cj}. Note that these restrictions are all due to supersymmetry: in a non-supersymmetric theory, the potential is not constrained and the theory admits de Sitter or anti-de Sitter spaces as possible solutions \cite{Randjbar-Daemi:1982hi}.

\section{Ungauged Theories: Compactification on a Magnetized Torus}
\label{sec-7-2}

An attractive scenario for magnetic compactification has been proposed by Bachas \cite{Bachas:1995ik} in the contexts of 10D superstring theory and 6D supersymmetric gauge theory, and involves compactification of the theories in question on magnetized tori. The resulting theories have chiral spectra and exhibit spontaneous supersymmetry breaking as well as Nielsen-Olesen instabilities that trigger electroweak symmetry breaking. The problem with the proposal is that it does not correspond to a classical solution of the equations of motion of the underlying theory, mainly because flatness of the tori is incompatible with the existence of magnetic fields along their directions. In what follows, we give a brief review of this proposal in the 6D case.

To begin, we neglect gravity for the moment and we consider a $D=6$, $N=2$ supersymmetric gauge theory containing vector multiplets and hypermultiplets. We compactify the theory on $\MBFT^2$ by making the periodic identifications $x_4 \cong x_4 + 2 \pi R_4$ and $x_5 \cong x_5 + 2 \pi R_5$. We then consider a $\mathrm{U}(1)_M$ subgroup of the gauge group, we let $Q$ denote the corresponding charge operator and we turn on the vector potential
\be
\label{e-7-2-1}
A_4 = a_4 ,\qquad\qquad A_5 = a_5 + H x_4
\ee
where $a_{4,5}$ and $H$ are constants. This configuration corresponds to a constant magnetic field
\be
\label{e-7-2-2}
F_{45} = H .
\ee
In order for the vector potential (\ref{e-7-2-1}) to be well-defined on the torus, $A_5(x_4)$ and $A_5(x_4 + 2 \pi R_4)$ must be related by a gauge transformation, $A_5(x_4 + 2 \pi R_4) = A_5(x_4) - \ii U^{-1} \partial_5 U$. A suitable choice for $U$ is given by $U = e^{2 \pi \ii Q H R_4 x_5}$. Single-valuedness of $U$ as $x_5$ goes around by $2 \pi R_5$ then demands that $2 \pi Q H R_4 R_5$ be an integer. We thus have the quantization condition
\be
\label{e-7-2-3}
Q H = \frac{n}{2 \pi R_4 R_5}.
\ee

Let us then consider the mass spectrum of the 4D effective theory. Starting from the case $H=0$, where the vector potential (\ref{e-7-2-1}) corresponds to two Wilson lines, the 4D masses of all charged fields are given by the usual Kaluza-Klein formula, including the shifts due to $a_{4,5}$. Thus, we have
\be
\label{e-7-2-4}
M^2 = \left(\frac{n_4}{R_4} - Q a_4 \right)^2 + \left(\frac{n_5}{R_5} - Q a_5 \right)^2
\ee
This formula is valid for both bosons and fermions and implies that all charged fluctuations, including the vector bosons carrying $\mathrm{U}(1)_M$ charges, are massive when $a_{4,5} \ne 0$. This is an example of gauge symmetry breaking by Wilson lines, with the \emph{vev}s $a_{4,5}$ playing the role of the continuous moduli. Turning to the case $H \ne 0$, one first difference is that $a_{4,5}$ are no longer moduli labelling inequivalent vacua; since the fluctuations of $A_{4,5}$ correspond to Goldstone bosons of a spontaneously broken symmetry, the associated \emph{vev}s are irrelevant. So, without loss of generality, we may consider the case $a_{4,5} = 0$. Then, the squared masses of the charged scalars are given by the eigenvalues of the internal Klein-Gordon operator
\be
\label{e-7-2-5}
- D_m D^m = ( \ii \partial_4 )^2 + ( \ii \partial_5 + Q H x_4 )^2
\ee
Since the quantity on the RHS is just the Hamiltonian for a 2D particle of mass $1/2$ in a constant magnetic field $H$, the squared masses are immediately read off from the famous Landau formula, with the result
\be
\label{e-7-2-6}
M_0^2 = (2k+1) | Q H | ; \qquad\qquad k=0,1,2,\ldots
\ee
where $k$ labels the Landau levels, each level having a degeneracy of order $Q n$. Eq. (\ref{e-7-2-6}) clearly implies that all charged scalars are massive. Passing to the charged fermions, their squared masses are given by the eigenvalues of the internal Dirac operator squared,
\be
\label{e-7-2-7}
- \Gamma^m D_m \Gamma^n D_n = - D_m D^m + Q F_{45} ( - \ii \Gamma^{45} ) =  - D_m D^m + 2 Q H S^{(1/2)},
\ee
where $S^{(1/2)} = -\frac{\ii}{2} \Gamma^{45}$ is the internal helicity operator with eigenvalues $\pm \frac{1}{2}$. This leads to the Landau formula for fermions, 
\be
\label{e-7-2-8}
M_{1/2}^2 = (2k+1) | Q H | + 2 Q H S^{(1/2)}; \qquad\qquad k=0,1,2,\ldots
\ee
where the degeneracies are as before. Eq. (\ref{e-7-2-8}) is in fact valid for charged particles of any spin, provided that we replace $S^{(1/2)}$ with the appropriate helicity operator. Now, for a 6D spinor subject to a Weyl condition $\Gamma_7 \psi = \pm \psi$, its 4D chirality is related to its internal helicity according to $2 S^{(1/2)} = - \ii \Gamma^{45} = \pm \Gamma_5$. Considering for definiteness the case $Q H > 0$, let us then examine the masses of the 4D spinors obtained by reduction of a 6D positive-chirality spinor. At the lowest Landau level, we see that the spinor with negative 4D chirality, $2 S^{(1/2)} = -1$, is \emph{massless} while the spinor with positive 4D chirality, $2 S^{(1/2)} = 1$, has a squared mass given by $2 Q H$. The apparent puzzle of a chiral spinor with nonzero mass is resolved once we pass to the next Landau level where we find a negative-chirality spinor with squared mass $2 Q H$; this spinor pairs with the positive-chirality spinor from the previous Landau level to form a massive Dirac spinor. Continuing this construction and including degeneracies, we see that the fermionic spectrum contains $Q n$ massless negative-chirality spinors plus an infinite tower of massive states, each one containing $Q n$ Dirac spinors of squared mass $2(k+1) |Q H|$. 

The scenario thus described provides a very attractive setup for supersymmetry breaking. However, it also suffers from some potential problems, described below.

\begin{enumerate}
\item \emph{Tachyonic instabilities}. Consider the mass formula for the scalar fluctuations of the $\mathrm{U}(1)_M$ gauge boson. Denote these fluctuations by $\phi_{5,6}$ and arrange them into the charged combinations $\phi_\pm = \frac{1}{\sqrt{2}} ( \phi_5 \mp \ii \phi_6 )$. Their masses are determined from the appropriate generalization of (\ref{e-7-2-8}) for a spin--1 field, given by
\be
\label{e-7-2-9}
M_1^2 = (2k+1) | Q H | + 2 Q H S^{(1)}; \qquad\qquad k=0,1,2,\ldots
\ee
where now the helicity operator $S^1$ has eigenvalues $\pm 1$ acting on $\phi_\pm$. So, for $Q H > 0$, the squared mass of the negative-helicity mode $\phi_-$ is given by $-2 |Q H|$ and thus the spectrum contains $Q n$ tachyonic states. As first pointed out by Nielsen and Olesen, the existence of such states renders the vacuum unstable. However, this problem can be remedied by (i) modding out the torus by a $\MBBZ_2$ symmetry that projects out the nonzero-helicity modes, (ii) adding more magnetic fields in other internal directions such that the masses of the tachyonic modes are shifted to zero or positive values or (iii) adding Wilson lines along internal cycles so that the 6D vector bosons are massive in the first place.

\item \emph{Gravitational/dilatonic tadpoles}. Another problem in this model arises once we include gravity. To see this, try embed the model into a $D=6$, $N=2$ supergravity theory and consider a constant-dilaton configuration. Since the torus is a flat manifold, Einstein's equations require that $T_{mn}$ be equal to zero. However, this cannot be so since $T_{mn}$ contains a contribution from the magnetic field and thus the theory has a gravitational tadpole. In a Poincar\'e supergravity where a potential for the dilaton is absent, the RHS of the dilaton equation of motion also contains a positive term and thus the configuration has a dilatonic tadpole. A potential resolution of this problem, given in \cite{Bachas:1995ik}, is that the classical supergravity (or string theory) vacuum may be modified by radiative corrections according to the Coleman-Weinberg mechanism. In such a scenario, the classical tadpoles described above are assumed to be cancelled by quantum corrections and the flat $\mathbf{M}^4 \times \MBFT^2$ configuration is identified with the ``true'' vacuum of the theory. However, this reasoning still lacks a rigorous justification.
\end{enumerate}

\section{Gauged Theories: Compactification on a Magnetized Sphere}
\label{sec-7-3}

In this section, we turn to the gauged theory, where the gauge group includes a $\mathrm{U}(1)_R$ factor. As noted earlier on, this theory admits a spontaneously compactification to four-dimensional spacetime with a magnetic monopole background and picks $\mathbf{M}^4 \times \mathbf{S}^2$ as its unique maximally-symmetric ground state. The properties of the resulting theories depend of the embedding of the magnetic $\mathrm{U}(1)_M$ in the gauge group. This section is devoted to examining the properties of such solutions in the context of the original Salam-Sezgin model and the anomaly-free theories discussed in Chapter 5.

\subsection{The Salam-Sezgin model}
\label{sec-7-3-1}

The simplest model of $D=6$ gauged supergravity admitting a spontaneous compactification of the sort discussed above is the Salam-Sezgin model \cite{Salam:1984cj}, where the gauge group of the theory is just the $\mathrm{U}(1)_R$ R-symmetry subgroup and a magnetic monopole is turned on in that direction. Although it is clear that this theory can only serve as a toy model (among other things, it is anomalous), it illustrates some of the basic principles of such compactifications. For this reason, we present the model below.

To begin the construction of the model, we first impose the requirement that the 4D spacetime be maximally symmetric, which implies that (i) all background fields must depend only on the two internal dimensions and (ii) all field strength components with spacetime indices must vanish. From condition (i) it follows that the $\mu\nu$ component of the Einstein equation (\ref{e-7-1-4}) gives $R_{\mu\nu} = 0$ i.e. that the 4D spacetime is Ricci-flat, the simplest choice being Minkowski space. Note that there is no fine-tuning involved in this construction: the 6D cosmological constant does not appear in the RHS of the Einstein equation due to the dilaton field equation and so the 4D cosmological constant is \emph{self-tuned} to zero. From condition (ii) it follows that all components of $G_3$ vanish, i.e. we must take $B_2=0$ or pure gauge, and that the only nonvanishing components of $F_2$ can be the internal components $F_{mn}$. Furthermore, assuming a constant dilaton and taking the hyperscalars to sit at the minimum of their potential where they all vanish, the dilaton field equation requires a non-vanishing $F_{mn}$ to compensate the contribution of the potential. Taking $F_{mn} \sim \epsilon_{mn}$, the RHS of the $mn$ component of the Einstein equation corresponds to a positive cosmological constant leading to an internal space of constant positive curvature, the simplest choice being an $\mathbf{S}^2$. Accordingly, the gauge field triggering this compactification is a magnetic monopole field, lying in a $\mathrm{U}(1)_M$ group which, in the present model, is identified with the $\mathrm{U}(1)_R$ gauge group. To summarize, the Salam-Sezgin ansatz is the following
\bea 
\label{e-7-3-1}
ds_6^2 &=& \eta_{\mu\nu} dx^\mu dx^\nu + R^2 ( d \vartheta^2 + \sin^2 \vartheta d \varphi^2 ), \nn\\
B_2 &=& 0, \nn\\
\phi &=& \phi_0 = \textrm{constant}, \nn\\
A_\pm &=& \frac{n}{2} ( \cos \vartheta \mp 1 ) d \varphi, \nn\\
\varphi^\alpha &=& 0.
\eea 
Here, $R$ is the $\mathbf{S}^2$ radius, while $\vartheta$ and $\varphi$ are the $\mathbf{S}^2$ polar angles. Also, $A_+$ and $A_-$ correspond to the potentials on the northern- and southern-hemisphere patches of the $\mathbf{S}^2$ which, on the equator, are related by a gauge transformation parameterized by $U=e^{ \ii n \varphi}$. In order for $U$ to be single-valued as $\varphi$ changes by $2 \pi$, $n$ must therefore be an integer. 

In a flat basis, the nonvanishing components of the Ricci tensor are
\be 
\label{e-7-3-2}
R_{\bar{4} \bar{4}} = R_{\bar{5} \bar{5}} = \frac{1}{R^2} ,
\ee
while those of the field strength are given by
\be 
\label{e-7-3-3}
H \equiv F_{\bar{4} \bar{5}} = \frac{n}{2 R^2} ,
\ee
Since the monopole ansatz solves its equations of motion, the only nontrivial equations to be satisfied are the $mn$ components of the Einstein equations and the dilaton field equation. These are given by
\be
\label{e-7-3-4}
\frac{n^2}{2 g_1^2 R^2} e^{\phi_0} = 1, \qquad e^{2\phi_0} = \frac{4 g_1^4 R^4}{n^2}, 
\ee
Solving the second equation for $e^{\phi_0}$ and inserting the result into the first equation, we find that the monopole charge is quantized as
\be
\label{e-7-3-5}
n = \pm 1,
\ee
from which it also follows the that the $\mathbf{S}^2$ radius is fixed in terms of the dilaton \emph{vev} and the $\mathrm{U}(1)_R$ coupling according to
\be
\label{e-7-3-6}
R^2 = \frac{e^{\phi_0}}{2 g_1^2}.
\ee

Let us now examine the supersymmetry of the background. For the ansatz (\ref{e-7-3-1}), the only nontrivial supersymmetry equations are the $\delta \psi_M$ and $\delta \lambda$ equations, which are respectively written as
\be
\label{e-7-3-7}
\left( \partial_M + \frac{1}{2} \omega_{M \bar{4} \bar{5}} \Gamma^{\bar{4} \bar{5}} + \ii A_M \right)  \epsilon = 0, \qquad ( \Gamma^{\bar{4} \bar{5}} F_{\bar{4} \bar{5}} + g_1^2 e^{-\phi_0} C^i  T_i ) \epsilon = 0.
\ee
Starting from the $\delta \lambda$ equation, we plug $F_{\bar{4} \bar{5}}$ and $C^i$ from (\ref{e-7-3-3}) and (\ref{e-7-1-18}) respectively and we write it in the form
\be
\label{e-7-3-8}
\Gamma^{\bar{4} \bar{5}} \epsilon = \ii n \epsilon.
\ee
Since $\epsilon$ satisfies $\Gamma_7 \epsilon = - \epsilon$, we have $\Gamma^{\bar{4} \bar{5}} \epsilon = - \ii \Gamma_5 \epsilon$ and Eq. (\ref{e-7-3-8}) is actually written as
\be
\label{e-7-3-9}
( \Gamma_5 + n ) \epsilon = 0.
\ee
Therefore, the $\delta \lambda$ equation fixes the 4D chirality of $\epsilon$ in terms of the monopole charge, implying that half of the 6D supersymmetries survive. Turning to the $\delta \psi_M$ equation, we plug $A_M$ from (\ref{e-7-3-1}), we note that the only nonvanishing spin connection component is
\be
\label{e-7-3-10}
\omega_{\varphi \bar{4} \bar{5}} = 1 - \cos \vartheta,
\ee
and we use again $\Gamma^{\bar{4} \bar{5}} \epsilon = - \ii \Gamma_5 \epsilon$ to find
\be
\label{e-7-3-11}
\partial_\mu \epsilon = 0 ,\qquad \partial_\vartheta \epsilon = 0 ,\qquad \left[ \partial_\varphi - \frac{\ii}{2} (1-\cos \vartheta)(\Gamma_5 + n) \right] \epsilon = 0
\ee
Provided that (\ref{e-7-3-9}) holds, the $\mathrm{U}(1)_R$ connection cancels the spin connection in the last of (\ref{e-7-3-11}) and thus the $\delta \psi_M$ equation reduces to $\partial_M \epsilon = 0$ and is solved by a constant spinor. Therefore, the background preserves half the 6D supersymmetries and leads to a 4D theory with $N=1$ supersymmetry.

Finally, a very interesting, recently discovered property of the Salam-Sezgin model is its uniqueness. Indeed, in \cite{Gibbons:2003di}, the authors relaxed the simplifying assumptions of the Salam-Sezgin model imposing only the requirements for (i) a maximally symmetric 4D spacetime and (ii) a compact, non-singular 2D internal space. Considering the appropriate ansatz for the metric
\be
\label{e-7-3-12}
\dd s_6^2 = e^{2 A(y)} \tilde{g}_{\mu\nu}(x) dx^\mu dx^\nu + g_{mn}(y) dy^m dy^n
\ee
where $\tilde{g}_{\mu\nu}(x)$ is the 4D Minkowski, dS or AdS metric and $A(y)$ is the warp factor, the authors proved that the Salam-Sezgin solution is in fact the \emph{only} possible solution.

\subsection{General embeddings}
\label{sec-7-3-2}

As stressed earlier on, the Salam-Sezgin model is only to be regarded as a toy model since its gauge group is just $\mathrm{U}(1)_R$, it is anomalous and it gives a vectorlike 4D spectrum. To obtain more interesting theories, one must try to embed this model into a more realistic and quantum-mechanically consistent theory, such as the anomaly-free gauged theories discussed in \S\ref{sec-5-6}. Doing so, there is the possibility of embedding the $\mathrm{U}(1)_M$ group corresponding to the magnetic monopole into any gauge group factor and not just identifying it with $\mathrm{U}(1)_R$. This possibility remedies one of the problems of the Salam-Sezgin model, namely it gives a chiral spectrum in 4D. However, in the generic case, it also raises issues concerning the classical stability of the monopole solutions. In what follows, we will present the general form of such solutions and then examine the zero-mode spectrum and classical stability of these theories.

\subsubsection{Equations of motion and supersymmetry}

We start by considering one of the anomaly-free theories of \S\ref{sec-5-6}, letting $\MG = \prod_x \MG_x \times \mathrm{U}(1)_R$ be the gauge group. The ansatz for the compactification is similar to (\ref{e-7-3-1}), but with the vector potential now replaced by
\be
\label{e-7-3-13}
A_\pm = \frac{n}{2} Q ( \cos \vartheta \mp 1 ) d \varphi,
\ee 
where $Q$ is the $\mathrm{U}(1)_M$ generator, that may in general be a linear combination of all commuting generators of $\MG$. The potentials $A_+$ and $A_-$ on the two patches should now be connected by a gauge transformation parameterized by $U=e^{ \ii n Q \varphi}$. Single-valuedness of $U$ as $\varphi$ changes by $2 \pi$ requires the quantity $n q_{\min}$, where $q_{\min}$ is the minimal $\mathrm{U}(1)_M$ charge in the theory, be an integer. This results in a set of quantization conditions on the coefficients defining $Q$ in terms of the generators in the Cartan subalgebra of $\MG$.

For simplicity, let us consider the case where $Q$ is identified with a single generator of a given $\MG_x$ factor and let $g_x$ be the corresponding gauge coupling. Now, the nonvanishing field-strength components are given by
\be 
\label{e-7-3-14}
H \equiv F_{\bar{4} \bar{5}} = \frac{n}{2 R^2} Q ,
\ee
What remains is to satisfy the $mn$ components of the Einstein equations and the dilaton field equation. These are now given by
\be
\label{e-7-3-15}
e^{\phi_0} = \frac{2 g_x^2 R^2}{n^2}, \qquad e^{2\phi_0} = \frac{4 g_x^2 g_1^2 R^4}{n^2}, 
\ee
Taking the square of the first equation and comparing with the second equation, we find that the two gauge couplings should be related by
\be
\label{e-7-3-16}
g_x = \pm n g_1.
\ee
Plugging this into the first of (\ref{e-7-3-15}), we see that the $\mathbf{S}^2$ radius is expressed in terms of the dilaton \emph{vev} and the $\mathrm{U}(1)_R$ coupling as before,
\be
\label{e-7-3-17}
R^2 = \frac{e^{\phi_0}}{2 g_1^2}.
\ee
So, in contrast to the Salam-Sezgin case, the monopole charge $n$ is no longer fixed by the field equations but can acquire any integer value. Moreover, we also have a certain type of tuning between the couplings $g_x$ and $g_1$.

Another departure from the Salam-Sezgin model is that now supersymmetry is completely broken. This most easily seen by considering the $\delta \psi_M$ equation which now takes the form
\be
\label{e-7-3-18}
\left( \partial_M + \frac{1}{2} \omega_{M \bar{4} \bar{5}} \Gamma^{\bar{4} \bar{5}} \right)  \epsilon = 0,
\ee
since the gravitino no longer couples to the monopole background. In terms of components, we have
\be
\label{e-7-3-19}
\partial_\mu \epsilon = 0 ,\qquad \partial_\vartheta \epsilon = 0 ,\qquad \left[ \partial_\varphi + \frac{1}{2} (1-\cos \vartheta) \Gamma^{\bar{4} \bar{5}} \right] \epsilon = 0
\ee
from which we see that we cannot satisfy the second and third equations simultaneously.

\subsubsection{Zero-mode spectrum}

We next turn to a brief discussion of the zero-mode spectrum of the resulting compactifications. Starting from the fermions, we first note that the free Dirac operator on has no zero modes on $\mathbf{S}^2$ and therefore the 4D fermionic zero modes necessarily originate from the 6D fermions carrying monopole charge. In the case where the monopole is embedded in one of the $\MG_x$ factors, the 6D fields that can give rise to 4D massless chiral fermions are contained among the associated gauginos and among the hyperinos that are charged under this group; the gravitino, tensorino and the rest of the gauginos will of course be massive. The number of these chiral zero modes is counted using index-theorem formulas. Turning to the bosons, the squared mass of each one of the lightest hyperscalar fluctuations will receive two contributions, one being proportional to the associated eigenvalue of $\frac{\partial^2 V}{\partial \varphi^\alpha \partial \varphi^\beta}$ at $\varphi^\alpha=0$ and the other being proportional to $\hat{\MD}^2$ where $\hat{\MD}$ is the covariant derivative acting on the hyperscalar fluctuations in the background of the monopole vector potential(s). The first contribution will make all hyperscalars massive. The second contribution, if the monopole charges do not add up to zero, will be a positive quantity proportional to $1/R^2$, where $R$ is the radius of $\mathbf{S}^2$. Furthermore in the case of a nonzero net monopole charge of the hyperscalar even the leading (lightest) $D=4$ scalar modes resulting from it will belong to a nontrivial irreducible representation of the Kaluza-Klein $\mathrm{SU}(2)$. We shall comment on the masses of some other bosonic modes below.

\subsubsection{Classical stability}

Let us finally turn to the issue of classical stability. A compactification is said to be classically stable if the squared masses of the 4D excitations of all fields are positive or zero. In the opposite case, there exist tachyonic states of negative mass squared which render the compactification unstable under small perturbations. In the present case, such states come from the scalar components of vector fields that carry magnetic charge, as in the toroidal compactification of Section \ref{sec-7-1}. To be specific, let $\phi_m$ be one of the excitations of the vector potential tangent to $\mathbf{S}^2$ and charged with respect to $\mathrm{U}(1)_M$. This vector has the components $\phi_{\pm}$ with respect to a complex basis in the tangent space of $\mathbf{S}^2$. We also have the reality condition $\phi = \phi^{\dagger}$. As a Lie-algebra-valued vector we can write $\phi = U_+^r T_r + W_+^r T_r^\dag$, where $U$ and $W$ are complex fields and the $T$'s are among the charged generators of the gauge group.  In order to be able to write down a general formula which can be applied for any model of this kind, denote the $\mathrm{U}(1)_M$ charge of $U$ or $W$ by $q$. The mass spectrum of $D=4$ spin-zero fields resulting from such a $D=6$ object is given by \cite{Randjbar-Daemi:1983bw} 
\be 
\label{e-7-3-20}
R^2 M^2 = \ell (\ell+1) - (\lambda -1)^2, 
\ee 
where 
\be 
\label{e-7-3-21}
\ell = |\lambda|, |\lambda|+1, \ldots ,\qquad \lambda =  1 + \frac{n}{2} q
\ee 
$n$ being the monopole charge. From these relations, it is easily seen that for all those fields for which $ nq \leq -2$ there is a tachyon.

In the case where tachyonic modes are present, one may adopt the point of view that they are welcome in the context of an effective theory as they are natural candidates for 4D Higgs fields. The quartic term in the potential for such fields will come from the self-coupling of the 6D gauge fields and their \emph{vev} will break the 4D gauge group at the Kaluza-Klein scale. Such an origin for the Higgs fields has been considered before as a possible solution to the hierarchy problem. In order for this interpretation to be complete, one needs to look for new stable solutions of the 6D field equations which would correspond to the minimum of the potential for the tachyons interpreted as Higgs fields. These solutions will necessarily break the spherical symmetry and their construction may give a geometrical origin to the Higgs mechanism. It will be interesting to find such solutions.

However, conventional phenomenology usually requires the absence of such tachyonic states from the spectrum. From Eqs. (\ref{e-7-3-20}) and (\ref{e-7-3-21}), we see that the condition for a compactification to be free of tachyonic modes is that the  embedding of the monopole in the gauge group must be such that $|nq|=1$ for all charged excitations, where $nq$ is understood as the sum of the individual $nq$'s over all the monopole directions with respect to which the corresponding excitation is charged. Choosing our conventions so that the minimal charge in the theory is $q_{\min} = \pm 1$, we need then to have $n=\pm 1$ and $q=\pm 1$ for \emph{all} charged fields. In the case where $\mathrm{U}(1)_M = \mathrm{U}(1)_R$ as in the Salam-Sezgin model, these conditions are automatically satisfied due to (\ref{e-7-3-5}) and to the fact that all fields charged under $\mathrm{U}(1)_R$ have unit charge. However, in the remaining cases where $\mathrm{U}(1)_M$ is embedded into the $\MG_x$ factors, this requirement imposes serious constraints on the allowed groups where the monopole can be embedded and on the allowed fermion representations with respect to these groups. One case where a stable compactification is guaranteed to exist is when the gauge group factor $\MG_x$ has a maximal-subgroup decomposition $\MG_x \supset \MH_x \times \mathrm{U}(1)$ with $\MH_x$ simple and where all fermions charged under this group transform in the adjoint. For if this is the case, the corrsponding decomposition of the adjoint of $\MG_x$,
\be 
\label{e-7-3-22}
\MA \to \MA_0 + \MR_{-1} + \overline{\MR}_1 + \mathbf{1}_0,
\ee 
shows that for all charged excitations we have $q=\pm 1$ and so the compactification is stable if $n=\pm 1$. However, in compactifications where $\MG_x$ does not have a maximal-subgroup decomposition of the above form or where there are fermions transforming in representations other than the adjoint, the resulting configurations are generically unstable.

\subsection{The $E_7 \times E_6 \times \mathrm{U}(1)_R$ model}
\label{sec-7-3-3}

Our first example of the above construction is the compactification of the $E_7 \times E_6 \times \mathrm{U}(1)_R$ model as discussed in \cite{Randjbar-Daemi:1985wc}. In the compactification discussed there, the monopole was embedded in the $E_6$ factor of the gauge group. Since the hyperinos are not charged under this group, the only charged fluctuations of fermions are those of the $E_6$ gauginos which, of course, transform in the adjoint $\mathbf{78}$. Embedding $\mathrm{U}(1)_M$ in $E_6$ according to the decomposition
\be 
\label{e-7-3-23}
E_6 \supset \mathrm{SO}(10) \times \mathrm{U}(1)_M,
\ee
decomposing the adjoint of $\mathbf{78}$ according to the branching rule \cite{Slansky:1981yr}
\be 
\label{e-7-3-24}
\mathbf{78} \to \mathbf{45}_0 + \mathbf{16}_{-1} + \mathbf{\overline{16}}_{1} + \mathbf{1}_{0}, 
\ee 
and discarding neutral fields, we see that the only $\mathrm{SO}(10) \times \mathrm{U}(1)_M$ representations that may give rise to fermion zero modes on $\mathbf{S}^2$ are
\be 
\label{e-7-3-25}
\mathbf{16}_{-1} + \mathbf{\overline{16}}_{1}, 
\ee
In 4D, the unbroken gauge group is $\mathrm{SO}(10) \times \mathrm{U}(1)_M \times \mathrm{SU}(2)_{KK}$ where $\mathrm{SU}(2)_{KK}$ is the Kaluza-Klein group originating from the $\mathbf{S}^2$ isometries. The number of chiral fermions is found to be $2 |n|$, i.e. the 4D massless chiral fermions form $2 |n|$ families of $\mathrm{SO}(10)$ transforming in the $|n|$--dimensional irreducible representation $\mathbf{n}$ of $\mathrm{SU}(2)_{KK}$. The representations of the chiral fermions under the $\mathrm{SO}(10) \times \mathrm{U}(1)_R \times \mathrm{SU}(2)_{KK}$ symmetry is then
\be 
\label{e-7-3-26}
(\mathbf{16} , \mathbf{n})_{1} + (\mathbf{16}, \mathbf{n})_{-1}, 
\ee 
with the subscripts here denoting the $\mathrm{U}(1)_R$ charges.

From the $\mathrm{U}(1)_R$ charge assignments in (\ref{e-7-3-26}), we immediately verify that the 4D theory is free of $\mathrm{U}(1)_R$ triangle anomalies. Furthermore, absence of $\mathrm{SO}(10)$ and $\mathrm{SU}(2)_{KK}$ anomalies in the 4D theory is guaranteed, since these groups are safe. Finally, according to the stability criteria stated in the end of \S\ref{sec-7-3-2}, the compactification described above is stable only if $n=\pm 1$, which corresponds to two families of chiral fermions.

\subsection{The $E_7 \times G_2 \times \mathrm{U}(1)_R$ model}
\label{sec-7-3-4}

Let us next turn to the compactifications of the $E_7 \times G_2 \times \mathrm{U}(1)_R$ model. A first example is given by embedding $\mathrm{U}(1)_M$ in $E_7$ according to the maximal-subgroup
decomposition 
\be 
\label{e-7-3-27}
E_7 \supset E_6 \times \mathrm{U}(1). 
\ee
Using the branching rules
\bea 
\label{e-7-3-28}
\mathbf{56} &\to& \mathbf{27}_1 + \mathbf{\overline{27}}_{-1} + \mathbf{1}_{3} + \mathbf{1}_{-3}, \nn\\
\mathbf{133} &\to& \mathbf{78}_0 + \mathbf{27}_{-2} +
\mathbf{\overline{27}}_{2} + \mathbf{1}_{0}, 
\eea 
we see that $q_{\min} = 1$ so that $n$ is an integer. Discarding neutral fields, we see that the fermion representations under $E_6 \times G_2 \times \mathrm{U}(1)_M$ which can give rise to fermion zero modes on $\mathbf{S}^2$ are 
\be 
\label{e-7-3-29}
(\mathbf{27},\mathbf{14})_1 +
(\mathbf{\overline{27}},\mathbf{14})_{-1} +
(\mathbf{1},\mathbf{14})_{3} + (\mathbf{1},\mathbf{14})_{-3}, 
\ee
for the $E_7$ hyperinos and 
\be
\label{e-7-3-30} 
(\mathbf{27},\mathbf{14})_{-2} + (\mathbf{\overline{27}},\mathbf{14})_{2}, 
\ee 
for the $E_7$ gauginos. The unbroken gauge group in $D=4$ is $E_6\times G_2\times \mathrm{U}(1)_R\times \mathrm{SU}(2)_{KK}$. The chiral fermions originate from the $\mathbf{27}$'s and the $\mathbf{\overline{27}}$'s. We can regard all the $D=4$ fermions as left-handed Weyl spinors. The chiral fermions originating from the decomposition of $\mathbf{56}$ of $E_7$ then are 
\be 
\label{e-7-3-31} 
2 (\mathbf{\overline{27}}, \mathbf{14}, \mathbf{n})_{0} ,
\ee 
while the fermions originating from the decomposition of the adjoint of $E_7$ produce 
\be 
\label{e-7-3-32} 
(\mathbf{27} , \mathbf{14}, \mathbf{2n})_{1} + (\mathbf{27}, \mathbf{14}, \mathbf{2n})_{-1}, 
\ee 
with the subscripts here denoting the $\mathrm{U}(1)_R$ charges.

As a second example, we consider the successive maximal-subgroup decompositions 
\be 
\label{e-7-3-33} 
E_7 \supset \mathrm{SO}(12) \times \mathrm{SU}(2)
\supset \mathrm{SO}(10) \times \mathrm{SU}(2) \times \mathrm{U}(1), 
\ee 
and we identify the last $\mathrm{U}(1)$ factor with $\mathrm{U}(1)_M$. Using the branching rules
\bea 
\label{e-7-3-34} 
\mathbf{56} &\to& (\mathbf{12},\mathbf{2}) + (\mathbf{32},\mathbf{1}), \nn\\
\mathbf{133} &\to& (\mathbf{1},\mathbf{3}) +
(\mathbf{32'},\mathbf{2}) + (\mathbf{66},\mathbf{1}), 
\eea 
for $E_7 \supset \mathrm{SO}(12) \times \mathrm{SU}(2)$ and 
\bea 
\label{e-7-3-35} 
\mathbf{12}  &\to& \mathbf{1}_1 + \mathbf{1}_{-1} + \mathbf{10}_0, \nn\\
\mathbf{32}  &\to& \mathbf{16}_1 + \overline{\mathbf{16}}_{-1}, \nn\\
\mathbf{32'}\!\! &\to& \mathbf{16}_{-1} + \overline{\mathbf{16}}_1, \nn\\
\mathbf{66}  &\to& \mathbf{1}_0 + \mathbf{10}_2 + \mathbf{10}_{-2}
+ \mathbf{45}_{0}, 
\eea 
for $\mathrm{SO}(12) \supset \mathrm{SO}(10) \times \mathrm{U}(1)$, we see again that, $q_{\min} = 1$ and $n = \mathrm{integer}$. Discarding neutral fields, we find that the muliplets of $\mathrm{SO}(10) \times \mathrm{SU}(2) \times G_2 \times \mathrm{U}(1)_M$ which can have fermion zero modes on $\mathbf{S}^2$ are
\be 
\label{e-7-3-36} 
(\mathbf{1},\mathbf{2},\mathbf{14})_1 + (\mathbf{1},\mathbf{2},\mathbf{14})_{-1} + (\mathbf{16},\mathbf{1},\mathbf{14})_1 + (\mathbf{\overline{16}},\mathbf{1},\mathbf{14})_{-1}, 
\ee 
for the hyperinos and 
\be 
\label{e-7-3-37} 
(\mathbf{16},\mathbf{2},\mathbf{14})_{-1} + (\overline{\mathbf{16}},\mathbf{2},\mathbf{14})_1 + (\mathbf{10},\mathbf{1},\mathbf{14})_{2} + (\mathbf{10},\mathbf{1},\mathbf{14})_{-2}, 
\ee
for the $E_7$ gauginos. Now, the unbroken gauge group in $D=4$ is $G= \mathrm{SO}(10) \times \mathrm{SU}(2)\times G_2\times \mathrm{SU}(2)_{KK}\times \mathrm{U}(1)_R$. The
spectrum of the $D=4$ chiral fermions is given by 
\be
\label{e-7-3-38} 
2(\mathbf{\overline {16}}, \mathbf{1}, \mathbf{14},\mathbf{n})_0, 
\ee 
and 
\be 
\label{e-7-3-39} 
(\mathbf{16},\mathbf{2},\mathbf{14},\mathbf{n})_1 + (\mathbf{16}, \mathbf{2}, \mathbf{14}, \mathbf{n})_{-1}. 
\ee
It is clearly seen that the spectrum in both cases is free from all chiral anomalies, because $E_6$ and $\mathrm{SO}(10)$ are safe groups in $D=4$ and the $\mathrm{U}(1)_R$ couplings are obviously vectorlike. It
is also seen that there is no value of $n$ which produces a realistic spectrum. One can study other embeddings with the aim of reducing the gauge group and the number of families. For example, the group $G_2$ can be broken completely by the embedding of a monopole in an $\mathrm{SU}(2)$ subgroup of $G_2$ relative to which the
branching is $\mathbf{14} = \mathbf{3} + \mathbf{11}$. By itself this will produce only a vectorlike theory in $D=4$ with an unbroken group $E_7 \times \mathrm{SU}(2)_{KK}\times \mathrm{U}(1)_R$. However, combined with other monopoles in the manner described above, one can break the group down to $\mathrm{SO}(10)\times \mathrm{SU}(2)_{KK}$; however, the number of families will still be large. Finally, according to the stability criteria of \S\ref{sec-7-3-2}, none of the above embeddings and, in fact, no other embedding is stable.
\appendix
\chapter{Notation and conventions}
\label{app-a}

Our main notational conventions are summarized as follows. 

\begin{itemize}
\item \emph{Metric}. We use the ``mostly-plus'' Minkowski signature $(-,+,\ldots,+)$ throughout the thesis, with the exception of certain sections of Chapter 3 and Appendix C, where we follow the standard conventions in the anomaly/index theory literature and work in the Euclidean. 

\item \emph{Physical constants}. We use physical units where $\hbar = c = 1$. The gravitational coupling constant $\kappa$ is defined in terms of Newton's constant as $\kappa^2 = 4 \pi G_N$, so that the canonical normalization of the Einstein term is $\frac{1}{4 \kappa^2} R$. In many cases, we set $\kappa = 1$. 

\item \emph{Spacetime indices}. In general, we denote world indices by $\mu,\nu,\ldots$ and tangent-space flat indices by $a,b,\ldots$. In the context of compactification, we will let $M,N,\ldots$, $\mu,\nu,\ldots$ and $m,n,\ldots$ denote world indices of the high-dimensional spacetime, the physical spacetime and the internal space respectively and $A,B,\ldots$, $a,b,\ldots$ and $\bar{m},\bar{n},\ldots$ denote the corresponding tangent-space indices. 

\item \emph{Covariant derivatives and curvatures}. Throughout the thesis, $D_\mu$ stands for the various GCT- and gauge-covariant derivatives acting on tensors, forms and spinors, the specific form being always clear from the context. Covariant derivarives that also involve composite connections will be denoted by $\MD_\mu$. The Riemann tensor is defined according to the convention
\be
\label{e-a-1}
[ D_\mu , D_\nu ] V_\rho = R_{\mu\nu\rho\sigma} V^\sigma.
\ee
The Ricci tensor and scalar are defined as
\be
\label{e-a-2}
R_{\mu\nu} =  R^\rho_{\phantom{\rho}\mu\rho\nu}, \qquad R = R^\mu_{\phantom{\mu}\mu}
\ee
Also, $R^{ab}$ will stand for the curvature 2--form
\be
\label{e-a-3}
R^{ab} =  \frac{1}{2} R_{\mu\nu}^{\phantom{\mu\nu} ab} dx^\mu dx^\nu
\ee
whose components are given by
\be
\label{e-a-4}
R_{\mu\nu}^{\phantom{\mu\nu} ab} = e^a_\rho e^b_\sigma R_{\mu\nu}^{\phantom{\mu\nu} \rho\sigma}.
\ee

\item \emph{Group theory}. Antihermitian generators are used throughout. The structure constants of an algebra are defined by
\be
\label{e-a-5}
[ T_I, T_J ]  = f_{IJ}^{\phantom{IJ} K} T_K.
\ee
In the case of gauge theories we adopt the conventional normalization, e.g. for $\mathrm{SU}(2)$ we set $T_I = - i \sigma_I /2$. However, in order to follow standard conventions in the anomaly literature, the traces involving Lie-algebra-valued quantities, e.g. $X= X^I T_I$, are \emph{defined} by
\be
\label{e-a-6}
\tr X^2  = - X^I X_I,
\ee
in the fundamental representation.

\end{itemize}
\chapter{Gamma Matrices and Spinors in Diverse Dimensions}
\label{app-b}
\thispagestyle{empty}

\section{Gamma Matrices in Diverse Dimensions}
\label{sec-b-1}

\subsection{The Clifford algebra}
\label{sec-b-1-1}

The Clifford algebra in an {\it even} number of dimensions, $D=2k+2$, is generated by a set of $D$ matrices $\Gamma^a$ which obey the anticommutation relations
\be
\label{e-b-1-1}
\{ \Gamma^a , \Gamma^b \} = 2 \eta^{ab}.
\ee
Any matrix of dimension equal to that of the $\Gamma^a$ can be expanded in the complete basis generated by
\be
\label{e-b-1-2}
1 , \Gamma^a , \Gamma^{ab} , \Gamma^{abc},\ldots, \Gamma^{a_1 \ldots a_{D}},
\ee
where
\be
\label{e-b-1-3}
\Gamma^{a_1 \ldots a_n} \equiv \Gamma^{[ a_1} \ldots \Gamma^{a_n ]}
\ee
is the fully antisymmetrized product of $n$ gamma matrices. One may define
\be
\label{e-b-1-4}
\Gamma = (-i)^{\frac{D-2}{2}} \Gamma^{0} \ldots \Gamma^{D-1},
\ee
which is easily seen to satisfy
\be
\label{e-b-1-5}
\Gamma^2 = 1 \qquad,\qquad \{ \Gamma , \Gamma^a \} = 0.
\ee
In the case of an {\it odd} number of dimensions, $D=2k+3$, the $D$--dimensional Clifford algebra is satisfied by the $2k+2$ matrices $\Gamma^{0}, \ldots ,\Gamma^{2k+1}$ of the even-dimensional case {\it plus} $\Gamma^{D-1} \equiv \pm \Gamma$.

The above definitions hold for flat space (or in an orthonormal frame in curved space). Curved-space gamma matrices are obtained from flat-space ones by raising indices with the inverse vielbein
\be
\label{e-b-1-6}
\Gamma^\mu = e^\mu_a \Gamma^a,
\ee
and satisfy the Clifford algebra
\be
\label{e-b-1-7a}
\{ \Gamma^\mu,\Gamma^\nu \} = 2 g^{\mu\nu}.
\ee

\subsection{Gamma matrix identities}
\label{sec-b-1-2}

From the Clifford algebra, one can derive numerous identities involving products or contractions of gamma matrices. Since these are indispensable when constructing supergravity theories, we will give a list of the most important ones below.

\begin{enumerate}
\item{\bf Duality relations.} Using the Clifford algebra (\ref{e-b-1-1}), the properties of the $\epsilon$--tensor and the definition (\ref{e-b-1-4}) of $\Gamma$, it is easily shown that an antisymmetrized product of $n$ gamma matrices can be expressed in terms of products of $D-n$ gamma matrices according to
\be
\label{e-b-1-7}
\Gamma^{a_1 \ldots a_n} =
  \begin{cases}
    \frac{ - \ii^k (-1)^{(D-n)(D-n+1)/2} }{(D-n)!} \epsilon^{a_1 \ldots a_n a_{n+1} \ldots a_D} 
     \Gamma_{a_{n+1} \ldots a_D} \Gamma, & D = 2k+2, \\
    \frac{ - \ii^k (-1)^{(D-n)(D-n-1)/2} }{(D-n)!} \epsilon^{a_1 \ldots a_n a_{n+1} \ldots a_D} 
     \Gamma_{a_{n+1} \ldots a_D}, & D =2k+3.
  \end{cases}
\ee
\item{\bf Gamma matrix products.} Since the antisymmetrized matrices in (\ref{e-b-1-2}) form a basis, any gamma-matrix product can be expressed as a linear combination of such matrices. Using the Clifford algebra (\ref{e-b-1-1}), one can easily establish the relations
\bea
\label{e-b-1-8}
\Gamma^a \Gamma^{b_1 \ldots b_n} &=& \Gamma^{a b_1 \ldots b_n} + n \eta^{a [ b_1} \Gamma^{b_2 \ldots b_n]} \nn\\
\Gamma^{b_1 \ldots b_n} \Gamma^a &=& (-1)^n \Gamma^{a b_1 \ldots b_n} + (-1)^{n+1} n \eta^{a [ b_1} \Gamma^{b_2 \ldots b_n]},
\eea
from which it also follows that
\be
\label{e-b-1-9}
[ \Gamma^a , \Gamma^{b_1 \ldots b_n} ] =
  \begin{cases}
    2 n \eta^{a [ b_1} \Gamma^{b_2 \ldots b_n]} & n = 0 \mod 2, \\
    2 \Gamma^{a b_1 \ldots b_n} & n = 1 \mod 2.
  \end{cases}
\ee
Then, by repeated application of (\ref{e-b-1-8}), one can derive the following identities
\bea
\label{e-b-1-10}
\Gamma^a \Gamma^{bc} &=& \Gamma^{a b c} + \eta^{a b} \Gamma^c - \eta^{a c} \Gamma^b, \nn\\
\Gamma^{bc} \Gamma^a &=& \Gamma^{a b c} - \eta^{a b} \Gamma^c + \eta^{a c} \Gamma^b, \nn\\
\Gamma^a \Gamma^{bcd} &=& \Gamma^{a b c d} + \eta^{a b} \Gamma^{c d} - \eta^{a c} \Gamma^{b d} + \eta^{a d} \Gamma^{b c}, \nn\\
\Gamma^{bcd} \Gamma^a &=& - \Gamma^{a b c d} + \eta^{a b} \Gamma^{c d} - \eta^{a c} \Gamma^{b d} + \eta^{a d} \Gamma^{b c}, \nn\\
\Gamma^{ab} \Gamma^{cd} &=& \Gamma^{a b c d} + \eta^{a d} \Gamma^{b c} - \eta^{a c} \Gamma^{b d} + \eta^{b c} \Gamma^{a d} - \eta^{b d} \Gamma^{a c} + \eta^{a d} \eta^{b c} - \eta^{a c} \eta^{b d}, \nn\\
\Gamma^a \Gamma^{bcde} &=& \Gamma^{a b c d e} + \eta^{a b} \Gamma^{c d e} - \eta^{a c} \Gamma^{b d e} + \eta^{a d} \Gamma^{b c e} - \eta^{a e} \Gamma^{b c d}, \nn\\
\Gamma^{bcde} \Gamma^a &=& \Gamma^{a b c d e} - \eta^{a b} \Gamma^{c d e} + \eta^{a c} \Gamma^{b d e} - \eta^{a d} \Gamma^{b c e} + \eta^{a e} \Gamma^{b c d}, \nn\\
\Gamma^{ab} \Gamma^{cde} &=& \Gamma^{a b c d e} - \eta^{a c} \Gamma^{b d e} + \eta^{a d} \Gamma^{b c e} - \eta^{a e} \Gamma^{b c d} + \eta^{b c} \Gamma^{a d e} - \eta^{b d} \Gamma^{a c e} + \eta^{b e} \Gamma^{a c d} \nn\\
&& + ( \eta^{a d} \eta^{b c} - \eta^{a c} \eta^{b d} ) \Gamma^e - ( \eta^{a e} \eta^{b c} - \eta^{a c} \eta^{b e} ) \Gamma^d + ( \eta^{a e} \eta^{b d} - \eta^{a d} \eta^{b e} ) \Gamma^c, \nn\\
\Gamma^{cde} \Gamma^{ab} &=& \Gamma^{a b c d e} + \eta^{a c} \Gamma^{b d e} - \eta^{a d} \Gamma^{b c e} + \eta^{a e} \Gamma^{b c d} - \eta^{b c} \Gamma^{a d e} + \eta^{b d} \Gamma^{a c e} - \eta^{b e} \Gamma^{a c d} \nn\\
&& + ( \eta^{a d} \eta^{b c} - \eta^{a c} \eta^{b d} ) \Gamma^e - ( \eta^{a e} \eta^{b c} - \eta^{a c} \eta^{b e} ) \Gamma^d + ( \eta^{a e} \eta^{b d} - \eta^{a d} \eta^{b e} ) \Gamma^c. 
\eea

\item{\bf Gamma matrix contractions.} One often encounters products of the above type where two or more indices are contracted. The resulting expressions are found by contracting the appropriate expressions on the RHS of (\ref{e-b-1-10}). One finds
\bea
\label{e-b-1-11}
\Gamma_a \Gamma^a &=& D, \nn\\
\Gamma_a \Gamma^b \Gamma^a &=& (2-D) \Gamma^b, \nn\\
\Gamma_a \Gamma^{bc} \Gamma^a &=& (D-4) \Gamma^{bc}, \nn\\
\Gamma_a \Gamma^{bcd} \Gamma^a &=& (6-D) \Gamma^{bcd}, \nn\\
\Gamma_a \Gamma^{ab} = \Gamma^{ba} \Gamma_a &=& (D-1) \Gamma^b, \nn\\
\Gamma_a \Gamma^{abc} = \Gamma^{bca} \Gamma_a &=& (D-2) \Gamma^{bc}, \nn\\
\Gamma_a \Gamma^{abcd} = \Gamma^{bcda} \Gamma_a &=& (D-3) \Gamma^{bcd}, \nn\\
\Gamma_{ba} \Gamma^{ac} &=& (D-2) \Gamma_b^{\phantom{b} c} + (D-1) \delta_b^{\phantom{b} c}, \nn\\
\Gamma_{ab} \Gamma^{ab} &=& - D (D-1), \nn\\
\Gamma_{ab} \Gamma^{abc} = \Gamma^{cab} \Gamma_{ab} &=& - (D-1)(D-2) \Gamma^c.
\eea

\end{enumerate}

\section{Representations of the Clifford Algebra}
\label{sec-b-2}

\subsection{Spinors of $SO(D-1,1)$}
\label{sec-b-2-1}

Let us now construct the representations of the Clifford algebra (\ref{e-b-1-1}). To this end, it is convenient to group the gamma matrices into lowering and raising operators as follows
\bea
\label{e-b-2-1}
\gamma^0 = \frac{1}{2} ( - \Gamma^0 + \Gamma^1 ) &,&\quad \gamma^{0 \dag} = \frac{1}{2} ( \Gamma^0 + \Gamma^1 )\nn\\
\gamma^i = \frac{1}{2} ( \Gamma^{2i} - i \Gamma^{2i+1} ) &,&\quad \gamma^{i \dag} = \frac{1}{2} ( \Gamma^{2i} + i \Gamma^{2i+1} ) ;\qquad i=1,\ldots,k.
\eea
The operators thus defined satisfy the algebra
\be
\label{e-b-2-2}
\{ \gamma^i , \gamma^{j \dag} \} = \delta^{ij}, \qquad,\qquad \{ \gamma^i , \gamma^j \} = \{ \gamma^{i \dag} , \gamma^{j \dag} \} = 0,
\ee
i.e. the algebra of $k+1$ uncoupled fermionic oscillators. The representations of this algebra are found in the usual way. We first define the Clifford vacuum $\psi_0 $ as the state that is annihilated by all lowering operators,
\be
\label{e-b-2-3}
\gamma^i \psi_0  = 0.
\ee
and we construct the remaining states in the representation by acting on this vacuum with the raising operators. Since each operator can act at most one time, a generic state will have the form
\be
\label{e-b-2-4}
\psi_s  = (\gamma^{k \dag})^{s_k + \frac{1}{2}} \ldots (\gamma^{0 \dag})^{s_0 + \frac{1}{2}} \psi_0 ,
\ee
where $s_i = \pm \frac{1}{2}$ and $s = (s_0,\ldots,s_k)$.

To verify that these states form a spinor representation of the Lorentz group, we introduce the matrices
\be
\label{e-b-2-5}
\Sigma^{ab} = \frac{1}{2} \Gamma^{ab}.
\ee
which, by virtue of the Clifford algebra (\ref{e-b-1-1}), satisfy the $SO(D-1,1)$ algebra
\be
\label{e-b-2-6}
[\Sigma^{ab},\Sigma^{cd}] = \eta^{ad}\Sigma^{bc} - \eta^{ac}\Sigma^{bd} - \eta^{bd}\Sigma^{ac} + \eta^{bc}\Sigma^{ad},
\ee
and can be identified with the Lorentz generators acting on the representation defined by the states (\ref{e-b-2-4}). Given these matrices, we can construct the following maximal set of commuting operators
\be
\label{e-b-2-7}
S_i = \ii^{\delta_{i,0}+1} \Sigma^{2i,2i+1} = - \frac{\ii^{\delta^{i0}+1}}{2} \Gamma^{2i} \Gamma^{2i+1} = \frac{1}{2} [ \gamma^{i \dag} , \gamma^i ] = \gamma^{i \dag} \gamma^i - \frac{1}{2}.
\ee
which are the analog of the spin operators in four dimensions. The action of $S_i$ on the state $\psi_s $ yields
\be
\label{e-b-2-8}
S_i \psi_s  = s_i \psi_s ,
\ee
and, since $s_i = \pm \frac{1}{2}$, the representation defined by $\psi_s$ is a {\it spinor} representation of the Lorentz group, called a {\it Dirac representation}. Its real dimension equals $2^{k+2} = 2^{\lfloor \frac{D}{2} \rfloor + 1}$. 

\subsection{The Weyl and Majorana conditions}
\label{sec-b-2-2}

The spinor representations thus defined are not, in general, minimal irreducible representations of the Lorentz group. As we will see, there are two types of projection which one may impose to reduce the independent components of a Dirac spinor. Their existence and their type depends on the spacetime dimension.

\subsubsection{Weyl spinors}

The first projection is the Weyl projection, which is possible in an even number of dimensions. In this case, one may easily prove that $\Gamma$ commutes with the Lorentz generators,
\be
\label{e-b-2-9}
[ \Gamma , \Sigma^{ab} ] = 0.
\ee
Therefore, the Dirac spinor representation is divided into two subspaces, classified by the eigenvalue of $\Gamma$ (called {\it chirality} and equal to $\pm 1$), both of which transform irreducibly under the Lorentz group. These are called \emph{Weyl spinors}, they are defined by
\be
\label{e-b-2-10}
\Gamma \psi_\pm = \pm \psi_\pm,
\ee
and they have real dimension $2^{\lfloor \frac{D}{2} \rfloor}$. In contrast, in the case of an odd number of dimensions where the Clifford algebra is satisfied by the $D-1$ matrices $\Gamma^{0}, \ldots ,\Gamma^{D-2}$ of the even-dimensional case plus $\Gamma^{D-1} \equiv \pm \Gamma$, one may verify that $\Gamma$ does not commute with $\Sigma^{a,D-1}$ and so the representation is irreducible and the notion of a Weyl spinor does not exist.

\subsubsection{Majorana spinors}

The second projection is the Majorana projection. To discuss it, we begin by defining the charge conjugation matrix $C$ by the requirement that the Majorana conjugate of $\psi$, $\widetilde{\psi} = \psi^T C$, satisfies the same equation as the usual Dirac adjoint $\bar{\psi} = \psi^\dag \Gamma^0$. For the case of massless spinors, this requirement easily leads to the relation
\be
\label{e-b-2-11}
\Gamma^{a T} = \alpha C \Gamma^{a} C^{-1} ;\qquad \alpha = \pm 1.
\ee
Using this relation, one can prove some important properties of $C$. First, writing $\Gamma^a = (\Gamma^{aT})^T$ and iterating (\ref{e-b-2-12}) twice, we find $\Gamma^a = ( C^{-1} C^T )^{-1} \Gamma^{a} ( C^{-1} C^T )$, which implies that $C^{-1} C^T$ is a multiple of the unit matrix, $C^T = \beta C$. Taking the transpose of this, we also find $C = \beta C^T$ and so
\be
\label{e-b-2-12}
C^T = \beta C ;\qquad \beta = \pm 1,
\ee
i.e. \emph{$C$ can be either symmetric or antisymmetric}. Second, using (\ref{e-b-2-13}) and the fact that $\Gamma^a$ are either hermitian or antihermitian, we can write $\Gamma^a = (C^* C) \Gamma^{a} (C^* C)^{-1}$ 
which tells us that $C^* C$ is a multiple of the unit matrix. Combining this with (\ref{e-b-2-12}) we find that the same must hold for $C^\dag C$. Since $C^\dag C$ is positive-definite, the proportionality constant must be greater than zero and thus we can choose our normalization so that
\be
\label{e-b-2-13}
C^\dag C = 1,
\ee
i.e. \emph{we can choose $C$ to be unitary}. Third, (\ref{e-b-2-11}) and (\ref{e-b-2-12}) imply that
\be
\label{e-b-2-14}
(C \Gamma^{a_1 \ldots a_n})^T = (-1)^{\frac{n(n-1)}{2}} \alpha^n \beta C \Gamma^{a_1 \ldots a_n}.
\ee
Using this last relation, one can determine the allowed values for $\alpha$ and $\beta$ for any spacetime dimension by counting the allowed numbers of symmetric and antisymmetric Clifford algebra matrices according to (\ref{e-b-2-14}) and comparing them to their expected values. The results are shown on Table B.1. 
\begin{table}[!t]
\label{t-b-1}
\begin{center}
\begin{tabular}{|r||l|}
\hline
Dimension & $(\alpha,\beta)$ \\
\hline
$0 \mod 8$ & $(+1,+1),(-1,+1)$ \\
$1 \mod 8$ & $(+1,+1)$ \\
$2 \mod 8$ & $(+1,+1),(-1,-1)$ \\
$3 \mod 8$ & $(-1,-1)$ \\
$4 \mod 8$ & $(+1,-1),(-1,-1)$ \\
$5 \mod 8$ & $(+1,-1)$ \\
$6 \mod 8$ & $(+1,-1),(-1,+1)$ \\
$7 \mod 8$ & $(-1,+1)$ \\
\hline
\end{tabular}
\end{center}
\caption{Values of $\alpha$ and $\beta$ for various dimensions.}
\end{table}

After these preliminaries, let us examine the complex conjugation properties of gamma matrices and spinors. Using (\ref{e-b-2-11}) and the hermiticity properties of the gamma matrices, we find the complex conjugation relation
\be
\label{e-b-2-15}
\Gamma^{a *} = - \alpha B \Gamma^a B^{-1} ;\qquad B = C \Gamma^0.
\ee
Using (\ref{e-b-2-13}), we easily see that the matrix $B$ is unitary. It also satisfies
\be
\label{e-b-2-16}
B^* B = \alpha \beta
\ee
Now, using (\ref{e-b-2-15}), we see that the complex conjugation of the Lorentz generators $\Sigma^{ab}$ gives
\be
\label{e-b-2-17}
\Sigma^{ab*} = B \Sigma^{ab} B^{-1}.
\ee
This relation has important consequences. Consider a Lorentz transformation of a spinor $\psi$, $\delta \psi = - \frac{1}{2} \lambda_{ab} \Sigma^{ab} \psi$. Then, we easily verify that $B \psi$ transforms in exactly the same way as $\psi^*$,
\bea
\label{e-b-2-18}
\delta ( B \psi ) &=& - \frac{1}{2} \lambda_{ab} B \Sigma^{ab} \psi = - \frac{1}{2} \lambda_{ab} B \Sigma^{ab} B^{-1} ( B \psi ), \nn\\
\delta \psi^* &=& - \frac{1}{2} \lambda_{ab} \Sigma^{ab*} \psi^* = - \frac{1}{2} \lambda_{ab} B \Sigma^{ab} B^{-1} \psi^*.
\eea
Therefore, we can investigate whether we can impose a reality condition on $\psi$ by setting $\psi^*$ and $B \psi$ proportional to each other. This is the Majorana condition
\be
\label{e-b-2-19}
\psi^* = \gamma B \psi,
\ee
where $\gamma$ is another constant. Consistency of this relation, $(\psi^*)^*=\psi$, requires $| \gamma |^2 B^* B = 1$ and, since $B^* B = \pm 1$, we must have $| \gamma |^2 = 1$. So, the Majorana condition is possible when 
\be
\label{e-b-2-20}
\alpha \beta = +1.
\ee
The spinors satisfying the Majorana condition with $\alpha = -1$ are called \emph{Majorana spinors} while those that satisfy it with $\alpha = +1$ are called \emph{pseudoMajorana spinors}. Consulting the previous table, we see that Majorana spinors exist for $D=2,3,4 \mod 8$ while pseudoMajorana spinors exist for $D=0,1,2 \mod 8$.

In the cases where $\alpha \beta = -1$, one may still impose a reality condition on spinors if we have extended supersymmetry. In such a case, we have $N$ spinors, labelled by an index $A$, and we may replace (\ref{e-b-2-19}) by the modified Majorana condition
\be
\label{e-b-2-21}
\psi_i^* \equiv (\psi^A)^* = B \Omega_{AB} \psi^B,
\ee
where $\Omega$ is a unitary $N \times N$ matrix. Consistency of this condition yields the constraint $\Omega \Omega^* = - 1$ which, combined with unitarity, implies that
\be
\label{e-b-2-22}
\Omega^T = - \Omega.
\ee
Hence, $\Omega$ must be simultaneously antisymmetric and unitary, which in turns requires that the number of spinors should be even. Now, the existence of a multiplet of spinors suggests that there is an internal symmetry group rotating the spinors as
\be
\label{e-b-2-23}
\delta \psi^a = M^A_{\phantom{A} B} \psi^B.
\ee
where $M$ is an antihermitian $N \times N$ matrix, as required by unitarity. Using (\ref{e-b-2-21}) and (\ref{e-b-2-23}) to compute the corresponding transformation of $\psi_i^*$, we find the two alternative expressions $\delta \psi_A^* = B \Omega_{AB} \delta \psi^B = B \Omega_{AB} M^b_{\phantom{B} C} \psi^C$ and $\delta \psi_A^* = (M^A_{\phantom{A} B} \psi^B)^* = - B M^B_{\phantom{B} A} \Omega_{BC} \psi^C$. In order for the symmetry (\ref{e-b-2-23}) to respect the modified Majorana condition, the two expressions must be equal, i.e. we must have
\be
\label{e-b-2-24}
\Omega M + M^T \Omega = 0.
\ee
The simplest choice for $\Omega$ is the symplectic metric
\be
\label{e-b-2-25}
\Omega = \left( \begin{array}{cc} 0 & \mathbf{1} \\ - \mathbf{1} & 0 \end{array} \right),
\ee
and so $M$ must be a matrix that preserves this metric. These matrices are just the generators of the symplectic group $\mathrm{USp}(2 N)$. For that reason, condition (\ref{e-b-2-21}) is referred to as the symplectic Majorana condition and the spinors satisfying it are called \emph{symplectic Majorana spinors} if $\alpha=-1$ and \emph{symplectic pseudoMajorana spinors} if $\alpha=+1$. From the table, we see that symplectic Majorana spinors exist for $D=0,6,7 \mod 8$ while symplectic pseudoMajorana spinors exist in $D=4,5,6 \mod 8$.

\subsubsection{Majorana-Weyl spinors}

In even dimensions, one could ask whether the Weyl and Majorana conditions may be simultaneously imposed as independent conditions, further reducing the components of a spinor. In order for this to be possible, the two types of projections must be consistent with each other. To find the consistency condition, we use (\ref{e-b-2-10}) and (\ref{e-b-2-19}) and we find the two alternative expressions $(\Gamma \psi_\pm)^* = \gamma B \Gamma \psi_\pm$ and $(\Gamma \psi_\pm)^* = \gamma \Gamma^* B \psi_\pm$. These must be equal and so we must have $\Gamma^* = B \Gamma B^{-1}$. On the other hand, Eqs. (\ref{e-b-1-4}) and (\ref{e-b-2-15}) imply that, for even dimensions
\be
\label{e-b-2-26}
\Gamma^* = (-1)^{\frac{D-2}{2}} B \Gamma B^{-1}.
\ee
So, we see that the two conditions may be simultaneously imposed if $D=2 \mod 4$. So, in $D=2 \mod 8$, where both Majorana and Weyl conditions are available, there exist \emph{Majorana-Weyl spinors}. In a similar manner, one finds that in $D=6 \mod 8$ there exist \emph{symplectic Majorana-Weyl spinors}.

\subsubsection{Irreducible spinors}

\begin{table}[!t]
\label{t-b-2}
\begin{center}
\begin{tabular}{|r||l|l|}
\hline
$D$ & Spinor type & $d(D)$\\
\hline
$1$ & $\textrm{pM}$ & 1\\
$2$ & $\textrm{MW},\textrm{pMW}$ & 1\\
$3$ & $\textrm{M}$ & 2\\
$4$ & $\textrm{W},\textrm{M},\textrm{pSM}$ & 4\\
$5$ & $\textrm{D}=2 \times \textrm{pSM}$ & 8\\
$6$ & $\textrm{W}=2 \times \textrm{SMW} = 2 \times \textrm{pSMW}$ & 8\\
$7$ & $\textrm{D}=2 \times \textrm{SM}$ & 16\\
$8$ & $\textrm{W},\textrm{pM},\textrm{SM}$ & 16\\
$9$ & $\textrm{pM}$ & 16\\
$10$ & $\textrm{MW},\textrm{pMW}$ & 16\\
$11$ & $\textrm{M}$ & 32\\
\hline
\end{tabular}
\end{center}
\caption{Types and dimensions of minimal spinors in diverse dimensions.}
\end{table}

We are now in a position to summarize the types of irreducible spinors in various dimensions and their dimensions. The latter are immediately found by recalling that the real dimension of a Dirac spinor is $2^{\lfloor \frac{D}{2} \rfloor +1}$. Since this is reduced by half for a Weyl or Majorana spinor, we find that the real dimension of the minimal spinor for given $D$ is
\be
\label{e-b-2-27}
d (D) = 2^{\lfloor \frac{D}{2} \rfloor + 1 - W - M},
\ee
where $W$ ($M$) equals $1$ if the spinor is subject to a Weyl (Majorana-type) condition and $0$ otherwise. The minimal spinors available in diverse dimensions and their real dimensions are shown on Table B.2.
Here D, W, M, pM and SM refer to Dirac, Weyl, Majorana, pseudoMajorana and symplectic Majorana spinors respectively. Also, the entries on the table containing symplectic Majorana spinors describe the decomposition of the minimal spinor into components related by the symplectic Majorana condition. E.g. in $D=6$ Minkowski space, the minimal spinor is a Weyl spinor with 8 real components which can also be written as a pair of symplectic (pseudo)Majorana-Weyl spinors with four real components each, related by (\ref{e-b-2-21}).

\section{Six-Dimensional Spinors} 

Since the bulk of this thesis is devoted to six-dimensional supergravities, we find it useful to summarize some of the properties of gamma matrices and spinors specific to six dimensions. These properties are extensively used in Chapter 5.

In six dimensions, the gamma-matrix duality relation () takes the form
\be
\label{e-b-3-1}
\Gamma^{a_1 \ldots a_n} = \frac{(-1)^{\lfloor n/2 \rfloor}}{(6-n)!} \epsilon^{a_1 \ldots a_n a_{n+1} \ldots a_D} \Gamma_{a_{n+1} \ldots a_6} \Gamma_7.
\ee
Also, the first four of the gamma-matrix contraction identities () are given by
\be
\label{e-b-3-2}
\Gamma_a \Gamma^a = 6 ,\quad \Gamma_a \Gamma^b \Gamma^a = -4 \Gamma^b ,\quad \Gamma_a \Gamma^{bc} \Gamma^a = 2 \Gamma^{bc} ,\quad \Gamma_a \Gamma^{bcd} \Gamma^a =0.
\ee

In six dimensions, the minimal spinor is symplectic Majorana-Weyl. In general, such spinors form a $2n$-dimensional multiplet $\psi^a$, $a=1,\ldots,2n$, of $\mathrm{USp}(2n)$. In the $\mathrm{USp}(2)$ case, the indices of such spinors are lowered (raised) by the $\mathrm{USp}(2)$ invariant tensor $\epsilon_{AB}$ ($\epsilon^{AB}$) according to the ``NW-SE'' convention, i.e.
\be
\label{e-b-3-3}
\chi_A = \chi^B \epsilon_{BA} ,\qquad \chi^A = \epsilon^{AB} \chi_B ;\qquad \epsilon_{12} = \epsilon^{12} = - \epsilon^{21} = - \epsilon^{21} = 1,
\ee
and the inner product of two such spinors is defined as
\be
\label{e-b-3-4}
\bar{\chi} \psi \equiv \bar{\chi}^A \psi_A =  \bar{\chi}^A \psi^B \epsilon_{BA}.
\ee
Similar identities hold in the general $\mathrm{USp}(2n)$ case, with $\epsilon_{AB}$ replaced by the $\mathrm{USp}(2n)$ invariant tensor $\Omega_{ab}$.

Finally, six-dimensional spinors satisfy the Majorana-flip property
\be
\label{e-b-3-5}
\bar{\chi}^A \Gamma^{a_1 \ldots a_n} \psi^B = (-1)^{n+1} \bar{\psi}^B \Gamma^{a_n \ldots a_1} \chi^A
\ee
which, after contraction with $\epsilon_{AB}$ gives
\be
\label{e-b-3-6}
\bar{\chi} \Gamma^{a_1 \ldots a_n} \psi = (-1)^n \bar{\psi} \Gamma^{a_n \ldots a_1} \chi.
\ee

\chapter{Characteristic Classes and Index Theorems}
\label{app-c}
\thispagestyle{empty}

\section{Characteristic Classes}
\label{sec-c-1}

\subsection{Generalities}
\label{sec-c-1-1}
 
Consider a set $\{ \alpha_i \}$, $i=1,\ldots,n$ of $k \times k$ complex matrices taking values in the Lie algebra of a subgroup $G$ of $\mathrm{GL}(k,\mathbb{C})$. A symmetric polynomial is a $\mathbb{C}$--valued linear function of the $\alpha_i$'s which is symmetric under interchange of any two $\alpha_i$'s,
\be
\label{e-c-1-1}
P( \alpha_1, \ldots, \alpha_i, \ldots, \alpha_j, \ldots, \alpha_n) = P( \alpha_1, \ldots, \alpha_j, \ldots, \alpha_i, \ldots, \alpha_n ).
\ee 
An invariant or characteristic polynomial is a symmetric polynomial which is in addition invariant under a $G$--transformation of all the $\alpha_i$'s,
\be
\label{e-c-1-2}
P( \alpha^g_1, \ldots, \alpha^g_n ) \equiv P( g^{-1} \alpha_1 g, \ldots, g^{-1} \alpha_n g ) = P( \alpha_1, \ldots, \alpha_n )
\ee
These definitions can be extended to the case where each $\alpha_i$ is also a $p_i$--form on a manifold $M$,
\be
\label{e-c-1-3}
\alpha_i = \alpha_{i,\mu_1\ldots\mu_{p_i}} dx^{\mu_1} \ldots dx^{\mu_{p_i}},
\ee
with the understanding that 
\be
\label{e-c-1-4}
P( \alpha_1, \ldots, \alpha_n ) \equiv P( \alpha_{1,\mu_1\ldots\mu_{p_1}}, \ldots, \alpha_{n,\mu_n\ldots\mu_{p_n}} ) dx^{\mu_1} \ldots dx^{\mu_{p_1}} \ldots dx^{\nu_1} \ldots dx^{\nu_{p_n}}.
\ee 
The canonical example of an invariant polynomial is the symmetrized trace
\be
\label{e-c-1-5}
P( \alpha_1, \ldots, \alpha_n ) = \Str(\alpha_1 \ldots \alpha_n ) \equiv \frac{1}{n!} \sum_P \tr( \alpha_{P(1)} \ldots \alpha_{P(n)} ).
\ee
where, in the case of $p_i$--forms, the symmetrization is only carried out in the matrix part with the form part fixed. In the case where all $\alpha_i$'s are equal, $\alpha_i=\alpha$, we define an invariant polynomial of degree $n$ as
\be
\label{e-c-1-6}
P_n ( \alpha ) = P( \alpha^n ) \equiv P( \alpha, \ldots, \alpha ),
\ee
and the canonical example is the ordinary trace
\be
\label{e-c-1-7}
P_n ( \alpha ) = \tr \alpha^n.
\ee

Let us now take the matrix-valued form $\alpha$ to be the curvature 2--form $\Omega = \dd \omega + \omega^2$ associated with a connection $\omega$ on a fiber bundle $E$ over $M$; the pair $(\omega,\Omega)$ will be understood as $(A,F)$ in the case of a gauge theory and as $(\omega,R)$ in the case of gravity. An invariant polynomial $P_n (\Omega)$ is a $2n$--form on $M$. This polynomial has the following very important properties. First, it is \emph{closed}
\be
\label{e-c-1-8}
\dd P_n (\Omega) = 0,
\ee
and, second, \emph{its integrals are topological invariants}. The second property is equivalent to the statement that the difference of two polynomials $P_n (\Omega_1)$ and $P_n (\Omega_0)$ is exact, i.e.
\be
\label{e-c-1-9}
P_n (\Omega_1) - P_n (\Omega_0) = \dd Q_{2n-1} (\omega_1,\omega_0),
\ee
for some $(2n-1)$--form $Q_{2n-1}(\omega_1,\omega_0)$. The first property implies that $P_n(\Omega)$ defines a cohomology class on $M$. The second property further implies that this cohomology class is independent of the connection. Such a cohomology class is called a \emph{characteristic class}.

The $(2n-1)$--form $Q_{2n-1}(\omega_1,\omega_0)$ appearing in (\ref{e-c-1-9}) is explicitly given by
\be
\label{e-c-1-10}
Q_{2n-1} (\omega_1,\omega_0) = n \int_0^1 \dd t P(\omega_1-\omega_0,\Omega_t^{n-1}),
\ee
where
\be
\label{e-c-1-11}
\omega_t = \omega_0 + t (\omega_1 - \omega_0) ,\qquad \Omega_t = \dd \omega_t + \omega_t^2.
\ee
On a local chart where the bundle is trivial, we can set $\omega_0 = \Omega_0 = 0$. Then (\ref{e-c-1-9}) turns into
\be
\label{e-c-1-12}
P_n(\Omega) = \dd Q_{2n-1}(\omega,\Omega)
\ee
with
\be
\label{e-c-1-13}
Q_{2n-1} (\omega,\Omega) = n \int_0^1 \mathrm{d} t t^{n-1} P( \omega, ( \dd \omega + t \omega^2)^{n-1} ).
\ee
The form $Q_{2n-1} (\omega,\Omega)$ thus defined is called the \emph{Chern-Simons form} of $P_n(\Omega)$.

Another quantity of interest is the $Q^1_{2n-2} (\delta \omega,\omega,\Omega)$, called the \emph{descent} of $Q_{2n-1} (\omega,\Omega)$ and defined by
\be
\label{e-c-1-14}
Q^1_{2n-2} (\delta \omega,\omega,\Omega) = Q_{2n-1} (\omega + \delta \omega,\Omega) - Q_{2n-1} (\omega,\Omega),
\ee
i.e. as the variation of $Q_{2n-1} (\omega,\Omega)$ when we vary $\omega$ keeping $\Omega$ fixed. An explicit expression for this form can be found by using the so-called Cartan homotopy operator, and is given by
\be
\label{e-c-1-15}
Q^1_{2n-2} (\delta \omega,\omega,\Omega) = n(n-1) \int_0^1 \mathrm{d} t t^{n-2} (1-t) P( \delta \omega, \dd ( \omega , ( \dd \omega + t \omega^2 )^{n-2} ) ).
\ee

\subsection{Characteristic classes of complex vector bundles}
\label{sec-c-1-2}

For a complex vector bundle the connection and curvature, denoted by $A$ and $F$ respectively, take their values in $GL(k,\mathbb{C})$. By a suitable $GL(k,\mathbb{C})$ transformation, $F$ may be brought to the diagonal form
\be
\label{e-c-1-16}
\frac{\ii F}{2\pi} \to \diag(x_1,x_2,\ldots,x_k).
\ee 
The characteristic classes relevant for complex vector bundles are the following.

\subsubsection{Chern class}

The {\it Chern class} of a complex vector bundle is defined by
\be
\label{e-c-1-17}
\ccl(F) = \det \left( 1 + \frac{\ii F}{2\pi} \right),
\ee
The term of order $F^i$ in the expansion is called the {\it $i$--th Chern class} of $F$ and is denoted by $\ccl_i(F)$. A more explicit form of this relation in terms of the $x_i$'s is provided by the so-called \emph{splitting principle} and is given by
\be
\label{e-c-1-18}
\ccl(F) = \prod_{i=1}^k (1 + x_i),
\ee 
which provides an efficient method for computing Chern classes.

\subsubsection{Chern characters}

The {\it Chern character} of a complex vector bundle is defined by
\be
\label{e-c-1-19}
\cch(F) = \tr \exp \left( \frac{\ii F}{2\pi} \right),
\ee
The $i$-th term of the expansion is called the {\it $i$--th Chern character} of $F$
\be
\label{e-c-1-20}
\cch_i(F) = \frac{1}{i!} \tr \left( \frac{\ii F}{2\pi} \right)^i.
\ee
In terms of the $x_i$'s, we have
\be
\label{e-c-1-21}
\cch(F) = \sum_{j=1}^k e^{x_j} ,\qquad
\cch_i(F) = \frac{1}{i!} \sum_{j=1}^k (x_j)^i.
\ee

\subsection{Characteristic classes of real vector bundles}
\label{sec-c-1-3}

For a real vector bundle, the connection and curvature, denoted by $\omega$ and $R$ respectively, are real antisymmetric matrices of $GL(k,\mathbb{R})$. Now, the curvature $F$ cannot be diagonalized but it can be brought to a skew-diagonal form. In the cases of interest $k$ is an even integer, $k=2r$, and we can write
\be
\label{e-c-1-22}
\frac{R}{2\pi} \to \left( 
\begin{array}{ccccc}
0 & x_1 \\
-x_1 & 0 \\
&& \ddots \\
&&& 0 & x_r \\
&&& -x_r & 0 \\
\end{array}
\right).
\ee
The characteristic classes relevant for real vector bundles are the following.

\subsubsection{Pontrjagin classes}

The {\it total Pontrjagin class} of $R$ is defined by
\be
\label{e-c-1-23}
p(R) = \det \left( 1 + \frac{R}{2\pi} \right),
\ee
Since the curvature $R$ is skew symmetric, this expression is even in $R$. Hence, the expansion of $p(R)$ in powers of $R$ has the form
\be
\label{e-c-1-24}
p(R) = \sum_{i=0}^r p_i (R) ,
\ee
where $p_i (R)$ is of order $2i$ in $R$. Obviously, the series terminates at $i=r$. The quantity $p_i (R)$ is called the {\it $i$--th Pontrjagin class} of the bundle. In terms of the $x_i$'s, we have
\be
\label{e-c-1-25}
p(R) = \prod_{i=0}^r ( 1 + x_i^2 ) ,\qquad p_i(R) = \sum_{j_1 < j_2 \ldots < j_i}^r x_{j_1}^2 \ldots x_{j_i}^2.
\ee

\subsubsection{$A$-roof genus}

The {\it $A$--roof genus} (or Dirac genus) is defined by
\be
\label{e-c-1-26}
\widehat{A}(R) = \prod_{i=1}^r \frac{x_i/2}{\sinh(x_i/2)} .
\ee
Its expansion in terms of Pontrjagin classes reads
\be
\label{e-c-1-27}
\widehat{A}(R) = 1 - \frac{1}{24} p_1 + \frac{1}{5760} (7 p_1^2 - 4 p_2) + \frac{1}{967680} ( -31 p_1^3 + 44 p_1 p_2 - 16 p_3) + \ldots,
\ee
while the first terms of its expansion in terms of curvature invariants has the form
\bea
\label{e-c-1-28}
\widehat{A}(R) &=& 1 + \frac{1}{(4\pi)^2} \frac{1}{12} \tr R^2 + \frac{1}{(4\pi)^4} \left[ \frac{1}{360} \tr R^4 + \frac{1}{288} (\tr R^2)^2 \right] \nonumber\\ &&+ \frac{1}{(4\pi)^6} \left[ \frac{1}{5760} \tr R^6 + \frac{1}{4320} \tr R^4 \tr R^2 + \frac{1}{10368} (\tr R^2)^3 \right] + \ldots .
\eea

\subsubsection{Hirzebruch polynomial}

The {\it Hirzebruch polynomial} (or $L$--polynomial) is defined by
\be
\label{e-c-1-29}
L(R) = \prod_{i=1}^r \frac{x_i}{\tanh x_i}.
\ee
Its expansion in terms of Pontrjagin classes reads
\be
\label{e-c-1-30}
L(R) = 1 + \frac{1}{3} p_1 + \frac{1}{45} (- p_1^2 + 7 p_2) + \frac{1}{945} ( 2 p_1^3 - 13 p_1 p_2 + 62 p_3) + \ldots,
\ee
while the first terms of its expansion in terms of curvature invariants has the form
\bea
\label{e-c-1-31}
L(R) &=& 1 - \frac{1}{(2\pi)^2} \frac{1}{6} \tr R^2 + \frac{1}{(2\pi)^4} \left[ - \frac{7}{180} \tr R^4 + \frac{1}{72} (\tr R^2)^2 \right] \nonumber\\ &&+ \frac{1}{(2\pi)^6} \left[ - \frac{31}{2835} \tr R^6 + \frac{7}{1080} \tr R^4 \tr R^2 - \frac{1}{1296} (\tr R^2)^3 \right] + \ldots
\eea

\section{The Descent Equations}
\label{sec-c-2}

An important set of relations between Chern-Simons forms, their descents and related quantities can be derived through the following formal construction. We consider the space of Lie-algebra-valued gauge transformations $U(\theta^\alpha,x)$, which is characterized by a set of parameters $\{ \theta^\alpha \}$. These parameters can be thought of as extra coordinates which can be ``wedged" with the spacetime coordinates $\{ x^\mu \}$ in the usual way. Accordingly, we can define an appropriate exterior derivative $\sss$ along the gauge orbits through
\be
\label{e-c-1-32}
\sss U = \frac{\partial U}{\partial \theta^\alpha} d \theta^\alpha.
\ee
By definition, this operator anticommutes with $\dd$,
\be
\label{e-c-1-33}
\{ \sss, \dd \} U =  \sss \frac{\partial U}{\partial x^\mu} \dd x^\mu + \dd \frac{\partial U}{\partial \theta^\alpha} d \theta^\alpha = \frac{\partial^2 U}{\partial x^\mu \partial \theta^\alpha} (\dd x^\mu \dd \theta^\alpha + \dd \theta^\alpha \dd x^\mu) = 0,
\ee
it is nilpotent,
\be
\label{e-c-1-34}
\sss^2 U = 0
\ee
and it satisfies the usual rules of exterior differentiation. This construction enables us to define generalized ``forms" in the extended space spanned by $\{ x^\mu , \theta^\alpha \}$. The operator $\sss$ increases the degree of forms in $\theta$--space by one as does $\dd$ for the degree of forms in $x$-space. Also, the operator 
\be
\label{e-c-1-35}
\Delta = \dd + \sss
\ee
can be thought of as the analog of $\dd$ in the extended space and is also nilpotent. 

Now, let $\omega$ and $\Omega$ be a reference connection and its curvature, independent of $\{ \theta^\alpha \}$. We define a new connection $\hat{\omega}$ and its associated curvature $\hat{\Omega}$ through the gauge transformation 
\be
\label{e-c-1-36}
\hat{\omega} = U^{-1} \omega U + U^{-1} \dd U ,\qquad \hat{\Omega} = U^{-1} \Omega U
\ee
The new connection depends on $\{ \theta^\alpha \}$ through the group element $U$ and thus it makes sense to speak of the action of $\sss$ on it. From the above, we easily find 
\bea
\label{e-c-1-37}
\sss \hat{\omega} = - U^{-1} \sss U \hat{\omega} - \hat{\omega} U^{-1} \sss U - \dd (U^{-1} \sss U).
\eea
If we define 
\be
\label{e-c-1-38}
\hat{v} = U^{-1} \sss U
\ee
(which is a 1-form in $\theta$--space and a scalar in $x$--space), we can write (\ref{e-c-1-37}) as
\be
\label{e-c-1-39}
\sss \hat{\omega} = - \dd \hat{v} - \{ \hat{\omega},\hat{v} \}.
\ee
Similarly, we can easily find that the action of $\sss$ on $\hat{\Omega}$ and $\hat{v}$ is given by
\be
\label{e-c-1-40}
\sss \hat{\Omega} = [\hat{\Omega},\hat{v}] ,\qquad \sss \hat{v} = - \hat{v}^2.
\ee
Therefore, we find that $\sss$ implements a gauge transformation of $\hat{\omega}$ parameterized by the anticommuting parameter $\hat{v}$, which itself transforms under $\sss$. This is just the familiar BRS transformation of Yang-Mills theory. For that reason, $\sss$ is referred to as the \emph{BRS operator} and $\hat{v}$ is called the \emph{ghost}. From the above, we can easily verify that $s$ is nilpotent on $\omega$, $\Omega$ and $v$, as expected. 

To derive the relations of interest, we introduce yet another connection by
\be
\label{e-c-1-42}
\bar{\omega} = \hat{\omega} + \hat{v},
\ee
which is a 1--form in the extended space. The associated field strength is defined by using the operator $\Delta$ and is easily seen to be equal to $\hat{\Omega}$,
\be
\label{e-c-1-43}
\bar{\Omega} = \Delta \bar{\omega} + \bar{\omega}^2 = d \hat{\omega} + \hat{\omega}^2 + (\sss \hat{\omega} + \dd \hat{v} +\{ \hat{\omega},\hat{v}\}) + (\sss \hat{v} + \hat{v}^2) = \hat{\Omega},
\ee
since both terms inside parentheses vanish. This tells us that, for any invariant polynomial $P_n$, we have $P_n(\hat{\Omega}) = P_n(\bar{\Omega})$. Expressing both sides of this equation through Chern-Simons forms and using (\ref{e-c-1-43}), we find $\dd Q_{2n-1}(\hat{\omega},\hat{\Omega}) = \Delta Q_{2n+1}(\bar{\omega},\hat{\Omega})$, that is,
\be
\label{e-c-1-46}
\dd Q_{2n-1}(\hat{\omega},\hat{\Omega}) = (\dd + \sss) Q_{2n-1}(\hat{\omega}+\hat{v},\hat{\Omega}).
\ee
Now, the form $Q_{2n-1}(\hat{\omega}+\hat{v},\hat{\Omega})$ can be expanded in powers of $\hat{v}$ as follows
\be
\label{e-c-1-47}
Q_{2n-1}(\hat{\omega}+\hat{v},\hat{\Omega}) = Q_{2n-1}(\hat{\omega},\hat{\Omega}) + Q^1_{2n-2}(\hat{v},\hat{\omega},\hat{\Omega}) + \ldots + Q^{2n-1}(\hat{v},\hat{\omega},\hat{\Omega}).
\ee
where the superscript and the subscript stand for the form degree in $\theta$--space and $x$--space respectively. Plugging (\ref{e-c-1-47}) into (\ref{e-c-1-46}) and equating terms of the same degree in $\theta$--space, we arrive at the so-called \emph{Stora-Zumino descent equations},
\bea
\label{e-c-1-48}
&\sss Q_{2n-1}(\hat{\omega},\hat{\Omega}) + \dd Q^1_{2n-2}(\hat{v},\hat{\omega},\hat{\Omega}) = 0, \nn\\ 
&\sss Q^1_{2n-2}(\hat{v},\hat{\omega},\hat{\Omega}) + \dd Q^2_{2n-3}(\hat{v},\hat{\omega},\hat{\Omega}) = 0, \nn\\
&\vdots \nn\\
&\sss Q^{2n-1}(\hat{v},\hat{\omega},\hat{\Omega}) = 0.
\eea
These relations play an important role in the discussion of gauge/gravitational anomalies.

\section{Index Theorems}
\label{sec-c-3}

Index theorems are theorems relating the index of an elliptic operator (an analytical quantity) with a certain invariant constructed from characteristic classes (a topological quantity). As such, they provide a direct connection between local properties of a manifold and its topological properties. Here, we will give some general definitions and we will present the relevant index theorems without proof. The original work of Atiyah and Singer leading to theorems of this type is presented in the series of papers \cite{Atiyah:1968mp,Atiyah:1968rj,Atiyah:1968ih,Atiyah:1971ws,Atiyah:1971rm}.

To define the index of a operator, consider two fiber bundles $E$ and $F$ over $M$. A differential operator $\Delta$ can be defined as an operator that maps sections in $\Gamma(M,E)$ to sections in $\Gamma(M,F)$,
\be
\label{e-c-3-1}
\Delta \quad:\quad s_E \in \Gamma(M,E) \to \Delta s_E \in \Gamma(M,F). 
\ee
If both $E$ and $F$ are equipped with well-defined inner products, denoted by $\langle  , \rangle_E$ and $\langle , \rangle_F$ respectively, then, it is possible to define the adjoint $\Delta^\dag$ of $\Delta$ as an operator mapping sections in $\Gamma(M,F)$ to sections in $\Gamma(M,E)$
\be
\label{e-c-3-2}
\Delta^\dag \quad:\quad s_F \in \Gamma(M,F) \to \Delta^\dag s_F \in \Gamma(M,E),
\ee
satisfying the requirement
\be
\label{e-c-3-3}
\langle s_F, \Delta s_E \rangle_F = \langle \Delta^\dag s_F , s_E \rangle_E.
\ee
The {\it index} of $\Delta$ is defined as the number of zero modes of $\Delta$ minus the number of zero modes of $\Delta^\dag$,
\be
\label{e-c-3-4}
\ind \Delta \equiv \dim \ker \Delta - \dim \ker \Delta^\dag. 
\ee 

According to the {\it Atiyah-Singer index theorems}, for certain operators $\Delta$, the index (\ref{e-c-3-4}) is actually a {\it topological quantity} expressed in terms of characteristic classes related to the base manifold and the fiber space. To state the Atiyah-Singer index theorem for the cases of interest, we consider a curved $2n$--dimensional manifold $M_{2n}$ on which we define a field carrying a spinor index, extra indices of some representation $\MR$ of $\mathrm{SO}(2n)$ with generators $\{ T_{\MR}^{ab} \}$ and internal gauge indices of some representation $\MS$ of a gauge group $\MG$. We then introduce the generalized positive-chirality Dirac operator $\ii \fslash \MD$ appropriate for this type of spinor (here, chirality refers to the explicit spinor index, not to possible extra $\mathrm{SO}(2n)$ spinor indices). Denoting the spaces of positive- and negative-chirality spinors by $\Delta_+$ and $\Delta_-$ respectively and the representation spaces of $\mathrm{SO}(2n)$ and $\MG$ by $R$ and $S$ respectively, we see that $\ii \fslash \MD$ defines a mapping $\ii \fslash \MD : \Delta_+ \times R \times S \to \Delta_- \times R \times S$, while the adjoint operator $( \ii \fslash \MD )^\dag$ defines a mapping $( \ii \fslash \MD )^\dag : \Delta_- \times R \times S \to \Delta_+ \times R \times S$. The Atiyah-Singer index theorem states that the index of the operator $\ii \fslash \MD$ is given by the quantity
\be
\label{e-c-3-5}
\ind (\ii \fslash \MD) = \int_{M_{2n}} \left[ \widehat{A}(R) \tr \exp \left( \frac{\ii}{4\pi} R_{ab} T_{\MR}^{ab} \right) \cch_{\MS}(F) \right]_\mathrm{vol} ,
\ee
involving the $A$-roof genus $\widehat{A}(R)$ of the spacetime manifold and the Chern character $\cch(F)$ associated with the connection $A$ on the bundle. Here, the subscript ``$\mathrm{vol}$'' implies that only the term proportional to the volume form should be retained.

In our study of anomalies, we will encounter three special cases of (\ref{e-c-3-5}). The first case occurs when the spinor in question is a usual spin-1/2 field $\psi_\alpha$ with no extra $\mathrm{SO}(2n)$ indices. Then, the relevant operator is the usual Dirac operator $\ii \fslash D_+ = \ii \fslash D P_+$ where $A$ the connection on the fiber. In this case, the exponential in () equals unity and the Atiyah-Singer index theorem reads
\be
\label{e-c-3-6}
\ind ( \ii \fslash D ) = \int_{M_{2n}} \left[ \widehat{A}(M) \cch(F) \right]_\mathrm{vol}.
\ee
The second case occurs where the spinor field is a spin $3/2$ gravitino $\psi^\mu_\alpha$ that carries an extra vector index and may be also coupled to the gauge field $A$. The appropriate operator is the Rarita-Schwinger operator $i \fslash D_{RS}$. In this case, the relevant representation of $\mathrm{SO}(2n)$ is the vector representation in which $R_{ab} (T^{ab})_{cd} = 2 R_{cd}$. Thus, our index theorem reads
\be
\label{e-c-3-7}
\ind ( \ii \fslash D_{RS} ) = \int_{M_{2n}} \left[ \widehat{A}(R) \tr \exp \left( \frac{iR}{2\pi} \right) \cch(F) \right]_\mathrm{vol}.
\ee
Finally, we will also need to consider the case of a bispinor field $\phi_{\alpha\beta}$ which carries an extra spinor index and does not couple to $A$. The corresponding operator is denoted by $\ii \fslash D_{\phi}$ and the relevant $\mathrm{SO}(2n)$ representation is the spinor representation where $R_{ab} (T^{ab})_{\alpha\beta} = \frac{1}{2} R_{ab} (\Gamma^{ab})_{\alpha\beta}$. Hence, we arrive at the index theorem
\be
\label{e-c-3-8}
\ind ( \ii \fslash D_{\phi} ) = \int_{M_{2n}} \left[ \widehat{A}(M) \tr \exp \left( \frac{\ii}{8\pi} R_{ab} \Gamma^{ab} \right) \right]_\mathrm{vol} .
\ee
To obtain a compact representation of this relation, one may choose an appropriate basis for the gamma matrices and such that $\frac{\ii}{8 \pi} R_{ab} \Gamma^{ab} = \frac{1}{2} \bigotimes_{i=1}^n \sigma^3 x_i$. Then, the trace in (\ref{e-c-3-8}) is equal to $2^n \prod_{i=1}^n \cosh (x_i / 2)$ and the integrand in (\ref{e-c-3-8}) is just the volume term of the Hirzebruch polynomial $L(R)$, 
\be
\label{e-c-3-9}
\left[ \widehat{A}(R) \tr \exp \left( \frac{\ii}{8\pi} R_{ab} \Gamma^{ab} \right) \right]_\mathrm{vol} = 2^n \prod_{i=1}^n \left[ \frac{x_i/2}{\tanh(x_i/2)} \right]_\mathrm{vol} = \left[ \prod_{i=1}^n \frac{x_i}{\tanh x_i} \right]_\mathrm{vol} = \left[ L(R) \right]_\mathrm{vol}
\ee
Therefore, the index theorem can be written in the compact form
\be
\label{e-c-3-10}
\ind ( \ii \fslash D_{\phi} ) = \int_{M_{2n}} \left[ L(R) \right]_\mathrm{vol}.
\ee

\chapter{Anomaly Coefficients}
\label{app-d}
\thispagestyle{empty}

\section{Second and Fourth Indices}
\label{sec-d-1}

The anomaly-related coefficients $a$, $b$ and $c$, introduced in Chapter \ref{chap-5}, are computed in terms of the second and fourth indices of the respective groups. To define them, we consider a simple group $\MG$ and we let $\MR$, $\MF$ and $\MA$ be a generic representation, the fundamental and the adjoint respectively. The $n$--th index $\ell_n(\MR)$ of $\MR$ is defined in terms of the symmetrized trace of the product of $n$ generators. In particular the second and fourth indices are determined by
\begin{equation}
\label{e-d-1}
\Str_{\MR} T_I T_J = \ell_2(\MR) d_{IJ},
\end{equation}
and
\begin{equation}
\label{e-d-2}
\Str_{\MR} T_I T_J T_K T_L = \ell_4( \MR ) d_{IJKL}  + \frac{3}{2 + \dim \MA } \ell_2( \MR )^2 \left[ \frac{\dim \MA}{\dim \MR} - \frac{1}{6} \frac{\ell_2( \MA )}{\ell_2 ( \MR )} \right] d_{(IJ}d_{KL)}.
\end{equation}
where $d^{I_1 \ldots I_n}$ are the invariant symmetric tensors of $\MG$ subject to the orthogonality conditions $d^{I_1 \ldots I_m} d_{I_1 \ldots I_m \ldots I_n} = 0$ for $m<n$; their normalization is determined by fixing the values of $\ell_n ( \MF )$. The normalization of second-order indices is irrelevant for our purposes while the normalization of fourth-order indices can be fixed by setting $\ell_4(\MF) = 1$ for all groups.

\begin{table}[!t]
\label{t-d-1}
\begin{center}
\begin{tabular}{|c||c|c|c|}
\hline
Group & Irrep $\MR$ & $b_{\MR}$ & $c_{\MR}$ \\
\hline
\hline
$E_8$ & $\mathbf{248}$ & $1/100$ & $1$ \\
\hline
      & $\mathbf{56}$ & $1/24$ & $1$ \\
\cline{2-4}
$E_7$ & $\mathbf{133}$ & $1/6$ & $3$ \\
\cline{2-4}
      & $\mathbf{912}$ & $31/12$ & $30$ \\
\hline
      & $\mathbf{27}$ & $1/12$ & $1$ \\
\cline{2-4}
      & $\mathbf{78}$ & $1/2$ & $4$ \\
\cline{2-4}
$E_6$ & $\mathbf{351}$ & $55/12$ & $25$ \\
\cline{2-4}
      & $\mathbf{351'}$ & $35/6$ & $28$ \\
\cline{2-4}
      & $\mathbf{650}$ & $10$ & $50$ \\
\hline
      & $\mathbf{26}$ & $1/12$ & $1$ \\
\cline{2-4}
$F_4$ & $\mathbf{52}$ & $5/12$ & $3$ \\
\cline{2-4}
      & $\mathbf{273}$ & $49/12$ & $21$ \\
\cline{2-4}
      & $\mathbf{324}$ & $23/4$ & $27$ \\
\hline
      & $\mathbf{7}$ & $1/4$ & $1$ \\
\cline{2-4}
$G_2$ & $\mathbf{14}$ & $5/2$ & $4$ \\
\cline{2-4}
      & $\mathbf{27}$ & $27/4$ & $9$ \\
\cline{2-4}
      & $\mathbf{64}$ & $38$ & $32$ \\
\hline
        & $\mathbf{3}$ & $1/2$ & $1$ \\
\cline{2-4}
$\textrm{SU}(3)$ & $\mathbf{8}$ & $9$ & $6$ \\
\cline{2-4}
        & $\mathbf{6}$ & $17/2$ & $5$ \\
\hline
$\textrm{SU}(2)$ & $\mathbf{2}$ & $1/2$ & $1$ \\
\cline{2-4}
        & $\mathbf{3}$ & $8$ & $4$ \\
\hline
\end{tabular}
\end{center}
\caption{The coefficients $b$ and $c$ for groups with no fourth-order invariants.}
\end{table}

\section{Anomaly Coefficients}
\label{sec-d-2}

Using the above definitions, we turn to the computations of the anomaly coefficients. Starting from the $c$--coefficients, we consider an algebra element $X=X^I T_I$ and we use (\ref{e-d-1}) for the representations $\MR$ and $\MF$ to find
\begin{equation}
\label{e-d-3}
\tr_{\MR} X^2 = \ell_2(\MR) (X^I)^2 ,\qquad \tr_{\MF} X^2 = \ell_2(\MF) (X^I)^2
\end{equation}
where we use the notation $(X^I)^n \equiv d_{I_1 \ldots I_n} X^{I_1} \ldots X^{I_n}$. So, we have
\begin{equation}
\label{e-d-4}
c_{\MR} =  \frac{\ell_2 ( \MR )}{\ell_2 ( \MF )} .
\end{equation}
To compute the $a$-- and $b$--coefficients, we first consider the case where $\MR$ has no fourth-order Casimirs so that $\ell_4(\MR) = 0$. Then Eq. (\ref{e-d-2}) leads to
\begin{equation}
\label{e-d-5}
\tr_{\MR} X^4 = \frac{3}{2 + \dim \MA } \ell_2( \MR )^2 \left[ \frac{\dim \MA}{\dim \MR} - \frac{1}{6} \frac{\ell_2( \MA )}{\ell_2 ( \MR )} \right] ((X^I)^2)^2 ,
\end{equation}
and so, $a_{\MR}=0$ and
\begin{equation}
\label{e-d-6}
b_{\MR} = \frac{3}{2 + \dim \MA } \frac{\ell_2( \MR )^2}{\ell_2( \MF )^2} \left[ \frac{\dim \MA}{\dim \MR} - \frac{1}{6} \frac{\ell_2( \MA )}{\ell_2 ( \MR )} \right] .
\end{equation}
We next consider the case where $\MR$ possesses fourth-order Casimirs. Then, using (\ref{e-d-2}) for the representations $\MR$ and $\MF$, we find
\begin{equation}
\label{e-d-7}
\tr_{\MR} X^4 = \ell_4( \MR ) (X^I)^4  + \frac{3}{2 + \dim \MA } \ell_2( \MR )^2 \left[ \frac{\dim \MA}{\dim \MR} - \frac{1}{6} \frac{\ell_2( \MA )}{\ell_2 ( \MR )} \right] ((X^I)^2)^2 ,
\end{equation}
and
\begin{equation}
\label{e-d-8}
\tr_{\MF} X^4 = (X^A)^4  + \frac{3}{2 + \dim \MA } \ell_2( \MF )^2 \left[ \frac{\dim \MA}{\dim \MF} - \frac{1}{6} \frac{\ell_2( \MA )}{\ell_2 ( \MF )} \right] ((X^I)^2)^2 ,\qquad
\end{equation}
Solving (\ref{e-d-8}) for $(X^I)^4$, substituting in (\ref{e-d-7}) and using the second of (\ref{e-d-3}), we find
\begin{equation}
\label{e-d-9}
a_{\MR} = \ell_4( \MR ) ,
\end{equation}
\begin{equation}
\label{e-d-10}
b_{\MR}  =  \frac{3}{2 + \dim \MA } \left\{ \frac{\ell_2( \MR )^2}{\ell_2( \MF )^2} \left[ \frac{\dim \MA}{\dim \MR} - \frac{1}{6} \frac{\ell_2( \MA )}{\ell_2 ( \MR )} \right]  -  \ell_4(\MR) \left[ \frac{\dim \MA}{\dim \MF} - \frac{1}{6} \frac{\ell_2( \MA )}{\ell_2 ( \MF )} \right] \right\}.
\end{equation}
\begin{table}[!t]
\label{t-d-2}
\begin{center}
\begin{tabular}{|c||c|c|c|c|}
\hline
Group & Irrep $\MR$ & $a_{\MR}$ &$b_{\MR}$ & $c_{\MR}$\\
\hline
\hline
& $\mathbf{N}$ & $1$ & $0$ & $1$\\
\cline{2-5}
$\textrm{SU}(N)$ & $\mathbf{N^2-1}$ & $2N$ & $6$ & $2N$\\
\cline{2-5}
& $\textstyle{\mathbf{\frac{N(N+1)}{2}}}$ & $N+8$ & $3$ & $N+2$\\
\cline{2-5}
& $\textstyle{\mathbf{\frac{N(N-1)}{2}}}$ & $N-8$ & $3$ & $N-2$\\
\hline
\hline
& $\mathbf{N}$ & $1$ & $0$ & $1$ \\
\cline{2-5}
$\textrm{SO}(N)$ & $\textstyle{\mathbf{\frac{N(N-1)}{2}}}$ & $N-8$ & $3$ & $N-2$\\
\cline{2-5}
& $\mathbf{2^{\lfloor \frac{N+1}{2} \rfloor - 1}}$ & $- 2^{\lfloor \frac{N+1}{2} \rfloor -5} $ & $3 \cdot 2^{\lfloor \frac{N+1}{2} \rfloor -7} $ & $2^{\lfloor \frac{N+1}{2} \rfloor - 4} $\\
\hline
\hline
& $\mathbf{2N}$ & $1$ & $0$ & $1$\\
\cline{2-5}
$\textrm{USp}(2N)$ & $\mathbf{N(2N+1)}$ & $2N+8$ & $3$ & $2N+2$\\
\cline{2-5}
& $\mathbf{N(2N-1)-1}$ & $2N-8$ & $3$ & $2N-2$\\
\hline
\end{tabular}
\end{center}
\caption{The coefficients $a$, $b$ and $c$ for for groups with fourth-order invariants.}
\end{table}

From these expressions, one may determine all the group-theoretical coefficients of interest using the values of the indices $\ell_2(\MR)$ and $\ell_4(\MR)$ which are tabulated e.g. in \cite{Slansky:1981yr}, \cite{McKay}, \cite{vanRitbergen:1998pn}. The values of $b_{\MR}$ and $c_{\MR}$ for groups with no fourth-order Casimirs are listed on Table D.1. The values of $a_{\MR}$, $b_{\MR}$ and $c_{\MR}$ for groups having fourth-order Casimirs are listed on Table D.2.

\backmatter

\begin{small}
\bibliographystyle{unsrt}
\bibliography{biblio}
\end{small}
}

\end{document}